\documentclass[3p,12pt,eqsecnum]{article}

\usepackage{etoolbox}
\pdfoutput=1

\usepackage{lmodern}
\usepackage[T2A,T1]{fontenc}
\usepackage{mathtools}
\usepackage[utf8]{inputenc}
\usepackage{bm}
\usepackage{graphicx}
\usepackage{amsmath}
\usepackage{amssymb}
\usepackage{color}
\usepackage{xifthen}
\usepackage{shuffle}
\usepackage{multirow}
\usepackage{subcaption}
\usepackage[titletoc]{appendix}
\usepackage{cancel}
\usepackage{hyperref}
\usepackage[margin=1in]{geometry}
\usepackage[numbers,sort&compress]{natbib}
\usepackage{textcomp}
\usepackage{array}  

\usepackage{afterpage}
\usepackage{tikz}
\usepackage{longtable}
\usepackage{fancyhdr}
\usepackage{caption}
\usepackage{calc}
\def\@oddfoot{\footnotesize\itshape {}}

\def\be{\begin{equation}}
\def\ee{\end{equation}}
\def\D{\Delta}
\def\e{\epsilon}
\def\G{\Gamma}
\def\Dphi{{\Delta_\phi}}
\def\s{\sigma}

\usepackage{amsmath}
\numberwithin{equation}{section}
\renewcommand\theequation{{\thesection.}\arabic{equation}}%

\newlength\imheight
\newlength\imwidth	 
\def\diagsizetwo{0.15\textwidth} 
\def\diagsize{0.19\textwidth}

\newcommand{\num}[1]{
	\settoheight\imheight{\includegraphics{#1}}
	\settowidth\imwidth{\includegraphics{#1}}
	c \ifthenelse{ \lengthtest{\imheight < 0.9 \imwidth}}
	{  \left( 
		\parbox{\widthof{\scalebox{.96}{\includegraphics[width=\diagsize]{#1}}}}{\includegraphics[width=\diagsize]{#1}}
	}
	{ \left(		\parbox{\widthof{\scalebox{.96}{\includegraphics[height=\diagsizetwo]{#1}}}}{\includegraphics[height=\diagsizetwo]{#1}}		
	} 
	\right)}

\newcommand{\col}[1]{
	\settoheight\imheight{\includegraphics{#1}}
	\settowidth\imwidth{\includegraphics{#1}}
	c \ifthenelse{ \lengthtest{\imheight < 0.9 \imwidth}}
	{  \left( 
		\parbox{\widthof{\scalebox{.96}{\includegraphics[width=\diagsize]{#1}}}}{\includegraphics[width=\diagsize]{#1}}
		}
	{ \left(		\parbox{\widthof{\scalebox{.96}{\includegraphics[height=\diagsizetwo]{#1}}}}{\includegraphics[height=\diagsizetwo]{#1}}	
	} 
	\right)}

\def\draftnote#1{{}} 

\DeclareMathAlphabet\mathbfcal{OMS}{cmsy}{b}{n}

\newbox\charbox
\newbox\slabox
\def\s#1{{      
        \setbox\charbox=\hbox{$#1$}
        \setbox\slabox=\hbox{$/$}
        \dimen\charbox=\ht\slabox
        \advance\dimen\charbox by -\dp\slabox
        \advance\dimen\charbox by -\ht\charbox
        \advance\dimen\charbox by \dp\charbox
        \divide\dimen\charbox by 2
        \raise-\dimen\charbox\hbox to \wd\charbox{\hss/\hss}
        \llap{$#1$} }}

\usepackage{afterpage}
\usepackage{longtable}
\usepackage{fancyhdr}

\begin{document}

	\thispagestyle{empty}

	\begin{center}

		{ \bf 
\LARGE Selected Topics in Analytic Conformal Bootstrap: A Guided Journey
			
			\vspace{0.25cm}
			
		}
		\bigskip\vspace{0.4cm}

	{\large Agnese Bissi}\\[3mm]
	
		{ \it  
			Department of Physics and Astronomy, \\
			  Uppsala University, Box 516, SE-751 20 Uppsala, Sweden} \\ [7mm]
			  
	 {\large Aninda Sinha}\\[3mm]

                        {\it Centre for High Energy Physics, Indian Institute of Science, \\
                         C.V. Raman Avenue, Bangalore 560012, India} \\ [7mm]
                         
          {\large Xinan Zhou}\\[3mm]
          
                         {\it Kavli Institute for Theoretical Sciences,\\
                               University of Chinese Academy of Sciences, Beijing 100190, China, and}\\[3mm]
                         {\it Princeton Center for Theoretical Science, \\
                                   Princeton University, Princeton, New Jersey 08544, USA}\\[-1mm]
		\let\thefootnote\relax\footnote{E-mails:  agnese.bissi@physics.uu.se, asinha@iisc.ac.in, xinan.zhou@ucas.ac.cn.}
                  
	\end{center}
	
	\medskip

\setcounter{footnote}{0}

{\centering{\section*{\small Abstract}}}
This review aims to offer a pedagogical introduction to the analytic conformal bootstrap program via a journey through selected topics. We review analytic methods which include the large spin perturbation theory, Mellin space methods and the Lorentzian inversion formula. These techniques are applied to a variety of topics ranging from large-$N$ theories, to the epsilon expansion and holographic superconformal correlators, and are demonstrated in a large number of explicit examples.

\vspace{2.6cm}

\begin{center}
{\it Invited review for Physics Reports}
\end{center}

\newpage
\tableofcontents

\newpage

\markboth{1\quad INTRODUCTION}{}
\section{Introduction}\label{sec_intro}
A ``bootstrap'' method or process is one that is self-generating or self-sustaining. As such, the bootstrap philosophy in quantum field theory refers to an ambitious program to use only basic symmetries and consistency conditions such as Poincar\'e invariance, unitarity, crossing symmetry and analyticity to constrain observables like the S-matrix elements \cite{heisenberg}. In the 1960s, the bootstrap program was pursued with the hope of understanding the strong interactions \cite{chew}. In the 1970s, a similar program was initiated to understand the physics of second order phase transitions, described by quantum field theories with conformal symmetries, {\it i.e.,} Conformal Field Theories (CFTs). This program is called the Conformal Bootstrap \cite{Ferrara:1973yt, Polyakov:1974gs}. In addition to the familiar Poincar\'e symmetries, CFTs enjoy scale symmetry as well as special conformal symmetries. These extra symmetries completely fix the structure of two- and three-point correlators  \cite{bigfatyellow}. One of the goals of the conformal bootstrap is to constrain the dynamical content appearing in four-point correlators in CFTs.

Conformal symmetry allows one to classify operators annihilated by the special conformal generators as ``primaries''. There are an infinite class of operators called ``descendants'' which are derivatives of these primary operators. The central idea of the conformal bootstrap program is to fix the operator product expansion (OPE) of any pair of local primary operators in the theory. Once this is accomplished, any $n$-point correlation function of local operators can be recursively calculated, at least in principle. In addition to conformal invariance, one uses crossing symmetry in a judicious manner. In the context of Euclidean correlators,  crossing symmetry arises due to operator associativity. This leads to the notion of different channels, which in an overlapping region of convergence are set equal, leading to the so-called crossing conditions.  Naively, these are an infinite number of conditions and finding any consistent solution seems to be a Herculean task. In fact, while the idea of the conformal bootstrap framework has been around since the 1970s, the main success it encountered, until recently, was restricted to two dimensional CFTs \cite{bpz, bigfatyellow}. In 2008, the work of \cite{rrtv} introduced a new numerical paradigm in the game. This paradigm enables us to extract, arguably, some of the most numerically accurate critical exponents for the 3d Ising model \cite{3dising1, 3dising2, 3disingkos}. In addition to this flagship result, numerical methods have enabled a systematic study of ``islands'' of CFTs allowed by unitarity and crossing symmetry. These developments have been recently reviewed in \cite{numrev}.

In  addition to these remarkable numerical results, it is worthwhile to develop analytic tools. There are several reasons for this. First, establishing potentially universal results for generic CFTs would require an analytic handle. Second, there is a plethora of results, both old and new, that the Feynman diagrammatic approach has produced; one would like to see how the bootstrap method compares to the successes of the diagrammatic approach. Finally, it is important to identify and establish techniques that  can produce results that are hard using other established methods.

\begin{figure}[ht]
\centering
\includegraphics[width=\textwidth]{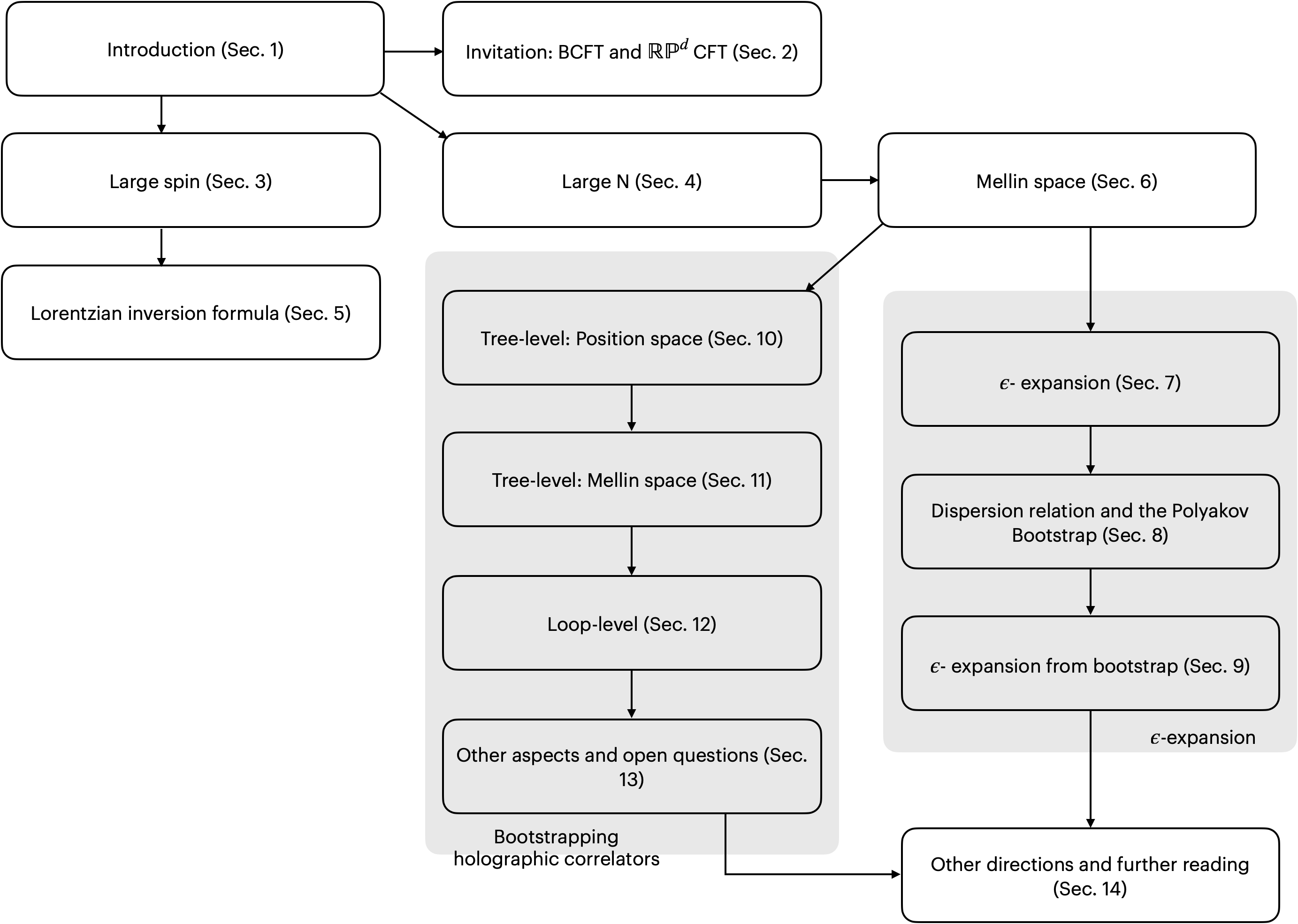}
\caption{The relation of sections in this review.}
    \label{fig:flow}
\end{figure}

In this present review, we will guide the reader on a journey through certain selected topics covering modern techniques in {\it analytic} conformal bootstrap in spacetime dimensions $d\geq 3$. The road map of the journey that we will take the reader on is depicted in Figure \ref{fig:flow}. It begins with an ``appetizer'' section \ref{Sec:BCFT}, which discusses boundary conformal field theories (BCFT). These are CFTs in the presence of a boundary or a co-dimension 1 defect. In nature, such systems may occur at the surface of a crystal. The two-point functions in such a scenario carry dynamical information, both of the bulk properties and of the new data due to the presence of a boundary. For technical reasons (lack of positivity in the so-called ``bulk channel''), setting up numerics in this scenario is hard. However, BCFTs allow for a rich phase structure corresponding to different boundary conditions. It is therefore important to develop analytic techniques. For our purpose, the case of BCFTs also serves as a simplifying example where we clarify some of the general ideas used in the analytic methods. Kinematically, this setup is very similar to CFTs placed on a real projective space. We will therefore also discuss analytic techniques for real projective space CFTs in the same section. 

We then discuss three possible routes. The first route begins in section \ref{Sec:largespin} and discusses large spin perturbation theory (LSPT). This is arguably the standard example in any discussion of the analytic conformal bootstrap. The main idea here is to reproduce contributions of certain known operators in one channel in the crossing equation in terms of the other channel. Typically, this needs an infinite number of operator contributions. One can argue that to reproduce the contribution of the identity operator, there have to be generalized free field  (GFF) operators in the spectrum. This is done by analyzing the large spin tail of such contribution. As we will review, this strategy works when there is a twist gap between the identity operator and other operators in the spectrum. We will study this canonical example in some detail and show how one can further go on to deriving leading order anomalous dimensions for the GFF spectrum. A natural continuation of this route is to discuss the now-famous Lorentzian inversion formula. This remarkable formula enables us to express the OPE coefficients as a convolution of the so-called double discontinuity of the position space correlator against an analytically continued (in spin) conformal block. This formula can then be used in the context of AdS/CFT to extract information about tree-level and loop-level AdS Witten diagrams.

Both the second and the third routes embark on perturbing away from the GFF spectrum (section \ref{Sec:largeN}). The perturbation parameter, by anticipating a connection with the AdS/CFT correspondence, is generically denoted by $1/N$, where $N$ is related to the central charge and taken to be large. Calculations along these routes are facilitated by a transition to Mellin space (section \ref{Sec:MellinFormalism}). Using Mellin techniques one can either continue the journey by discussing correlators in the $\epsilon$-expansion (the second route) or in the $1/N$ expansion (the third route). 

In the second route (sections \ref{Sec:epsilon}-\ref{Sec:epsilonbootstrap}), the $\epsilon$-expansion makes contact with the Wilson-Fisher fixed point \cite{Wilson:1971dc} and extracts the anomalous dimensions of certain operators in a perturbative expansion in $\epsilon$ where the spacetime dimension is written as $d=4-\epsilon$. Quite remarkably, not only can all the results of the famous Wilson-Kogut review \cite{Wilson:1973jj} be reproduced, but one can also easily get novel results for OPE coefficients which are difficult to calculate using the diagrammatic approach. In order to extract OPE data analytically, it is convenient to use Polyakov's 1974 seminal idea \cite{Polyakov:1974gs}, where he postulated that the bootstrap equations can be solved analytically by starting with a basis that is manifestly crossing symmetric. As we will review, this approach, in modern parlance, is tied with the crossing symmetric Witten diagrams in AdS space. The crossing symmetric AdS Witten diagrams provide a convenient kinematical basis for expanding the Mellin space correlator. Since the basis is crossing symmetric, constraints arise on demanding consistency with the OPE, leading to the so-called Polyakov conditions. This needs a discussion of crossing symmetric dispersion relations (section \ref{Sec:dispersionPolyakov}) which enables one to fix the so-called contact term ambiguities. 

In the third route (sections \ref{Sec:PositionSpace}-\ref{Sec:openproblemscorrelators}), we discuss efficient modern techniques to compute holographic correlators in various top-down string theory/M-theory models. We will focus on the regime where the bulk dual descriptions are weakly coupled and local. The basic observables are holographic correlators which correspond to on-shell scattering amplitudes in AdS. From these objects we can extract analytic data of the strongly coupled boundary theories by performing standard CFT analysis. The models which we will consider include the paradigmatic example of the strongly coupled 4d $\mathcal{N}=4$ super Yang-Mills, which is dual to IIB supergravity on $AdS_5\times S^5$, along with others preserving a certain amount of supersymmetry. Due to the presence of a compact internal manifold in these models, the Kaluza-Klein reduced effective theory in AdS contains infinitely many particles. The extreme complexity of the bulk effective action together with the proliferation of curved-space diagrams render the standard diagrammatic expansion method practically useless beyond just a few simplest cases. However, as we will see, using symmetries and consistency conditions allows us to fix the correlators completely and therefore circumvents these difficulties. After a brief review of the superconformal kinematics in section \ref{Subsec:scfkinematics}, we will discuss in detail three complementary bootstrap methods to compute tree-level correlators (sections \ref{Sec:PositionSpace}, \ref{Sec:MellinSpace}). These methods yield all four-point tree-level correlators of arbitrary Kaluza-Klein modes in all maximally superconformal theories, and reveal remarkable simplicity and structures hidden in the Lagrangian description. We also discuss various extensions: higher-point correlators (section \ref{Subsec:5ptfunctions}), correlators corresponding to super gluon scattering in AdS (section \ref{Subsec:supergluons}), and loop-level correlators (section \ref{Sec:loops}).  The results and techniques which we will review in this part of the review also bear great resemblance with the on-shell scattering amplitude program in flat space, as we will point out along the way.

In organizing this review, we have presented the material in a way such that these routes are relatively independent and can be read separately. We also accompanied the discussions with many pedagogical examples. All the journeys along these different routes end with a brief discussion of open questions in this research area. We also conclude in section \ref{Sec:omissions} with a discussion of further reading material which covers a broader range of topics. Where possible, we will delegate lengthy formulas and algebraic steps to the appendices. The third appendix (appendix \ref{App:WittenDiagrams}) also constitutes a self-contained review of various properties of Witten diagrams which make appearances at multiple places in this review. We will assume some familiarity with CFTs on the part of the reader. For introductory material on this topic, we refer the reader to \cite{bigfatyellow,slavaepfl,Simmons-Duffin:2016gjk, numrev}. For introductory material on the  AdS/CFT correspondence we refer the reader to \cite{Aharony:1999ti,DHoker:2002nbb,Penedones:2016voo}. 
There will be special functions like Gauss and generalized hypergeometric functions used in several places. Most of these functions are inbuilt in $\mathtt{Mathematica}$. For authoritative references, we ask the reader to consult \cite{AnAsRo99, 10.5555/1830479}.

\newpage
\markboth{2\quad OVERTURE: BOOTSTRAP WITH TWO-POINT FUNCTIONS}{}
\section{Overture: Bootstrap with two-point functions}\label{Sec:BCFT}
This section serves as an appetizer for the reader to get a taste of the kind of analytic conformal bootstrap techniques which we will present in the review. To this end, we would like to choose systems which are as simple as possible (yet still nontrivial). One toy example that comes to mind is the one dimensional CFT where the simplest nontrivial observables are the four-point functions. Though conceptually closer to the higher dimensional case as it deals with the same kind of observables, the application of 1d CFTs is quite limited. Therefore, we choose to investigate instead two closely related but perhaps less familiar setups, namely, CFTs with a conformal boundary and CFTs on  real projective space. These setups are equally simple compared to CFT$_1$ but can be discussed in arbitrary spacetime dimensions. This gives them a wider range of physical applicability. In particular, BCFTs have important applications in various condensed matter systems. Therefore, we believe that the greater effort needed to get acquainted with these new CFT systems is justified and will be rewarding in the end. The most noticeable feature of these setups is that conformal symmetry is only partially preserved. But as a result, there are new observables. The simplest nontrivial observables are the two-point functions. We will use these two-point functions to demonstrate the power of analytic conformal bootstrap without too much technical complexity. Note that this section is structured to be independent from the other sections. Therefore, if the reader wishes to go directly to the three routes of the review, skipping it will not affect their understanding.

 The rest of this section is organized as follows. In Section \ref{Subsec:BCFTkinematics} we introduce the setups and discuss the kinematics. In Section \ref{Subsec:analyticmethods} we review analytic bootstrap methods for studying two-point functions. In Section \ref{Subsec:otherbackgrounds} we give a short discussion of CFTs in other backgrounds. As we already mentioned, the two setups which we will study in this section are also interesting in their own right. For readers who are interested in learning more about these topics, we refer them to the original papers. An incomplete sampling of the literature on BCFTs from the bootstrap perspective includes \cite{Liendo:2012hy,Gliozzi:2015qsa,Liendo:2016ymz,Rastelli:2017ecj,Bissi:2018mcq,Mazac:2018biw,Kaviraj:2018tfd,Dey:2020lwp,Dey:2020jlc,Giombi:2020rmc,Giombi:2021cnr}. For works on CFTs on real projective space, see \cite{Nakayama:2015mva,Verlinde:2015qfa,Nakayama:2016cim,Nakayama:2016xvw,Hasegawa:2016piv,Hogervorst:2017kbj,Hasegawa:2018yqg,Giombi:2020xah,Wang:2020jgh,Tsiares:2020ewp,Nakayama:2021muk}.

\subsection{Kinematics}\label{Subsec:BCFTkinematics}
To discuss the conformal symmetry these systems preserve, it is most convenient to use the embedding space formalism. We can represent each point $x^\mu\in\mathbb{R}^{d-1,1}$ by a null ray in the embedding space $\mathbb{R}^{d,2}$
\begin{equation}
P^A\;,\quad A=1,2,\ldots, d+2\;,\quad P\cdot P=0\;,\quad P\sim \lambda P\;.
\end{equation}
Operators are defined on the null rays with the condition 
\begin{equation}\label{opsca}
\mathcal{O}_\Delta(\lambda P)=\lambda^{-\Delta}\mathcal{O}_\Delta(P)\;.
\end{equation}
Let us choose the signature of the embedding space to be $(-,+,-,+,\ldots,+)$. Then we can choose a particular $\lambda$ to parameterize the null vector as 
\begin{equation}\label{Pemb}
P^A=\big(\frac{1+x^2}{2},\frac{1-x^2}{2},x^\mu \big)
\end{equation}
where $x^2=x^\mu x_\mu$. Conformal group transformations correspond to $SO(d,2)$ rotations on $P^A$. Their actions on $x^\mu$ are obtained by further rescaling  $P^1+P^2$ of the rotated embedding vector to 1. 

Let us now introduce two fixed vectors
\begin{equation}
N_b=(0,0,0,\ldots,1)\;,\quad\quad N_c=(1,0,0,\ldots,0)\;,
\end{equation}
which will correspond to two different systems. Either vector partially breaks the conformal group. In the first case, the surface with $N_b\cdot P=0$ gives rise to a planar boundary located at $x_d=0$. We will often denote $x_d$ as $x_\perp$ as is common in the BCFT literature. Therefore, this case is related to boundary CFTs, and the residual conformal symmetry is $SO(d-1,2)$. This $SO(d-1,2)$ symmetry is just the conformal group of the $d-1$ dimensional boundary.   
Note that we can also perform a conformal transformation to change the planar boundary into a sphere. This can be accomplished by choosing the fixed vector $\tilde{N}_b=(0,1,0,\ldots,0)$, and the boundary is a unit sphere centered at $x=0$. Now we consider the second case. The symmetry group preserving the vector $N_c$ is $SO(d,1)$. Using this vector, we can define a transformation 
\begin{equation}\label{invP}
P\to -2(P\cdot N_c)N_c-P\;,
\end{equation}
which upon rewriting in the form (\ref{Pemb}) by rescaling gives the conformal inversion transformation
\begin{equation}\label{invx}
x^\mu\to -\frac{x^\mu}{x^2}\;.
\end{equation}
Upon identifying $x^\mu\sim -x^\mu/x^2$, we obtain the real projective space $\mathbb{RP}^d$.\footnote{A more familiar definition of the real projective space is to take the $\mathbb{Z}_2$ quotient of a sphere $S^d$
\begin{equation}
\mathbf{X}^2=1\;,\quad \mathbf{X}\in\mathbb{R}^{d+1}\;,\quad \mathbf{X}\sim-\mathbf{X}\;.
\end{equation}
Since we are considering CFTs, we can perform a Weyl transformation to map it to the flat space by $x^\mu=\frac{X^\mu}{1-X^{d+1}}$, $\mu=1,\ldots,d$ and $d s^2_{\mathbb{R}^d}=\frac{(1+x^2)^2}{4}ds^2_{S^d}$. In two dimensions, this is also known as a crosscap.} To consider CFTs on this quotient space, we also need to identify the operators inserted at points related by inversion. For scalar operators, we have
\begin{equation}\label{invopid}
\mathcal{O}^\pm_\Delta(x)\leftrightarrow \pm x^{2\Delta} \mathcal{O}^\pm_\Delta(x')\;,\quad x'^\mu=-\frac{x^\mu}{x^2}\;.
\end{equation}
where we have two choices for the parity of the operator. 

Let us now discuss correlators of local operators. For these systems, the simplest correlators are one-point functions. Because there are residual Lorentz symmetries, spinning operators cannot have one-point functions and we only need to consider scalar operators. The only invariants which we can construct from the embedding vector $P$ and the fixed vectors are $P\cdot N_b$ and $P\cdot N_c$. Moreover, using the scaling behavior (\ref{opsca}) we can fix the one-point functions up to an overall constant. For BCFTs this gives
\begin{equation}
\langle O_\Delta \rangle_b=\frac{a_{b,\Delta}}{(2P\cdot N_b)^\Delta}=\frac{a_{b,\Delta}}{(2x_\perp)^\Delta}\;,
\end{equation} 
and for $\mathbb{RP}^d$ CFTs we have
\begin{equation}
\langle O_\Delta^+ \rangle_c=\frac{a_{c,\Delta}}{(-2P\cdot N_c)^\Delta}=\frac{a_{c,\Delta}}{(1+x^2)^\Delta}
\end{equation} 
for operators with $+$ parity and zero for the other choice.\footnote{When we perform a Weyl transformation and map it to $S^d$, the one-point functions are just constants. Identifying operators on antipodal points with a minus sign forces their expectation values to be zero.} The coefficients $a_{b,\Delta}$ and $a_{c,\Delta}$ are new CFT data defining the theories.\footnote{To see the coefficients correspond to new data, let us try to absorb them by changing the normalization of the operators. However, this would change the normalization of two-point functions. Note that when the points are very close to each other, we can ignore the presence of the boundary or the identification under inversion. The limiting two-point functions to should approach those in the CFT in infinite flat space with the same normalization.} 

Let us move on to two-point functions. In this case, one can construct cross ratios which are invariant under the residual conformal symmetry and independent rescalings of the embedding vectors. The cross ratio for the BCFT case is 
\begin{equation}\label{defcrxi}
\xi=\frac{(-2P_1\cdot P_2)}{(2N_b\cdot P_1)(2N_b\cdot P_2)}=\frac{(x_1-x_2)^2}{4x_{1,\perp}x_{2,\perp}}\;,
\end{equation} 
and the cross ratio for the real projective space case is 
\begin{equation}\label{defcreta}
\eta=\frac{(-2P_1\cdot P_2)}{(-2N_c\cdot P_1)(-2N_c\cdot P_2)}=\frac{(x_1-x_2)^2}{(1+x_1^2)(1+x_2^2)}\;.
\end{equation} 
The two-point functions can be written as functions of the cross ratios after extracting a kinematic factor 
\begin{eqnarray}
\langle \mathcal{O}_{\Delta_1}(x_1)\mathcal{O}_{\Delta_2}(x_2)\rangle_b&=&\frac{\mathcal{G}(\xi)}{|2N_b\cdot P_1|^{\Delta_1}|2N_b\cdot P_2|^{\Delta_2}}=\frac{\mathcal{G}(\xi)}{|2x_{1,\perp}|^{\Delta_1}|2x_{2,\perp}|^{\Delta_2}}\;,\\
\langle \mathcal{O}^\pm_{\Delta_1}(x_1)\mathcal{O}^\pm_{\Delta_2}(x_2)\rangle_c&=&\frac{\mathcal{G}^\pm(\eta)}{(-2N_c\cdot P_1)^{\Delta_1}(-2N_c\cdot P_2)^{\Delta_2}}=\frac{\mathcal{G}^\pm(\eta)}{(1+x_1^2)^{\Delta_1}(1+x_2^2)^{\Delta_2}}\;.
\end{eqnarray}
Note that for two-point functions to be nonzero in real projective space CFTs, the two operators must have the same parity so that the two-point function is neutral under the parity $\mathbb{Z}_2$. We can expand the two-point functions in the limits of operator product expansion (OPE), and the contributions are organized by the residual conformal symmetry into conformal blocks. We look at these two cases separately.

\begin{figure}
  \centering
\begin{subfigure}{0.45\textwidth}
 \centering
  \includegraphics[width=0.7\linewidth]{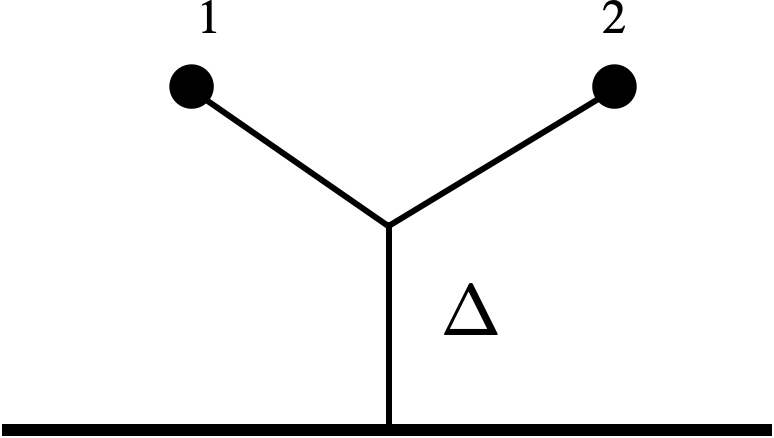}
  \caption{Bulk channel.}
  \label{fig:BCFTOPEB}
\end{subfigure}
\begin{subfigure}{0.45\textwidth}
  \centering
  \includegraphics[width=0.7\linewidth]{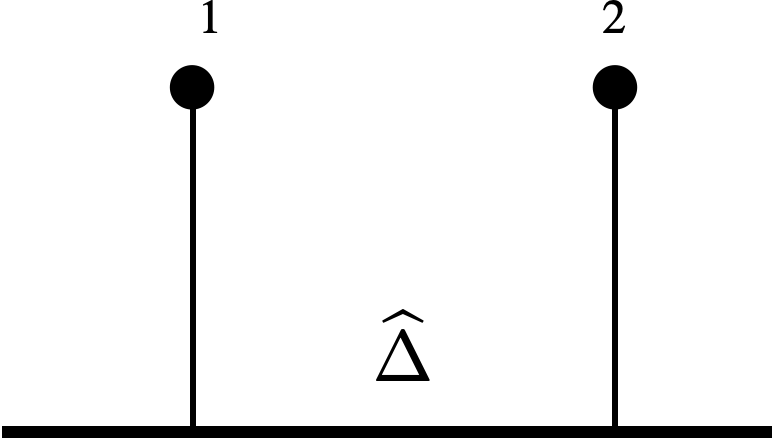}
  \caption{Boundary channel.}
  \label{fig:BCFTOPEbdr}
\end{subfigure}
\caption{Two OPE limits in a BCFT two-point function. The horizontal lines represent the conformal boundary.}
\label{fig:de}
\end{figure}

In BCFTs we have two distinct OPEs. The first one is usually referred to as the {\it bulk channel} OPE (Figure \ref{fig:BCFTOPEB}) where the two operators are taken to be close to each other
\begin{equation}\label{OPEB}
\mathcal{O}_{\Delta_1}(x_1)\mathcal{O}_{\Delta_2}(x_2)=\frac{\delta_{12}}{(x_1-x_2)^{2\Delta_1}}+\sum_k C_{12k} D[x_1-x_2,\partial_{x_2}]\mathcal{O}_{\Delta_k}(x_2)
\end{equation}
where $C_{12k}$ are OPE coefficients and the differential operators $D[x_1-x_2,\partial_{x_2}]$ are determined by conformal symmetry. For simplicity, we have only displayed in the OPE the scalar operators which contribute to the two-point function. Using this OPE, two-point functions can be expressed as an infinite sum of one-point functions.The contribution of each primary operator and its descendants can be resummed into a bulk-channel conformal block \cite{McAvity:1995zd}
\begin{equation}\label{BCFTbulkg}
g_{b,\Delta}^{\rm bulk}(\xi)=\xi^{\frac{\Delta-\Delta_1-\Delta_2}{2}}{}_2F_1\big(\frac{\Delta+\Delta_1-\Delta_2}{2},\frac{\Delta+\Delta_2-\Delta_1}{2};\Delta-\frac{d}{2}+1;-\xi\big)\;,
\end{equation}
and the two-point function can be written as 
\begin{equation}
\mathcal{G}(\xi)=\delta_{12}\xi^{-\Delta_1}+\sum_k \mu_{b,12k}\, g_{b,\Delta_k}^{\rm bulk}(\xi)\;,
\end{equation}
with $\mu_{b,12k}=a_{b,\Delta_k}C_{12k}$. The second OPE is the so-called {\it boundary channel} OPE (Figure \ref{fig:BCFTOPEbdr}) where operators are taken near the boundary and expressed in terms of operators living on the boundary at $x_\perp=0$
\begin{equation}
\mathcal{O}_\Delta(x)=\frac{a_{b,\Delta}}{|2x_\perp|^\Delta}+\sum_l \rho_l C[x]\widehat{O}_{\widehat{\Delta}_l}(x)\;.
\end{equation}
Here $\rho_l$ are OPE coefficients and the differential operators $C[x]$ are fixed by conformal symmetry. Using this OPE we can write the two-point function as an infinite sum of two-point functions on the boundary which are fixed by the residual conformal symmetry. The contribution of each operator is resummed into a {\it boundary channel} conformal block \cite{McAvity:1995zd}
\begin{equation}\label{BCFTbdrg}
g_{b,\widehat{\Delta}}^{\rm boundary}(\xi)=\xi^{-\widehat{\Delta}}{}_2F_1\big(\widehat{\Delta},\widehat{\Delta}-\frac{d}{2}+1;2\widehat{\Delta}+2-d;-\frac{1}{\xi}\big)\;.
\end{equation}
In terms of the boundary channel conformal blocks, we can write the two-point function as 
\begin{equation}
\mathcal{G}(\xi)=a_{b,\Delta}^2\delta_{12}+\sum_l \rho_{1,l}\rho_{2,l} g_{b,\widehat{\Delta}_l}^{\rm boundary}(\xi)\;.
\end{equation}
Similar to four-point conformal blocks in CFTs without boundaries, the bulk channel and the boundary channel conformal blocks are also more conveniently computed as the eigenfunctions of conformal Casimir operators \cite{Liendo:2012hy}. The two ways of expanding two-point functions are equivalent, and the equivalence gives rise to the BCFT crossing equation
\begin{equation}
\delta_{12}\xi^{-\Delta_1}+\sum_k \mu_{b,12k}\, g_{b,\Delta_k}^{\rm bulk}(\xi)=a_{b,\Delta}^2\delta_{12}+\sum_l \rho_{1,l}\rho_{2,l} g_{b,\widehat{\Delta}_l}^{\rm boundary}(\xi)\;.
\end{equation}
Here in both channels we have explicitly singled out the identity operator exchange. We can also absorb them into the sums by extending the sums to include operators with dimension zero.

\begin{figure}
\centering
\includegraphics[width=0.35\textwidth]{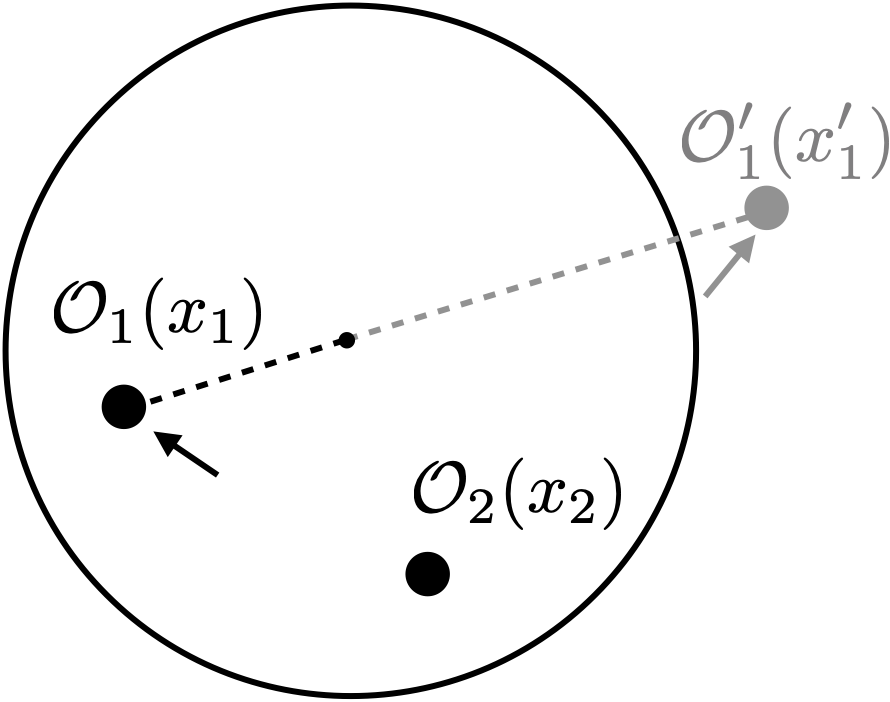}
\caption{Two OPE channels in real projective space CFTs. The circle represents the unit sphere at $x=0$, and $\mathcal{O}'_1(x'_1)$ is the inversion image of $\mathcal{O}_1(x_1)$ with respect to the unit sphere.}
    \label{fig:RPdOPE}
\end{figure}

In real projective space CFTs, the situation is slightly different. We still have the bulk channel OPE (\ref{OPEB}), which allows us to express the two-point function as a sum of one-point functions. The contribution of an operator resums into the conformal block \cite{Nakayama:2016xvw}
\begin{equation}\label{RPdCFTg}
g_{c,\Delta}(\eta)=\eta^{\frac{\Delta-\Delta_1-\Delta_2}{2}}{}_2F_1\big(\frac{\Delta+\Delta_1-\Delta_2}{2},\frac{\Delta+\Delta_2-\Delta_1}{2};\Delta-\frac{d}{2}+1;\eta\big)\;,
\end{equation}
and the two-point function can be written as 
\begin{equation}\label{RPdcfdecom}
\mathcal{G}^\pm(\eta)=\delta_{12}\eta^{-\Delta_1}+\sum_k \mu_{c,12k}\, g_{c,\Delta_k}(\eta)
\end{equation}
where $\mu_{c,12k}=a_{c,\Delta_k}C_{12k}$. On the other hand, we no longer have the boundary channel OPE since there is no boundary.\footnote{For this reason, there are a lot more data in the BCFT case which are associated to the operators living on the boundary.} Instead, we can move $\mathcal{O}_2$ towards the inversion image of $\mathcal{O}_1$. Due to the operator identification \ref{invopid}, we can apply the same OPE (\ref{OPEB}). This gives rise to a new channel which we will refer to as the {\it image channel}. These two OPE channels are illustrated in Figure \ref{fig:RPdOPE}. The image channel conformal blocks are given by \cite{Nakayama:2016xvw}
\begin{equation}\label{RPdCFTgmirror}
\bar{g}_{c,\Delta}(\eta)=(1-\eta)^{\frac{\Delta-\Delta_1-\Delta_2}{2}}{}_2F_1\big(\frac{\Delta+\Delta_1-\Delta_2}{2},\frac{\Delta+\Delta_2-\Delta_1}{2};\Delta-\frac{d}{2}+1;1-\eta\big)\;,
\end{equation}
and the two-point function can be written as 
\begin{equation}
\mathcal{G}^\pm(\eta)=\delta_{12}(1-\eta)^{-\Delta_1}+\sum_k \mu_{c,12k}\, \bar{g}_{c,\Delta_k}(\eta)\;.
\end{equation}
The conformal blocks in the two channels can also be obtained from solving Casimir equations. Equating these two conformal block decompositions, we arrive at the following crossing equation
\begin{equation}\label{RPdcrossingeqn}
\delta_{12}\eta^{-\Delta_1}+\sum_k \mu_{c,12k}\, g_{c,\Delta_k}(\eta)=\pm\big(\delta_{12}(1-\eta)^{-\Delta_1}+\sum_k \mu_{c,12k}\, \bar{g}_{c,\Delta_k}(\eta)\big)
\end{equation}
where $\pm$ is the common parity of the two operators. 

\begin{figure}
\centering
\includegraphics[width=0.6\textwidth]{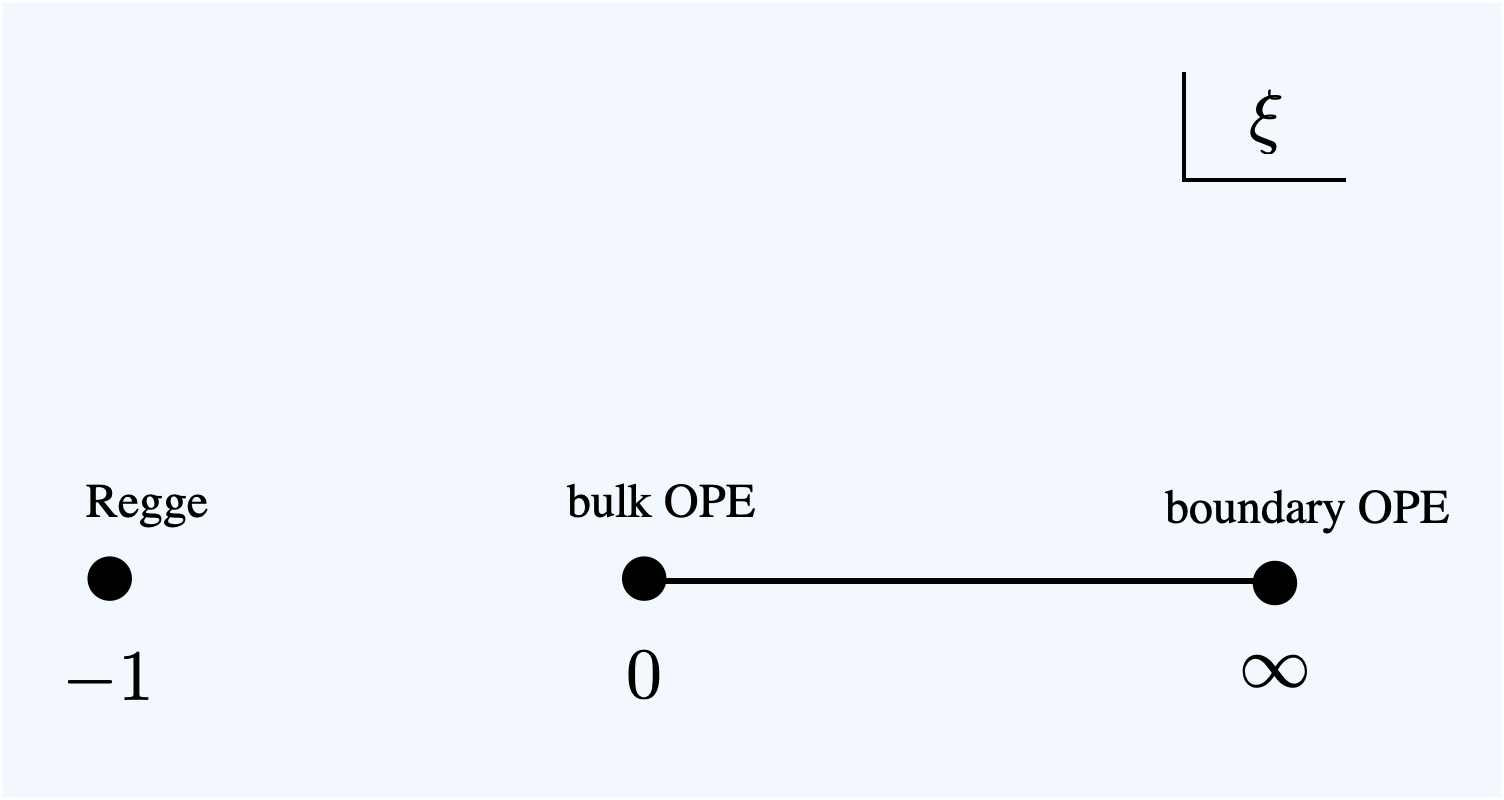}
\caption{The $\xi$-plane for a BCFT two-point function. Three interesting points on this plane are the bulk channel OPE limit ($\xi=0$), the boundary channel OPE limit ($\xi=\infty$), and the Regge limit ($\xi=-1$).}
    \label{fig:xiplane}
\end{figure}

\begin{figure}
\centering
\includegraphics[width=0.35\textwidth]{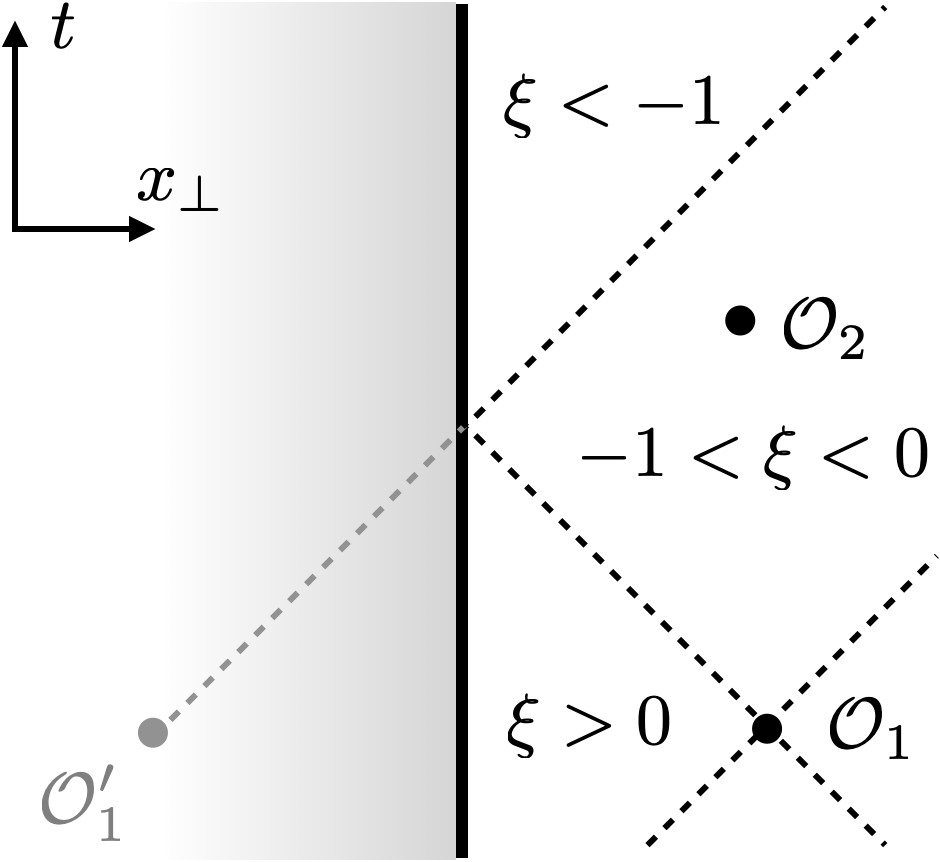}
\caption{The Regge limit of a BCFT two-point function in Lorentzian spacetime. The vertical line represents the conformal boundary, and time is one of the dimensions of the boundary. The Regge limit $\xi=-1$ is reached when $\mathcal{O}_2$ approaches the lightcone of the image of $\mathcal{O}_1$.}
    \label{fig:xiLorentzian}
\end{figure}

Finally, let us comment that we can complexify the cross ratios, and study the analytic property of correlators on the complex plane. For BCFTs, the complex $\xi$-plane is shown in Figure \ref{fig:xiplane}. The two special points $\xi=0$ and $\xi=\infty$ correspond to the bulk channel and boundary channel OPE limits respectively. In Euclidean spacetime, the cross ratio is restricted to the semi-infinite real axis $\xi\in[0,\infty)$. However, there is another interesting point at $\xi=-1$ which can be reached via analytic continuation. In Lorentzian signature,  $\xi=-1$ corresponds to one operator approaching the lightcone of the other operator's image with respect to the boundary (Figure \ref{fig:xiLorentzian}). This limit is referred to as the Regge limit. In a unitary BCFT, one can prove that the growth of the two-point function in the Regge limit is bounded by the exchange of the operator with the lowest conformal dimension in the bulk channel. The proof takes advantage of the so-called $\rho$ coordinate \cite{Hogervorst:2013sma}, and the positivity of the conformal block decomposition coefficients in the boundary channel. Details of the proof can be found in Appendix A of \cite{Mazac:2018biw}.

\begin{figure}
\centering
\includegraphics[width=0.6\textwidth]{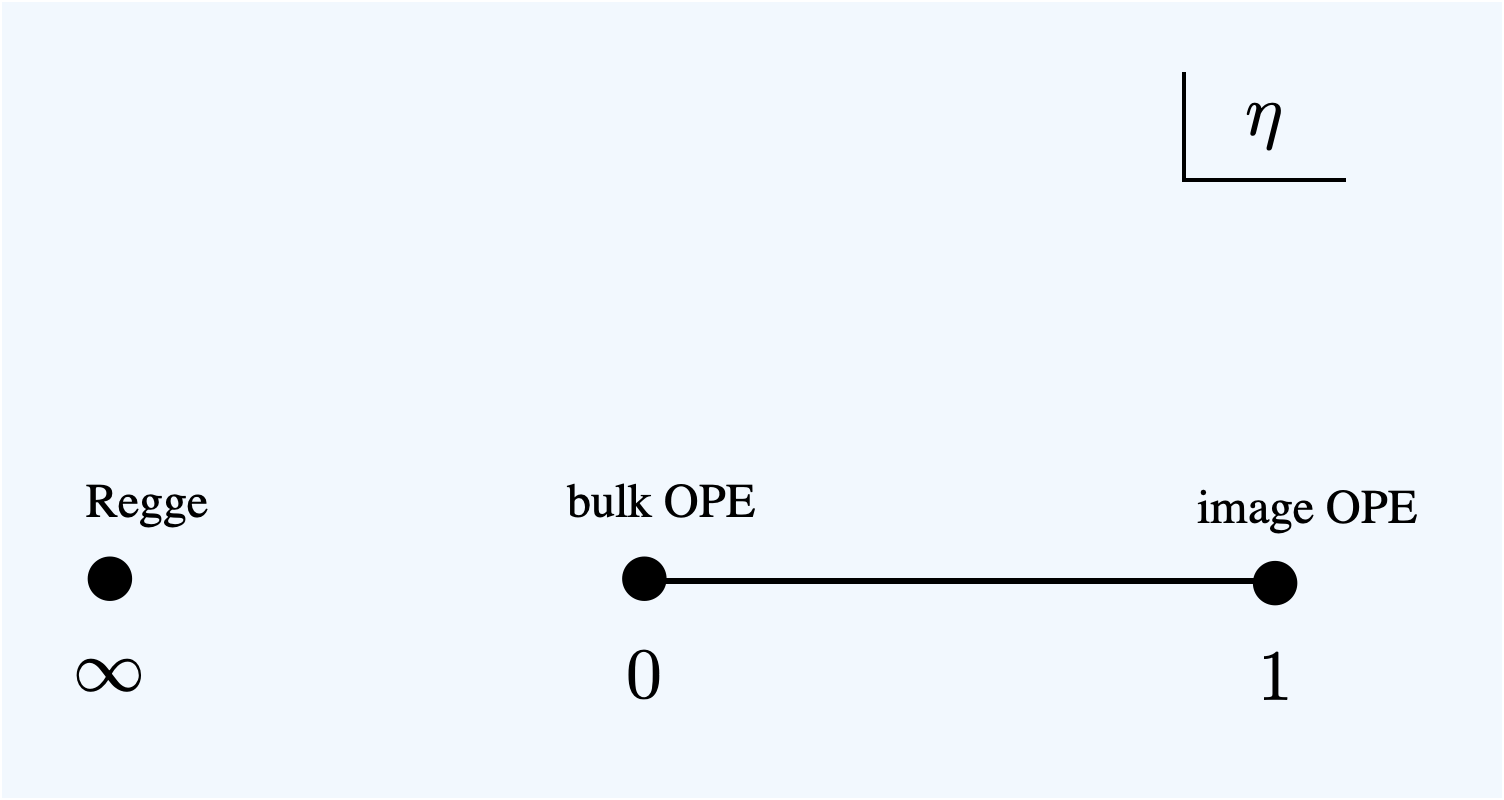}
\caption{The $\eta$-plane for a real projective space CFT two-point function. Three interesting points on this plane are the bulk channel OPE limit ($\eta=0$), the image channel OPE limit ($\eta=1$), and the Regge limit ($\eta=\infty$).}
    \label{fig:etaplane}
\end{figure}

The complex $\eta$-plane for real projective space CFTs is shown in Figure \ref{fig:etaplane}. There are also three points of special interest. The points $\eta=0$ and $\eta=1$ respectively correspond to the bulk channel OPE and the image channel OPE limits, and $\eta\in[0,1]$ for Euclidean space. The point $\eta=\infty$ plays a similar role as the Regge limit in BCFT two-point functions \cite{Giombi:2020xah}, and can only be reached via analytic continuation in Euclidean signature. However, unlike in the BCFT case, there is no analogue of the boundary channel where the conformal block decomposition coefficients are positive. Therefore, one cannot adapt the proof for BCFTs to prove boundedness of two-point functions in the Regge limit.   

\subsection{Analytic methods}\label{Subsec:analyticmethods}
In this subsection we discuss analytic methods for BCFTs \cite{Kaviraj:2018tfd,Mazac:2018biw} and real projective space CFTs \cite{Giombi:2020xah} which are based on ``analytic functionals''. Such functional methods were originally introduced for four-point functions in 1d CFTs \cite{Mazac:2016qev,mazacpaulos1,mazacpaulos2}, and later generalized to higher dimensions in \cite{Paulos:2019gtx,MRZ19,Caron-Huot:2020adz}. While the level of technical sophistication varies greatly in these different setups, the essential ideas remain the same. Here we will exploit the simpler kinematics of two-point functions to demonstrate the main features of such an approach. 

To help the reader navigate through this subsection, let us give below a quick summary of these features and also point out the connections. We will argue that the ``double-trace'' conformal blocks, from both the direct and the crossed channels, form a new basis for expanding the correlators. These double-trace conformal blocks are associated with special product operators of which the conformal dimensions are the sums of the elementary building operators. This should be contrasted with the standard conformal block decomposition which exploits only one channel at a time and does not require the spectrum to be discrete. The dual of the double-trace conformal blocks are the analytic functionals. Their actions on the crossing equation turn it into sum rules for the CFT data. We will develop this functional approach both from a dispersion relation, and by exploiting the structure of Feynman diagrams (Witten diagrams) \cite{Witten:1998qj} in certain holographic setups. The first argument can be viewed as a toy example of the CFT dispersion relation for four-point functions \cite{Carmi:2019cub}. The second argument is closely related to Polyakov's original version of the conformal bootstrap \cite{Polyakov:1974gs}, which will be reviewed later in Section \ref{Sec:dispersionPolyakov}. As we will see, the Witten diagrams also give rise to another set of basis which are essentially those used in \cite{Polyakov:1974gs}. Moreover, the sum rules from the functionals are just a modern paraphrase of the consistency conditions imposed by Polyakov in his approach.

\subsubsection{Real projective space CFTs}

\begin{figure}
\centering
\includegraphics[width=0.6\textwidth]{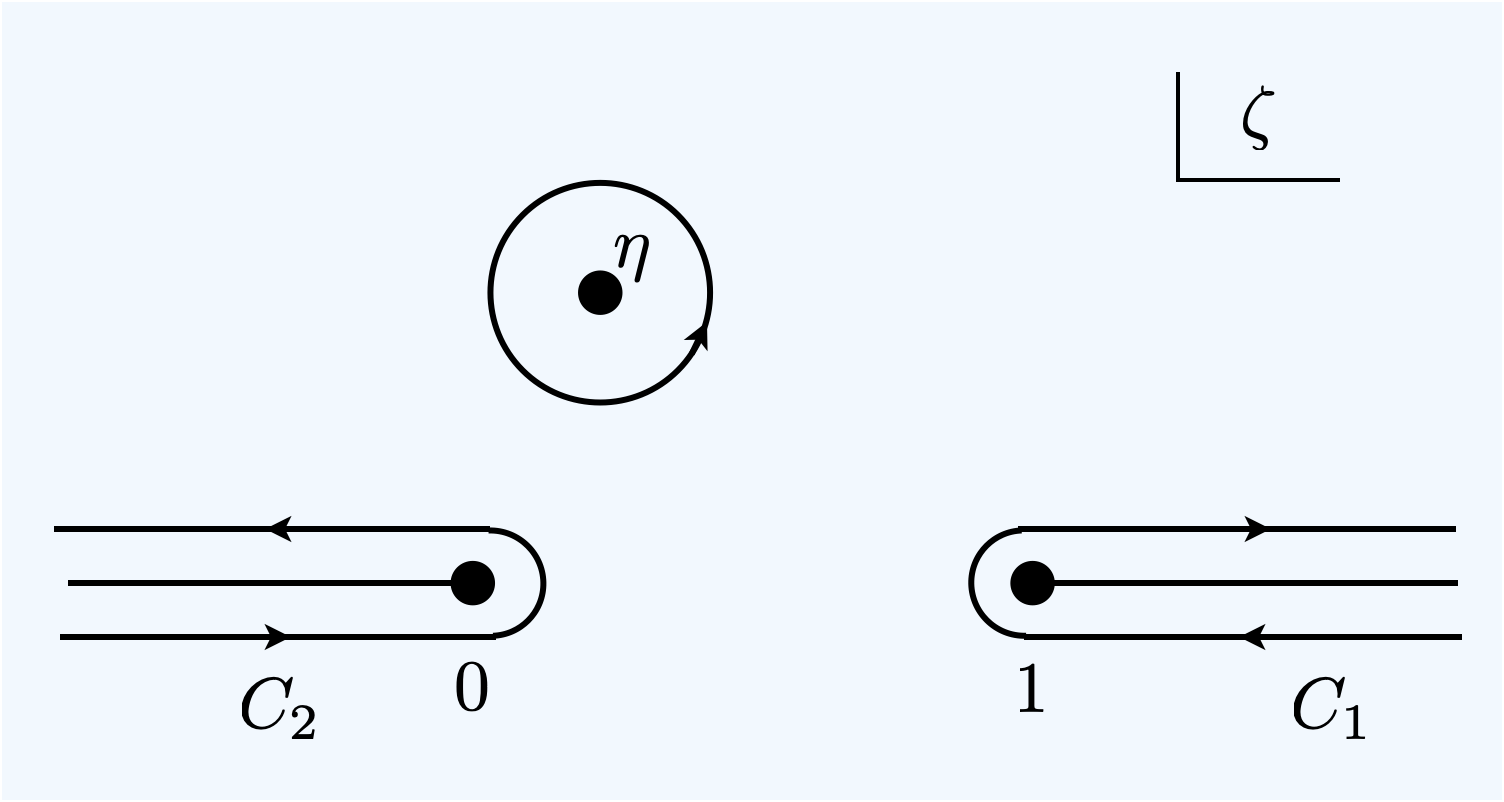}
\caption{Deformation of contours in the dispersion relation.}
    \label{fig:RPdcontours}
\end{figure}

Let us first consider a simplified example of a two-point functions in a 2d real projective CFT where the external dimensions are equal $\Delta_1=\Delta_2=\Delta_\varphi$, following the discussion in  \cite{Giombi:2020xah}. The two-point function can be written as 
\begin{equation}
\mathcal{G}(\eta)=\oint \frac{d\zeta}{2\pi i} \frac{\mathcal{G}(\zeta)}{\zeta-\eta}
\end{equation}
using Cauchy's integral formula. To lighten the notation we will suppress the parity choice $\pm$ of the operators in this subsection. We can deform the contour to wrap it around the two branch cuts $[1,\infty)$, $(-\infty,0]$ as in Figure \ref{fig:RPdcontours}. Assuming that the two-point function has the following behavior in the Regge limit
\begin{equation}\label{Reggebounded}
|\mathcal{G}(\eta)|\lesssim |\eta|^{-\epsilon}\;,\quad \eta\to \infty\;,
\end{equation}
where $\epsilon$ is an infinitesimal positive number, we can drop the contribution from the arcs at infinity. In other words, no subtraction is needed in this case. The two-point function becomes
\begin{equation}
\mathcal{G}(\eta)=\mathcal{G}_1(\eta)+\mathcal{G}_2(\eta)
\end{equation}
where 
\begin{equation}\label{G1G2int}
\begin{split}
\mathcal{G}_1(\eta)=&\int_{C_1} \frac{d\zeta}{2\pi i} \frac{\mathcal{G}(\zeta)}{\zeta-\eta}=\int_1^\infty  \frac{d\zeta}{2\pi i} \frac{{\rm Disc}_1[\mathcal{G}(\zeta)]}{\zeta-\eta}\;,\\
\mathcal{G}_2(\eta)=&-\int_{C_2} \frac{d\zeta}{2\pi i} \frac{\mathcal{G}(\zeta)}{\zeta-\eta}=\int_{-\infty}^0  \frac{d\zeta}{2\pi i} \frac{{\rm Disc}_2[\mathcal{G}(\zeta)]}{\zeta-\eta}\;,
\end{split}
\end{equation}
and
\begin{equation}
\begin{split}
{\rm Disc}_1[\mathcal{G}(\zeta)]=&\mathcal{G}(\zeta+i0^+)-\mathcal{G}(\zeta-i0^+)\;,\quad \zeta\in(1,\infty)\;,\\
{\rm Disc}_2[\mathcal{G}(\zeta)]=&\mathcal{G}(\zeta+i0^+)-\mathcal{G}(\zeta-i0^+)\;,\quad \zeta\in(-\infty,0)\;.
\end{split}
\end{equation}
The two functions $\mathcal{G}_1(\eta)$ and $\mathcal{G}_2(\eta)$ are related by crossing symmetry
\begin{equation}\label{RPdG1G2}
\mathcal{G}_2(\eta)=\pm \mathcal{G}_1(1-\eta)\;.
\end{equation}
Here we have also assumed that the integrals converge, {\it i.e.}, ${\rm Disc}_2[\mathcal{G}(\zeta)] \sim \zeta^{-a}$ with $a<1$ as $\zeta \to 0$. To proceed, let us define a function 
\begin{equation}
k_h(\eta)=\eta^h {}_2F_1(h,h;2h,\eta)
\end{equation}
which has the following orthonormal property
\begin{equation}
\oint_{|\eta|=\epsilon} \frac{d\eta}{2\pi i}\eta^{-2}k_{x+n}(\eta)k_{1-x-m}(\eta)=\delta_{nm}\;.
\end{equation}
We note that the conformal blocks with $d=2$, $\Delta_1=\Delta_2=\Delta_\varphi$ are related to $k_h(\eta)$ by 
\begin{equation}
g_{c,\Delta}(\eta)=\eta^{-\Delta_\varphi} k_{\frac{\Delta}{2}}(\eta)\;.
\end{equation}
We will now show that the two-point function $\mathcal{G}(\eta)$ can be decomposed in terms of a special class of conformal blocks with dimensions $\Delta_n^{\rm d.t.}=2\Delta_\varphi+2n$ in {\it both} OPE channels. Here the superscript ${\rm d.t.}$ stands for {\it double-trace} as $\Delta_n^{\rm d.t.}$ is the conformal dimension of a double-trace operator of the schematic form $:\varphi\square^n\varphi:$. These are operators which appear universally in the mean field theory, and their dimensions are just the sums of the dimensions of the building blocks.\footnote{The operator $\varphi$ has dimension $\Delta_\varphi$ and $\square$ has dimension 2. Therefore, $:\varphi\square^n\varphi:$ has dimension $2\Delta_\varphi+2n$. The terminology ``double-trace'' is borrowed from gauge theory to denote the fact such an operator is made of {\it two} ``single-trace'' operators $\phi$. Here ``trace'' refers to the trace over gauge group indices because a single-trace operator in gauge theories has the form ${\rm tr}(X_1X_2\ldots X_n)$, with operators $X_i=X_i^aT^a$ in the adjoint representation of the gauge group. For the moment, these terminologies can just be regarded as names if they are not familiar to the reader.} To show this, we note that the kernel in the Cauchy integral admits the following expansion in terms double-trace conformal blocks
\begin{equation}\label{Ckernalindt}
\frac{1}{\zeta-\eta}=\sum_{n=0}^\infty H_n(\zeta) g_{c,\Delta_n^{\rm d.t.}}(\eta)
\end{equation}
with coefficients which are functions of $\zeta$. These coefficients can be computed using the orthonormal property of $k_h(\eta)$
\begin{equation}
H_n(\zeta)=\oint_{|\eta|=\epsilon}\frac{d\eta}{2\pi i} \frac{\eta^{\Delta_\varphi-2}}{\zeta-\eta}k_{1-\Delta_\varphi-n}(\eta)\;,
\end{equation}
and gives
\begin{equation}
H_n(\zeta)=\frac{(-4)^{-n}(\Delta_\varphi)_n(2\Delta_\varphi-1)_n}{n!(\Delta_\varphi-\frac{1}{2})_n}\zeta^{-1}{}_3F_2(1,-n,2\Delta_\varphi+n-1;\Delta_\varphi,\Delta_\varphi;\zeta^{-1})\;.
\end{equation}
Inserting (\ref{Ckernalindt}) into (\ref{G1G2int}), we find that $\mathcal{G}_1(\eta)$ can be expanded in terms of double-trace conformal blocks
\begin{equation}
\mathcal{G}_1(\eta)=\sum_{n=0}^\infty r_{n,1}\, g_{c,\Delta_n^{\rm d.t.}}(\eta)\;,
\end{equation}
with
\begin{equation}\label{rn1}
r_{n,1}=\int_{C_1}\frac{d\zeta}{2\pi i} H_n(\zeta)\mathcal{G}(\zeta)\;.
\end{equation} 
To get this result, we have assumed that we can exchange the order of the integral and the infinite sum. However, to avoid being overly technical in this introductory section, we will not discuss when this assumption is valid. Now using crossing symmetry (\ref{RPdG1G2}), we find that $\mathcal{G}_2(\eta)$ can be expanded in terms of double-trace conformal blocks in the image channel
\begin{equation}
\mathcal{G}_2(\eta)=\sum_{n=0}^\infty r_{n,2}\, \bar{g}_{c,\Delta_n^{\rm d.t.}}(\eta)\;,
\end{equation}
where $r_{n,2}=\pm r_{n,1}$. This proves that any two-point function $\mathcal{G}(\eta)$, suitably bounded in the Regge limit as in (\ref{Reggebounded}), can be decomposed as a linear combination of double-trace conformal blocks $\{g_{c,\Delta_n^{\rm d.t.}}(\eta),\bar{g}_{c,\Delta_n^{\rm d.t.}}(\eta)\}$ from both channels. Note this is quite different from the standard conformal block decomposition where we use only one OPE channel and the conformal dimensions of the conformal blocks are not forced to take discrete values. 

The conclusions we reached in this simple example in fact generalize to the general case. Let us consider a two-point function in a $d$-dimensional real projective space CFT with dimensions $\Delta_1$ and $\Delta_2$. If the two-point function satisfies the boundedness condition (\ref{Reggebounded}), then a basis is given by the conformal blocks in the bulk channel and the image channel
\begin{equation}
\{g_{c,\Delta_n^{\rm d.t.}}(\eta),\bar{g}_{c,\Delta_n^{\rm d.t.}}(\eta)\}\;,\quad\quad n=0,1,2,\ldots
\end{equation}
where 
\begin{equation}
\Delta_n^{\rm d.t.}=\Delta_1+\Delta_2+2n\;.
\end{equation}
With this basis of functions, we can define a dual basis whose elements are the {\it functionals}
\begin{equation}
\{\omega_{c,n},\bar{\omega}_{c,n}\}\;,\quad\quad n=0,1,2,\ldots\;.
\end{equation}
These functional are defined to have the following orthonormal action on the basis vectors
\begin{equation}\label{omegaong}
\begin{split}
&\omega_{c,m}(g_{c,\Delta_n^{\rm d.t.}})=\delta_{nm}\;,\quad\quad \omega_{c,m}(\bar{g}_{c,\Delta_n^{\rm d.t.}})=0\;,\\
&\bar{\omega}_{c,m}(g_{c,\Delta_n^{\rm d.t.}})=0\;,\quad\quad \bar{\omega}_{c,m}(\bar{g}_{c,\Delta_n^{\rm d.t.}})=\delta_{nm}\;.
\end{split}
\end{equation}

To fully specify these functionals, we need to know how they act on a generic conformal block, {\it i.e.}, computing
\begin{equation}\label{omegaaction}
\omega_{c,n}(g_{c,\Delta})\;,\quad \omega_{c,n}(\bar{g}_{c,\Delta})\;,\quad \bar{\omega}_{c,n}(g_{c,\Delta})\;,\quad \bar{\omega}_{c,n}(\bar{g}_{c,\Delta})\;,
\end{equation}
for a general conformal dimension $\Delta$. Let us consider decomposing a conformal block in the above double-trace basis
\begin{equation}\label{gcdecomp}
g_{c,\Delta}(\eta)=\sum_n M_n\, g_{c,\Delta_n^{\rm d.t.}}(\eta)+\sum_n N_n\, \bar{g}_{c,\Delta_n^{\rm d.t.}}(\eta)\;.
\end{equation}
Acting on it with the basis functionals and using the orthonormal relation (\ref{omegaong}), we find 
\begin{equation}
M_n=\omega_{c,n}(g_{c,\Delta})\;,\quad \quad N_n=\bar{\omega}_{c,n}(g_{c,\Delta})\;.
\end{equation}
Similarly, the actions $\omega_{c,n}(\bar{g}_{c,\Delta})$, $\bar{\omega}_{c,n}(\bar{g}_{c,\Delta})$ appear in the decomposition coefficients of the image channel conformal block $\bar{g}_{c,\Delta}$. Once we know the actions of these functionals, we can act with them on the crossing equation of two-point functions (\ref{RPdcrossingeqn}) to systematically extract the constraints on the CFT data in the form of {\it sum rules}\footnote{Here we have absorbed the identity exchange into the infinite sum. Moreover, we have assumed that we are allowed to swap the infinite summation with the action of the functionals. However, this may not always be true. For a detailed discussion on this swapping subtlety, see \cite{Qiao:2017lkv,mazacpaulos2}.}
\begin{equation}\label{PRdsumrule}
\sum_k \mu_{c,12k}\, \omega_n(g_{c,\Delta_k})=\pm \sum_k \mu_{c,12k}\, \omega_n(\bar{g}_{c,\Delta_k})\;.
\end{equation}
In the 2d example considered above, these actions can be computed as contour integrals (\ref{rn1}) with $\mathcal{G}(\eta)$ taken to be a conformal block. However, these coefficients can also be computed in a different way, by considering a seemingly unrelated problem of conformal block decomposition of tree-level Witten diagrams in AdS space, as we now explain. 

\begin{figure}
\centering
\includegraphics[width=0.55\textwidth]{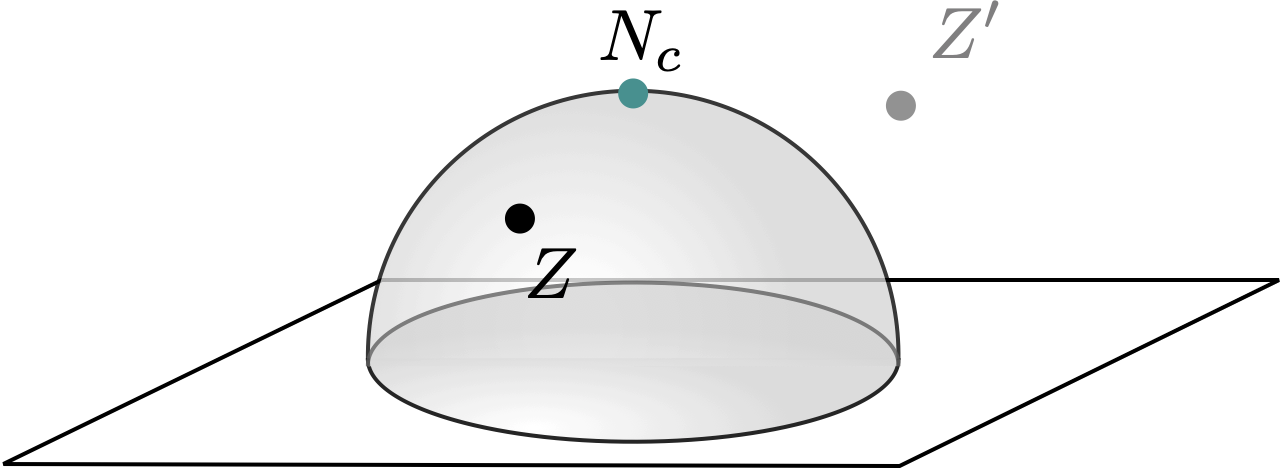}
\caption{The quotient AdS space obtained by identifying points related by the inversion with respect to the unit radius hemisphere. The point $Z'$ is the inversion image of $Z$. The north pole $N_c$ of the hemisphere is invariant under inversion.}
    \label{fig:qAdS}
\end{figure}

We consider the following simple setup that realizes the kinematics of a real projective space CFT in AdS space. We first extend the inversion (\ref{invx}) in $\mathbb{R}^d$ to $AdS_{d+1}$ by
\begin{equation}\label{AdSinv}
z_0\to \frac{z_0}{z_0^2+\vec{z}^2}\;,\quad \quad \vec{z}\to -\frac{\vec{z}}{z_0^2+\vec{z}^2}
\end{equation}
where $z=(z_0,\vec{z})$ are the Poincar\'e coordinates of AdS and $z_0$ is the radial direction. Note that at the conformal boundary $z_0=0$, (\ref{AdSinv}) reduces to (\ref{invx}). This transformation can also be obtained from (\ref{invP}) by replacing the embedding space vector $P$ by the embedding space vector $Z$ of an AdS point
\begin{equation}
Z^A=\frac{1}{z_0}\bigg(\frac{1+z_0^2+\vec{z}^2}{2},\frac{1-z_0^2-\vec{z}^2}{2},\vec{z}\bigg)\;.
\end{equation}
Geometrically, (\ref{AdSinv}) corresponds to an inversion with respect to a unit radius hemisphere located at $z_0=0$, $\vec{z}=0$, as is illustrated in Figure \ref{fig:qAdS}. The kinematics of real projective space CFTs can be realized in the quotient space $AdS_{d+1}/\mathbb{Z}_2$ which is defined by identifying points under the inversion (\ref{AdSinv})
\begin{equation}
z_0\leftrightarrow \frac{z_0}{z_0^2+\vec{z}^2}\;,\quad \quad \vec{z}\leftrightarrow -\frac{\vec{z}}{z_0^2+\vec{z}^2}\;.
\end{equation}
Note that (\ref{AdSinv}) has a special fixed point at $z_0=1$, $\vec{z}=0$, which corresponds to the north pole of the hemisphere. In fact, written in terms of the embedding space coordinates, this point is nothing but the fixed vector $N_c$. 

\begin{figure}
\centering
\includegraphics[width=\textwidth]{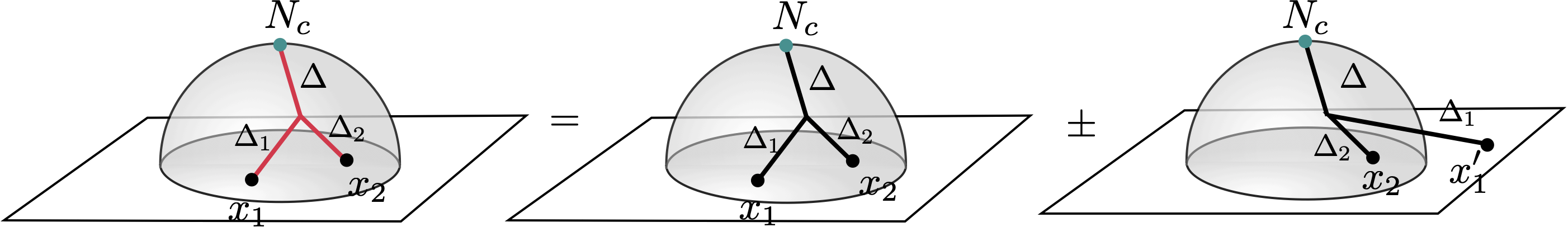}
\caption{Tree-level Witten diagrams. On the LHS, we have a single exchange Witten diagram in the quotient AdS space. On the RHS, we use the method of images to express it in terms of the exchange Witten diagram in the full AdS space and its image diagram where one boundary point $x_1$ has been moved to its inversion image $x'_1$.}
    \label{fig:RPdWD}
\end{figure}

Let us now consider scalar fields on this quotient AdS space, and require the fields to have the same value at points related by inversion 
\begin{equation}\label{Phicondi}
\Phi(Z)=\pm \Phi(Z')\;,
\end{equation}
up to a sign which corresponds to the parity of the dual CFT operator. Here $Z'$ is the inversion image of $Z$. We further assume that the effective action of these fields contains a cubic term $\int dZ \Phi_1(Z) \Phi_2(Z) \Phi(Z)$, and a linear term $\Phi(N_c)$ that is localized at the fixed point $N_c$. We can then consider an exchange Witten diagram $V^{\rm exchange}_{\Delta}$ in the quotient AdS space as is shown on the LHS of Figure \ref{fig:RPdWD}. The propagators in this diagram need to be consistent with the condition (\ref{Phicondi}) on the hemisphere. The insertion point of the cubic vertex is integrated over the quotient AdS space, while the end $N_c$ is held fixed. By using the method of images, we can express this exchange diagram in terms of exchange Witten diagrams defined on the full AdS space without taking the quotient (Figure \ref{fig:RPdWD})
\begin{equation}\label{VqAdS}
V^{\rm exchange}_{\Delta}(P_1,P_2)=W^{\rm exchange}_{\Delta}(P_1,P_2)\pm (x_1^2)^{-\Delta_1}\bar{W}^{\rm exchange}_{\Delta}(P_1,P_2)\;.
\end{equation}
Here, the Witten diagrams
\begin{equation}\label{Wcexchange}
W^{\rm exchange}_{\Delta}(P_1,P_2)=\int_{AdS_{d+1}}d^{d+1}Z G^\Delta_{BB}(N_c,Z)G^{\Delta_1}_{B\partial}(Z,P_1)G^{\Delta_2}_{B\partial}(Z,P_2)\;,
\end{equation}
\begin{equation}
\bar{W}^{\rm exchange}_{\Delta}(P_1,P_2)=W^{\rm exchange}_{\Delta}(P'_1,P_2)\;,
\end{equation}
are defined with the standard AdS bulk-to-boundary propagator
\begin{equation}
G^{\Delta}_{B\partial}(Z,P)=\frac{1}{(-2Z\cdot P)^\Delta}\;,
\end{equation}
and the bulk-to-bulk propagator satisfying 
\begin{equation}
\big(\square_Z-\Delta(\Delta-d)\big)G^\Delta_{BB}(Z,W)=\delta(Z,W)\;.
\end{equation}
The integration region of the cubic vertex insertion points is the entire $AdS_{d+1}$ space. There appears to be two more diagrams where the sources are inserted at $(P'_1,P'_2)$ and $(P_1,P'_2)$. but one can show that they are the same as the two diagrams above. Let us also extract the kinematic factor from (\ref{VqAdS}), and then we have 
\begin{equation}
V^{\rm exchange}_{\Delta}(P_1,P_2)=\frac{1}{(1+x_1^2)^{\Delta_1}(1+x_2^2)^{\Delta_2}}\left(\mathcal{W}^{\rm exchange}_{\Delta}(\eta)\pm \bar{\mathcal{W}}^{\rm exchange}_{\Delta}(\eta)\right)\;,
\end{equation}
with 
\begin{equation}\label{WWbar}
\bar{\mathcal{W}}^{\rm exchange}_{\Delta}(\eta)=\mathcal{W}^{\rm exchange}_{\Delta}(1-\eta)\;.
\end{equation}

After this long detour into AdS, let us finally get to our point: the functional actions (\ref{omegaaction}) can be extracted from the conformal block decomposition coefficients of the Witten diagrams $\mathcal{W}^{\rm exchange}_{\Delta}(\eta)$, $\bar{\mathcal{W}}^{\rm exchange}_{\Delta}(\eta)$. One can show that  
both diagrams obey the boundedness condition (\ref{Reggebounded}) in the Regge limit. Moreover, under conformal block decomposition the exchange Witten diagram $\mathcal{W}^{\rm exchange}_{\Delta}(\eta)$ is comprised of a single-trace conformal block and infinitely many double-trace conformal blocks in the same channel
\begin{equation}
\mathcal{W}^{\rm exchange}_{\Delta}(\eta)=A\, g_{c,\Delta}(\eta)+\sum_{n=0}^\infty A_n\, g_{c,\Delta_n^{\rm d.t.}}(\eta)\;,
\end{equation} 
and infinitely many double-trace conformal blocks in the crossed channel
\begin{equation}
\mathcal{W}^{\rm exchange}_{\Delta}(\eta)=\sum_{n=0}^\infty B_n\, \bar{g}_{c,\Delta_n^{\rm d.t.}}(\eta)\;.
\end{equation}
We are stating here these decomposition properties merely as facts to avoid going into unnecessary technicalities. But they follow directly from a study of these integrals and the details of the analysis can be found in \cite{Giombi:2020xah}. Comparing these two expansions with (\ref{gcdecomp}), one finds that the functional actions can be expressed in terms of the conformal block decomposition coefficients of exchange Witten diagrams as
\begin{equation}
\omega_{c,n}(g_{c,\Delta})=-\frac{A_n}{A}\;,\quad \quad \bar{\omega}_{c,n}(g_{c,\Delta})=\frac{B_n}{A}\;.
\end{equation}
As was shown in \cite{Giombi:2020xah}, one can explicitly evaluate the exchange Witten diagram integral (\ref{Wcexchange}) in terms of hypergeometric functions, and recursively compute all the conformal block decomposition coefficients. Here we do not give the explicit expressions of these coefficients, and refer the reader to \cite{Giombi:2020xah} for details. Similarly, the image diagram decomposes as 
\begin{equation}
\bar{\mathcal{W}}^{\rm exchange}_{\Delta}(\eta)=\sum_{n=0}^\infty B_n\, g_{c,\Delta_n^{\rm d.t.}}(\eta)=A\, \bar{g}_{c,\Delta}(\eta)+\sum_{n=0}^\infty A_n\, \bar{g}_{c,\Delta_n^{\rm d.t.}}(\eta)\;,
\end{equation}
which follows from the crossing relation (\ref{WWbar}). From these identities we find 
\begin{equation}\label{oobarfWbar}
\omega_{c,n}(\bar{g}_{c,\Delta})=\bar{\omega}_{c,n}(g_{c,\Delta})\;,\quad \bar{\omega}_{c,n}(\bar{g}_{c,\Delta})=\omega_{c,n}(g_{c,\Delta})\;.
\end{equation}
All in all, these Witten diagrams give us an efficient holographic method to obtain these functionals. 

In fact, there is a further use of these Witten diagrams. As we now show, they also furnish a new basis of functions to decompose conformal correlators. The decomposition reads
\begin{equation}\label{RPdWDexp}
\mathcal{G}(\eta)=\sum_k \frac{\mu_{c,12k}}{A}(\mathcal{W}_{\Delta_k}^{\rm exchange}(\eta)\pm\bar{\mathcal{W}}_{\Delta_k}^{\rm exchange}(\eta))
\end{equation}
where we sum over the same spectrum appearing  in the conformal block decomposition (\ref{RPdcfdecom}) and $\mu_{c,12k}$ are the same coefficients. To prove it, we expand both $\mathcal{W}_{\Delta_k}^{\rm exchange}(\eta)$ and $\bar{\mathcal{W}}_{\Delta_k}^{\rm exchange}(\eta)$ in the $\eta\to 0$ channel. This gives
\begin{equation}
\mathcal{G}(\eta)=\sum_k \mu_{c,12k} g_{c,\Delta_k}(\eta)+\sum_k\sum_{n=0}^\infty \mu_{c,12k}\left(-\omega_{c,n}(g_{\Delta_k})\pm \bar{\omega}_{c,n}(g_{c,\Delta_k})\right)g_{c,\Delta_n^{\rm d.t.}}\;.
\end{equation}
Interchanging the order of the sums and using (\ref{oobarfWbar}), we find the second term vanishes when the sum rules (\ref{PRdsumrule}) are used. The expansion in terms of Witten diagrams then reduces to the conformal block decomposition in the bulk channel. While this new expansion is very similar to the conformal block expansion, we must note the important difference that it exploits building blocks from both channels at the same time. Basis of this kind first appeared in the original work of Polyakov \cite{Polyakov:1974gs}. Finally, we can also reverse the logic starting from (\ref{RPdWDexp}). Requiring that double-trace conformal blocks vanish in the conformal block decomposition gives rise to sum rules (\ref{PRdsumrule}).

\vspace{0.5cm}

\noindent{\bf An application}

\vspace{0.2cm}

\noindent The zeros in the functional actions (\ref{omegaong}) at the double-trace conformal dimensions can considerably simplify the sum rules (\ref{PRdsumrule}) if the theory spectrum contains such operators. The simplest (and almost trivial) example is the mean field theory. However, we can also consider CFTs which can be viewed as small perturbations around the mean field theory. This special feature of the analytic functionals therefore makes them particularly suitable for studying such theories. As a simple application, let us show how to use functionals to bootstrap the one-point function coefficients of the $O(N)$ model on real projective space. We will only outline the computation, and refer the reader to \cite{Giombi:2020xah} for the explicit details. 

The CFT of interest is the Wilson-Fisher fixed point of the Lagrangian theory 
\begin{equation}
S=\frac{\Gamma(\frac{d}{2}-1)}{4\pi^{\frac{d}{2}}}\int d^dx \bigg(\frac{1}{2}(\partial_\mu\varphi^I)^2+\frac{\lambda}{4}(\varphi^I\varphi^I)^2\bigg)\;,\quad I=1,\ldots,N\;,
\end{equation}
at $d=4-\epsilon$ dimension. We consider the $\langle\varphi^I\varphi^J\rangle$ two-point function. To order $\epsilon^2$, the only operators that can be exchanged are the identity and the double-trace operators $[\varphi\varphi]_n=\varphi\square^n\varphi$, and we parameterize the deviations from the mean field theory values as follows
\begin{equation}
\mu_{\varphi\varphi n}=\mu^{(0)}_{\varphi\varphi n}+\epsilon \mu^{(1)}_{\varphi\varphi n}+\epsilon^2 \mu^{(2)}_{\varphi\varphi n}\;,\quad \Delta_\varphi=\frac{d}{2}-1+\epsilon^2\gamma_\varphi^{(2)}\;,\quad \Delta_{[\varphi\varphi]_n}=\Delta_n^{\rm d.t.}+\epsilon\gamma_n^{(1)}+\epsilon^2\gamma_n^{(2)}\;.
\end{equation}
We have used the well known fact that the anomalous dimension of $\varphi$ starts at $\epsilon^2$. 

To proceed, we act on crossing equation with the functionals and expand the sum rules in powers of $\epsilon$
\begin{equation}
\omega_{c,n}(g_{c,0})+\sum_n \mu_{\varphi\varphi n}\, \omega_{c,n}(g_{c,\Delta_{[\varphi\varphi]_n}})=\pm \big(\omega_{c,n}(\bar{g}_{c,0}) +\sum_n \mu_{\varphi\varphi n}\,\omega_{c,n}(\bar{g}_{c,\Delta_{[\varphi\varphi]_n}})\big)+\mathcal{O}(\epsilon^3)\;.
\end{equation}
At the zeroth order, we have just the mean field theory and one can check that the sum rules 
\begin{equation}
\omega_{c,n}(g_{c,0})+\mu_{\varphi\varphi n}^{(0)}=\pm \omega_{c,n}(\bar{g}_{c,0})
\end{equation}
gives the correct mean field theory coefficients
\begin{equation}
\mu_{\varphi\varphi n}^{(0)}=\pm \delta_{n,0}\;.
\end{equation}
Moreover, one finds $\omega_n(g_{c,0})=0$. Using these results in the next order and we find that the sum rules at $\mathcal{O}(\epsilon)$ are given by
\begin{equation}
\mu_{\varphi\varphi n}^{(1)}\pm \big[\omega_{c,n}(g_{c,\Delta_0^{\rm d.t.}+\epsilon \gamma_0^{(1)}})\big]_{\mathcal{O}(\epsilon)}=\pm \big[\omega_{c,n}(\bar{g}_{c,0})\big]_{\mathcal{O}(\epsilon)}+\big[\omega_{c,n}(\bar{g}_{c,\Delta_0^{\rm d.t.}+\epsilon \gamma_0^{(1)}})\big]_{\mathcal{O}(\epsilon)}\;.
\end{equation}
Solving these equations, we find
\begin{equation}
\mu_{\varphi\varphi 0}^{(1)}=-\frac{\gamma_0^{(1)}}{2}\;,\quad \mu_{\varphi\varphi 1}^{(1)}=\frac{\gamma_0^{(1)}}{4}\;,\quad \mu_{\varphi\varphi\, n\geq 2}^{(1)}=0\;,
\end{equation}
which agrees with \cite{Hasegawa:2018yqg}. Note that the fact that only finitely many coefficients are nonzero at this order is very useful. It implies that at the next order the sum rules will continue to have only finitely many terms. Explicitly, we find at $\mathcal{O}(\epsilon^2)$
\begin{equation}
\begin{split}
&\mu_{\varphi\varphi n}^{(2)}\pm \big[\omega_{c,n}(g_{c,\Delta_0^{\rm d.t.}+\epsilon \gamma_0^{(1)}+\epsilon^2\gamma_0^{(2)}})\big]_{\mathcal{O}(\epsilon^2)}+\mu_{\varphi\varphi 0}^{(1)}\big[\omega_{c,n}(g_{c,\Delta_0^{\rm d.t.}+\epsilon \gamma_0^{(1)}})\big]_{\mathcal{O}(\epsilon)}\\&+\mu_{\varphi\varphi 1}^{(1)}\big[\omega_{c,n}(g_{c,\Delta_1^{\rm d.t.}+\epsilon \gamma_1^{(1)}})\big]_{\mathcal{O}(\epsilon)}=\pm\big[\omega_{c,n}(\bar{g}_{c,0})\big]_{\mathcal{O}(\epsilon^2)}+\big[\omega_{c,n}(\bar{g}_{c,\Delta_0^{\rm d.t.}+\epsilon \gamma_0^{(1)}+\epsilon^2\gamma_0^{(2)}})\big]_{\mathcal{O}(\epsilon^2)}\\
&\pm \mu_{\varphi\varphi 0}^{(1)} \big[\omega_{c,n}(\bar{g}_{c,\Delta_0^{\rm d.t.}+\epsilon \gamma_0^{(1)}})\big]_{\mathcal{O}(\epsilon)}\pm \mu_{\varphi\varphi 1}^{(1)} \big[\omega_{c,n}(\bar{g}_{c,\Delta_1^{\rm d.t.}+\epsilon \gamma_1^{(1)}})\big]_{\mathcal{O}(\epsilon)}\;.
\end{split}
\end{equation}
From these equations, we can solve the coefficients $\mu_{\varphi\varphi n}^{(2)}$ in terms of the bulk data of anomalous dimensions. After using their values in the $O(N)$ model, we find, for example
\begin{equation}
\begin{split}
&\mu_{\varphi\varphi 0}=\pm 1-\frac{N+2}{2(N+8)}\epsilon-\frac{3(N+2)(2N+6\pm(N+8))}{2(N+8)^3}\epsilon^2+\mathcal{O}(\epsilon^3)\;,\\
&\mu_{\varphi\varphi 1}=\frac{N+2}{4(N+8)}\epsilon-\frac{(N+2)}{4(N+8)^2}\big(\frac{76+N(N+10)}{2(N+8)}\pm(N-2)\big)\epsilon^2+\mathcal{O}(\epsilon^3)\;.
\end{split}
\end{equation}
These results are consistent with the analytic results obtained from large $N$ analysis \cite{Giombi:2020xah}, and also with the numerical bootstrap results \cite{Nakayama:2016cim} which considered $\epsilon=1$, $N=1$. Proceeding to $\mathcal{O}(\epsilon^3)$ and higher orders however is difficult. Due to the fact that all $\mu_{\varphi\varphi n}^{(2)}$ are nonzero, the functional sum rules at the next order inevitably contain infinitely many terms, making them difficult to solve analytically.

\subsubsection{Boundary CFTs}
Two-point functions in BCFTs also admit a similar functional treatment \cite{Kaviraj:2018tfd,Mazac:2018biw}, which is closely related to mean field theories with boundaries. Analogous to the choice of parity in the real projective CFT case, here one can choose either Neumann or Dirichlet boundary conditions for the associated mean field theory. For definiteness, we will only discuss the Neumann boundary condition case here. The Dirichlet case is similar and its discussion can be found in \cite{Kaviraj:2018tfd}. 

We start with the conformal block decomposition of the mean field theory two-point function with Neumann boundary condition
\begin{equation}
\langle \varphi(x_1) \varphi(x_2) \rangle_{\rm Neumann}=\frac{1}{(2x_{1,\perp})^{\Delta_\varphi}(2x_{2,\perp})^{\Delta_\varphi}}\big(\xi^{-\Delta_\varphi}+(\xi+1)^{-\Delta_\varphi}\big)\;.
\end{equation}
In the bulk channel, we find infinitely many double-trace operators with dimensions $\Delta_n^{\rm d.t.}=2\Delta_\phi+2n$, $n=0,1,\ldots$. In the boundary channel, we find an infinite tower of boundary modes $\widehat{\varphi}_n$ with dimensions $\widehat{\Delta}_n=\Delta_\varphi+2n$, $n=0,1,\ldots$. If we had considered the Dirichlet boundary condition, we would have found a different tower with dimensions $\widehat{\Delta}_n=\Delta_\varphi+2n+1$. 

Let us now consider a two-point function with $\Delta_1\neq \Delta_2$. We will also make a technical assumption that the two-point function satisfies the following boundedness condition in the Regge limit
\begin{equation}\label{BCFTsRbound}
|\mathcal{G}(\xi)|\lesssim |\xi+1|^{-\frac{\Delta_1+\Delta_2-1}{2}+\epsilon}\;,\quad \quad \xi\to-1^+\;,
\end{equation}
for some $\epsilon>0$. This behavior was referred to as {\it Regge super-boundedness} in \cite{Mazac:2018biw}, and here we assume it to simplify the discussion. The claim is that the following set of conformal blocks in {\it both} bulk and boundary channels, which are closely related to the mean field theory spectrum, furnishes a basis for Regge super-bounded functions
\begin{equation}\label{BCFTbasis}
\begin{split}
&g_{b,\Delta_n^{\rm d.t.}}^{\rm bulk}\;,\quad \text{with}\quad \Delta_n^{\rm d.t.}=\Delta_1+\Delta_2+2n\;,\quad n=0,1,\ldots\;,\\
&g_{b,\widehat{\Delta}_n^i}^{\rm boundary}\;,\quad \text{with}\quad \widehat{\Delta}_n^i=\Delta_i+2n\;,\quad n=0,1,\ldots\;,\; i=1,2\;.
\end{split}
\end{equation}
The dual basis is given by the set of functionals $\{\omega_n, \widehat{\omega}_n^{(i)}\}$ defined by the orthonormal relations
\begin{equation}
\begin{split}
&\omega_m(g_{b,\Delta_n^{\rm d.t.}}^{\rm bulk})=\delta_{mn}\;,\quad\quad \omega_m(g_{c,\widehat{\Delta}_n^i}^{\rm boundary})=0\;,\\
&\widehat{\omega}_m^{(j)}(g_{b,\Delta_n^{\rm d.t.}}^{\rm bulk})=0\;,\quad\quad \widehat{\omega}_m^{(j)}(g_{c,\widehat{\Delta}_n^i}^{\rm boundary})=\delta_{mn}\delta_{ij}\;.
\end{split}
\end{equation}
Similar to the real projective CFT case, a convenient way to see that $\{g_{b,\Delta_n^{\rm d.t.}}^{\rm bulk}, g_{b,\widehat{\Delta}_n^i}^{\rm boundary}\}$ provides a basis is to use holography. It also allows us to obtain the actions of the dual functionals. 

\begin{figure}
  \centering
\begin{subfigure}{0.45\textwidth}
 \centering
  \includegraphics[width=0.7\linewidth]{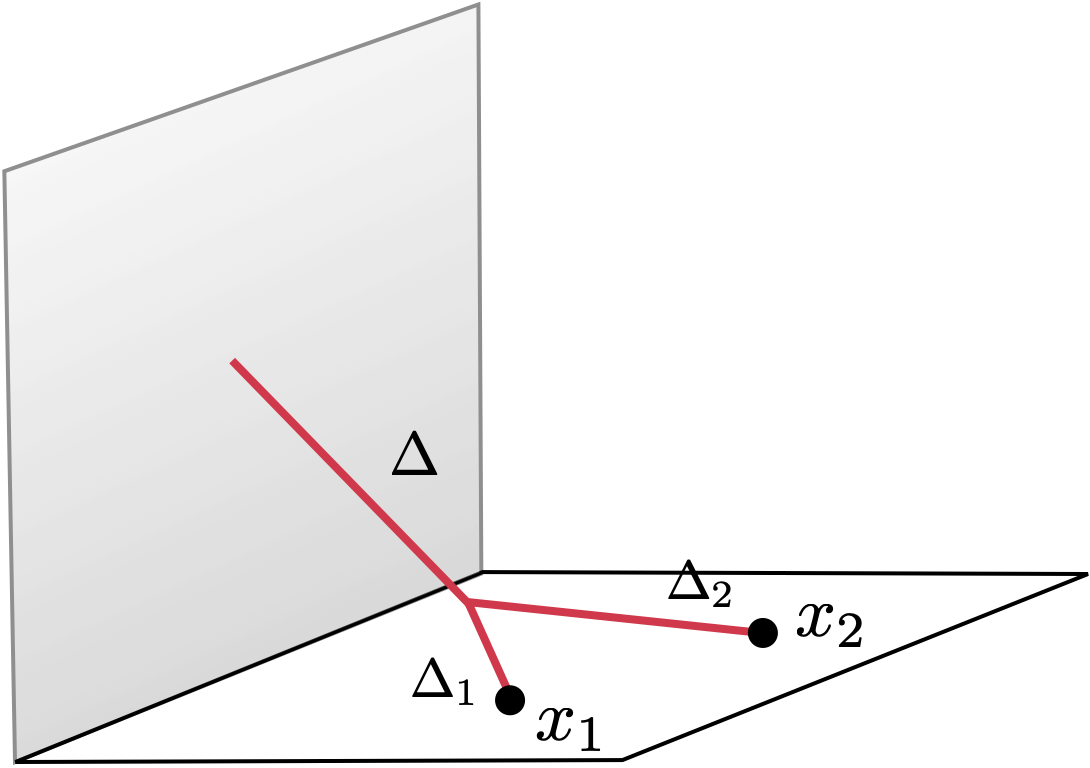}
  \caption{Bulk channel diagram.}
  \label{fig:hAdSbulk}
\end{subfigure}
\begin{subfigure}{0.45\textwidth}
  \centering
  \includegraphics[width=0.7\linewidth]{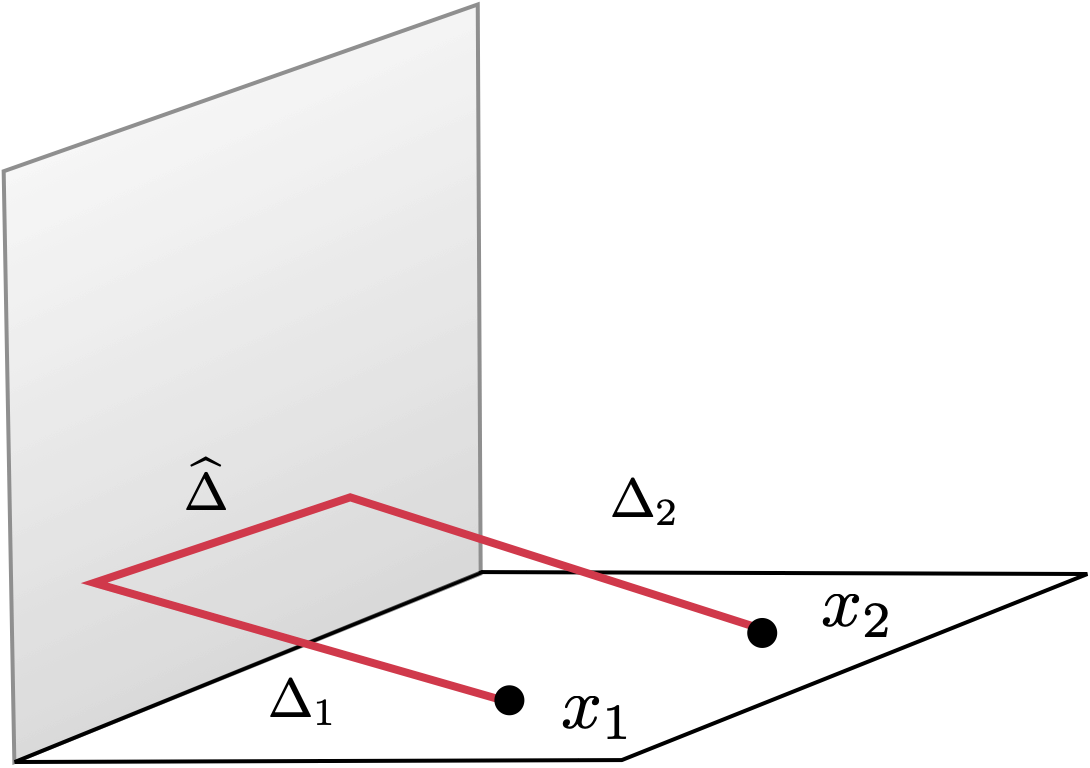}
  \caption{Boundary channel diagram.}
  \label{fig:hAdSboundary}
\end{subfigure}
\caption{Exchange Witten diagrams in the half AdS space. Here  spacetime stops after the vertical wall.}
\label{fig:hAdSWD}
\end{figure}

\begin{figure}
  \centering
\begin{subfigure}{0.63\textwidth}
 \centering
  \includegraphics[width=\linewidth]{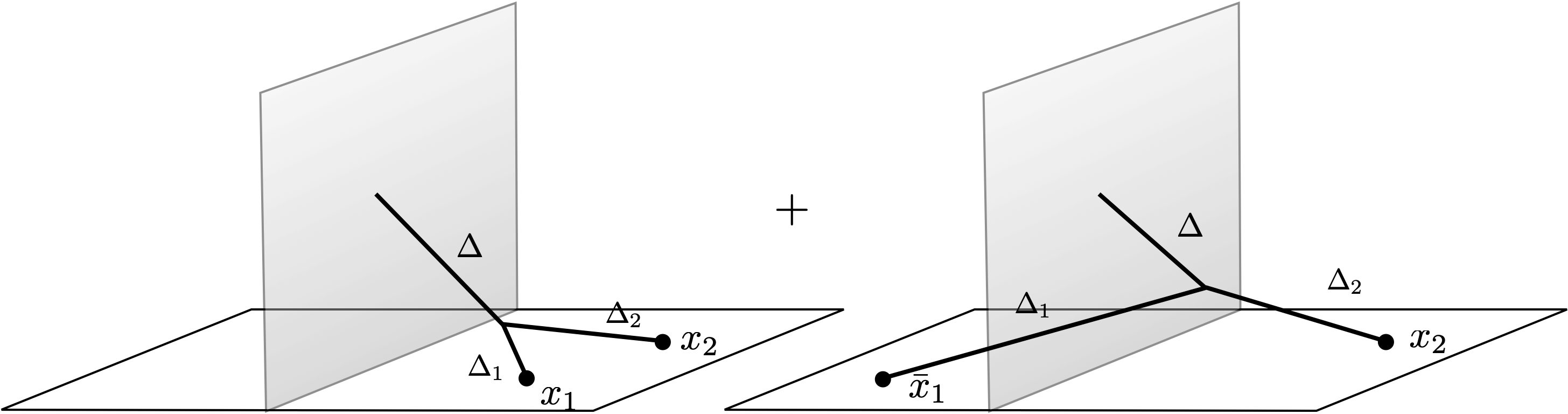}
  \caption{Bulk and mirror channel diagrams.}
  \label{fig:ICFTbulk}
\end{subfigure}
\begin{subfigure}{0.34\textwidth}
  \centering
  \includegraphics[width=\linewidth]{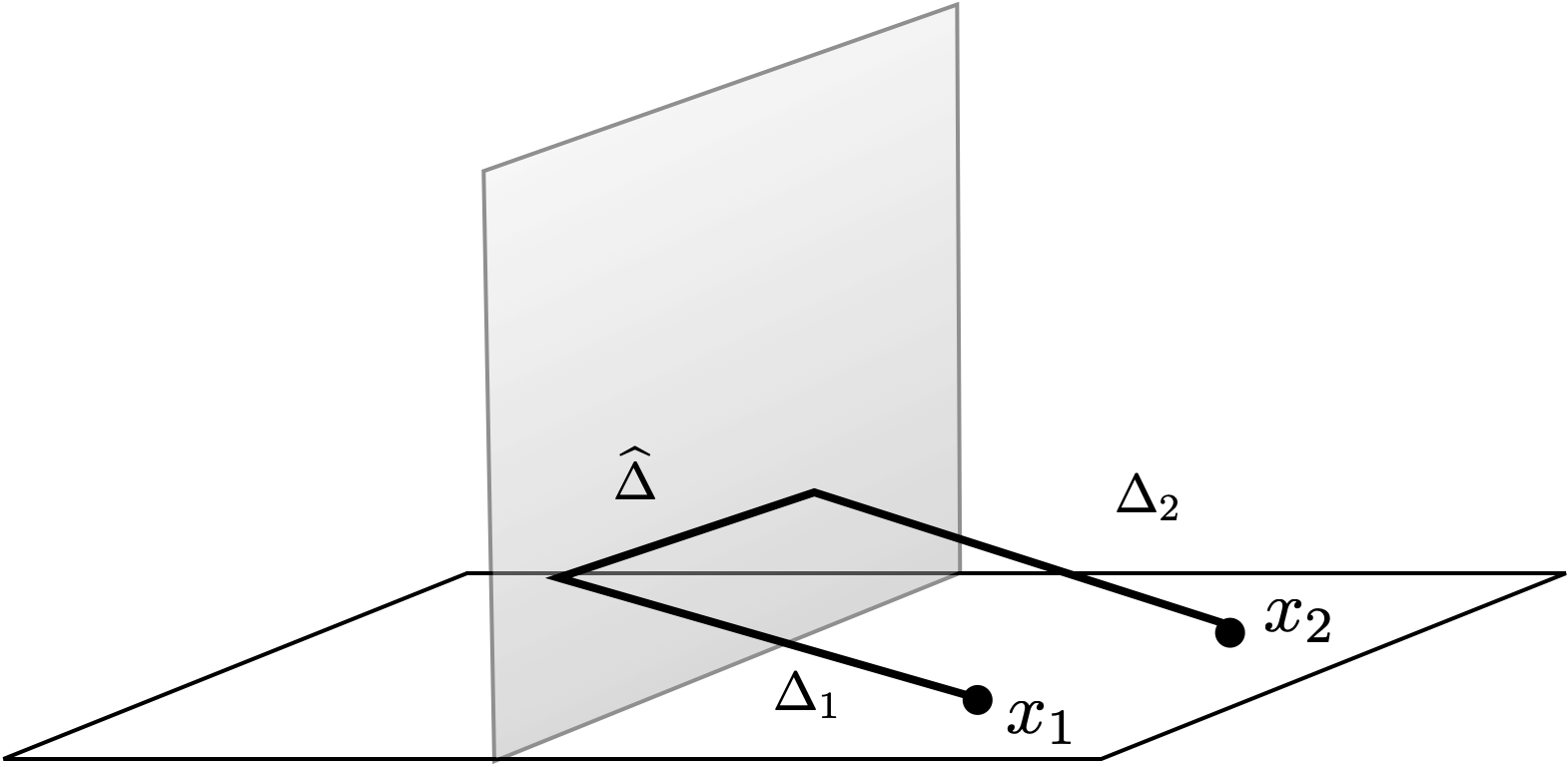}
  \caption{Boundary channel diagram.}
  \label{fig:ICFTboundary}
\end{subfigure}
\caption{Exchange Witten diagrams in the full AdS space. Here the $AdS_d$ subspace is just an interface which hosts localized degrees of freedom and does not back-react to the geometry.}
\label{fig:ICFTWD}
\end{figure}

Let us consider the following holographic setup where we take half of the $AdS_{d+1}$ space by requiring $z_\perp\geq 0$. This amounts to extending the boundary of the BCFT at $x_\perp=0$ into a wall in $AdS_{d+1}$. The mean field theory boundary condition is also extended by requiring scalar fields in the half AdS space to obey Neumann boundary condition on the wall. We can then consider the following two types of diagrams: the bulk channel exchange Witten diagram \ref{fig:hAdSbulk} and the boundary channel exchange Witten diagram \ref{fig:hAdSboundary}. Here both the bulk-to-bulk and the bulk-to-boundary propagators need to obey the Neumann boundary condition at the $AdS_d$ subspace $z_\perp=0$. In the bulk channel diagram, the cubic vertex insertion point is integrated over the half $AdS_{d+1}$ space, and the other end of the bulk-to-bulk propagator is integrated over the entire $AdS_d$ wall. In the boundary channel diagram, the bulk-to-bulk propagator lives in $AdS_d$ and both vertex insertion points are integrated over $AdS_d$. Again, by using the method of images, we can express these diagrams in terms of diagrams defined in the full $AdS_{d+1}$ (Figure \ref{fig:ICFTWD}).  In this new setup, we have a $AdS_d$ probe brane located at $z_\perp=0$ which is just an interface. There are localized degrees of freedom living on this subspace, but the brane does not back-react to the geometry.  The half AdS space diagram \ref{fig:hAdSbulk} is equivalent to the sum of a bulk channel exchange diagram in the full AdS space and its mirror diagram in which $\mathcal{O}_1$ is inserted at $-x_{1,\perp}$. These two diagrams are shown in Figure \ref{fig:ICFTbulk}, and we denote them by $\mathcal{W}^{\rm bulk}_\Delta$, $\mathcal{W}^{\rm mirror}_\Delta$ respectively. The integration over the cubic vertex insertion points is now over the entire $AdS_{d+1}$ space. On the other hand, $x_{1,\perp}\to-x_{1,\perp}$ in an $AdS_{d+1}$  boundary channel exchange diagram does not change its value. Therefore, doubling the space does not affect the boundary channel exchange Witten diagrams and the two diagrams \ref{fig:hAdSboundary} and \ref{fig:ICFTboundary} are the same. We denote \ref{fig:ICFTboundary} by $\mathcal{W}^{\rm boundary}_{\widehat{\Delta}}$.

\begin{figure}
\centering
\includegraphics[width=0.4\textwidth]{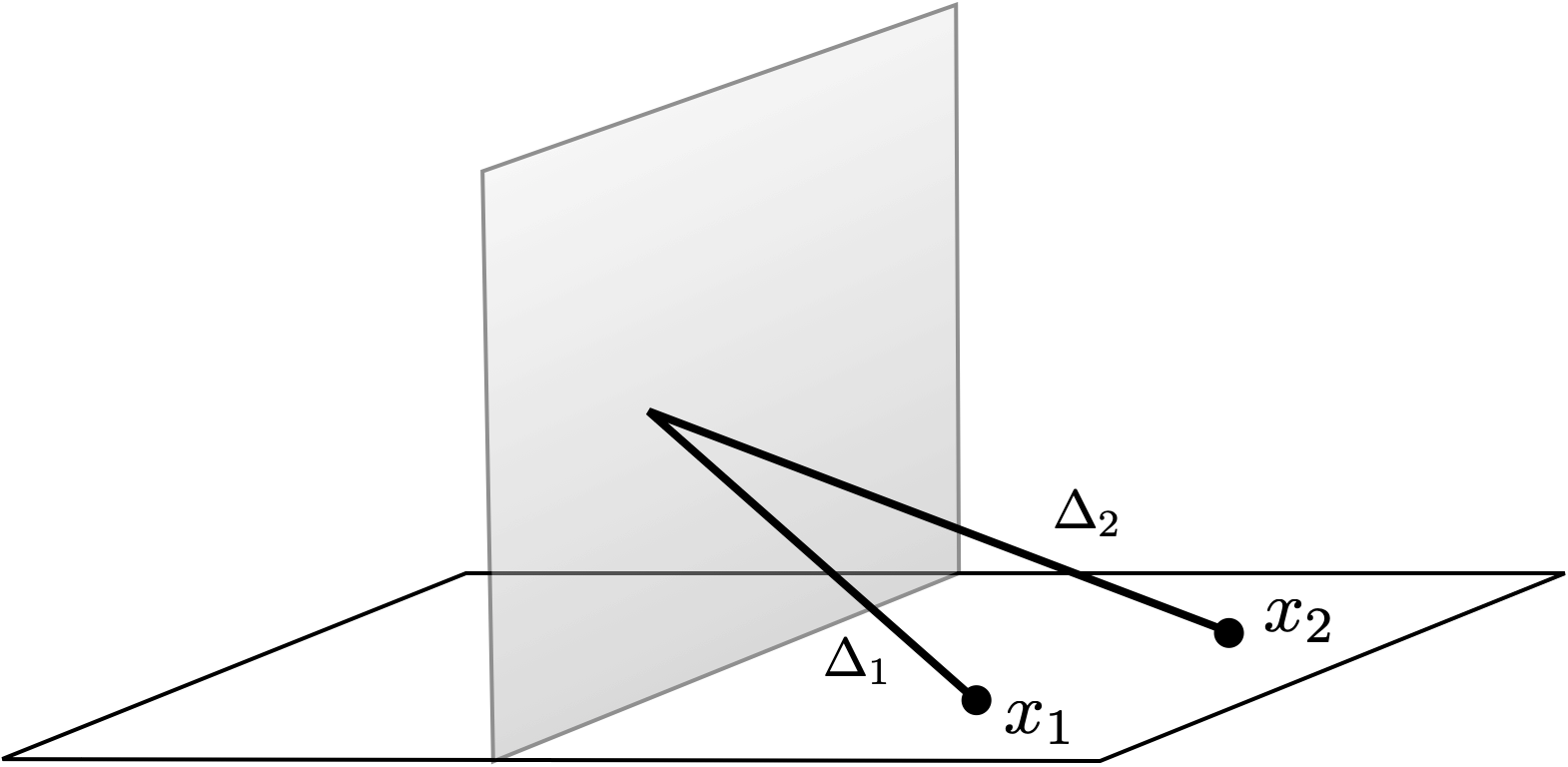}
\caption{The contact Witten diagram. This diagram is given by the product of two bulk-to-boundary propagators with the vertex point integrated over the $AdS_d$ brane.}
    \label{fig:ICFTcontact}
\end{figure}

The crucial property we need to make progress is how these Witten diagrams decompose into conformal blocks. Using for example the Mellin representation for BCFTs \cite{Rastelli:2017ecj}, one can show that $\mathcal{W}^{\rm bulk}_\Delta+\mathcal{W}^{\rm mirror}_\Delta$ decomposes into single-trace and double-trace conformal blocks in the bulk channel
\begin{equation}
\mathcal{W}^{\rm bulk}_\Delta(\xi)+\mathcal{W}^{\rm mirror}_\Delta(\xi)= E\, g_{b,\Delta}^{\rm bulk}(\xi)+\sum_{n=0}^\infty E_n\, g_{b,\Delta_n^{\rm d.t.}}^{\rm bulk}(\xi)\;,
\end{equation}
and only double-trace conformal blocks in the boundary channel
\begin{equation}
\mathcal{W}^{\rm bulk}_\Delta(\xi)+\mathcal{W}^{\rm mirror}_\Delta(\xi)= \sum_{n=0}^\infty\sum_{i=1,2} F_n^{(i)}\, g_{b,\widehat{\Delta}_n^i}^{\rm boundary}(\xi)\;,
\end{equation}
 Similarly, the boundary channel exchange diagram $\mathcal{W}^{\rm boundary}_{\widehat{\Delta}}$ decomposes as
\begin{equation}
\mathcal{W}^{\rm boundary}_{\widehat{\Delta}}(\xi)=K g_{b,\widehat{\Delta}}^{\rm boundary}(\xi)+\sum_{n=0}^\infty\sum_{i=1,2} K_n^{(i)}\, g_{b,\widehat{\Delta}_n^i}^{\rm boundary}(\xi)
\end{equation}
in the boundary channel, and 
\begin{equation}
\mathcal{W}^{\rm boundary}_{\widehat{\Delta}}(\xi)=\sum_{n=0}^\infty L_n\, g_{b,\Delta_n^{\rm d.t.}}^{\rm bulk}(\xi)
\end{equation}
in the bulk channel. Equating the two decompositions in each case, we find that conformal blocks $g_{b,\Delta}^{\rm bulk}(\xi)$, $g_{b,\widehat{\Delta}}^{\rm boundary}(\xi)$ with arbitrary conformal dimensions $\Delta$, $\widehat{\Delta}$ can be expanded in terms of the double-trace conformal blocks (\ref{BCFTbasis}). This almost leads to our claim that (\ref{BCFTbasis}) is a basis. However, we need to check if the Regge behaviors of the Witten diagrams satisfy the condition (\ref{BCFTsRbound}). One can show that as $\xi\to -1^+$, these diagrams behave as \cite{Mazac:2018biw}
\begin{equation}
|\mathcal{W}^{\rm bulk}_\Delta(\xi)+\mathcal{W}^{\rm mirror}_\Delta(\xi)|\sim |\xi+1|^{-\frac{\Delta_1+\Delta_2-1}{2}}\;,\quad |\mathcal{W}^{\rm boundary}_{\widehat{\Delta}}(\xi)|\sim |\xi+1|^{-\frac{\Delta_1+\Delta_2-3}{2}}\;.
\end{equation}
Therefore, only $\mathcal{W}^{\rm boundary}_{\widehat{\Delta}}$ is Regge super-bounded and $\mathcal{W}^{\rm bulk}_\Delta+\mathcal{W}^{\rm mirror}_\Delta$ is only {\it Regge bounded} in the parlance of \cite{Mazac:2018biw}. To see why this point is important, we note that there is another Regge-bounded diagram $\mathcal{W}^{\rm contact}$ (Figure \ref{fig:ICFTcontact}) which decomposes into only $\{g_{b,\Delta_n^{\rm d.t.}}^{\rm bulk}, g_{b,\widehat{\Delta}_n^i}^{\rm boundary}\}$ in both channels
\begin{equation}
\mathcal{W}^{\rm contact}(\xi)=\sum_{n=0}^\infty R_n\, g_{b,\Delta_n^{\rm d.t.}}^{\rm bulk}(\xi)=\sum_{n=0}^\infty\sum_{i=1,2} S_n^{(i)}\, g_{b,\widehat{\Delta}_n^i}^{\rm boundary}(\xi)\;.
\end{equation}
This implies a linear relation among the basis vectors. However, we can avoid this relation by insisting that we are in the smaller space of functions defined by (\ref{BCFTsRbound}). It turns out that there is a unique combination of the exchange diagrams and the contact diagram
\begin{equation}
\mathcal{V}^{\rm bulk}_\Delta(\xi)=\mathcal{W}^{\rm bulk}_\Delta(\xi)+\mathcal{W}^{\rm mirror}_\Delta(\xi)+\theta\, \mathcal{W}^{\rm contact}(\xi)
\end{equation}
such that $\mathcal{V}^{\rm bulk}_\Delta$ has improved Regge behavior $|\xi+1|^{-\frac{\Delta_1+\Delta_2-3}{2}}$ and is therefore super-bounded. Then in this Regge super-bounded space a basis is given by (\ref{BCFTbasis}). Moreover, the actions of the dual functionals can be read off from the conformal block decomposition coefficients of the combination $\mathcal{V}^{\rm bulk}_\Delta$ and $\mathcal{W}^{\rm boundary}_{\widehat{\Delta}}$
\begin{equation}
\begin{split}
&\omega_m(g_{b,\Delta}^{\rm bulk})=-\frac{1}{E}(E_m+\theta R_m)\;,\quad \widehat{\omega}^{(j)}_m(g_{b,\Delta}^{\rm bulk})=\frac{1}{E}(F_m^{(j)}+\theta S_m^{(j)})\;,\\
& \omega_m(g_{b,\widehat{\Delta}}^{\rm boundary})=\frac{L_m}{K}\;,\quad \widehat{\omega}^{(j)}_m(g_{b,\widehat{\Delta}}^{\rm boundary})=-\frac{K_m^{(j)}}{K}\;.
\end{split}
\end{equation}
Similar to the real projective case, one can also show that $\mathcal{V}^{\rm bulk}_\Delta$ and $\mathcal{W}^{\rm boundary}_{\widehat{\Delta}}$ form a Polyakov style basis for expanding correlators.

The above discussion assumed $\Delta_1\neq \Delta_2$. However, the story of the equal weight case $\Delta_1=\Delta_2=\Delta_\varphi$ is similar and requires only minor modifications. In this case the two towers of boundary channel conformal blocks in the basis (\ref{BCFTbasis}) become degenerate, but the degeneracy can be compensated by turning one tower into derivative conformal blocks
\begin{equation}
\begin{split}
&g_{b,\Delta_n^{\rm d.t.}}^{\rm bulk}\;,\quad \text{with}\quad \Delta_n^{\rm d.t.}=2\Delta_\varphi+2n\;,\quad n=0,1,\ldots\;,\\
&g_{b,\widehat{\Delta}_n}^{\rm boundary}\;,\partial g_{b,\widehat{\Delta}_n}^{\rm boundary}\;,\quad \text{with}\quad \widehat{\Delta}_n=\Delta_\varphi+2n\;,\quad n=0,1,\ldots\;,
\end{split}
\end{equation}
Here $\partial g_{b,\widehat{\Delta}}^{\rm boundary}=\partial_{\widehat{\Delta}} g_{b,\widehat{\Delta}}^{\rm boundary}$. This basis again can be found by examining the conformal block decomposition of Witten diagrams with equal external weights, where the derivative conformal blocks are related to anomalous dimensions. The dual functional basis is then defined to be $\{\omega_n,\widehat{\omega}_n,\widetilde{\omega}_n\}$, which acts on the basis vectors $\{g_{b,\Delta_n^{\rm d.t.}}^{\rm bulk},g_{b,\widehat{\Delta}_n}^{\rm boundary},\partial g_{b,\widehat{\Delta}_n}^{\rm boundary}\}$ in the orthonormal way. Their actions on general conformal blocks (and their derivatives) can be read off from the conformal block decomposition coefficients of $\mathcal{V}^{\rm bulk}_\Delta$ and $\mathcal{W}^{\rm boundary}_{\widehat{\Delta}}$.

Finally, the functionals discussed in this subsection can be applied to a variety of analytic bootstrap problems. For example, \cite{Kaviraj:2018tfd} used the functionals to recover the Wilson-Fisher BCFT data to order $\epsilon^2$. In \cite{Mazac:2018biw}, the functionals were used to study a deformation of the mean field theory which interpolates the Neumann and Dirichlet boundary conditions. These applications are similar to the $O(N)$ model example we studied in the real projective space CFT subsection, and therefore will not be further discussed. We refer the reader to the original papers for the details.

\subsection{CFTs on other backgrounds}\label{Subsec:otherbackgrounds}
The two situations we reviewed in this section can be viewed more generally as special cases of CFTs on backgrounds which are not conformally equivalent to (empty) $\mathbb{R}^d$. There has been a lot of progress in applying bootstrap techniques to study such CFTs. 

Closely related to boundary CFTs are CFTs with conformal defects of various codimensions. There is a vast literature on this topic in the context of conformal bootstrap, see, {\it e.g.}, \cite{Billo:2013jda,Gaiotto:2013nva,Gliozzi:2015qsa,Bianchi:2015liz,Billo:2016cpy,Gadde:2016fbj,Giombi:2017cqn,Soderberg:2017oaa,Lauria:2017wav,Lemos:2017vnx,Liendo:2018ukf,Isachenkov:2018pef,Lauria:2018klo,Liendo:2019jpu,Wang:2020seq,Drukker:2020swu,Komatsu:2020sup,Lauria:2020emq,Gimenez-Grau:2020jvf,Giombi:2021uae,Gimenez-Grau:2021wiv,Barrat:2021yvp}. Another important background is $\mathbb{R}^{d-1}\times S^1$ and is related to CFTs at finite temperature. There the simplest nontrivial observable is also the two-point function, and the Kubo-Martin-Schwinger condition is cast into a crossing equation. Therefore, the situation is quite similar to the cases of BCFTs and CFTs on real projective space. For works in this direction, see \cite{El-Showk:2011yvt,Iliesiu:2018fao,Manenti:2019wxs,Alday:2020eua,Dodelson:2020lal}.




\newpage
\markboth{3\quad LARGE SPIN ANALYTIC BOOTSTRAP}{}
\section{Large spin analytic bootstrap}\label{Sec:largespin}
In this section we would like to discuss how crossing symmetry, the structure of the OPE and basic properties of the conformal blocks imply the presence of operators with large spins, and how to characterize them. These developments are based on \cite{Alday:2007mf, Komargodski:2012ek, Fitzpatrick:2012yx}. For reader's convenience, we also offer a quick review of some basic concepts of CFT in Section \ref{Subsec:basicCFTreview}. However, for the readers who already have a working knowledge of CFT, this subsection can be safely skipped.
\subsection{Important concepts: A lightning review}\label{Subsec:basicCFTreview}
In this subsection, we will briefly summarize the important concepts needed in order to understand the rest of the review which deals with mostly four-point functions.  For readers who have read Section \ref{Sec:BCFT}, they will already find great familiarity with these concepts. Nevertheless, we will still go through them due to their essential importance and also to set up the notations that we will use in the review. It should be noted that this subsection is not intended to be a pedagogical introduction to CFT since these basic concepts have already been discussed in great detail in many excellent reviews \cite{numrev, slavaepfl,Penedones:2016voo,Simmons-Duffin:2016gjk}. Our discussion will be concise, and the reader is referred to these references for further details. For this subsection, we will focus on external scalar operators.

\begin{itemize}
    
\item {\it Operator product expansion (OPE)}: The concept of OPE holds the center stage in the discussion of the conformal bootstrap. In quantum field theory, the idea of OPE enables us to replace the product of two operators which are close to each other by an infinite set of operators inserted at the midpoint. Unlike QFT, where OPE is asymptotic, in CFT the OPE has a finite radius of convergence. For scalar primary operators $\varphi_1(x)$, $\varphi_2(x)$, we have the following operator equation\footnote{We already encountered this OPE in (\ref{OPEB}).}
\begin{equation}\label{OPE}
   \varphi_1(x_1)\varphi_2(x_2)=\frac{\delta_{12}}{(x_1-x_2)^{2\Delta_{\varphi_1}}}+\sum_{\mathcal{O}} C_{12\mathcal{O}}\,D[x_1-x_2,\partial_{x_2}]\mathcal{O}(x_2)\,,
    \end{equation}
where the sum is over primary operators $\mathcal{O}$. $C_{12\mathcal{O}}$ are the OPE coefficients and $D[x_1-x_2,\partial_{x_2}]$ are differential operators whose form is fixed by conformal invariance. The goal of the bootstrap is to constrain the OPE coefficients as well as the spectrum (scaling dimensions) of primary operators that appear in the OPE. In the CFT literature, the operator spectrum and the OPE coefficients are often referred to as the CFT data.  If a theory is unitary then there are unitarity bounds that the scaling dimensions of operators have to obey, namely
\begin{eqnarray}\label{unit}
&&\Delta\geq \frac{d-2}{2}\,,\quad  \ell=0\,,\\
&&\Delta\geq d-2+\ell\,,\quad \ell>0\,,
\end{eqnarray}
where $\ell$ denotes the spin of the operator. The quantity $\tau\equiv\Delta-\ell$ is referred to as the twist of the operator.
\item {\it Four-point functions}: The spacetime dependence of  two- and three-point functions are completely fixed by conformal invariance. Starting at four points, however, there are quantities which are invariant under all conformal transformations.\footnote{These statements are easy to see in the embedding space formalism introduced in Section \ref{Subsec:BCFTkinematics}.} These are the conformal cross ratios\footnote{They are the analogues of the cross ratios $\xi$ and $\eta$ for BCFTs and real projective space CFTs defined in (\ref{defcrxi}) and (\ref{defcreta}).}
    \begin{equation}
U=\frac{x_{12}^2 x_{34}^2}{x_{13}^2 x_{24}^2}\,,\quad V=\frac{x_{14}^2 x_{23}^2}{x_{13}^2 x_{24}^2}\,,
\end{equation}
As a result, conformal symmetry can only determine a four-point function up to an arbitrary function of $U$ and $V$. For example, we can write the correlation function of four identical scalar primary operators $\varphi$ with dimension $\Delta_\varphi$ as 
\begin{equation}
    \langle \varphi(x_1) \varphi(x_2) \varphi(x_3) \varphi(x_4) \rangle=\frac{1}{\left(x^2_{12}x^2_{34}\right)^{\Delta_\varphi}} \mathcal{G}(U,V)\,.
\end{equation}

\item {\it Conformal blocks:} Four-point functions can be deconstructed by using the OPE. Performing the OPE \eqref{OPE} for $\varphi(x_1)$ and $\varphi(x_2)$ we reduce the four-point function to a sum of three-point functions which are fixed by conformal symmetry up to the OPE coefficients. Equivalently, we can perform \eqref{OPE} for  $\varphi(x_1)$, $\varphi(x_2)$ and $\varphi(x_3)$, $\varphi(x_4)$ to reduce the four-point function as a sum of two-point functions of operators which are contained in both OPEs. In other words, the four-point function can be interpreted as the sum of infinitely many operator exchanges. The contribution to the four-point function from exchanging a conformal primary operator and its conformal descendants is known as a conformal block $g_{\Delta,\ell}(U,V)$.\footnote{Recall that we had similar notions for BCFTs and real projective CFTs. Depending on the OPE which we use, we have the bulk channel conformal block (\ref{BCFTbulkg}) and the boundary channel conformal block (\ref{BCFTbdrg}) for BCFTs. Similarly, we have the bulk channel conformal block (\ref{RPdCFTg}) and the image channel conformal block (\ref{RPdCFTgmirror}) for real projective space CFTs.} It can be obtained by directly resumming these contributions contained in the RHS of \eqref{OPE} for a specific primary operator $\mathcal{O}$. But more efficiently, the conformal block can be obtained as the eigenfunction of the bi-particle quadratic conformal Casimir operator. Explicit expressions for $g_{\Delta, \ell}(U,V)$ in any spacetime dimensions can be found in \cite{Dolan:2011dv} and they have a closed form expression in even spacetime dimensions. Using conformal blocks, we can write the decomposition of the four-point function more explicitly as follows
\begin{equation}\label{ope12}
\mathcal{G}(U,V)=1+\sum_{\Delta,\ell} a_{\Delta,\ell} U^{\frac{\Delta-\ell}{2}}g_{\Delta, \ell}(U,V)\;.
\end{equation}
Here we have separated out the contribution of the identity operator, whose presence we shall assume. The coefficients $a_{\Delta,\ell}=C^2_{\varphi \varphi \mathcal{O}_{\Delta,\ell}}$ are the squares of the OPE coefficients. For unitary theories $a_{\Delta,\ell}$ are positive as the OPE coefficients are real. It should be noted that the OPE coefficients depend on the normalizations which one chooses for the conformal blocks
$g_{\Delta, \ell}(U,V)$. For a survey of different normalizations used in the literature, see \cite{numrev}.

\item{\it Crossing equation:} In \eqref{ope12}, we made a particular choice of applying the OPE \eqref{OPE} to $\varphi(x_1)$, $\varphi(x_2)$ and $\varphi(x_3)$, $\varphi(x_4)$. We could have also used the OPE for $\varphi(x_1)$, $\varphi(x_4)$ and $\varphi(x_2)$, $\varphi(x_3)$ instead. Equating the two cases leads to the following crossing equation
\begin{equation}
\mathcal{G}(U,V)=\left( \frac{U}{V}\right)^{\Delta_\varphi}\mathcal{G}(V,U)\,.
\end{equation}
Note that the crossing equation does not obviously follow from the conformal block decomposition (\ref{ope12}). Instead, they together impose infinitely many constraints on the CFT data and form the cornerstone of the Numerical Conformal Bootstrap. The goal of the Analytic Conformal Bootstrap program is to develop analytic techniques to extract information from these equations.

\item{\it Generalized Free Fields:} In this review, we will frequently refer to generalized free fields (GFF) or the mean field theories (MFT) which constitute the simplest examples of conformal theories. These theories also arise as the leading order approximation in the expansion of certain small parameters. GFF theories exhibit similar features as free theories. For example, if we consider the four-point function of identical scalars with scaling dimension $\Delta_\varphi$, the exchanged spectrum consists of operators with dimensions
\begin{equation}
    \Delta_{n,\ell}=2\Delta_{\varphi}+2n+\ell\,,
\end{equation}
and spin $\ell$. These operators are the normal ordered products with the schematic form $:\varphi\square \partial^\ell \varphi:$ and their conformal dimensions are just the engineering dimensions. The motivation behind these operators is that one can use Wick contraction to get their contribution. In what we will discuss, we will consider corrections to the scaling dimensions of these operators (anomalous dimensions) and also their OPE coefficients. These small corrections are subleading in the expansion parameter.\footnote{We encountered examples of GFF and studied perturbations around them in Section \ref{Sec:BCFT}. The discussions of four-point functions will be similar in spirit.}

\item{\it Holographic correlators:} Much of the discussion that will follow is motivated by the AdS/CFT correspondence \cite{Maldacena:1997re,Witten:1998qj,Gubser:1998bc}. This correspondence is an equivalence between a specific string theory (or M-theory) in anti de Sitter (AdS) space in $d+1$ dimensions and a CFT in $d$ dimensions. Correlation functions in the CFT are mapped to scattering amplitudes in AdS space under this duality. In the bulk, the Feynman diagrams with external points anchored at the boundary of the AdS space are referred to as Witten diagrams. Similar to Feynman diagrams in flat space, we can classify Witten diagrams according to their topologies. For example, Witten diagrams relevant for four-point functions at tree level can be either contact diagrams or exchange diagrams.\footnote{In Section \ref{Subsec:analyticmethods}, we have seen similar Witten diagrams in more complicated setups, and they played an important role in the construction of the functional approach.}
\end{itemize}

\subsection{Euclidean vs Lorentzian}
\begin{figure}[ht]
	\centering
  \includegraphics[width=0.8\linewidth]{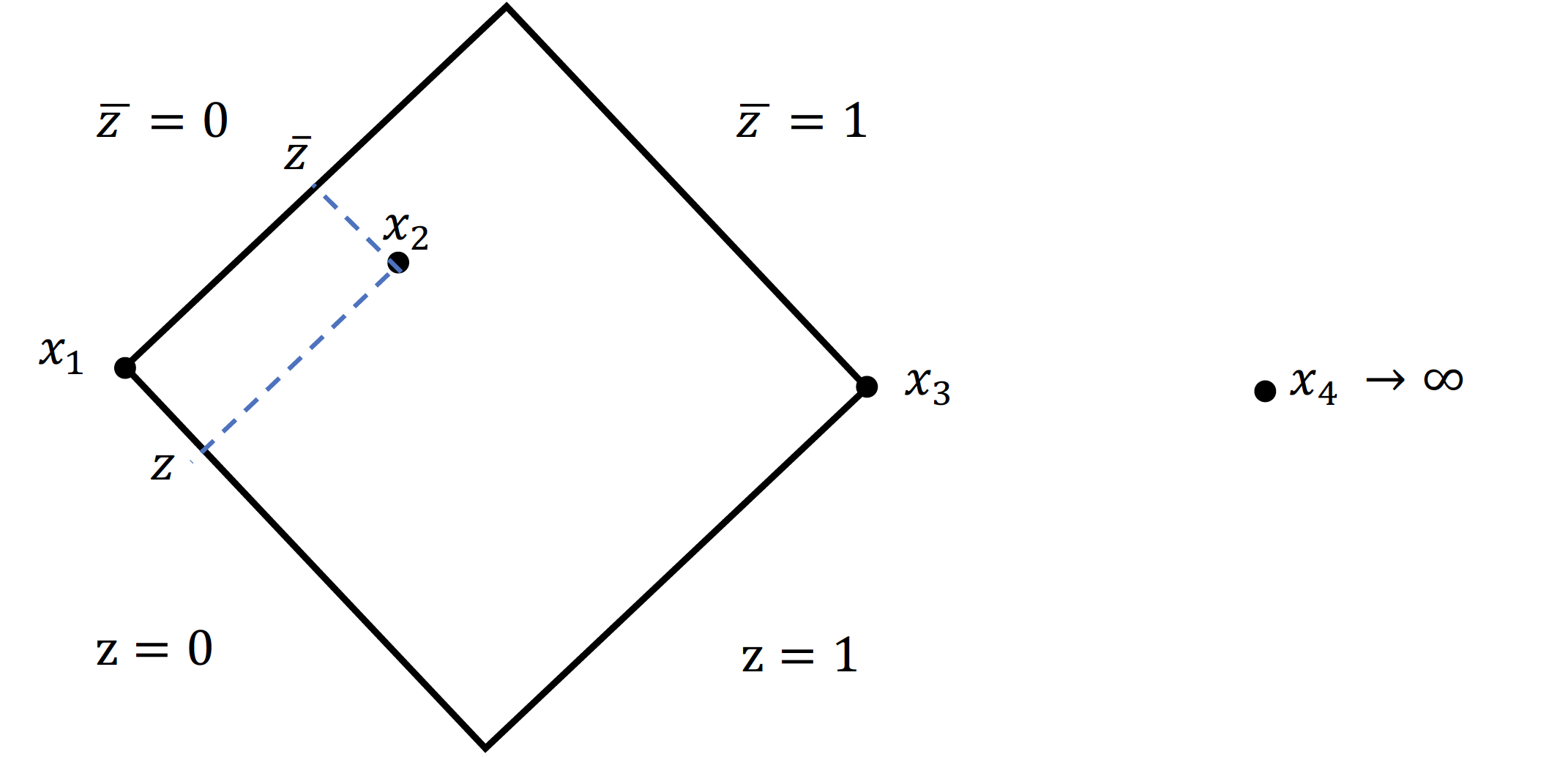}
	\caption{Kinematics in Lorentzian signature.}
	\label{figlorent}
\end{figure}

In this section we need to discuss some important differences when discussing CFTs in Euclidean or Lorentzian kinematics. In the Lorentzian case, it is possible to define the so called lightcone limit, which amounts to sending $x^2 \to 0$ while being on the lightcone. This is realised because it is possible to send one of the lightcone coordinates to zero while keeping at least another one fixed. Let us write the conformal cross ratios 
as
\begin{equation}
U= z \bar z\,,\quad V=(1-z)(1-\bar z)\,.    
\end{equation}
While in Euclidean signature $z^*=\bar{z}$, in the Lorentzian case $z$ and $\bar{z}$ are independent from each other. If we consider a four-point function in a space-like configuration in Minkowski signature, it is possible to use conformal symmetry to set the coordinates of the four points to be one at the origin $x_1=(0,0)$, one at $x_2=(z,\bar{z})$, one at  $x_3=(1,1)$ and $x_4$ is sent to infinity along both directions, see Fig.\ref{figlorent}. 
Then the lightcone limit amounts to taking $z$ small with $\bar{z}$ fixed. An interesting limit is the so called double lightcone limit, in which we send $z\to 0$, and then with $\bar{z} \to 1$, where $z \ll 1-\bar{z} \ll 1$.  The study of the conformal block decomposition, or of the OPE, in the Lorentzian regime necessarily probes the operators with small twists and large spins. \footnote{Notice that in Lorentzian signature the value of the spin is continuous, differently from the Euclidean counterpart. Despite the fact that we deal with local operator having integer spin, this is essential in the context of the Lorentzian inversion formula \cite{Caron-Huot:2017vep} which we will review in section \ref{Sec:inve}, see also \cite{Kravchuk:2018htv}.} This is exactly the spirit of the following section. Throughout the section, we will also use $U$, $V$ interchangeably with $z$, $\bar{z}$.
\subsection{Necessity of a large spin sector}
In this section we would like to study the regime of large spins and understand how crossing symmetry applied to the four-point function of a scalar operator $\varphi$ constrains the CFT data in this regime. Let us start with the simplest example of generalised free fields in four space-time dimensions\footnote{The generalisation of this discussion to generic even dimensions is straightforward, see for instance\cite{Fitzpatrick:2012yx}.}, which are the dual of free field theories in AdS. We will study the four-point correlator of four identical scalars of dimension $\Delta_\varphi$ in such a theory. Correlators in mean field theory are given by the sum of two-point contractions, giving
\begin{eqnarray} \label{MFT}
\nonumber\langle \varphi(x_1) \varphi(x_2) \varphi(x_3) \varphi(x_4) \rangle&=& \frac{1}{\left(x^2_{12}x^2_{34}\right)^{\Delta_\varphi}}+ \frac{1}{\left(x^2_{13}x^2_{24}\right)^{\Delta_\varphi}}+\frac{1}{\left(x^2_{14}x^2_{23}\right)^{\Delta_\varphi}}\\
\nonumber&=& \frac{1}{\left(x^2_{12}x^2_{34}\right)^{\Delta_\varphi}}\left( 1+ U^{\Delta_\varphi}+ \left(\frac{U}{V}\right)^{\Delta_\varphi} \right)\\
&=& \frac{1}{\left(x^2_{12}x^2_{34}\right)^{\Delta_\varphi}} \mathcal{G}(U,V)
\end{eqnarray}
where in the last line we have defined $\mathcal{G}(U,V)$ for later convenience. 
If we decompose the above correlator in conformal blocks, assuming that we are taking the OPE of $\varphi(x_1) \varphi(x_2)$ together with $\varphi(x_3) \varphi(x_4)$, we obtain that
\begin{equation}
\mathcal{G}(U,V)=1+\sum_{\Delta,\ell} a_{\Delta,\ell} U^{\frac{\Delta-\ell}{2}}g_{\Delta, \ell}(U,V)\;.
\end{equation}
We observe that in addition to exchanging the identity operator, there is a tower of intermediate operators of the form $\varphi \partial^{2n}\partial_{\mu_1}\dots \partial_{\mu_\ell}\varphi=[\varphi, \varphi]_{n,\ell}$ being exchanged. Their dimensions are $\Delta_{n,\ell}=2\Delta_\varphi+2n+\ell$ and the corresponding $a_{\Delta,\ell}$'s read \footnote{Notice that we use the following normalisation for the four dimensional conformal blocks $g_{\Delta, \ell}(z,\bar{z})=\frac{2^{-\ell}}{(z-\bar{z})}\left((-z)^{\ell} z k_{\Delta+\ell}(z)k_{\Delta-\ell-2}(\bar{z}) -(z \leftrightarrow \bar{z})\right)$ and with $k_a(z)={}_2 F_1\left(\frac{a}{2},\frac{a}{2},a,z\right)$. }
\begin{equation} \label{MFTa}
\begin{split}
 a^{\text{MF}}_{n,\ell}={}&\frac{2^{\ell+1} (\ell+1) (\ell+2 (\Delta_\varphi+n-1)) \Gamma (n+\Delta_\varphi-1)^2 }{(\Delta_\varphi-1)^2 n! \Gamma (\Delta_\varphi-1)^4 \Gamma (\ell+n+2)}\\
   {}&\times\frac{\Gamma
   (n+2\Delta_\varphi-3) \Gamma (\ell+n+\Delta_\varphi)^2 \Gamma (\ell+n+2 \Delta_\varphi-2)}{\Gamma (2 n+2 \Delta_\varphi-3) \Gamma (2 \ell+2 n+2 \Delta_\varphi-1)}\;.
   \end{split}
\end{equation}
We can set up this problem more abstractly and consider the constraints of crossing symmetry which are 
\begin{equation}
\mathcal{G}(U,V)=\left( \frac{U}{V}\right)^{\Delta_\varphi}\mathcal{G}(V,U)\,.
\end{equation}
This relation implies that we can decompose both sides in conformal blocks, leading to 
\begin{eqnarray} \label{cross}
1+ \sum_{\tau, \ell}a_{\tau, \ell}U^{\frac{\tau}{2}}g_{\tau, \ell}(U,V)=\left( \frac{U}{V}\right)^{\Delta_\varphi} \left(1+ \sum_{\tau, \ell}a_{\tau, \ell}V^{\frac{\tau}{2}}g_{\tau, \ell}(V,U)\right)\,,
\end{eqnarray}
where we have introduced the conformal twist $\tau=\Delta-\ell$. Before proceeding, it is useful to discuss some properties of the conformal blocks \cite{Dolan:2000ut,Dolan:2003hv,Dolan:2011dv}. While we are showing them explicitly only for four-dimensional conformal blocks, such properties are much more general and can be easily generalised to any dimension. We discuss three properties of the conformal blocks that will be relevant later on.
\begin{itemize}
\item {\it Small $U$ limit:} This limit is already explicit and it is controlled by the twist of the operator. Specifically, the conformal block behaves as
\begin{equation}
U^{\frac{\tau}{2}}g_{\tau, \ell}(U,V) \xrightarrow{U \ll 1}  -2^{-\ell} U^{\tau/2} (1-V)^{\ell} {}_2F_1 \left( \frac{\tau}{2}+\ell ,\frac{\tau}{2}+\ell ,\tau+2\ell,1-V \right)+\dots\;.
 \end{equation}
This limit has to be understood as $z \to 0$, for any value of $\bar{z}$.
\item {\it Small $V$ limit:} This limit is more subtle. We will discuss at length this limit later, but the structure is as follows 
\begin{equation} \label{vdiv}
g_{\tau, \ell}(U,V) \xrightarrow{V \ll 1}  a(U,V) \log(V) + b(U,V) 
 \end{equation}
where $a(U,V)$ and $b(U,V)$ admit a regular series expansion in the small $U$ and $V$ limits, meaning $z \ll 1-\bar{z} \ll 1$. In particular, the relation above should be understood as meaning that a small $V$ expansion of a single conformal block does not contain any power-law divergence and the only divergence appearing is logarithmic. 
\item {\it Casimir operator:} The conformal blocks are eigenfunctions of the quadratic and quartic Casimir operators of the conformal group, whose eigenvalues depend on the twist and spin of the intermediate operator. Specifically to four dimensions, we have
\begin{eqnarray}
\mathcal{D}_2 \left(U^{\frac{\tau}{2}}g_{\tau, \ell}(U,V) \right)&=& \frac{1}{2} ((\ell+\tau -4) (\ell+\tau )+\ell (\ell+2)) \left(U^{\frac{\tau}{2}}g_{\tau, \ell}(U,V) \right)\;,\\
\mathcal{D}_4 \left(U^{\frac{\tau}{2}}g_{\tau, \ell}(U,V) \right)&=& \ell (\ell+2) (\ell+\tau -3) (\ell+\tau -1) \left(U^{\frac{\tau}{2}}g_{\tau, \ell}(U,V) \right)
\end{eqnarray}
where 
\begin{eqnarray} \label{casimirdef}
{\cal D}_2 &=&D+\bar D + 2 \frac{z \bar z}{z-\bar z}\left( (1-z) \partial -(1-\bar z) \bar \partial \right) \label{quadrcas}\;, \\
{\cal D}_4 &=& \left( \frac{z \bar z}{z-\bar z}\right)^{2} (D-\bar D)\left( \frac{z-\bar z}{z \bar z}\right)^{2}(D-\bar D)\;.
\end{eqnarray}
Here $D=(1-z)z^2 \partial^2 - z^2 \partial$.
\end{itemize}

After this digression, let us come back to \eqref{cross}. By taking the limit of $U \ll 1$ on both sides of the relation, we note that there is a potential paradox. In particular, we observe that
\begin{eqnarray}
1&\sim &\frac{U^{\Delta_\varphi}}{V^{\Delta_\varphi}}\sum_{\tau, \ell} a_{\tau, \ell} V^{\tau/2}g_{\tau, \ell}(V,U)\;,\\
\implies \frac{1}{U^{\Delta_\varphi}} &\sim& \frac{1}{V^{\Delta_\varphi}}\sum_{\tau, \ell} a_{\tau, \ell} V^{\tau/2}g_{\tau, \ell}(V,U)\;. \label{largespin}
\end{eqnarray}
The LHS has a divergence $U^{-\Delta_{\varphi}}$ as $U \to 0$ while each conformal block on the RHS, following \eqref{vdiv}, has a logarithmic divergence. Then the question becomes: how is it possible to reproduce a power-law divergence with a sum of logarithmic divergences? This is only possible by having an infinite sum of conformal blocks on the RHS, with twist $\tau=2\Delta_\varphi$. This is the case because the sum does not converge for all real $U$. In particular, when $\sqrt{U} <0$ the sum diverges and by analytically continuing the sum to the region of convergence, it can be seen that it contains a power-law behaviour which fixes the problem. The next step is to understand if there are any parameters controlling such divergence. It is possible to study the limit of large $\tau=2 \Delta_{\varphi}+2n$, at fixed $\ell$, and it is possible to see that 
\begin{equation}
U^{\tau/2} g_{\tau, \ell} (U,V) \xrightarrow{U,V \ll 1}  U^{\tau/2} V^{\tau/2} +\dots\;.
\end{equation}
Moreover, $a_{\tau, \ell}$ for large $\tau$ are bounded \cite{Pappadopulo:2012jk}, ensuring that the sum for small $U$ and $V$ converges.

The limit of large spin $\ell$ and fixed $\tau$ is instead different. Let us study it in a more detailed way. We would like to study the RHS of \eqref{largespin}. In particular, if we consider the small $V$ limit, again in the regime in which $z\ll 1-\bar{z}\ll 1$, of this term we have
\begin{equation}
\sum_{\ell} \left( -\frac{1}{2}\right)^{\ell} a_{\tau, \ell} V^{\tau/2-\Delta_{\varphi}} \left( 1-U\right)^{\ell} {} _2F_1\left(\ell+\frac{\tau}{2}, \ell+\frac{\tau}{2},2\ell+\tau,1-U\right)\;.
\end{equation}
In this sum, most of the contribution comes from the region of $U$ goes to zero, when the spin $\ell$ is large. Thus we can make the following change of coordinates 
\begin{equation} 
\ell=\frac{x}{\sqrt{U}},
\end{equation}
where $x$ is a constant that does not depend on $U$. In this way, we can replace the sum with an integral over the parameter $x$. At the same time, we also consider the integral representation of the hypergeometric function
\begin{equation} \label{hyper}
{} _2F_1\left(a,b,c,q\right) = \int_{0}^{1} dt \frac{\Gamma(c)}{\Gamma(b)\Gamma(c-b)}t^{b-1}(1-t)^{c-b-1}(1-t q)^{-a}\;.
\end{equation}
To start with, we would like to see how the example of generalised free field works. Thus we can use $a^{\text{MF}}_{\tau, \ell}$ as the squared OPE coefficients and by combining all the pieces together and performing the change of coordinates $t \to 1-t \sqrt{U}$ in the $U\to 0$ limit we obtain\footnote{Notice that there is a factor of $\frac{1}{2}$ comes from the fact that we are summing only over even spins.}
\begin{eqnarray} \label{transf}
\nonumber && \frac{4 V^{\tau/2} U^{-\Delta_{\varphi}}  \Gamma \left(\frac{\tau }{2}-1\right)^2 \Gamma \left({-\Delta_{\varphi}}+\frac{\tau }{2}-3\right)}{({-\Delta_{\varphi}}-1)^2 \Gamma
   ({\Delta_{\varphi}}+1)^4 \Gamma (\tau -3) \Gamma \left(\frac{1}{2} (\tau -2 {-\Delta_{\varphi}})+1 \right)}
\displaystyle \int_0^{\infty} d x  x^{2 {-\Delta_{\varphi}}-1}\text{K}_0(2 x)   \\
&&=\frac{V^{\tau/2} U^{-\Delta_{\varphi}} \Gamma (\Delta_{\varphi})^2 \Gamma \left(\frac{\tau }{2}-1\right)^2 \Gamma \left(\Delta_{\varphi}+\frac{\tau }{2}-3\right)}{(\Delta_{\varphi}-1)^2 \Gamma
   (\Delta_{\varphi}-1)^4 \Gamma (\tau -3) \Gamma\left(\frac{1}{2} (\tau -2 \Delta_{\varphi})+1\right)},
\end{eqnarray}
where the function $\text{K}_0$ is the modified Bessel function of the second kind. If we combine this result with \eqref{largespin} we can see that it has several interesting features. We have proven that the tail of large spin of the sum in \eqref{largespin} is essential to reproduce the divergence as $U \to 0$ that we were studying. In particular, we see that in order to reproduce the leading terms in a small $U,V$ expansion, the CFT under study needs to have infinitely many operators with twist that accumulates at $\tau=2 \Delta_{\varphi}$. Remarkably
\begin{equation}
\frac{U^{-\Delta_{\varphi}} \Gamma (\Delta_{\varphi})^2 \Gamma \left(\frac{\tau }{2}-1\right)^2 \Gamma \left(\Delta_{\varphi}+\frac{\tau }{2}-3\right)}{(\Delta_{\varphi}-1)^2 \Gamma
   (\Delta_{\varphi}-1)^4 \Gamma (\tau -3) \left(\frac{1}{2} (\tau -2 \Delta_{\varphi})\right)!}
\xrightarrow{\tau=2 \Delta_{\varphi}} V^{\tau/2} U^{-\Delta_{\varphi}}\;,
\end{equation}
so it exactly reproduce the LHS of \eqref{largespin}. In addition, at any order in $V$ we need to have the same behaviour and thus the twist should accumulate around $\tau=2 \Delta_{\varphi}+2n$, with $n$ being an integer. To sum up: in this particular regime, where $z \ll 1-\bar{z}\ll1$, the leading contribution in the direct channel is controlled by operators with small twist and it is mapped in the crossed channel to the large spin contribution \footnote{Notice that when we mention the small $U$ and $V$ limit in the following sections we refer to this particular kinematical regime.}. An interesting remark is that the regime of $U$, $V$ going both to zero can only be reached in Minkowski spacetime. With this starting point it is also possible to study several cases, and in particular it is possible to see how to study corrections around large spins \cite{Fitzpatrick:2012yx,Komargodski:2012ek} \footnote{It is also possible to apply similar techniques to higher point functions, see \cite{Bercini:2020msp,Antunes:2021kmm}}. In addition, in \cite{Alday:2015eya} it has been shown that at any order in the perturbative series in large spin $\ell$ it is possible to compute all the terms in such expansion of the squared of the three-point functions and of the dimension away from the degenerate point by matching all the divergences in the direct and crossed channels.
\subsubsection{Anomalous dimensions at large spin}
The starting point is the situation that we reviewed in the previous subsection, in particular we consider a setup in which there exists a family of operators of a given twist, that is unbounded in the spin. In this large spin regime, there is a family of operators whose twist is independent on the spin, as soon as we consider finite values of the spin the operators start gaining an anomalous dimension and this degeneracy is lifted. We parametrise this perturbation with the anomalous dimension $\gamma_{\ell}$ that we require to be small. Explicitly, we write
\begin{equation}
\Delta_{n,\ell}=2\Delta_\varphi+2n+\ell+\gamma_{n,\ell}\,,\quad \gamma_{0,\ell}\equiv \gamma_\ell\,.    
\end{equation}
The case $\gamma_{\ell}=0$ corresponds to the case of the previous subsection. Now we would like to understand the constraints coming from crossing symmetry, unitarity and the structure of the conformal block decomposition on the correction to the anomalous dimensions.  In order to do so, we need to explore more orders in the small $U$ and $V$ expansion. In particular, in the OPE $\varphi \varphi$ there will be an operator with twist $\tau_{min}$ and spin $\ell_0$ with associated squared OPE coefficient $a_{\tau_{min},\ell_0}$ and thus the expansion in small $U$ reads
\begin{equation}
{\cal G}(U,V) =1+ a_{\tau_{min},\ell_0} U^{\frac{\tau_{min}}{2}} (V-1)^{\ell_0} ~_2F_1(\ell_0 + \tau_{min}/2,\ell_0+\tau_{min}/2,2\ell_0+\tau_{min},1-V) +\ldots\;.
\end{equation}
Crossing symmetry then implies a term of the form
\begin{align} \label{cr11}
{\cal G}(U,V) &= \frac{U^{\Delta_{\varphi}}}{V^{\Delta_{\varphi}}} \left( 1+ a_{\tau_{min},\ell_0} V^{\frac{\tau_{min}}{2}} (U-1)^{\ell_0} ~_2F_1(\ell_0 + \frac{\tau_{min}}{2},\ell_0+\frac{\tau_{min}}{2},2\ell_0+\tau_{min},1-U) +...\right) \nonumber \\
& =  \frac{U^{\Delta_{\varphi}}}{V^{\Delta_{\varphi}}} \left( 1+ a_{\tau_{min},\ell_0} V^{\frac{\tau_{min}}{2}} \left( \alpha \log U+ \beta+\ldots \right) +\ldots\right)\,,
\end{align}
where $\alpha,\beta$ are related to $a(U,V)$ and $b(U,V)$  in \eqref{vdiv}. The dots stand for more suppressed powers in $U$ and $V$. As discussed in
\cite{Fitzpatrick:2012yx,Komargodski:2012ek, Alday:2013cwa,Alday:2015eya}, crossing symmetry together with the structure of conformal blocks imply that the powers of $V$ multiplying $V^{\tau_{min}/2}$ are integers. In a small $U$ limit, we have then
\begin{eqnarray}
\label{crosslarge}
\nonumber &&\sum_{\ell} a_{\ell} U^{{\Delta_{\varphi}}+\gamma_\ell/2}(1-V)^\ell ~_2F_1\left(\Delta_{\varphi}+\ell+\frac{\gamma_{\ell}}{2},\Delta_{\varphi}+\ell+\frac{\gamma_{\ell}}{2},2(\Delta_{\varphi}+\ell+\frac{\gamma_{\ell}}{2}),1-V\right)\\
&&=\frac{U^{\Delta_{\varphi}}}{V^{\Delta_{\varphi}}} \left( 1+ a_{\tau_{min},\ell_0} V^{\frac{\tau_{min}}{2}} \left( \alpha \log U+ \beta+\ldots \right)+\ldots\right)\;.
\end{eqnarray}
The divergence in  $\frac{U^{\Delta_{\varphi}}}{V^{\Delta_{\varphi}}}$ fixes the behavior of $a_{\ell}$ to be the same as the one of $a_{0,\ell}^{\text{MF}}$ at large $\ell$, as we have already discussed.  In order to study the consequences of (\ref{crosslarge}) having only integer powers of $V$ times $V^{\tau_{min}/2-\Delta_{\varphi}}$, let us make a few remarks. The main idea is to go through similar steps compared to the previous subsection to obtain equations constraining the OPE data.
Firstly, it is convenient to rescale the squared OPE coefficient in the following way
\begin{equation}
a_{\ell}=\frac{2^{\ell+1}\Gamma \left(\ell+\frac{\gamma_{\ell} }{2}+\Delta_{\varphi} \right)^2 \Gamma \left(\ell+\frac{\gamma_{\ell} }{2}+2 \Delta_{\varphi} -1\right)}{\Gamma (\Delta_{\varphi})^2 \Gamma
   \left(\ell+\frac{\gamma_{\ell} }{2}+1\right) \Gamma \left(2 \ell+\frac{\gamma_{\ell} }{2}+2 \Delta_{\varphi} -1\right)} \hat{a}_{\ell}\;.
\end{equation}
At leading order in the expansion, we have $\gamma_{\ell}=0$, $\hat{a}_{\ell}=1$. This rescaling makes the manipulation in \eqref{transf} less lengthy. 
The second insight resides in the usage of the quadratic Casimir operator \eqref{quadrcas}. In particular, since we are working in a small $U$ expansion, we need to compute the limit of the Casimir operator in  \eqref{quadrcas} and the corresponding eigenvalues can be written as 
\begin{equation}
J^2=(\ell+\Delta_{\varphi} +\gamma_\ell/2)(\ell+\Delta_{\varphi} +\gamma_\ell/2-1)\;.
\end{equation}
The most interesting point is that if we act with the Casimir operator on the RHS of \eqref{cr11} it increases the power divergence as $V \to 0$ of the equation, and correspondingly it follows that the LHS has an enhanced behaviour for large $\ell$. This is crucial, and allows us to act repeatedly to the crossing equation to explore more and more divergences as $V \to 0$, and probe subleading corrections of the CFT data in the large spin limit. To do so, we can rewrite both  $\gamma_{\ell}$ and $\hat{a}_{\ell}$ as functions of $J$, and expand them in inverse powers of $J$. Then we see that the leading behaviour at large $J$ is fixed by the divergence $V^{\frac{\tau_{min}}{2}-\Delta_{\varphi}}$  to be
\begin{align}
\gamma_\ell &= \frac{c_1}{J^{\tau_{min}}}+\ldots\;,\\
\hat a_\ell &= 1+  \frac{d_1}{J^{\tau_{min}}}+\ldots
\end{align}
where the coefficients $c_1,d_1$ can be fixed in terms of $\alpha,\beta$ in \eqref{crosslarge}. The subleading corrections depend on the value of $\tau_{min}$. For instance  $\tau_{min}=2$, which corresponds to the presence of the stress tensor, has an expansion of the form 
\begin{align}\label{largejnonperturbative}
\gamma_\ell &= \frac{c_1}{J^2}+ \frac{c_2}{J^3}+ \frac{c_3}{J^4}+ \frac{c_4}{J^5}+\ldots\;, \\
\hat a_\ell &= 1+  \frac{d_1}{J^2}+ \frac{d_2}{J^3}+ \frac{d_3}{J^4}+ \frac{d_4}{J^5}+\ldots\;. 
\end{align}
To fix the coefficients $c_i$ and $d_i$, we plug the expressions \eqref{largejnonperturbative} into \eqref{crosslarge} and we follow a similar procedure to the one in the previous section.  Now $J^2$ has to scale as $V^{-2}$, and then using \eqref{hyper} and the same scalings as previously, we end up with integral relations containing $\gamma_J$ and $\hat{a}_J$. These expansions are valid also to subleading order as $V\to 0$, and by requiring such expansion not to have half-integer divergent powers of $V$ we find arbitrarily many relations for the coefficients $c_i$ and $d_i$.  The final results can be summarised by saying that the expansion of $\gamma(J)$ for large $J$ contains only even powers of $1/J$, and the expansion of $\hat a(J)\left(1 - \frac{\sqrt{1+4 J^2}}{4J} \gamma'(J)\right)$ for large $J$ contains only even powers of $1/J$. These expansions provide all orders in the large spin expansion.

The same technology, as presented in \cite{Alday:2015eya}, can be used in the case of perturbative theories. In this case the minimal twist that appears in the crossed channel is generically $\tau_0$ but the same analysis can be carried over. It is also possible to use similar methods to compute anomalous dimensions \cite{Kaviraj:2015cxa, Kaviraj:2015xsa} to operators with leading dimension $\Delta=2\Delta_\varphi+2n+\ell$, with $n\neq 0$ and $n\ll \ell$. 
In the next subsection we will discuss how to construct these corrections using a more powerful technology, which is based on the simple observations that we made so far.

\subsection{Twist conformal blocks and large spin perturbation theory}\label{Subsec:TCB}
In this section we are going to review  \cite{Alday:2016njk}, from which the definition of the twist conformal blocks stems. This approach builds on what was described in the previous sections and most importantly, provides an algebraic way of solving the constraints of crossing symmetry which was also observed in \cite{Alday:2015ewa} \footnote{An analysis of the accuracy of the resummation of the large spin expansion down to low spins has been performed in \cite{Alday:2015ota,Simmons-Duffin:2016wlq}}. The plan of this section is to lightly review the abstract construction of the twist conformal blocks and their properties. The main aim is, in section \ref{Sec:largeN}, using this technology to study one of the most interesting applications of this method which are theories admitting a large central charge expansion. We will again restrict our discussion to the $d=4$ case.

The main idea of \cite{Alday:2016njk} is to introduce a family of functions  $H^{(\rho)}_{\tau}(U,V)$, called twist conformal blocks, that can be easily expanded both around small $U$ and $V$ in the Lorentzian regime, differently from the conformal blocks. 
\subsubsection{Degenerate point}
As we have seen in the case of generalised free fields,there are infinitely many double-trace operators whose spins are not bounded. We will then define 
\begin{equation}
\sum_{\ell} a^{(0)}_{\tau, \ell} U^{\tau/2} g_{\tau, \ell}(U,V)=H^{(0)}_{\tau}(U,V)
\end{equation}
where we have introduced the notation  $a^{(0)}_{\tau, \ell}=a^{\text{MF}}_{\tau, \ell}$. The properties of the functions $H^{(0)}_{\tau}(U,V)$ are\footnote{Notice that we are specifying our discussion to the four dimensional case. But with minor modification it can be extended to any number of spacetime dimensions, as it is discussed in  \cite{Alday:2016njk}. }
\begin{itemize}
\item $H^{(0)}_{\tau}(U,V)  \xrightarrow{U \to 0} U^{\tau/2} $.
\item $H^{(0)}_{\tau}(U,V)  \xrightarrow{V \to 0} V^{-\Delta_{\varphi}} $.
\item  $\mathcal{H}_{\tau} H^{(0)}_{\tau}(U,V)=\frac{\tau}{4} (\tau -6) (\tau -4) (\tau -2)   H^{(0)}_{\tau}(U,V) $ where $\mathcal{H}_{\tau}$ is a combination of the Casimir operators of the conformal group given by $\mathcal{H}_{\tau}=\mathcal{D}_4-\mathcal{D}_2^2+\left( (\tau -6) \tau +6\right)\mathcal{D}_2$.
\end{itemize}
The first two properties come from the fact that we can decompose these objects into conformal blocks both in the direct and the crossed channels. In particular, they need to reproduce the identity conformal block in the crossed channel. On the other hand, the last property resides on the fact that since $H^{(0)}_{\tau}(U,V)$ does not depend on the spin, it has to be the eigenfunction of a specific differential operator whose eigenvalues do not depend on the spin either. 

Now the idea is to use the second property to write down an expansion for the twist conformal blocks, and then plug it into the differential equation given by the Casimir and use the first property to fix the boundary conditions.

\subsubsection{Large spin perturbation}
When we consider corrections to the regime of infinite spin, we need to consider the following contributions
\begin{equation}
\sum_{\ell} a_{\tau,\ell}^{(0)} \frac{U^{\tau/2}}{J_{\tau,\ell}^{2m}} g_{\tau,\ell}(U,V) = H^{(m)}_{\tau}(U,V)\;.
\end{equation}
In particular, we define the following Casimir operator
\begin{equation}
\mathcal{C}_\tau ={\cal D}_2 +\frac{1}{4} \tau(6-\tau)\;.
\end{equation}
This Casimir operator is slightly different from the one in \eqref{quadrcas}, and its eigenvalues can be easily computed as
\begin{equation}
J^2_{\tau,\ell} = \frac{1}{4}(2\ell+\tau)(2\ell+\tau-2)\;.
\end{equation}
We will assume that the dimensions and the squared of the OPE coefficients have the following structure in the expansion in inverse powers of $J$ around large spins $\ell$. The functions $H^{(m)}_{\tau}$ are not eigenfunctions of the quadratic Casimir $\mathcal{C}_\tau$. But its action defines a recurrence relation which relates twist conformal blocks associated to a given twist $\tau$ but with different values of $m$
\begin{equation}
\label{recurrence}
{\cal C}_\tau H^{(m+1)}_{\tau}(U,V)= H^{(m)}_{\tau}(U,V)\;.
\end{equation}
Analogously to the small $U$ and $V$ limits of $H^{(0)}_{\tau}(U,V)$, it is possible to see that
\begin{itemize}
\item $H^{(m)}_{\tau}(U,V)  \xrightarrow{U \to 0} U^{\frac{\tau}{2}}$.
\item $H^{(m)}_{\tau}(U,V)  \xrightarrow{V \to 0}  V^{-\Delta_\varphi+m}$.
\end{itemize}
Note that when $\Delta_\varphi-m$ is an integer, it is also possible to get a  $\log^2 V$. By using this properties and expanding in different regimes, it is possible to compute these functions.

We would like to end this section with some remarks on this method. While in spirit its applicability is generic, it is best suited when one considers a perturbation around a regime in which there is degeneracy, meaning that there are infinitely many operators with a given twist $\tau_0$ and unbounded spin. This situation is present in several interesting expansions, for instance in the $\epsilon$-expansion  or the $1/N$ expansion that we are going to review in the next section. In these situations, it is possible to construct the twist conformal blocks, for any value of $\tau_0$ and $\Delta_{\varphi}$ by using the recurrence relation and fixing the coefficients of the $V \to 0$ expansion. In particular, in all the considered cases, while $\tau_0$ and $\Delta_{\varphi}$ are generically not integer, there are combinations of them that are integers and lead to perturbative series in $U,V \to 0$ which can be expanded in the precise meaning discussed in this section \footnote{For a detailed analysis of an example in which the dimension $\Delta_{\varphi}$ is completely generic, we refer the reader to \cite{Alday:2017gde}.}. This can be seen as a limitation of this set of techniques, which can be overcome by introducing a fully non perturbative setup to express the CFT data as an integral over a specific function encoding the $V \to 0$ divergence of the correlator as will be discussed in Sec.\ref{Sec:inve}.

\newpage
\markboth{4\quad LARGE N}{}
\section{Large N}\label{Sec:largeN}

In this section we are going to review the perturbative expansion around large $N$, which corresponds to the limit of the large central charge. This is a setup which is most interesting when studying holographic theories, where $N$ plays the role of the degrees of freedom. In particular this study has been pioneered to understand the family of large $N$ CFTs that can have weakly coupled and local gravity duals. For concreteness, in this section we will implicitly identify $N$ with the rank of $SU(N)$ gauge group in four dimensional gauge theories. However, the discussion of large $N$ expansion is universal and $N$ can take other meanings. It should be noted that the expansion powers may differ depending on the context. The main reference of this topic is \cite{Heemskerk:2009pn}. We will review the content and results of the paper and also discuss the expansion at subleading orders  \cite{Aharony:2016dwx}. 

\subsection{Setup}
We consider a setup in which we have a generic CFT with a large $N$ expansion and a large gap in conformal dimensions. Holographically, this corresponds to a local quantum field theory in AdS with a large mass gap. More precisely, we assume that there exists a ``single-trace'' type\footnote{Here and below, when writing ``single-trace'', ``double-trace'', {\it etc}, we are borrowing the terminology from gauge theories. It should be noted, however, that in a generic large $N$ theory we do not necessarily need to have the notion of traces. Roughly speaking, we may think of single-trace and double-trace as single-particle and double-particle in AdS space.  } of scalar field  $\varphi$ which has a fixed dimension $\Delta_{\varphi}$. We consider the four-point function 
\begin{equation}
\langle \varphi(x_1)  \varphi(x_2)   \varphi(x_3)   \varphi(x_4) \rangle=\frac{\mathcal{G}(U,V)}{x_{12}^{2\Delta_{\varphi}} x_{34}^{2\Delta_{\varphi}}}\;,
\end{equation}
and its large $N$ expansion which takes the form
\begin{equation}
\mathcal{G}(U,V)=\mathcal{G}^{(0)}(U,V)+\frac{1}{N^2}\mathcal{G}^{(1)}(U,V)+\frac{1}{N^4}\mathcal{G}^{(2)}(U,V)+\cdots\;.
\end{equation}
The displayed first three orders of the expansion will respectively correspond to the disconnected, tree-level and one-loop level contributions in AdS.
We will assume that the OPE content of $\varphi  \varphi$ is 
\begin{equation}
\varphi  \varphi= 1+\varphi+T_{\mu \nu}+[\varphi \varphi]_{n,\ell}+[T T]_{n,\ell}+[\varphi T]_{n,\ell}+[\varphi \varphi \varphi]_{n,\ell}+\dots
\end{equation}
where the dots denote higher-traces operators. The stress tensor $T_{\mu\nu}$ is dual to the graviton field in AdS.

\subsection{Leading order: $N^0$}
To simplify even further the setup, we can assume that there is a $\mathbb{Z}_2$ symmetry which will allow for only double-trace operators $[\varphi \varphi ]_{n, \ell}$. Notice however that as we have seen, double-trace operators are necessary since the identity operator in one channel requires their presence in the crossed channel. We assume also that at this order in $N$ the stress-tensor is not present\footnote{Notice that the $\mathbb{Z}_2$ symmetry forbids the presence of double trace operators of the form $[\varphi T]_{n, \ell}$}. 

At this order in $N$, the only contributions come from the disconnected part of the four-point correlator, thus practically this is a mean field theory correlator \eqref{MFT}. For completeness, let us reproduce it
\begin{equation}
\mathcal{G}^{(0)}(U,V)= 1+ \left( \frac{U}{V} \right)^{\Delta_{\varphi}}+U^{\Delta_{\varphi}}\;.
\end{equation}
The OPE data are the ones discussed in \eqref{MFTa}. In particular, the intermediate operators are double-trace operators (besides the unit operator) with dimensions and squared OPE coefficients 
\begin{eqnarray}
&&\Delta^{(0)}_{n,\ell}=2 \Delta_{\varphi}+ 2n +\ell\;,\\
&&a_{n,\ell}^{(0)}=a_{n,\ell}^{\text{MF}}= \frac{2^{\ell+1} (\ell+1) (\ell+2 (\Delta_\varphi+n-1)) \Gamma (n+\Delta_\varphi-1)^2 }{(\Delta_\varphi-1)^2 n! \Gamma (\Delta_\varphi-1)^4} \nonumber\\ &&\quad\quad\quad\quad\quad\quad\times\frac{\Gamma
   (n+2\Delta_\varphi-3) \Gamma (\ell+n+\Delta_\varphi)^2 \Gamma (\ell+n+2 \Delta_\varphi-2)}{ \Gamma (\ell+n+2)
   \Gamma (2 n+2 \Delta_\varphi-3) \Gamma (2 \ell+2 n+2 \Delta_\varphi-1)}\;.
\end{eqnarray}
As we have discussed, even if we did not know the structure of the four-point correlator, we could have arrived at this answer by using the fact that the identity operator is exchanged in one channel. Under crossing, this generates a power law divergence that requires an infinite number of double-trace operators in the OPE. 

\subsection{First order: $N^{-2}$} 
We would like to understand how to fix the corrections to the OPE data at order $N^{-2}$ . In particular notice that crossing symmetry should be satisfied at each order. Also, there are two scenarios that can be studied now. One situation is to consider corrections to the OPE data, in the absence of any other operators appearing at order $N^{-2}$. The other situation is to consider the corrections to the OPE data of the double-trace operators in the presence of a new operator appearing at order $N^{-2}$. It requires the OPE coefficient of the new operator to scale as $N^{-1}$. 
\subsubsection{Absence of new operators}
Let us study the first scenario, which has been extensively analysed in \cite{Heemskerk:2009pn}. In this case the OPE expansion looks like
\begin{equation}
\varphi  \varphi = 1+ [\varphi \varphi].
\end{equation}
Thus we will focus on the correction to the dimensions of the double-trace (or double-twist) operators and to their squared three-point functions.  They can be expanded to this order as 
  \begin{eqnarray}
  \label{expansions}
 \Delta_{n,\ell} &=&\Delta_{n,\ell}^{(0)}+ \frac{1}{N^2} \gamma_{n,\ell}^{(1)} + \cdots,\\
 a_{n,\ell} &=& a_{n,\ell}^{(0)}+ \frac{1}{N^2} a_{n,\ell}^{(1)} + \cdots\;.
 \end{eqnarray}
If we insert them into the four-point function and expand to order $N^{-2}$, we obtain
\begin{align} \label{expN}
{\cal G}^{(1)}(U,V) = \sum_{n,\ell} U^{\Delta_{\varphi}+n}  \left(a^{(1)}_{n,\ell} + \frac{1}{2} a^{(0)}_{n,\ell} \gamma^{(1)}_{n,\ell} \left(\log U+\frac{\partial}{\partial n}\right)\right) g_{2\Delta_{\varphi}+2n,\ell}(U,V)\;.
\end{align}
Crossing symmetry would require then a term of the form
\begin{equation}
{\cal G}^{(1)}(V,U)=\frac{U^{\Delta_{\varphi}}}{V^{\Delta_{\varphi}}}\sum_{n,\ell} V^{\Delta_{\varphi}+n}  \left(a^{(1)}_{n,\ell} + \frac{1}{2} a^{(0)}_{n,\ell} \gamma^{(1)}_{n,\ell} \left(\log V+\frac{\partial}{\partial n}\right)\right) g_{2\Delta_{\varphi}+2n+\ell,\ell}(V,U)\;.
\end{equation}
Let us study the limit of small $V$. Different from the previous order, there is no power-law divergence due to the fact that we have only double-trace operators. The consequence of this simple observation is striking, there is no need for infinitely many operators with large spins since they would otherwise produce an enhanced divergence in the small $V$ limit. Thus the correction to the OPE data are different from zero only for a finite range of spins. In the language of twist conformal blocks, we have exactly the same structure. In particular, since there is no divergence in $V$, there cannot be any twist conformal block $H^{m}(U,V)$ with $m=0,1, \dots \Delta_{\varphi}-1$. On the other hand, since $\Delta_{\phi}$ is an integer, all the higher $m$ terms would produce terms of the form $(\log V)^2$ which are incompatible with crossing, due to the fact that the only possible logarithmic term is $\log(U)$. Another option would have been to have $H^{m}(V,U)$ but those are also absent due to the fact that there is no divergence in $V$ to allow for them.  
Thus it is possible to state that 
\begin{eqnarray}
a^{(1)}_{n, \ell} & \neq 0 \qquad \forall \quad \ell=0,2, \dots, L\;,\\
\gamma^{(1)}_{n, \ell} & \neq 0 \qquad \forall \quad \ell=0,2, \dots, L\;.
\end{eqnarray}
The precise structure of this solution can be found by studying the small $U$ and $V$ limits of the crossing equations, and using projectors to isolate the contribution of only a finite number of spins. In particular there are $ \frac{(L+2)(L+4)}{8}$ undetermined constants for each spin $L$, this means that the structure of the conformal block decomposition together with crossing symmetry is not enough to fix completely the OPE data. The details can be found in \cite{Heemskerk:2009pn}. As an example, one finds
\begin{equation}
\gamma^{(1)}_{n, 0}=\alpha \frac{\left(2 \Delta_{\varphi} -1\right)\left(n+1\right)\left(2 \Delta_{\varphi} +n-3\right)\left( \Delta_{\varphi}+n -1\right)}{\left( \Delta_{\varphi}-1\right)\left( 2\Delta_{\varphi}+2n -3\right) \left( 2\Delta_{\varphi}+2n -1\right)}
\end{equation}
where $\alpha$ is an unfixed parameter corresponding to the freedom we discussed before. Generically, for the squared OPE coefficient it is possible to find a derivative relation, meaning that 
\begin{equation}
a^{(1)}_{n,\ell}=\frac{1}{2}\partial_n \left(a^{(0)}_{n,\ell} \gamma^{(1)}_{n, \ell} \right) \qquad \ell=0,2, \dots L\;.
\end{equation}
We can now make contact with the AdS physics. In particular, these solutions correspond to quartic vertex of the kind $\varphi^4$, $\varphi^2 \nabla^2 \varphi^2$ and so on.\footnote{Here we are abusing the notation slightly by using $\varphi$ to denote both the CFT operator and the field in AdS. Notice also that there are no cubic vertices since we are in the simplest setup where we imposed a $\mathbb{Z}_2$ symmetry.}  
\begin{figure}[ht]
	\centering
  \includegraphics[width=0.3\textwidth, height=0.3\textwidth]{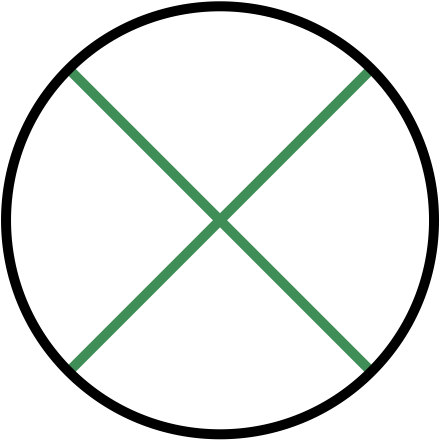}
	\caption{The quartic tree-level contact diagram, which is the only one at order $N^{-2}$ in the case where only double-trace operators are exchanged.}
	\label{figvertex}
\end{figure}

Also in this case, we can count how many interactions with $2L$ derivatives are present which contribute a spin up to $L$ and we have exactly the same number $ \frac{(L+2)(L+4)}{8}$. Notice that the results that we have presented in this section are valid when $\Delta_\varphi$ is an integer. 
\subsubsection{Presence of new operators}
The situation changes when the OPE contains another operator $\varphi_{\tau,s}$ which contributes to the four-point function at  order $N^{-2}$
\begin{equation}
\varphi  \varphi =1+[ \varphi \varphi]+\frac{1}{N}\varphi_{\tau,s}\;.
\end{equation}
The new operator has conformal twist $\tau$ and spin $s$. As a result, corrections to the OPE data of the double-trace operators will depend also on the presence of $\varphi_{\tau,s}$. We are interested in understanding how crossing symmetry fixes the corrections of the form \eqref{expansions}. The situation is different compared to the previous case. In particular, the four-point function receives a contribution corresponding to the conformal block associated with the exchange of the new operator 
\begin{equation}
a_{\tau, s} U^{\tau/2}g_{\tau,s}(U,V)
\end{equation}
where $a_{\tau, s}$ is the squared three-point function coefficient $\langle \varphi \varphi \varphi_{\tau,s}\rangle^2 $. If we use crossing, we observe that it requires the presence of a term of the form
\begin{equation}
\mathcal{G}^{(1)}(U,V) =\frac{U^{\Delta_{\varphi}}}{V^{\Delta_{\varphi}-\frac{\tau}{2}}} a_{\tau, s} g_{\tau,s}(V,U)+\dots
\end{equation}
This already signals that for any positive non-integer $\Delta_\varphi-\frac{\tau}{2}$  the corrections to the CFT data need to have an infinite support in the spin. This is because, differently from the previous section, there is a divergence as $V\to 0$ that needs to be reproduced by the divergent part of this sum
\begin{equation}
\mathcal{G}^{(1)}(U,V)= \frac{1}{2}\sum_{n,\ell}a_{n,\ell}^{(0)}\gamma_{n,\ell}U^{\Delta_{\varphi}+n}g_{n,\ell}(U,V) \log U +\dots
\end{equation}
where the dots denote terms which are analytic as $z$ goes to zero. This means that 
\begin{equation}\label{crosin}
\frac{1}{2}\sum_{n,\ell}a_{n,\ell}^{(0)}\gamma_{n,\ell}U^{\Delta_{\varphi}+n}g_{n,\ell}(U,V) \sim  \frac{U^{\Delta_{\varphi}}}{V^{\Delta_{\varphi}-\frac{\tau}{2}}}a_{\tau,s}g_{\tau,s}(V,U)|_{ \log U}
\end{equation}
where this equation means that we need to consider the divergence on the LHS as $V \to 0$. To control this problem, we need to construct the twist conformal blocks and in particular
\begin{equation} \label{crossingle}
\sum_{m, n} B_{m,n}H^{(m)}_{n}(U,V)|_{div}=\frac{U^{\Delta_{\varphi}}}{V^{\Delta_{\varphi}-\frac{\tau}{2}}} a_{\tau, s} g_{\tau,s}(V,U)|_{\log U}
\end{equation}
where it is assumed that $\gamma_{n,\ell}=2\sum_m \frac{B_{mn}}{J^{2m}}$ and $J^2$ is the conformal spin.
\paragraph{Intermezzo on twist conformal blocks.}
Let us fill in more details regarding the twist conformal blocks. Similarly to Section \ref{Subsec:TCB}, we can define them as 
\begin{equation} \label{twconf}
H^{(m)}_{n}(z,\bar{z})=\sum_{\ell}a_{n,\ell} \frac{(z \bar{z})^{\Delta_{\varphi}+n}}{J^{2m}}g_{n,\ell}(z,\bar{z})
\end{equation}
where $J^2=(\ell+n+\Delta_{\varphi})(\ell+n+\Delta_{\varphi}-1)$. To solve the problem above we are interested in to the divergent contribution of such blocks in the limit in which $\bar{z}\to 1$. To this end, we can construct 
\begin{equation} \label{serex}
\sum_{n}H^{(0)}_n(z,\bar{z})=\left( \frac{z \bar{z}}{(1-z)(1-\bar{z})}\right)^{\Delta_{\varphi}}\;.
\end{equation}
In addition, the structure of the conformal blocks fixes the form to be
\begin{equation}
\sum_{n}H^{(0)}_n(z,\bar{z})=\frac{1}{\bar{z}-z} z^{\Delta_{\varphi}+n}F_{\Delta_{\varphi}+n-1}(z)\bar{H}^{(0)}_n(\bar{z})
\end{equation}
where $F_{\beta}(z)= {}_2F_1(\beta, \beta, 2\beta,z)$. By matching the series expansion of both sides of \eqref{serex}, it is possible to find the full structure for $\bar{H}^{(0)}_n(\bar{z})$ which reads 
\begin{equation}
\bar{H}^{(0)}_n(\bar{z})=\left(\frac{\bar{z}}{1-\bar{z}} \right)^{\Delta_\varphi}d_n(1+b_n (1-\bar{z}))\;.
\end{equation}
Here
\begin{equation}
d_n=-\frac{\sqrt{\pi } 2^{-2 \Delta_{\varphi} -2 n+4} \Gamma (n+\Delta_{\varphi} -1) \Gamma (n+2 \Delta_{\varphi} -3)}{\Gamma (\Delta_{\varphi} -1)^2 \Gamma (n+1) \Gamma \left(n+\Delta_{\varphi} -\frac{3}{2}\right)}\;,
\end{equation}
and
\begin{equation}
b_n=-\frac{(\Delta_{\varphi} -1)^2+n^2+(2 \Delta_{\varphi} -3) n}{(\Delta_{\varphi} -1)^2}\;.
\end{equation}
We can use then the recurrence relation to extrapolate this result to any positive $m$. The idea is to use the fact that the Casimir operator acts on the twist conformal blocks in the following way
\begin{equation}
\mathcal{C}_\tau H^{(m+1)}_n(z,\bar{z})=H^{(m)}_n(z,\bar{z})
\end{equation}
which, due to the factorization in $z$ and $\bar{z}$, leads to a recurrence relation for $\bar{H}^{(m)}_n(z,\bar z)$ as
\begin{equation}
\mathcal{D}\bar{H}^{(m+1)}_n(z,\bar z)=\bar{H}^{(m)}_n(z,\bar z)\;.
\end{equation}
Here $\mathcal{D}=\bar{z} \bar{D} \bar{z}^{-1}$ and $D$ is defined in \eqref{casimirdef}. For a fixed twist, it is possible to write down an expansion of the form 
\begin{equation}
\bar{H}^{(m)}_n(z,\bar z)=\left(\frac{\bar{z}}{1-\bar z} \right)^{\Delta_{\varphi}-m}h_0^{(m)}(1+h_1^{(m)}(1-\bar{z})+h_2^{(m)}(1-\bar{z})^2+\dots)\;,
\end{equation}
and the coefficients $h_n^{(m)}$ can be found iteratively.  

With this piece of information we can tackle the main problem \eqref{crossingle}. Due to the factorisation property of \eqref{twconf}, it is possible to see that also the functions $B_{mn}$ satisfy a similar equation. In particular, by inserting the expansion of the anomalous dimension in \eqref{crossingle} one gets
\begin{eqnarray} 
 \sum_{mn} B_{mn} F_{\Delta_{\varphi}+n-1}(z) \frac{z^{\Delta_{\varphi}+n}}{z-\bar z}\bar{H}_n^{(m)}(\bar z)&=&\frac{a_{\tau,s} (z \bar z)^{\Delta_{\varphi}}}{\left( (1-z)(1-\bar{z}) \right)^{\Delta_{\varphi}-\tau/2}} \label{crosag}\\
\nonumber &\times & \frac{(1-\bar z)^{s+1} F_{\tau/2+s}(1-\bar z) F_{\tau/2-1}(1- z)}{z-\bar z}\big|_{\log z}\;. 
\end{eqnarray}
Then we can factor out the dependence on $m$ in the following way
\begin{equation}
B_{mn}=\kappa_{\tau-2}(n) \rho^{(\tau+2s)}_m(J)-\kappa_{\tau+2s}(n) \rho^{(\tau-2)}_m(n)\;.
\end{equation}
Inputting this expression in \eqref{crosag} we get two decoupled equations for $\kappa$ and $\rho$ respectively
\begin{align} \label{finaleq}
\nonumber \sum_{n} \kappa_{\tau-2}(n) z^{n+\Delta_{\varphi}}F_{\Delta_{\varphi}+n-1}(z)&=\frac{1}{d_n}\frac{\Gamma(\tau-2)}{\Gamma^2\left(\frac{\tau-2}{2} \right)}\frac{z^{\Delta_{\varphi}}}{(1-z)^{\Delta_{\varphi}}}(1-z)^{\tau/2}{}_2F_1\left(\frac{\tau-2}{2}, \frac{\tau-2}{2},1,z\right)\;,\\
\sum_m \rho_m^{(\tau+2s)}(n)\bar{H}^{(m)}_n (\bar{z})&= a_{\tau,s} d_n \frac{z^{\Delta_{\varphi}}}{(1-\bar z)^{\Delta_{\varphi}}}(1-\bar z)^{\tau/2+s+1}F_{\tau/2+s}(1-\bar z)\;.
\end{align}
Expanding order by order \eqref{finaleq} in $z$ and $1-\bar z$ it is possible to find all $\kappa$ and $\rho$, for any twist $\tau$ and spin $s$. Moreover, it is possible to show that crossing fixes the range for $m$ to be integer and  $m=\tau/2+s+1, \tau/2+s+2, \dots$. 

We give the solution for $\tau=2$. This case is important because it corresponds to, when $s=2$, the stress-energy tensor which is always present in consistent CFTs. The solution of $\eqref{finaleq}$ gives all the coefficients in the large $J$ expansion. For this case, the expansion can be resummed to give 
\begin{equation}\label{gamas}
\gamma_{n,\ell}^{as}=-a_{2,s}\frac{2 \kappa_{2+2s}(n)(\Delta_{\varphi}-1)^2}{(\ell+1)(\ell+2\Delta_{\varphi}+2n-2)}\;.
\end{equation}
The function $\kappa_{2+2s}(n)$ is a degree $2s$ polynomial in $n$. For the exchange of a scalar operator, with spin $s=0$, $\kappa_2(n)=1$ while for $s=2$ such as the stress-energy tensor we have
\begin{equation}
\kappa_6(n)=30\frac{6 n^4+12(2\Delta_{\varphi}-3)n^3+6 (5 \Delta_{\varphi}^2-14\Delta_{\varphi}+11)n^2+6(2\Delta_{\varphi}^3-7\Delta_{\varphi}^2+10\Delta_{\varphi}-6)n}{\Delta_{\varphi}^2(\Delta_{\varphi}-1)^2}+30\;.
\end{equation}
By plugging the solution for $\tau=2$ back into \eqref{crosin} it is possible to show that such anomalous dimension do not solve crossing. In fact, it needs to be supplemented with a correction $\gamma_{n,\ell}^{fin}$ which is different from zero only for $\ell=0,1,\dots, s$. The precise structure can be found in a very similar way as in \cite{Heemskerk:2009pn}. As an example, for the scalar exchange of twist two, {\it i.e.,} $\tau=2$ and $s=0$, this extra piece differs from zero only for $\ell=0$ and is given by 
\begin{equation}
\gamma_{n,0}^{fin}=\frac{1}{2}a_{2,0}\frac{(n+1)(2\Delta_{\varphi}+n-3)(\Delta_{\varphi}-1)^2}{(\Delta_{\varphi}+n-1)(2\Delta_{\varphi}+2n-3)(2\Delta_{\varphi}+2n-1)}\;.
\end{equation}
In this way we have that the full anomalous dimension is given by $\gamma_{n,\ell}^{as}+\gamma_{n,0}^{fin}$. In addition, it is always possible to add solutions truncated in the spin, which are crossing symmetric by themselves and for which \eqref{crosin} does not put any constraints. More examples of such anomalous dimensions can be found in \cite{Alday:2017gde}.

The corrections to the three-point functions $a_{n,\ell}^{(1)}$ can be found in a very similar way as the correction to the anomalous dimensions. The only difference is that they are not proportional to $\log U$, as it is clear from \eqref{expN}.

Quite nicely, the structure of $a_{n,\ell}^{(1)}$ is simple and it is given by 
\begin{equation}
a_{n,\ell}^{(1)}=\frac{1}{2}\partial_n \left(a_{n,\ell}^{(0)}\gamma_{n,\ell} \right)+a_{n,\ell}^{(0)} \hat{a}_{n,\ell}^{(1)}\;,
\end{equation}
and generically $\gamma_{n,\ell}=\gamma_{n,\ell}^{as}+\gamma_{n,\ell}^{fin}$ and $\hat{a}_{n,\ell}^{(1)}=0$ for $\Delta_{\varphi}=2,3,\dots ,\tau/2+1+s$.  This form is reminiscent of the situation in which there are no new operator, except for the extra piece. Explicit results for this term can be found in \cite{Alday:2017gde}.

\begin{figure}[ht]
	\centering
  \includegraphics[width=0.8\linewidth]{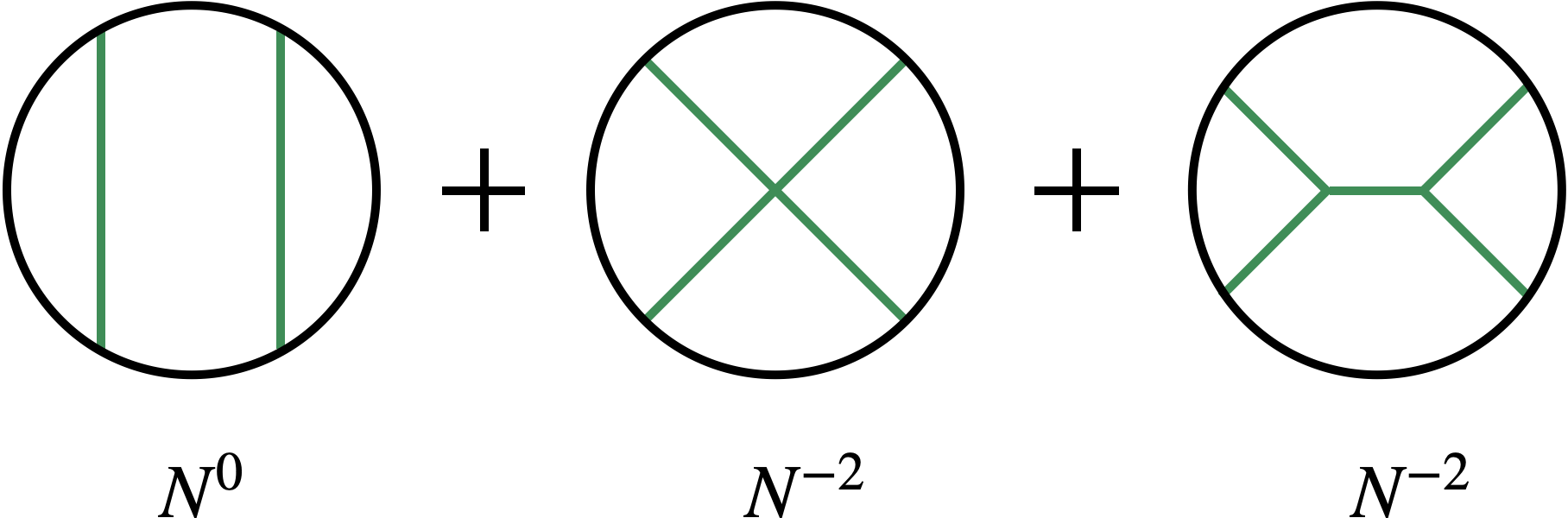}
	\caption{In the presence of a single-trace intermediate operator exchange, there are two contributions to the anomalous dimension and three-point functions of double-trace operators at order $N^{-2}$, that come from the quartic and cubic vertices depicted. }
	\label{figvertex}
\end{figure}

\subsection{Second order: $N^{-4}$}
In this section we would like to understand the corrections to order $N^{-4}$ of the anomalous dimension and of the squared three-point functions \cite{Aharony:2016dwx}. In particular, we assume that at order $N^{-2}$ there are no new operators, thus we only have corrections to the anomalous dimensions and OPE coefficients of double-trace operators which have support on finitely many spins. We also implicitly assume that there is only one operator with the same quantum numbers, which are the dimension and the spin. Thus the OPE data admit the following expansion
\begin{align}
\Delta&=2\Delta_{\varphi}+2n+\ell+\frac{1}{N^2}\gamma^{(1)}_{n,\ell}+\frac{1}{N^4}\gamma^{(2)}_{n,\ell}\;,\\
a_{n,\ell}&=a_{n,\ell}^{(0)}+\frac{1}{N^2}a_{n,\ell}^{(1)}+\frac{1}{N^4}a_{n,\ell}^{(2)}
\end{align}
where $\gamma^{(1)}_{n,\ell}\neq 0$ and $a_{n,\ell}^{(1)}$ for $\ell=0,2,\dots, L$. 
This expansion together with the conformal block decomposition imply that the correction to the four-point function at order $N^{-4}$ has the form
\begin{align} \label{crossing4}
 \mathcal{G}^{(2)}(U,V)=\sum_{n,\ell}U^{\Delta_{\varphi}+n} &\left(a_{n,\ell}^{(2)}+\frac{1}{2}a_{n,\ell}^{(0)}\gamma^{(2)}_{n,\ell}\left(\log U+ \frac{\partial}{\partial n} \right) \right.\\
\nonumber &\left. +\frac{1}{2}a_{n,\ell}^{(1)}\gamma^{(1)}_{n,\ell}\left( \log U+\frac{\partial}{\partial n}\right)\right.\\
\nonumber &\left.+\frac{1}{8}a_{n,\ell}^{(1)} \left(\gamma^{(1)}_{n,\ell}\right)^2 \left( \log^2 U+2 \log U \frac{\partial}{\partial n}+\frac{\partial^2}{\partial n^2}\right)\right)g_{2\Delta_{\varphi}+2n+\ell,\ell}(U,V)\;.
\end{align}
The corrections to the CFT data appearing at order $N^{-4}$ are only in the first line of the equation above while the remaining two lines pertain to corrections to order $N^{-2}$ and $N^{0}$ that we already determined from solving constraints from crossing at previous orders. Most importantly, due to the order of the perturbation, there is a logarithmic singularity $\log^2 U$, which correspondingly is mapped to $\log^2 V$ under crossing. This simple observation already signals the fact that $\gamma^{(2)}_{n,\ell}$ and $a_{n,\ell}^{(2)}$ need to be different from zero for arbitrarily large spins, because a finite number of conformal blocks can have a divergence which is at most $\log V$. Let us now analyse the problem in more detail. 
\begin{itemize}
\item The term proportional to $\log^2 U$ corresponds to 
\begin{equation} \label{rhssum}
\frac{1}{8}\sum_{n}\sum_{\ell=0}^{L} U^{\Delta_{\varphi}+n}a_{n,\ell}^{(0)}\left(\gamma_{n,\ell}^{(1)} \right)^2 g_{2\Delta_{\varphi}+2n+\ell,\ell}(U,V)= U^{\Delta_{\varphi}} \left(g_1(U,V) \log V+g_{2} (U,V) \right) \end{equation}
where the functions $g_1(U,V)$ and $g_2(U,V)$ can be expanded in positive integer powers in $U$ and $V$ and the presence of the $\log V$ makes manifest the fact that the sum over the spin is truncated, up to spin $L$. For later convenience we define $\lim_{V \to 0}g_1(U,V)=\tilde{g}_1(U)$.
\item Crossing symmetry implies that $\mathcal{G}^{(2)}(U,V)$ contains 
\begin{equation}
U^{\Delta_{\varphi}} \log^2 V \left(g_1(V,U) \log U+ g_2(V,U)\right)\;.
\end{equation}
This is the only term which contains a $\log^2 V$. 
\item The last two lines of \eqref{crossing4} involve only finite sums over the spin, so they cannot reproduce the $\log^2 V$ divergence. This implies that the only candidate is the first line of \eqref{crossing4}, in particular
\begin{equation} \label{fineq}
\frac{1}{2}\sum_{n,\ell}U^n a_{n,\ell}^{(0)} \gamma_{n,\ell}^{(2)} g_{2\Delta_{\varphi}+2n+\ell,\ell}(U,V) |_{\log^2 V}=g_1(V,U)\;,
\end{equation}
and similarly for $a_{n,\ell}^{(2)}$. This provides an equation for $\gamma_{n,\ell}^{(2)}$, which is given in terms of $g_1(V,U)$, a fully specified function once we know $\gamma^{(1)}_{n,\ell}$ and $ a_{n,\ell}^{(0)}$. 
\end{itemize}
In order to solve this equation one would need to compute the sum \eqref{rhssum}, which is an infinite sum over $n$. Instead of performing the sum directly, it is convenient to compute the contribution to the sum of a single conformal block and then sum these terms. From now on, we will consider the leading twist correction for the anomalous dimension $\gamma^{(2)}_{0,\ell}$. Despite its simplicity, this case already contains several interesting information that we can extract from this problem. To extract this contribution it is enough to focus on the leading $U \to 0$ term of the LHS of \eqref{fineq}. Conversely, \eqref{fineq} tells us that we should focus on the leading term as $V \to 0$ of \eqref{rhssum}, corresponding to $\tilde{g}_1(V)$. More compactly, the answer can be written as 
\begin{equation} \label{gammatwo}
\gamma^{(2)}_{0,\ell}=\frac{1}{8}\sum_{n,s}a_{n,s}^{(0)} \left(\gamma^{(1)}_{n,s} \right)^2 \gamma_{0,\ell}^{(2)}|_{(n,s)}
\end{equation}
where  $\gamma_{0,\ell}^{(2)}|_{(n,s)}$ denotes the contribution to the anomalous dimension for a single conformal block corresponding to exchanged operators with quantum number $n$, $s$. The procedure of computing $\gamma_{0,\ell}^{(2)}|_{(n,s)}$ is general but its final result depends on the quantum numbers. The idea is very similar to what we have seen in the previous section and it amounts to computing the contribution of a single conformal block to $\tilde{g}_1(U)$ and then acting with the Casimir operator to probe higher orders terms in $J$. With this piece of information we can insert the specific $\gamma^{(1)}_{n,s}$ and perform the sum.

Let us report the results for a specific case in which the operator $\varphi$ has dimension two and we have a $\varphi^4$ type interaction at order $N^{-2}$. The corresponding anomalous dimension is
\begin{equation}
\gamma^{(1)}_{n,0}= \frac{3(n+1)^3}{(1+2n)(3+2n)}\alpha\;,
\end{equation}
and $\alpha$ is a proportionality constant that cannot be fixed using solely crossing symmetry. By inserting this information in \eqref{gammatwo}, we find for the first few values of the spin that 
\begin{align}
\gamma_{0,0}^{(2)} & \rightarrow \text{divergent}\;,\\
\gamma_{0,2}^{(2)} & =\frac{2(174 \pi^2-1925)}{3465}\alpha^2\;,\\
\gamma_{0,4}^{(2)} & =\frac{150600 \pi^2-1520519}{2252250}\alpha^2\;,
\end{align}
and generically as a function of $J$ 
\begin{equation} \label{larges}
\gamma^{(2)}_{0,\ell}=-\frac{12}{J^4}\left(1+\frac{18}{5} \frac{1}{J^2}+\frac{96}{7} \frac{1}{J^4}+\frac{360}{7} \frac{1}{J^6}+\dots \right)\alpha^2\;.
\end{equation}
With these results at hand it is possible to reconstruct the full Mellin amplitude as the polar terms can be reconstructed with \eqref{larges} supplemented by crossing symmetry (the Mellin representation will be introduced in section \ref{Sec:MellinFormalism}). The presence of the divergence at spin zero could be worrisome but actually it is consistent with the expectations from its AdS interpretation. In particular, we expect any bulk loop diagram in AdS that can be considered in the specific setup to have UV divergences. Since the curvature of AdS can be ignored in the UV, these divergences have to behave in the same way as the flat space ones. This necessitates the presence of counterterms that are contact diagrams and need be included in the effective field theory description to make the CFT data finite. In fact, each local bulk term comes with an arbitrary coefficient, that is expected to be responsible of the cancellation of the divergences. In particular,  there is a divergent part which is precisely needed to cancel the divergence and an arbitrary finite part. From the CFT point of view, the same happens since each of the terms that we considered at order $N^{-2}$ comes with a coefficient that cannot be fixed with any consideration based on symmetries. 

There are two main lessons that can be learnt from solving crossing up to order $N^{-4}$:
\begin{itemize}
\item It is possible to reconstruct fully the one-loop answer using lower order CFT data, supplemented with crossing symmetry, the structure of the OPE and the singularity pattern.
\item The CFT analysis contains all the ingredients that are expected from the dual AdS picture. 
\end{itemize}
\begin{figure}[ht]
	\centering
  \includegraphics[width=0.8\linewidth]{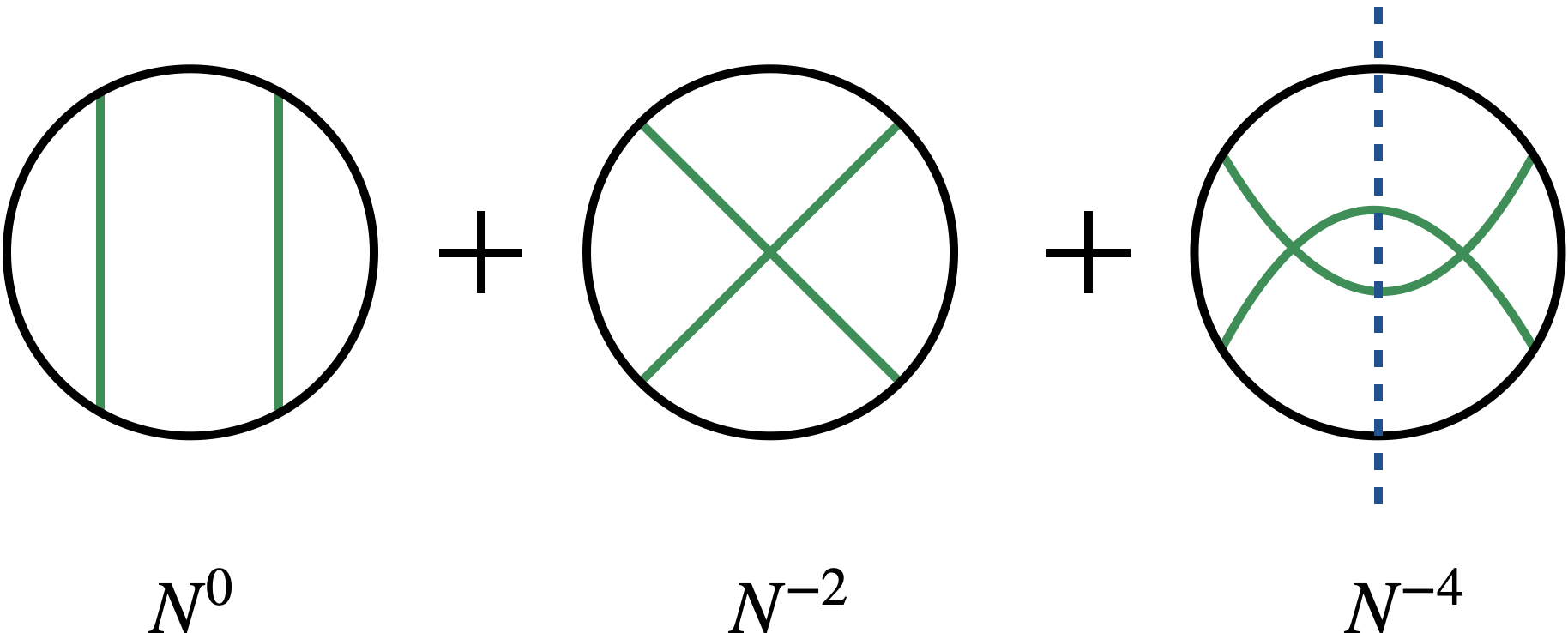}
	\caption{In the presence of only quartic vertices, this is the set of allowed Witten diagrams to the order $N^{-4}$. These diagrams are also supplemented by the crossing symmetric counterpart.}
	\label{figvloop}
\end{figure}

\newpage
\markboth{5\quad LORENTZIAN INVERSION FORMULA}{}
\section{Lorentzian inversion formula}\label{Sec:inve}
In this section we are going to review a method that goes under the name of Lorenzian inversion formula, first introduced by Caron-Huot in \cite{Caron-Huot:2017vep}. This approach serves as a proof of the fact that the large spin expansion is analytic, down to spin one. Thus the resummations that we presented in the previous sections are not accidental but are instead solidly based on this fact. More importantly, it provides us with an alternative way of computing OPE data from the singularities of the correlators. The plan of this section is to first present the main formula. Then we will motivate it and  discuss a few applications of this formula.

\subsection{Motivation and sketch of a proof}
For simplicity, let us consider the  correlator of  four identical scalar operators.\footnote{The discussion of \cite{Caron-Huot:2017vep} is valid for any external scalar operator.} The correlator can be decomposed into conformal blocks as
\begin{equation}
\mathcal{G}(z,\bar{z})=\sum_{\Delta, \ell}a_{\Delta, \ell} G_{\Delta, \ell}(z,\bar{z})
\end{equation}
where $G_{\Delta, \ell}(z,\bar{z})=(z \bar z)^{\frac{\Delta-\ell}{2}}g_{\Delta, \ell}(z, \bar z)$ and $g_{\Delta, \ell}(z, \bar z)$ is the conformal block associated to the exchange of an operator of dimension $\Delta$ and spin $\ell$. 
The OPE coefficients in a unitary CFT are not arbitrary. In particular, they are not independent of each other but give rise to an analytic function which fixes completely their structure. Based on that, it is possible to exploit the analytic properties of the correlator to extract the OPE data, such as the conformal dimensions and three-point function coefficients. The reason behind analyticity in spins resides in the fact that generically Euclidean physics needs to resum into a function which is sensible at high energy, so it is intimately related to causality. Following \cite{Caron-Huot:2017vep}, we would like to present a simple example which shows the logic behind the inversion formula. 

Let us consider a function
\begin{equation}
f(x)=\sum_{j=1}^{\infty}f_j x^j\;,
\end{equation}
with the properties that 
\begin{itemize}
    \item $f(x)$ is analytic in the whole complex plane, except for the branch cuts at real $x >1$,
    \item $\left|\frac{f(x)}{x}\right|\to 0$ as $x \to \infty$.
\end{itemize}
This allows us to use Cauchy's theorem to extract the coefficients $f_j$ as
\begin{equation}
    f_j=\frac{1}{2 \pi i} \oint \frac{dx}{x} x^{-j} f(x)\;.
\end{equation}
By deforming the contour and using the second property above we can write 
\begin{equation}
    f_j=\frac{1}{2 \pi} \int_{1}^{\infty} \frac{dx}{x} x^{-j} \text{Disc} f(x)
\end{equation}
where $\text{Disc} f(x)=-i[f(x(1+i0))-f(x(1-i0))]$. From this relation, it is clear that the coefficients $f_j$ are analytic for $\text{Re}(j)\geq 1$ and fully determined by the imaginary part of $f(x)$. 

In spirit, this is the same that happens for the Froissart-Gribov formula \cite{Gribov:1961ex,Donnachie:2002en}, which forms the foundation of the Regge theory by proving that the relativistic S-matrix is analytic in spins. In that case, the role of  $x^j$ in the simple example is played by Legendre polynomials, and the formula is proven by ``inverting'' these polynomials. It turns out that Euclidean CFTs admit a similar treatment and the idea of \cite{Caron-Huot:2017vep} was to adapt such a reasoning to the case of CFT four-point functions. In particular, it is possible to start with the usual Euclidean decomposition of the four-point function $G(z,\bar{z})$ into conformal blocks, or more precisely into conformal partial waves. This step has been achieved in \cite{Costa:2012cb}\footnote{Notice that we specialised this formula to $d=4$, the bound of integration is also related to $d$.}
\begin{equation}
    G(z,\bar{z})=\delta_{12}\delta_{34}+\sum_{\ell=0}^{\infty} \int_{2-i \infty}^{2+i \infty}\frac{d \Delta}{2 \pi i} c(\ell,\Delta)F_{\ell,\Delta}(z, \bar{z})
\end{equation}
where the first term is the contribution of the identity operator and the functions $F_{\ell,\Delta}$ is a single-valued combination of the conformal blocks $G_{\Delta, \ell}$ and their shadows $G_{d-\Delta, \ell}$. Notice that in this decomposition the spin $\ell$ takes integer values while the dimension $\Delta$ is continuous. Now the idea is to use the orthogonality of the functions $F_{\ell,\Delta}$ to invert such integral and obtain $c(\ell,\Delta)$.
This can be achieved and it reads
\begin{equation} \label{inveucl}
    c(\ell,\Delta)= N(\ell, \Delta) \int \frac{d^2 z}{z^2 \bar{z}^2} \left(\frac{z -\bar{z}}{z \bar{z}} \right)^2 F_{\ell,\Delta}(z, \bar{z}) G(z,\bar{z})
\end{equation}
The function $ N(\ell, \Delta)$ can be computed by using the behaviour of the functions $F_{\ell,\Delta}(z, \bar{z})$ around $z=0$. Notice that the expression \eqref{inveucl} is valid in the Euclidean signature, so $\bar{z}=z^*$ and the integration is over the complex Euclidean $z$ plane. The most important observation is that the conformal block decomposition is satisfied if the spectral function $c(\ell,\Delta)$ has poles and residues related to the conformal dimensions and OPE coefficients of the exchanged operator respectively.\footnote{There are subtleties related to the convergence of this integral and on the precise location of the shadow poles. We refer the interested reader to Section 3 and Appendix A of \cite{Caron-Huot:2017vep}} 

In order to make it Lorenzian, we need to introduce appropriate variables and do the following manipulations \cite{Hogervorst:2013sma}
\begin{equation}
    z=\frac{4\rho}{\left( 1+\rho\right)^2},\;\,\,\, w=\sqrt{\frac{\rho}{\bar{\rho}}}=e^{i \theta},\quad
    \int d^2 z \to \int_0^1 d|\rho| \oint \frac{d w}{w}
\end{equation}
Let us recall that for any $d$, the conformal blocks are eigenfunctions of the quadratic and quartic Casimir operators, and it is possible to see that generically solving the differential equations associated to these eigenvalue problems leaves us with 8 solutions. These solutions can be built starting from pure power laws in the configuration $0 \ll z \ll \bar{z} \ll 1$, and they are given by\footnote{The symmetries are $\ell \leftrightarrow 2-d-\ell$, $\Delta \leftrightarrow d-\Delta$ and $\Delta \leftrightarrow 1-\ell$. Thus by using them, it is possible to generate all the solutions.}
\begin{equation} 
    g_{\Delta, \ell}(z,\bar{z}) \sim z^{\frac{\Delta-\ell}{2}}\bar{z}^{\frac{\Delta-\ell}{2}}
\end{equation}
Then in principle the functions $F_{\ell, \Delta}$ are complicated linear combinations of these 8 solutions. In order to close the $w$-contour we would like to roughly decompose the function $F_{\ell, \Delta}=F^{+}(\ell,\Delta)+F^{-}(\ell,\Delta)$  such that 
\begin{equation}
    F^{+}(\ell,\Delta)\to w^{\ell} \quad  \text{as} \,\,w \to 0
\end{equation}
and 
\begin{equation}
    F^{-}(\ell,\Delta)\to w^{-\ell} \quad  \text{as} \,\,w \to \infty
\end{equation}
This is not straightforward and {\it a priori} it is not guaranteed to be possible. Quite remarkably, as shown in \cite{Caron-Huot:2017vep}, with some manipulations it is possible to find a precise linear combination which brings \eqref{inveucl} to its Lorenzian counterpart \eqref{inv}, properly integrated on the Lorentzian diamond. Now it becomes clear that we have a relation which gives the s-channel OPE data as an integral of the $\text{dDisc}$ which is convergent in the t-channel and a kernel which is essentially the Lorenzian counterpart of the conformal blocks, and it is convergent for spin larger than 1. What happens for spins smaller or equal than 1 is that the contribution of the arc in the $w$-plane cannot be dropped.\footnote{Notice that in the discussion presented here the spin of intermediate operators can only be even, since we started from a four-point function of identical scalar operators. Thus we can say that the inversion formula is valid for $j>0$.} This ends our sketch of the derivation of the Lorentzian inversion formula. We refer to the original paper \cite{Caron-Huot:2017vep} and to \cite{Simmons-Duffin:2017nub} for a more detailed and rigorous proof, and in the next section we give the explicit expression. 

\subsection{Main formula}
In this section we would like to write down all the ingredients of the Lorentzian inversion formula that can be used to invert the OPE decomposition, meaning that it gives the OPE data from the analytic structure of the four-point correlator. We first need to analytically continue to the Lorentzian regime and define the double-discontinuity 
\begin{equation}
\text{dDisc}\left[ \mathcal{G}(z,\bar{z})\right]=\mathcal{G}_{\text{Eucl}}(z,\bar{z})-\frac{1}{2}\mathcal{G}^{\circlearrowright}(z,\bar z)-\frac{1}{2}\mathcal{G}^{\circlearrowleft}(z,\bar z)\;.
\end{equation}
Here $\mathcal{G}_{\text{Eucl}}(z,\bar{z})$ is the Euclidean correlator and the other terms are the two possible analytic continuations around the branch point $\bar z=1$. We have all the ingredients to write down the inversion formula which reads
\begin{equation}
    c_{\ell,\Delta}= c^{(t)}_{\ell, \Delta}+(-1)^{\ell}c^{(u)}_{\ell, \Delta}
\end{equation}
where
\begin{equation} \label{inv}
c^{(t)}_{\ell, \Delta}=\frac{1}{4}k_{\frac{\Delta+\ell}{2}}\int_{0}^{1}d z d\bar{z} \left(\frac{z-\bar z }{z \bar z} \right)^2 \frac{G_{\ell+3,\Delta-3}(z, \bar{z})}{z^2 \bar z^2}\text{dDisc}\left[ \mathcal{G}(z,\bar{z})\right]\;,
\end{equation}
with $k_{\alpha}=\frac{\Gamma(\alpha)^4}{2 \pi^2 \Gamma(2 \alpha -1) \Gamma(2 \alpha)}$ and $c^{(u)}_{\ell, \Delta}$ has the same form as  $c^{(t)}_{\ell, \Delta}$ but with $x_1 \leftrightarrow x_2$. This relation is fully analytic in the spin, except the term $(-1)^{\ell}.$\footnote{In the case discussed here $(-1)^{\ell}=1$ since the spin of the intermediate operators is always even.}. 
The spectral functions $c_{\ell, \Delta}$ are related to  $a_{\Delta, \ell}$ In particular, it is related to the s-channel OPE data in this way
\begin{equation}
c_{\ell, \Delta} \xrightarrow[\Delta \to \Delta_k]{} \frac{a_{\Delta_k,\ell}}{\Delta_k-\Delta}\;.
\end{equation}

\subsection{Examples}
Let us analyse this formula \eqref{inv}, in particular in connection with the discussion about large spin reconstruction. The first point to make is that this formula is analytic up to spin one, so it is possible to invert the four-point correlator up to this value of the spin. The second is that it turns out that the information contained in the double discontinuity is the same as the one obtained when considering the singularities as $V \to 0$ of the correlators. To understand these points, let us list the double discontinuity of some useful functions:
\begin{align} \label{doub}
\text{dDisc}[\log(1-\bar{z})]&=0\;, \nonumber\\
\text{dDisc}[\log^2(1-\bar{z})]&=4\pi^2\;,\\
\text{dDisc}\left[\left( \frac{1-\bar z}{\bar z} \right)^p \right] &=\left( \frac{1-\bar z}{\bar z} \right)^p 2\sin ^2(\pi p)\;  \nonumber.
\end{align}
This set of functions are the ones that we have encountered in the previous sections, in particular in the discussion of large $N$ CFTs. We have seen that the functions that perform a singularity as $V \to 0$ are the last two in \eqref{doub} which appear at order $N^{-4}$ and $N^{0}$ respectively. Let us discuss more in details the last line. If we consider the four-point correlator introduced in \eqref{MFT}, we see that written in terms of $z$ and $\bar z$ the function has a non-vanishing double discontinuity only due to the presence of the term 
\begin{equation}
\left( \frac{z \bar z}{(1-z)(1-\bar z)}\right)^{\Delta_{\varphi}}\;.
\end{equation}
Thus we have that 
\begin{align}
\nonumber c_{\ell, \Delta}&=\frac{1+(-1)^{\ell}}{4}k_{\frac{\Delta+\ell}{2}}\int_{0}^{1}d z d\bar{z} \left(\frac{z-\bar z }{z \bar z} \right)^2 \frac{G_{\ell+3,\Delta-3}(z, \bar{z})}{z^2 \bar z^2}\text{dDisc}\left[ \left(\frac{z \bar z}{(1-z)(1-\bar z)}\right)^{\Delta_{\varphi}} \right] \\
\nonumber &=\frac{1+(-1)^{\ell}}{4}k_{\frac{\Delta+\ell}{2}}\int_{0}^{1}d z d\bar{z} \left(\frac{z-\bar z }{z \bar z} \right)^2 \frac{G_{\ell+3,\Delta-3}(z, \bar{z})}{z^2 \bar z^2}\left( \frac{z\bar z}{  (1-\bar z)(1- z)}\right)^{\Delta_{\varphi}} 2\sin ^2(\pi \Delta_{\varphi})\\
\nonumber &=\frac{2^{2-\Delta} (\Delta-2) \Delta_{\varphi} \left((-1)^\ell+1\right) (\ell+1) \Gamma (2-\Delta_{\varphi})^2 \Gamma (-\Delta_{\varphi}) \Gamma (-\Delta+\ell+4) }{\Gamma (1-\Delta_{\varphi}) \Gamma (\Delta_{\varphi})^2 \Gamma
   \left(\frac{1}{2} (-\Delta+\ell+4)\right)^2 \Gamma (\Delta+\ell-1) \Gamma \left(\frac{1}{2} (-\Delta-2 \Delta_{\varphi}+\ell+8)\right)} \\
   \times& \frac{\Gamma \left(\frac{\Delta+\ell}{2}\right)^2 \Gamma
   \left(-\frac{\Delta}{2}+\Delta_{\varphi}+\frac{\ell}{2}\right) \Gamma \left(\frac{1}{2} (\Delta+2 \Delta_{\varphi}+\ell-4)\right)}{ \Gamma \left(\frac{1}{2} (\Delta-2 \Delta_{\varphi}+\ell+4)\right)}\;.
\end{align}
We can see that 
\begin{equation}
\text{Res}_{\Delta=2\Delta_{\varphi}+2n+\ell}c_{\ell, \Delta} =a_{n,\ell}^{\text{MF}}\;.
\end{equation}
This confirms our previous observation that the presence of the term $\left(\frac{U}{V} \right)^{\Delta_{\varphi}}$ in the four-point function fully fixes the OPE data. Using the inversion formula, one recovers from it both the dimensions of the exchanged operators $\Delta=2\Delta_{\varphi}+2n+\ell$ which correspond to the poles, and the squared three-point functions coefficients which correspond to the residues at the pole. Let us also point out that the third equation in (\ref{doub}) is responsible for the double discontinuity of a conformal block where $p=\tau-2\Delta_\varphi$. Note that when $\tau=2\Delta_\phi+2n$ for $n\in \mathbb{Z}_{\geq 0}$, {\it i.e.}, when the exchanged operator is a double-trace operator, the double discontinuity vanishes and it does not contribute to the spectral function. This is a welcome feature of the Lorentzian inversion formula, in particular in applications to tree-level correlators in AdS where we only need to consider the contribution of single-trace operators.

\newpage
\markboth{6\quad MELLIN SPACE}{}
\section{Mellin space}\label{Sec:MellinFormalism}

\subsection{General comments}

The Mellin space formalism was introduced in \cite{Mack:2009mi,Penedones:2010ue} (see also \cite{Paulos:2011ie,Fitzpatrick:2011ia}) and is a natural language for discussing holographic correlators. In position space, these are rather complicated functions of the conformal cross ratios. However, in this formalism the analytic structure of holographic correlators becomes drastically simplified, and manifests the underlying scattering amplitude nature of  these objects. 

Consider the correlation function of $n$ scalar operators.\footnote{For spinning correlators, the Mellin formalism is more difficult to define. See \cite{Goncalves:2014rfa} for the case of $n$-point function with one spinning operator, and \cite{Sleight:2018epi, Chen:2017xdz} for correlators of four spinning operators. Mellin formalism can also be developed for boundary CFT \cite{Rastelli:2017ecj} and defect CFTs \cite{Goncalves:2018fwx}.} The correlator can be written as a multi-dimensional inverse Mellin transformation
\begin{equation}\label{defMellinnpt}
\langle \mathcal{O}_1(x_1)\ldots \mathcal{O}_n(x_n)\rangle=\int [d\delta_{ij}] \bigg(\prod_{i<j} (-2P_i\cdot P_j)^{-\delta_{ij}}\Gamma[\delta_{ij}]\bigg) \mathcal{M}(\delta_{ij}).
\end{equation}
Here we have used the embedding coordinates\footnote{The use of the embedding space manifests the conformal covariance but is not strictly necessary. One can also prove the statements within the physical space with coordinates $x^\mu$.} defined in Section \ref{Subsec:BCFTkinematics}
\begin{equation}
P^A=\big(\frac{1+x^2}{2},\frac{1-x^2}{2},x^\mu \big)\;,
\end{equation}
where the signature of the first two components are respectively $-$ and  $+$, and $x^\mu$ are the coordinates of the operators. This gives 
\begin{equation}
-2P_i\cdot P_j=x_{ij}^2\;.
\end{equation}
We can set 
\begin{equation}
\delta_{ij}=\delta_{ji}\;,\quad \delta_{ii}=-\Delta_i\;.
\end{equation}
Requiring the correlator to have the correct scaling behavior under $P_i\to \lambda_i P_i$ imposes the conditions 
\begin{equation}
\sum_{j=1}^n \delta_{ij}=0\;.
\end{equation}
The integration is over the independent $\delta_{ij}$ and along the imaginary axes.\footnote{Note that for $n=2$ and $n=3$, these constraints completely fix $\delta_{ij}$. There is no integral and the Mellin representation just gives the standard two- and three-point functions.} All the information of the correlator is transferred to the function $\mathcal{M}(\delta_{ij})$ which is defined to be the {\it Mellin amplitude}. 

To see why it is natural to call $\mathcal{M}(\delta_{ij})$ an amplitude, we note that the conditions on $\delta_{ij}$ can be automatically solved by parameterizing $\delta_{ij}$ using auxiliary flat-space momenta $\vec{p}_i$
\begin{equation}
\delta_{ij}=\vec{p}_i\cdot \vec{p}_j\;.
\end{equation} 
Here the momenta are conserved and on-shell
\begin{equation}
\sum_i \vec{p}_i=0\;,\quad \quad \vec{p}_i^2=-\Delta_i\;,
\end{equation}
with the squared masses replaced by the conformal dimensions. Therefore, the independent $\delta_{ij}$, which count the independent conformal cross ratios of the form $\frac{(P_i\cdot P_j)(P_k\cdot P_l)}{(P_i\cdot P_k)(P_j\cdot P_j)}$, are in correspondence to the independent Mandelstam variables defined from the auxiliary momenta. 

In fact, we can further show that the auxiliary momenta live in a $d+1$ dimensional spacetime by performing a simple counting. Consider first the CFT side. For $n$ points in a CFT$_d$, the number of independent cross ratios should equal to the number of independent parameters after using all nontrivial conformal symmetry actions. This is the number of all coordinates minus the dimension of the conformal group
\begin{equation}\label{cftcounting1}
\underbrace{n d}_{\text{coordinates of }n\text{ points}}-\underbrace{\frac{1}{2}(d+1)(d+2)}_{\text{dimension of }SO(d+1,1)}\;.
\end{equation} 
However, this counting is only correct for $n>d+2$ as for $n\leq d+2$ there is a nontrivial stability group. To see this explicitly, we can use a conformal transformation to send two points to $0$ and $\infty$. The remaining $n-2$ points define an $n-2$ dimensional hyperplane. The rotation group $SO(d+2-n)$ in directions orthogonal to the plane is a stability group, and we should subtract this group when it is nontrivial. Adding back its dimension, we get 
\begin{equation}\label{cftcounting2}
\frac{n(n-3)}{2}\;.
\end{equation}
Another way to understand this change of counting behaviors is that for $n>d+2$ an $M\times M$ matrices with elements $\{P_i\cdot P_j\}$ and $d+2<M\leq n$ are no longer of the full rank. There are additional relations given by $\det \{P_i\cdot P_j\}=0$. Let us now perform the counting of independent Mandelstam variables on the amplitude side. Here we consider $n$ on-shell momenta in $\mathbb{R}^{D-1,1}$. For $n>D+1$, we have similarly
\begin{equation}
\underbrace{n (D-1)}_{\text{coordinates of }n\text{ on-shell momenta}}-\underbrace{\frac{1}{2}D(D+1)}_{\text{dimension of the Poincar\'e group}}\;,
\end{equation} 
where we note that the on-shell condition eliminates one degree of freedom for each particle. For $n\leq D+1$, there is also a stability group which can be seen as follows. We can go to the frame in which the total momentum is zero
\begin{equation}
\sum_i \vec{p}_i=0\;.
\end{equation}
Then these momenta span an $n-1$ dimensional subspace, which remains invariant under an $SO(D-n+1)$ rotation group which is orthogonal to it. Adding back the dimension of the stability group, we again arrive at 
\begin{equation}
\frac{n(n-3)}{2}\;.
\end{equation}
Note that the answers from the two counting problems coincide precisely if $D=d+1$. This indicates that a correlation function in CFT$_d$ can be mapped into a scattering amplitude in a $d+1$ dimensional spacetime. Of course, this is not surprising as we know that the AdS/CFT correspondence is a way to establish such a relation.  

Before we move on, let us make a quick comment regarding (\ref{defMellinnpt}). In writing (\ref{defMellinnpt}), we are implicitly thinking that we are in the case of (\ref{cftcounting2}) where the spacetime dimension is sufficiently high with respect to the number of operators and we are free of the determinant relations. While (\ref{defMellinnpt}) remains a valid representation even when we are in the case of (\ref{cftcounting1}), one might imagine that there is an alternative formalism better suited for this situation with fewer Mellin variables. This is particularly relevant for the case of CFT$_1$ where  starting from the first nontrivial case with $n=4$ one encounters only the case (\ref{cftcounting1}), and the case (\ref{cftcounting2}) never shows up. On the other hand, it is also useful to think of this problem from the dual perspective. It is well known that remarkable properties of S-matrices such as integrability crucially rely on the special kinematics in 2d. The redundant parameterization in (\ref{defMellinnpt}), however, does not reflect these special kinematic features. It would therefore be of great interest to find another Mellin representation that is intrinsic to the CFT$_1$ kinematics. Relevant works on the 1d Mellin representation include \cite{fgsz,Bianchi:2021piu}, but it is not yet clear how to establish such a formalism for general $n$-point functions.

Let us get back to the definition (\ref{defMellinnpt}). Operators exchanged in a CFT correlator are manifested as poles in the Mellin formalism. To see this, let us consider the OPE
\begin{equation}
\mathcal{O}_1(x_1)\mathcal{O}_2(x_2)=\sum_k C_{12k}(x_{12}^2)^{\frac{\Delta_k-\Delta_1-\Delta_2}{2}}(\mathcal{O}_k(x_2)+hx_{12}^2 \partial^2\mathcal{O}_k(x_2)+\ldots)\;.
\end{equation}
Here we have restricted to the scalar operators for simplicity. The constant $h$ is fixed by conformal symmetry and multiplies the first of the descendant terms with the others collectively denoted by $\ldots$. We now perform this OPE in the $n$-point function and compare it with the Mellin representation. In the limit of $x_{12}^2\to 0$, it is convenient to integrate over $\delta_{12}$ by closing the contour to the left in the complex plane. In order to match the OPE, it is clear that the integrand of the inverse Mellin transformation must have poles at 
\begin{equation}
\delta_{12}=\frac{\Delta_1+\Delta_2-\Delta_k-2m}{2}\;,\quad m=0\;,1\;,2\;,\ldots.
\end{equation}
The residues of these poles are proportional to the product of the OPE coefficient $C_{12k}$ and the Mellin amplitude of the lower-point  correlator. In other words, OPE in the CFT correlator leads to factorization in the Mellin amplitude. The precise relation was derived in \cite{Goncalves:2014rfa}. Similar reasoning also applies to the general case where we exchange a spinning operator with dimension $\Delta_k$ and spin $\ell_k$. The corresponding poles are at
\begin{equation}\label{nptOPEpole}
\delta_{12}=\frac{\Delta_1+\Delta_2-(\Delta_k-\ell_k)-2m}{2}\;,\quad m=0\;,1\;,2\;,\ldots.
\end{equation}

Note that in the definition (\ref{defMellinnpt}),we have included a factor of Gamma functions which contain poles at integer locations. These poles correspond to the ``double-trace'' operators which are ubiquitous in holographic theories (in the strict central charge $c\to\infty$ limit they are just mean field theories). It turns out that separating out their contributions in this way is convenient  when considering holographic correlators, as we will explain in more detail in the next subsection in the four-point function context.  

\subsection{Four-point function case}
Since the focus of most of this review will be on four-point functions, here we spell out the details of the Mellin representation formalism for $n=4$.

In this case, the general definition (\ref{defMellinnpt}) reduces to 
\begin{equation}
\langle\mathcal{O}_1(x_1)\ldots \mathcal{O}_4(x_4)\rangle=\frac{1}{(x_{12}^2)^{\frac{\Delta_1+\Delta_2}{2}}(x_{34}^2)^{\frac{\Delta_3+\Delta_4}{2}}}\left(\frac{x_{14}^2}{x_{24}^2}\right)^a\left(\frac{x_{14}^2}{x_{13}^2}\right)^b \mathcal{G}(U,V)\;,
\end{equation}
with $a=\frac{1}{2}(\Delta_2-\Delta_1)$, $b=\frac{1}{2}(\Delta_3-\Delta_4)$, and 
\begin{equation}\label{defMellin4pt}
\begin{split}
\mathcal{G}(U,V)=&\int_{-i\infty}^{i\infty}\frac{dsdt}{(4\pi i)^2}U^{\frac{s}{2}}V^{\frac{t}{2}-\frac{\Delta_2+\Delta_3}{2}}\mathcal{M}(s,t)\,\Gamma(\tfrac{\Delta_1+\Delta_2-s}{2})\Gamma(\tfrac{\Delta_3+\Delta_4-s}{2})\\
&\quad\quad\quad\times \Gamma(\tfrac{\Delta_1+\Delta_4-t}{2})\Gamma(\tfrac{\Delta_2+\Delta_3-t}{2}) \Gamma(\tfrac{\Delta_1+\Delta_3-u}{2})\Gamma(\tfrac{\Delta_2+\Delta_4-u}{2})\;.
\end{split}
\end{equation}
Here to make it more symmetric we have also introduced the third Mandelstam variable $u$ satisfying 
\begin{equation}
s+t+u=\sum_{i=1}^4\Delta_i\;.
\end{equation} 
The integration contours of $s$ and $t$ run parallel to the imaginary axis and separate semi-infinite series of poles running to the left and to the right. Bose symmetry acts by permuting $s$, $t$, $u$ and the operator labels. Therefore the Mellin amplitude has the same symmetry properties as a flat-space amplitude. For example, for four identical operators the Mellin amplitude satisfies
\begin{equation}
\mathcal{M}(s,t)=\mathcal{M}(s,u)=\mathcal{M}(t,u)\;.
\end{equation} 

Let us consider again the OPE of the four-point function in the s-channel. Then (\ref{nptOPEpole}) gives rise to poles in $s$ and the Mellin amplitude takes the following form
\begin{equation}\label{4ptOPEinMellin}
\mathcal{M}(s,t)\supset \sum_{m=0}^{\infty} C_{12k}C_{34k} \frac{Q_{\ell_k,m}(t)}{s-(\Delta_k-\ell_k)-2m}\;.
\end{equation}
The numerators $Q_{\ell,m}(t)$ are {\it kinematic} polynomials of degree $\ell$ in $t$ and are known as the {\it Mack polynomials}. They can be obtained by, for example, matching the position space expressions of the conformal blocks after evaluating the Mellin integrals.  

Let us note that the above form (\ref{4ptOPEinMellin}) of OPE in Mellin space is reminiscent of flat-space tree-level scattering. Such a behavior strengthens the analogy between correlators and amplitudes which we argued about in the previous subsection. To further appreciate this analogy, let us examine the structure of tree-level Witten diagrams in Mellin space. These tree-level Witten diagrams are the leading corrections in the $1/N$-expansion of correlators in a local holographic theory. 

\begin{figure}
\centering
\includegraphics[width=0.7\textwidth]{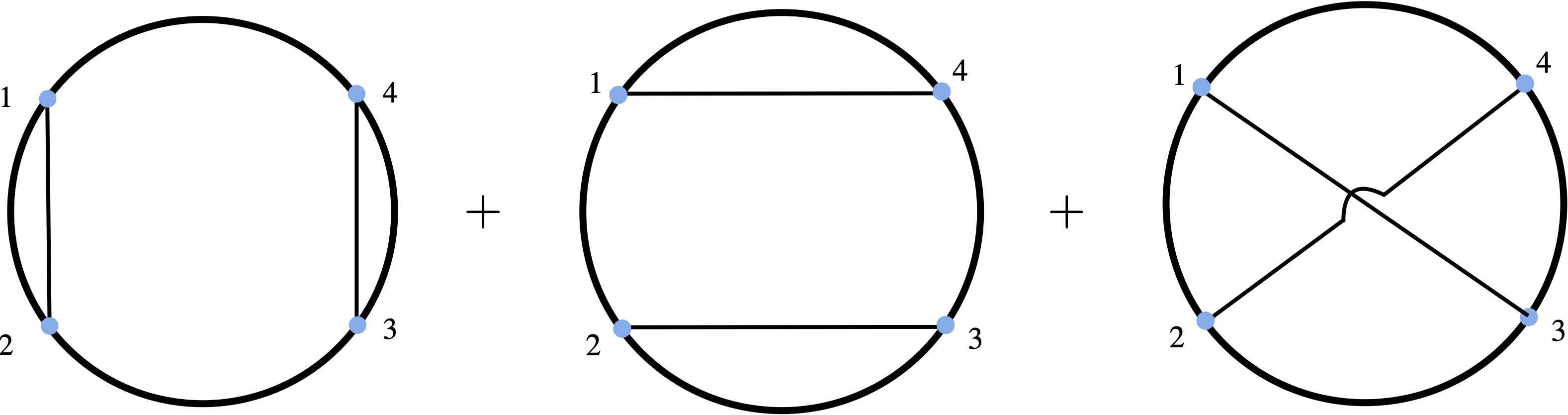}
\caption{Disconnected Witten diagrams.}
    \label{fig:Mellindisc}
\end{figure}

\begin{figure}
\centering
\includegraphics[width=0.85\textwidth]{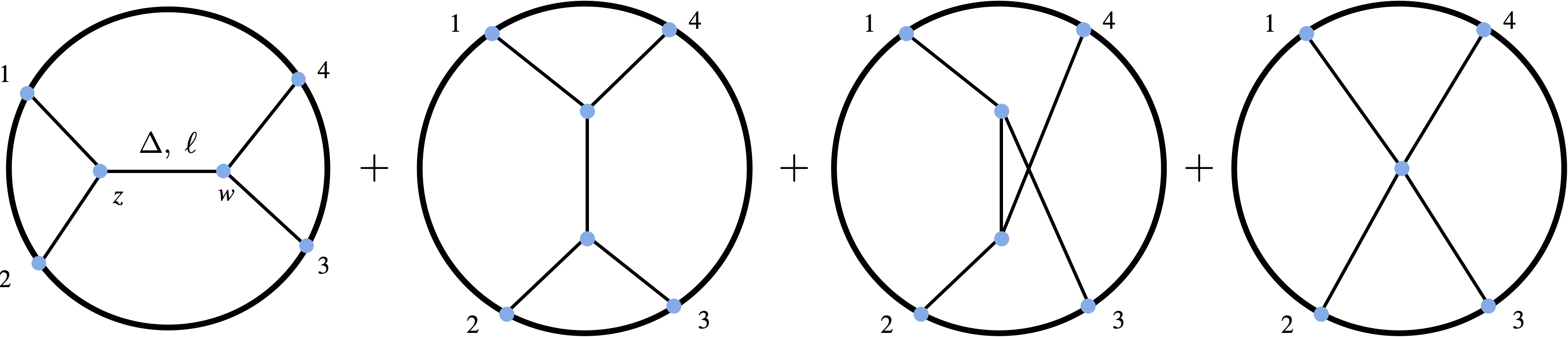}
\caption{Tree-level Witten diagrams.}
    \label{fig:Mellintrees}
\end{figure}

For simplicity, let us consider four identical operators of dimension $\Delta_\varphi$. The four-point function can be expanded in the following way
\begin{equation}
\mathcal{G}=\mathcal{G}_{\rm disc}+\frac{1}{N^2}\mathcal{G}_{\rm tree}+\frac{1}{N^4}\mathcal{G}_{\rm 1-loop}+\ldots\;.
\end{equation}
In the $N\to\infty$ limit, the leading order contribution to the correlator is given by the mean field theory, and is comprised of products of two-point functions
\begin{equation}
\mathcal{G}_{\rm disc}=1+U^{\Delta_\varphi}+\left(\frac{U}{V}\right)^{\Delta_\varphi}\;.
\end{equation}
This corresponds to the disconnected diagrams depicted in Figure \ref{fig:Mellindisc}. Decomposing it into conformal blocks we find only double-trace operators of the schematic form 
\begin{equation}
:\varphi \square^n \partial^\ell \varphi:\;,
\end{equation}
which have conformal dimension $\Delta_{n,\ell}=2\Delta_\varphi+2n+\ell$ and spin $\ell$. They correspond to poles in the integrand at $s=2\Delta_\varphi+2n$ which are precisely the poles of the s-channel Gamma functions. At the next order, we have connected tree-level diagrams depicted in Figure \ref{fig:Mellintrees}. These tree-level diagrams can further be divided into exchange diagrams and contact diagrams. They are built out of propagators following Feynman rules similar to those in flat space. For example, the contact diagram following from a quartic vertex without derivatives is defined as, see, {\it e.g.}, \cite{Witten:1998qj,DHoker:1999mqo,DHoker:2002nbb}
\begin{equation}\label{defWcon0der}
W_{\rm con}(x_i)=\int_{AdS_{d+1}} dz \prod_{i=1}^4 G_{B\partial}^{\Delta_\varphi}(x_i,z)\;,
\end{equation} 
and the s-channel exchange diagram of a scalar field with dimension $\Delta$ is defined as 
\begin{equation}
W_{\Delta,\ell=0}(x_i)=\int_{AdS_{d+1}} dz dw  G_{B\partial}^{\Delta_\varphi}(x_1,z)G_{B\partial}^{\Delta_\varphi}(x_2,z)G_{BB}^\Delta(z,w)G_{B\partial}^{\Delta_\varphi}(x_3,w)G_{B\partial}^{\Delta_\varphi}(x_4,w)\;.
\end{equation} 
Here $G_{B\partial}^{\Delta}(x,z)$ and $G_{BB}^{\Delta}(z,w)$ are the bulk-to-boundary and the bulk-to-bulk propagators in AdS respectively. Since these tree diagrams are corrections to the mean field theory correlator, the double-trace operators appearing in the OPE at the disconnected order will also appear at this order. Their presence is conveniently captured by the Gamma function factor. In addition, in the exchange diagrams there is a ``single-trace'' operator which is dual to the exchanged field in AdS. This requires the Mellin amplitude of an s-channel exchange diagram to contain the contribution (\ref{4ptOPEinMellin}). Since there are no other operators exchanged in the OPE at this order, we conclude that the Mellin amplitude of the exchange Witten diagram is just 
\begin{equation}
\mathcal{M}_{\Delta,\ell}(s,t)=\sum_{m=0}^{\infty} \frac{Q_{\ell,m}(t)}{s-(\Delta-\ell)-2m}+R_{\ell-1}(s,t)
\end{equation}
where $R_{\ell-1}(s,t)$ is a degree-$(\ell-1)$ polynomial free of poles and we have set $C_{12k}$ and $C_{34k}$ to 1 for convenience. By contrast, the contact diagrams contain only double-trace operators in the conformal block decomposition. Therefore, their Mellin amplitudes are regular in the Mandelstam variables. For example, the Mellin amplitude of the zero-derivative contact diagram (\ref{defWcon0der}) is just a constant. More generally, the Mellin amplitude of a contact diagram with $2L$ contracted derivatives in the quartic vertex is a degree-$L$ polynomial
\begin{equation}
\mathcal{M}_{\text{con, }2L\text{-der}}=P_{L}(s,t)\;.
\end{equation}

Evidently, the Mellin amplitudes of AdS tree-level diagrams are highly similar to the flat-space tree-level amplitudes. This similarity makes the Mellin formalism a very useful tool to study holographic correlators and allows us to apply many intuitions from flat-space scattering. 

\subsection{Flat-space limit}
We have considered scattering processes of a relativistic theory placed in an AdS space, and we have implicitly set its radius $R$ to be 1 when discussing the diagrams. Here let us restore the $R$ dependence and make it tunable. If we take $R$ to be much larger than any length scales in the theory, then clearly the curvature effects should be negligible. The AdS scattering amplitude should correspondingly reduce to the flat-space scattering amplitude in this limit.  The Mellin formalism provides a convenient way to extract the flat-space limit of AdS scattering amplitudes. 

The precise relation was given in \cite{Penedones:2010ue}. On the one hand, we have the Mellin amplitude $\mathcal{M}(\delta_{ij})$ of an $n$-point scalar correlator where the conformal dimension of each operator is $\Delta_i$. On the other hand, we have the scattering amplitude $\mathcal{T}_n$ of $n$ massless particles in flat space. The flat-space amplitude is effectively reproduced from the high-energy limit of the Mellin amplitude
\begin{equation}\label{MtoT}
\mathcal{T}_n(s_{ij})=\mathcal{N}R^{\frac{n(d-1)}{2}-d-1}\lim_{R\to\infty}\int_{-i\infty}^{i\infty} d\alpha e^\alpha \alpha^{\frac{d-2-\sum_i\Delta_i}{2}}\mathcal{M}\left(\delta_{ij}=-\frac{R^2s_{ij}}{4\alpha},\Delta_a=R m_a\right)\;.
\end{equation}
Here $s=2\vec{p}_i\cdot \vec{p}_j$ are the Mandelstam variables in the flat space, and $\mathcal{N}$ is an overall factor depending on the external dimensions $\Delta_i$. The dimensions $\Delta_a$ belong to exchanged internal fields. They scale linearly with $R$ if we wish to assign a nonzero mass $m_a$ in the flat-space limit.\footnote{This follows from the mass relation $m^2R^2=\Delta(\Delta-d)$ in the large $R$ limit.} Finally, the integration contour of $\alpha$ runs to the right of all poles in the Mellin amplitude and the branch cut from $\alpha^{\frac{d-2-\sum_i\Delta_i}{2}}$. The relation (\ref{MtoT}) was presented as a conjecture in \cite{Penedones:2010ue} and was checked in many explicit examples, including contact and exchange diagrams at tree level, and four-point one-loop amplitudes. It was also derived in \cite{Fitzpatrick:2011hu} using wavepackets where the scattering was limited to a small flat region of AdS.

\newpage
\markboth{7\quad THE EPSILON EXPANSION}{}
\section{The epsilon expansion}\label{Sec:epsilon}
\subsection{A brief review}
The epsilon expansion was introduced in \cite{Wilson:1971dc, Wilson:1973jj} as an approximate technique to compute critical exponents for the 3d Ising and $O(N)$ models. The idea is to start with the $O(N)$ model
\be
S=\sum_{i=1}^{N}\int d^{d} x\, \left(\partial_\mu \phi^i \partial^\mu \phi^i-m^2 \phi^i\phi^i  +\lambda (\phi^i\phi^i)^2\right)\,,
\ee
with $d=4-\epsilon$
and compute various scaling dimensions in this model at the fixed point, where $m$ and $\lambda$ are tuned appropriately, pretending $\epsilon$ to be small using the Feynman diagram approach \cite{Wilson:1973jj} to some loop order. Then at the end of the calculation either $\epsilon$ is set equal to unity or some resummation technique is used \cite{Kleinert:2001ax} to obtain physical answers. For reasons not completely well understood, the results are remarkably close to both Monte Carlo simulations of the 3d Ising/O(N) models as well as experimental measurements.

In terms of $\tau=\frac{T-T_c}{T_c}$, the specific heat $C\propto \tau^{-\alpha}$ defines the critical exponent $\alpha$. Similarly at the critical point $T=T_c$, the correlator is expected to behave like $\langle \phi(r)\phi(0)\rangle\propto r^{-d+2+\eta}$, defining the exponent $\eta$. Let us focus for now on $N=1$ which is relevant for the 3d Ising model. In terms of the scaling dimension $\D_\phi$ of $\phi$ and $\D_{\phi^2}$ of $\phi^2$, we have
\be
\alpha=2-\frac{d}{d-\D_{\phi^2}}\,,\quad \eta=2\D_\phi-d+2\,.
\ee
In a free theory $\D_\phi=1-\epsilon/2, \D_{\phi^2}=2\D_\phi=2-\epsilon$. The $\lambda (\phi^i \phi^i)^2$ interaction induces a flow to the Wilson-Fisher fixed point where operators get anomalous dimensions.
The results for $\D_\phi$ and $\D_{\phi^2}$ are \cite{Wilson:1973jj}:
\begin{eqnarray}\label{WF}
\D_\phi&=& 1-\frac{\e}{2}+\frac{\e^2}{108}+\frac{109 \e^3}{11664}+O(\e^4)\,,\\
\D_{\phi^2}&=&2-\frac{2\e}{3}+\frac{19\e^2}{162}+O(\e^3)\,.
\end{eqnarray}
The results for double field higher gradient operators of the form $\mathcal{O}_\ell=\phi\partial_{\mu_1}\cdots \partial_{\mu_\ell}\phi$ are also known and were worked out to $O(\e^2)$ in \cite{Wilson:1973jj}. Using the Feynman diagram approach, the $O(\e^4)$ anomalous dimensions of ${\mathcal O}_\ell$ \cite{gracey}, the $O(\e^7)$ anomalous dimension of $\phi^2$ and $O(\e^8)$ anomalous dimension of $\phi$ \cite{panzer, schnetz} have been worked out.  For a recent summary of the critical O(N) epsilon expansion data, refer to \cite{Henriksson:2022rnm}.

If we take the Wilson-Fisher results in eq.(\ref{WF}) above and naively substitute $\e=1$, we get $\D_{\phi}=0.5186$ which is in a remarkable agreement with numerical results which give 0.5181 as the answer to 4 significant figures \cite{numrev}. For $\phi^2$ we find $\D_{\phi^2}=1.45$ while numerical results give 1.41 to 2 decimal places. However, despite these encouraging findings, it is difficult to compute OPE coefficients using this approach. Also it is known that the $\e^k$ term in the anomalous dimension calculations grows as $k^{k+4} e^{-k}(\frac{\e}{3})^k$ \cite{Brezin:1976vw} necessitating the use of resummation techniques. Furthermore, a more crucial drawback is that this approach does not use the conformal symmetry of the critical point and is inherently perturbative. In what follows, we will review how conformal field theory techniques can be used to extract OPE data including OPE coefficients in the epsilon expansion.

\subsection{CFT derivation of leading order anomalous dimension}
In this section, we will review the derivation of the leading order anomalous dimension of the operators $\phi^4$ using CFT techniques using the elegant method of \cite{rychkovtan}. We will need this information in what follows. Following \cite{rychkovtan}, we assume that
\begin{enumerate}
\item The WF fixed point is invariant under the full conformal symmetry.
\item Each local operator in the free theory at $\e=0$ has a counterpart at the WF fixed point. In particular:
\be
\lim_{\e\rightarrow 0} V_n=\phi^n\,.
\ee
This enables us to refer to $\phi^n$ in the WF theory unambiguously. The conformal dimension of $V_n$ is denoted by $\D_n$. We will further define
\be
\D_n=n(1-\frac{\e}{2})+\gamma_n\,.
\ee
\item $V_3$ is a descendant. Namely
\be
\partial^2 V_1=\beta V_3\,.
\ee
Here $\beta=\beta(\e)$ will be fixed later. This equation also means that $\Delta_3=\Delta_1+2$.
\end{enumerate}
We will choose the normalizations such that 
\be
\langle \phi(x)\phi(0)\rangle=\frac{1}{|x|^2}\,,\quad \langle V_1(x) V_1(0)\rangle=\frac{1}{|x|^{2\Delta_1}}\,.
\ee
This will enable us to fix $\beta$. To do this we compare $\langle \partial^2 V_1(x)\partial^2 V_1(0)\rangle$ and $\langle V_3(x) V_3(0)\rangle$.  This leads to
\be\label{alf}
\beta=4 \left(\frac{\gamma_1}{3}\right)^{1/2}\,.
\ee
Consider first the OPE in the free theory
\be\label{phiop}
\phi^n(x)\times \phi^{n+1}(0)\supset (n+1)! |x|^{-2n}\left(\phi(0)+\frac{n}{2} |x|^2\phi^3(0)\right)\,.
\ee
The RHS is obtained using Wick contractions. Next we need the WF OPE
\be \label{Vop}
V_n(x)\times V_{n+1}(0)\supset \tilde f |x|^{\D_1-\D_n-\D_{n+1}}\left(1+q_1 x^\mu\partial_\mu+q_2 x^\mu x^\nu \partial_\mu \partial_\nu+q_3 x^2 \partial^2+\cdots\right)V_1(0)\,.
\ee
Here $q_1,q_2,q_3$ are fixed in terms of $\D_1, \Delta_n,\Delta_{n+1}$. By considering $\langle V_n(x) V_{n+1}(0) V_1(z)\rangle$ and matching with $\langle \phi^n(x)\phi^{n+1}(0)\phi(z)\rangle$ in the $\e\rightarrow 0$ limit, we will find that $\tilde f=(n+1)!+O(\e)$. Next using eq.(\ref{Vop}) we have
\be
\langle V_n(x) V_{n+1}(0) V_3(z)\rangle \approx (n+1)!|x|^{\delta}\left(1+q_1 x^\mu\partial_\mu+q_2 x^\mu x^\nu \partial_\mu \partial_\nu+q_3 x^2 \partial^2+\cdots\right)\langle V_1(0) V_3(z)\rangle\,,
\ee
where $\delta=\D_1-\D_n-\D_{n+1}.$
We have to match this in the limit $|x|\ll |z|$ with
\be
\langle \phi^n(x)\phi^{n+1}(0)\phi^3(z)\rangle \approx  (n+1)! \frac{n}{2} |x|^{-2n+2}\langle \phi^3(0)\phi^3(z)\rangle\,,
\ee
in the $\e\rightarrow 0$ limit which follows from eq.(\ref{phiop}). We consider first $n=1$ or $n\geq 4$. The key step is to match the $O(x^2)$ terms for which we need $q_3 \beta\rightarrow n/2$. Since one can show \cite{rychkovtan} that $q_3\approx (\gamma_{n+1}-\gamma_n+\gamma_1)/(16\gamma_1)$ and $\beta\sim O(\e)$, this would need $q_3$ to be singular in the $\e\rightarrow 0$ limit. This leads to $\gamma_1=O(\e^2)$. Writing $\gamma_1=\delta_1^{(2)}\e^2$ and $\gamma_n=\delta_n^{(1)}\e$ we find 
\be
(\delta_{n+1}^{(1)}-\delta_n^{(1)})\e = \frac{\beta}{6} n\,.
\ee
Using eq.(\ref{alf}), we conclude that $\beta=O(\e)$. 
One can further argue that this relation holds for $n=2,3$ as well and hence for all integer $n$. Matching $\delta_1^{(1)}=0$ and $\D_3-\D_1=2$ fixes 
\be\label{phin}
\delta_n^{(1)}=\frac{n(n-1)}{6}\,.
\ee
These agree with the Feynman diagram calculation of the anomalous dimensions of $\phi^n$ operators in the WF theory. 
In particular we have $\Delta_{4}=4+O(\e^2)$ which we will need in the next section. In order to go beyond leading order and also compute corrections to OPE coefficients, we will need to use bootstrap equations. Further applications of this technique to evaluate leading order anomalous dimensions in the $\phi^6$ theory in $2+\epsilon$ dimensions can be found in \cite{fep1}, for the Gross-Neveu model in $2+\epsilon$ dimensions can be found in \cite{fep2}, while $\phi^3$ theory in $6-\epsilon$ dimensions was examined in \cite{fep3}.

\newpage
\markboth{8\quad POLYAKOV BOOTSTRAP FROM DISPERSION RELATION}{}
\section{Polyakov Bootstrap from dispersion relation}\label{Sec:dispersionPolyakov}
In his seminal 1974 work \cite{Polyakov:1974gs}, Polyakov postulated a crossing symmetric way to solve the dynamical content of the conformal bootstrap program. In this paper, he looked at momentum space consistency conditions in the context of the leading order epsilon expansion, as well as a non-perturbative version of these conditions. In order to frame the non-perturbative conditions, Polyakov used a spectral function representation of the conformal correlator and argued that in order to have better convergence in the spectral variable, one needs to incorporate spurious double poles, corresponding to operators that are absent from the spectrum. Since crossing symmetry is in-built in this formalism, consistency conditions arise on demanding that the OPE does not include contributions from these spurious double poles. The modern incarnation of the Polyakov bootstrap was discussed in  \cite{Sen:2015doa, Gopakumar:2016wkt, Gopakumar:2016cpb, Dey:2017fab}. Mellin space was found to be suitable in understanding Polyakov's seminal paper in the language of exchange Witten diagrams. Note that these provide a convenient kinematical basis and we are not assuming any knowledge of the dual gravity theory.

\begin{figure}[b]
\centering
\includegraphics[width=0.9\textwidth]{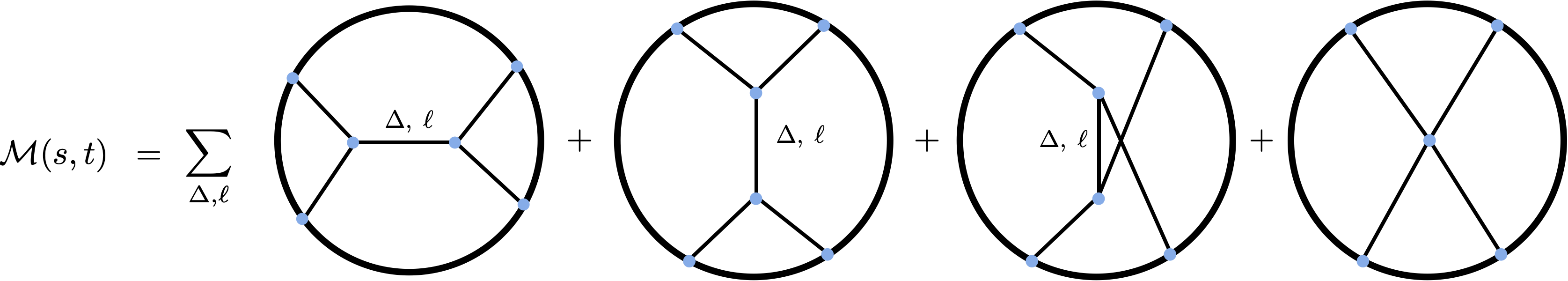}
\caption{Polyakov's 1974 \cite{Polyakov:1974gs} idea in its modern incarnation. The Mellin amplitude can be expanded in a basis of crossing symmetric AdS exchange Witten diagrams and contact diagrams. The crossing symmetric dispersion relation fixes this basis.}
    \label{fig:witttendiag_cs}
\end{figure}

We will focus on identical external scalars for which we have
\be
G(x_i)=\frac{1}{(x_{12}^2 x_{34}^2)^{\D_\phi}}{\mathcal G}(U,V)\,,
\ee
with
\be
{\mathcal G}(U,V)=\int_{-i\infty}^{i\infty}\frac{ ds dt}{(4\pi i)^2} U^{\frac{s}{2}}V^{\frac{t}{2}-\D_\phi} {\mathcal M}(s,t)\mu(s,t)\,,
\ee
where the measure factor $\mu$ is given by
\be
\mu(s,t)=\G^2(\D_\phi-\frac{s}{2})\G^2(\D_\phi-\frac{t}{2})\G^2(\D_\phi-\frac{u}{2})\,,
\ee
and 
\be
s+t+u=4\D_\phi\,.
\ee
We will call ${\mathcal M}(s,t)$ as the Mellin amplitude. 
The double poles in the measure factor, if not canceled, would correspond to operators in the spectrum with exact dimensions $\D=2\D_\phi+2n+\ell$. Since in generic non-supersymmetric CFTs, we expect operators to gain anomalous dimensions, such exact operators would be spurious. The original Polyakov conditions in \cite{Polyakov:1974gs} are then the cancellation of such contributions in the Mellin amplitude. In  \cite{Sen:2015doa, Gopakumar:2016wkt, Gopakumar:2016cpb}, $O(\e^2)$ anomalous dimension for the scalar $\phi^2$ operator, $O(\e^3)$ anomalous dimensions for the higher spin operators $\phi\partial_{a_1}\cdots \partial_{a_\ell}\phi$ as well as the corresponding OPE coefficients to one higher order in $\e$ were calculated. The anomalous dimensions were in perfect agreement with existing Feynman diagram calculations, while the OPE coefficients were new. For the stress tensor OPE, alternative arguments (see appendix B in \cite{Dey:2016mcs}) give rise to the same answer, giving credence to such calculations. Nevertheless, in spite of these successes, in \cite{Gopakumar:2018xqi}, it was realized that there are contact term ambiguities in the kinematical basis being used\footnote{The simpler case of 1d CFTs where there are no spins  was discussed in \cite{mazacpaulos1, mazacpaulos2}; for a Mellin space discussion see \cite{fgsz}.}. These ambiguities resulted in a mismatch at $O(\e^3)$ for the anomalous dimension of $\phi^2$ compared to the Feynman diagram results. Thus the question becomes how to fix such ambiguities. This requires understanding the non-perturbative existence of Mellin amplitudes.

The non-perturbative existence of Mellin amplitudes was discussed in detail in \cite{Penedones:2019tng}. The main criteria are analyticity in a sectorial domain (arg $[U]$, arg $[V]$)$\in \Theta_{CFT}$ and polynomial boundedness for ${\mathcal G}(U,V)$. By analysing ${\mathcal G}(U,V)$ in various limits and employing crossing symmetry, it was concluded that the integrals over $U,V$ leading to the definition of the Mellin amplitude, which run from $0$ to $\infty$, do not converge. It was shown that an improved convergence is possible on performing subtractions. In general, it was found that subtracting off the crossing symmetric contributions of operators with twists lying between the assumed twist-gap and the smallest twist accumulation point, would lead to convergence. Our interest is in the epsilon expansion to the first few orders, where subtracting off the disconnected contribution arising from the exchange of the identity operator is sufficient. After subtractions, one can write down fixed-$t$ dispersion relations for the Mellin amplitude, much like how one writes dispersion relations for flat space scattering amplitudes. The non-perturbative origin of the Polyakov conditions is subtle and has been explained in \cite{Penedones:2019tng, Caron-Huot:2020adz}. In \cite{Caron-Huot:2020adz}, these conditions originate from demanding consistency between dispersion relations and the $s$-channel OPE. For the epsilon expansion, the conclusion from such analyses is that we can continue to use the Polyakov conditions as discussed in \cite{Gopakumar:2016wkt, Gopakumar:2016cpb, Gopakumar:2018xqi}.

In order to make connection with Polyakov's original idea of a manifestly crossing symmetric approach, we need to start with a crossing symmetric dispersion relation. This was done in \cite{Gopakumar:2021dvg}. We will now summarize the derivation. For ease of notation, we will use
\be\label{stuvar}
s=2s_1+\frac{4\D_\phi}{3},\quad t=2s_2+\frac{4\D_\phi}{3},\quad u=2s_3+\frac{4\D_\phi}{3}\,,
\ee
so that we have $s_1+s_2+s_3=0$. Full crossing symmetry means invariance under the permutations of the $s_i$'s. In order to write a crossing symmetric dispersion relation, we use an old but forgotten idea given by Auberson and Khuri in 1972  \cite{Auberson:1972prg}. For QFT, this was resurrected in \cite{Sinha:2020win} and then developed for CFT in \cite{Gopakumar:2021dvg}. Rather than working with Mandelstam variables, we will use a different parametrization, namely
\be
s_k= a\left(1-\frac{(z_k-z)^3}{z^3-1}\right)\,,\quad {k=1,2,3}\,,
\ee
where $z_k=\exp(2\pi i (k-1)/3)$ are the cube-roots of unity. The parameter $a$ works out to be 
\be\label{adef}
a=\frac{s_1 s_2 s_3}{s_1 s_2+s_1 s_3+s_2 s_3}=\frac{y}{x}\,,{\qquad} y=-s_1 s_2 s_3\,,\quad x=-(s_1 s_2+s_1 s_3+s_2 s_3)\,,
\ee
so that in terms of the $s_i$'s, $a$ is manifestly crossing symmetric. 
The idea now is to write a dispersion relation in the variable $z$ keeping $a$ fixed. Notice that since \be s_1+s_2+s_3=0\,,\ee the above equation for fixed $a$ gives two roots for $s_2$ in terms of $s_1$, namely
\be
s_2^{\pm}=-\frac{s_1}{2}\left[1\mp (\frac{s_1+3a}{s_1-a})^{1/2}\right]\,.
\ee
In the fully crossing symmetric case of interest, both roots give the same result so we will work with $s_2^+$. 
In terms of the $z$ variable, the poles on the real $s_1$ axis gets mapped to the boundary of the disc $|z|=1$. The region where $|s_i|$'s are small is the neighborhood of $z=0$ while the Regge limit, for example, $s_1\rightarrow \infty$, keeping $s_2$ fixed gets mapped to $z\rightarrow z_2$.
In order to proceed, we have to make assumptions about the fall-off of ${\mathcal M}(s,t)$ as $|z|\rightarrow 1$. We will first assume that two subtractions suffice, i.e., ${\mathcal M}(s,t)\rightarrow o(s^2)$ for fixed-$t$.
The final form of this dispersion relation is given by \cite{Gopakumar:2021dvg}:
\be\label{disp}
{\mathcal M}(s_1,s_2)={\mathcal M}(0,0)+\frac{1}{\pi}\int_{\sigma}^\infty \frac{ds_1'}{s_1'}{\mathcal A}(s_1';s_2^+(s_1',a)) H_2(s_1'; s_1,s_2,s_3)\,,
\ee
where 
\be
H_2(s;s_1,s_2,s_3)= \left(\frac{s_1}{s-s_1}+\frac{s_2}{s-s_2}+\frac{s_3}{s-s_3}\right)\,,
\ee
is a manifestly crossing symmetric kernel. The lower limit of the integrand $\sigma$ is where the chain of poles in the Mellin variable $s_1$ starts. ${\mathcal A}(s_1;s_2)$ is the s-channel discontinuity. Since the Mellin amplitude is meromorphic, this will generally be a sum of delta functions. Denoting these poles by 
\be
\tau_k=\frac{\Delta-\ell}{2}+k-\frac{2\Dphi}{3}\,,
\ee
we can explicitly write
\be\label{poly}
{\mathcal M}(s_1,s_2)={\mathcal M}(0,0)+\sum_{\Delta,\ell,k}^\infty \frac{c_{\Delta,\ell}}{\tau_k}{\mathcal Q}_{\ell,k}^{(\Delta)}(a)H_2(\tau_k; s_1,s_2,s_3)\,,
\ee
where $c_{\Delta,\ell}={\mathcal N}_{\Delta,\ell} C_{\Delta,\ell}$, with $C_{\Delta,\ell}$'s being the OPE coefficient square, given in appendix (\ref{secConv}) and
\be
{\mathcal Q}_{\ell,k}^{(\D)}(a)=\mathcal{R}_{\D,\ell}^{(k)} P_{\Delta,\ell}(\tau_k, s_2^+(\tau_k,a))\,,
\ee
with $P_{\Delta,\ell}$'s being the Mack polynomials and $\mathcal{R}_{\D,\ell}^{(k)}$ being some normalization factors, whose explicit expressions can be found in appendix (\ref{secConv}).
We will refer to this as the Polyakov block expansion.

One can argue \cite{Gopakumar:2021dvg} that the conformal partial wave expansion converges in the neighbourhood of $a=0$ so that we can consider Taylor expanding around $a=0$. Now notice that in terms of $x,y$ defined in eq.(\ref{adef}), the kernel is
\be
H_2(s_1'; x,a)=\frac{x(2s_1'-3a)}{x a-x s_1'+ (s_1')^3}\,,
\ee
so that if we Taylor expand around $a=0$ followed by $x=0$, we will only get positive powers of $x$ and $y$. In other words, the kernel is ``local''. On the other hand, ${\mathcal A}$ is a function of $a$  and $s_1'$ only so that Taylor expanding around $a=0$ will generically lead to arbitrary powers of $a$ and hence would lead to inverse powers of $x$ in the expansion. Specifically, the form of the integrand is
\be
{\mathcal A}(s_1';s_2^+(s_1',a)) \times H_2(s_1'; x,a)=\left(\sum_{p=0}^\infty d_p  a^p\right)\times\left(\sum_{m=0}^\infty \sum_{n=0}^m a^n x^m c_{nm}\right)\,,
\ee
where $d_p, c_{nm}$ are functions of $s_1'$. Assuming the Mellin amplitude to be meromorphic, $d_p$ would be proportional to $\delta(s_1'-s_k)$ where $s_k$'s are the location of the $s_1$ poles. After integration over $s_1'$ we would get an expression with a sum over these poles.
As an example, let us consider $x^2$. We see that every term in $\left(\sum_{p=3}^\infty d_p a^p\right)$ when multiplied by $c_{02} x^2+c_{12} a x^2+ c_{22} a^2 x^2$ would lead to negative powers of $x$ after using $a=y/x$. Further for $p\leq 2$ in the sum, we would have $d_1 c_{22} a^3 x^2+ d_2 c_{12} a^3 x^2+d_2 c_{22} a^4 x^2$ which would also give negative powers of $x$. In a local theory, one should expect to see only positive powers of $x$ when expanded around $a=0$, $x=0$. Thus we expect that the sum over the Mellin poles would give a cancellation of such negative powers. Cancellation of negative powers of $x$ leads to the ``locality'' constraints  \cite{Gopakumar:2021dvg}. A non-trivial example worked out in \cite{Gopakumar:2021dvg} shows how this works for the 2d-Ising model. It was further argued in  \cite{Gopakumar:2021dvg} that these locality constraints are identical to the crossing symmetry conditions that one would impose in the fixed-$t$ dispersion relation.

In addition to these locality constraints, we impose the Polyakov conditions. For the purpose of epsilon expansion, we will use the original Polyakov conditions as discussed in \cite{Gopakumar:2018xqi}. These read
\be\label{con1}
{\mathcal M}(s_1=\frac{\Delta_\phi}{3}+p,s_2)=0\,,
\ee
and
\be\label{con2}
\partial_{s_1}{\mathcal M}(s_1=\frac{\Delta_\phi}{3}+p,s_2)=0\,.
\ee
The validity of these conditions in this context was established in \cite{Carmi:2020ekr}.
These have to hold for any integer $p\geq 0$ and for any $s_2$ inside the region shown in the figure.
\begin{figure}
\centering
\includegraphics[width=0.5\textwidth]{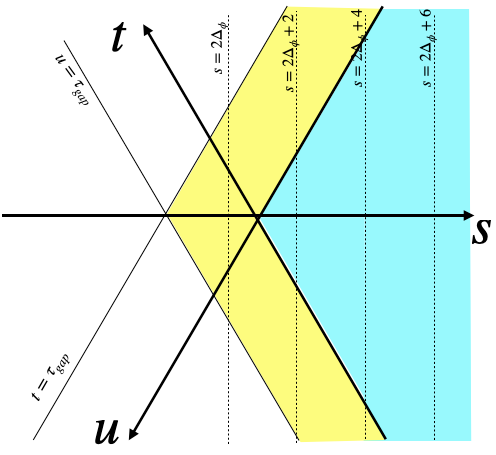}
\caption{Convergence regions in the Mellin-Mandelstam $stu$-plane eq.(\ref{stuvar}). The non-derivative condition converges in the yellow region (which overlaps with the blue region on the right). The derivative condition converges in the blue region. $\tau_{gap}$ is the minimum twist of the operator that appears in the $\phi\times\phi$ OPE. Figure adapted from \cite{Carmi:2020ekr}.}
    \label{fig:convreg}
\end{figure}

 Notice that in eq.(\ref{disp}), there is an unfixed constant ${\mathcal M}(0,0)$. Thus, to make use of the constraints eq.(\ref{con1}) we will work with subtracted equations in which this unfixed constant drops out. From a different perspective, this was also discussed in \cite{Gopakumar:2018xqi}. 
\newpage
\markboth{9\quad EPSILON EXPANSION FROM BOOTSTRAP}{}
\subsection{Witten diagram basis}
To facilitate an analytical investigation, we will need a change of basis. Namely, we will use a basis where the locality constraints are already imposed, i.e., each block in the basis will be local. This will make a connection with the Witten diagram basis expansion, originally envisaged in  \cite{Gopakumar:2016wkt, Gopakumar:2016cpb, Gopakumar:2018xqi}, with the contact term ambiguities discussed in \cite{Gopakumar:2018xqi} now fully fixed. Following \cite{Gopakumar:2021dvg}, we will now review the logic behind how this happens. Notice that each term in the Polyakov block expansion in eq.(\ref{poly}) contains ``non-local'' terms, which cancel on summing over the spectrum. The resulting cancellation conditions, which we dubbed as locality constraints, are analogous to the crossing symmetry constraints that arise in fixed-$t$ dispersion relation \cite{Penedones:2019tng} or the so-called odd-spin constraints \cite{Caron-Huot:2020adz} as shown in \cite{Gopakumar:2021dvg}. We can get the Witten diagram basis by re-expanding in a basis where the locality constraints have already been imposed. The price we pay for doing this is that, unlike the Polyakov blocks, the resulting ``Witten'' blocks will not be Regge bounded. Explicitly, we will write the Witten block expansion as
\be\label{witten}
{\mathcal M}(s_1, s_2)=\alpha_0+\sum_{\Delta,\ell,k}^\infty c_{\Delta,\ell}\left[ \sum_{i=s,t,u} M_{\Delta,\ell,k}^{(i)}(s_1,s_2)+M_{\Delta,\ell,k}^{(c)}(s_1,s_2)\right]\,,
\ee
where
\begin{eqnarray}
M_{\Delta,\ell,k}^{(s)}(s_1,s_2)&=&\mathcal{R}_{\D,\ell}^{(k)} \frac{P_{\D,\ell}(s_1,s_2)}{\tau_k-s_1}\,,\quad M_{\Delta,\ell,k}^{(t)}(s_1,s_2)=M_{\Delta,\ell,k}^{(s)}(s_2,s_3)\,,\nonumber\\ M_{\Delta,\ell,k}^{(u)}(s_1,s_2)&=&M_{\Delta,\ell,k}^{(s)}(s_3,s_1)\,,
\end{eqnarray}
and $M^{(c)}_{\Delta,\ell,k}(s_1, s_2)$ are crossing-symmetric contact terms which arise on demanding the equivalence between eq.(\ref{poly}) and eq.(\ref{witten}) after imposing the locality constraints. $\alpha_0$ is a constant that we cannot fix at this stage. A general expression for arbitrary $\ell$ for $M^{(c)}_{\Delta,\ell,k}(s_1,s_2)$ is difficult to find but explicit expressions can be worked out for a given spin. The simplest case spin-0 leads to
\be\label{sp0con}
M^{(c)}_{\Delta,0,k}=- 3\mathcal{R}_{\D,0}^{(k)} \frac{P_{\Delta,0}(0,0)}{\tau_k}\,,
\ee
with $P_{\Delta,0}(0,0)=1$, while
for spin 2, we explicitly find with $\tau_k=\frac{\Delta-2}{2}+k-\frac{2\Dphi}{3}$
\be\label{sp2con}
M^{(c)}_{\Delta,2,k}=- \mathcal{R}_{\D,2}^{(k)}\left[3 \frac{P_{\Delta,2}(0,0)}{\tau_k}+\frac{1}{4}\left(\frac{x}{\tau_k}+\frac{y}{\tau_k^2}\right)\right]\,.
\ee
For eq.(\ref{sp0con}), the sum over $k$ leads to ${}_3F_2$, which needs analytic continuation to be applicable for the epsilon expansion. Eq. (\ref{witten}) now is exactly the form used in \cite{Gopakumar:2018xqi} with the contact terms fixed. The $k$ sum in the non-zero spin contact terms is convergent and can be explicitly performed, but we will not do so here. 

\subsection{Polyakov conditions}
We will use the technology developed in \cite{Gopakumar:2018xqi} to analyse the Polyakov conditions in eq.(\ref{con1}) and eq.(\ref{con2}). In light of the rigorous results in \cite{Penedones:2019tng}, there will be some important differences. For starters, in \cite{Gopakumar:2018xqi} and previous similar discussions, the contribution of the identity operator was included in eq.(\ref{con2}). The identity operator contributes to the $p=0$ condition and the analytic findings give OPE expansion coefficients that are consistent. However, according to \cite{Penedones:2019tng}, the non-perturbative definition of the Mellin amplitude requires that this contribution be subtracted as the Mellin amplitudes corresponding to $U^{\Dphi}$ or $(U/V)^{\Dphi}$ do not exist. Nevertheless, we will follow what was done in \cite{Gopakumar:2018xqi}  and the earlier papers. One can think of the resulting condition as ensuring that the spurious $U^\Dphi$ and its crossing symmetric counterparts are absent in the expansion of the position space correlator around $U\sim 0$. Explicitly, in position space, when one expands the correlator around $U\sim 0, V\sim 1$, one does not expect powers like $U^{\Delta_\phi}$ and $U^{\Delta_\phi}\ln U$. Let us examine the conditions eq.(\ref{con1},\ref{con2}) a bit more. These conditions carry the label $s_2$. We could choose some special value of $s_2$ here and Taylor expand around this value. Alternatively, one could take a linear combination of the resulting conditions. A priori it is not clear which route is a better one. However, the epsilon expansion suggests that expanding these conditions in the basis of the so-called continuous Hahn polynomials, which are the Mellin space generalizations of the Gegenbauer polynomials, is a better thing to do. What happens here is that in the $s$-channel, up to some order in epsilon, only a single operator contributes. This is the route we will review.  The conditions in eq.(\ref{con1},\ref{con2}), after decomposing in the continuous Hahn polynomial basis, can be written as  \cite{Gopakumar:2018xqi} 
\begin{eqnarray}\label{qcond}
&&\Big(\sum_{\D,\ell}c_{\D, \ell}\big(q_{\D,\ell' | \ell}^{(s)}(s_1)+2 q_{\D,\ell' |\ell}^{(t)}(s_1)\big)+\sum_{n,m} a_{n,m}q^{(c)}_{n,m}(s_1)\Big) |_{s_1=\frac{\D_\phi}{3}+r}=0\,,\\
&&\Big( \sum_{\D, \ell}c_{\D, \ell}\partial_s\big(q_{\D,\ell' | \ell}^{(s)}(s_1)+2 q_{\D, \ell' | \ell}^{(t)}(s_1)\big) +2 \partial_s q_{\ell'}^{diss}(s_1)+\sum_{n,m} a_{n,m}\partial_{s_1} q^{(c)}_{n,m}(s_1)\Big)|_{s_1=\frac{\D_\phi}{3}+r}=0\,, \nonumber
\end{eqnarray}
where explicit expressions for $q_{\D,\ell' | \ell}^{(s,t)}(s_1)$ etc can be found in appendix (\ref{seclen}). For obvious reasons, we will refer to the first set of conditions as the non-derivative conditions and the second set as derivative conditions. $q^{diss}$ is the contribution from the identity operator and enters in only the $r=0$ derivative condition. Here $a_{0,0}$ will depend on $\alpha_0$ in eq.(\ref{witten}). Since $\alpha_0$ cannot be fixed at this stage, solving these equations will require us to work with subtracted equations where the contribution of both $\alpha_0$ as well as the spin-0 contact term in eq.(\ref{sp2con}) cancels out. In the next section, we will see how the Wilson-Fisher $\epsilon$-expansion results are produced using these equations.

\section{Epsilon expansion from bootstrap}\label{Sec:epsilonbootstrap}
We begin by reproducing the anomalous dimension results in Wilson-Kogut \cite{Wilson:1973jj} and summarise the new results for the OPE coefficients.
\subsection{Summary of steps}
For the epsilon expansion, we will put $d=4-\epsilon$ in the equations (\ref{qcond}) and solve these equations order by order in epsilon. We will focus on extracting the anomalous dimensions and OPE coefficients of the scalar and spin-2 operators. To have any hope of extracting useful analytic information, at least to the first few orders, we should be in a lucky situation where only a finite number of operators contribute to eq.(\ref{qcond}). In eq.(\ref{witten}), $c_{\Delta,\ell}={\mathcal N}_{\Delta,\ell} C_{\Delta,\ell}$, where the normalization ${\mathcal N}_{\Delta,\ell}$ is defined in appendix \ref{secConv} and $C_{\Delta,\ell}$ are the OPE coefficients defined so that near $U\sim 0, V\sim 1$, the contribution of an operator with conformal dimension $\Delta$ and spin $\ell$ gives $C_{\Delta,\ell}U^{\frac{\Delta-\ell}{2}}(1-V)^\ell$.  Now ${\mathcal N}_{\Delta,\ell}$ has double zeros at the location of MFT operators with dimensions $\Delta=2\Dphi+2n+\ell$. This observation will be crucial in calculating the first few orders in the epsilon expansion. To make progress, we will make the following assumptions:
\begin{enumerate}
\item We will assume the existence of a unique spin-2 stress tensor whose dimension is $\Delta_2=4-\epsilon$. 
\item We will assume that there are unique double field operators of leading dimension $\Delta_\ell=2\Dphi+\ell+O(\e)$ and $\Dphi=1+O(\epsilon)$. For convenience, we will denote the OPE coefficient squared of these operators by $C_\ell$.
\item The twist $\geq 4$ operators have their OPE coefficient squared beginning at $O(\epsilon^2)$. 
\end{enumerate}
Let us parametrize
\begin{eqnarray}
\Delta_\phi &=& 1+\sum_{n=1}^\infty \delta_\phi^{(n)}\e^n\,,\quad \Delta_0=2+\sum_{n=1}^\infty \delta_0^{(n)}\epsilon\,,\\
C_\ell &=&\sum_{n=0}^\infty C_\ell^{(n)}\epsilon^n\,, \quad \ell \geq 0.
\end{eqnarray}
In addition to these operators, we will also need the unique twist-4 scalar primary $\phi^4$ to solve the equations consistently. At the order we will be interested in, and with the equations we will use, the higher spin, twist-4 operators, and higher than twist-4 operators do not contribute. We will parametrise the $\phi^4$ operator by
\be
\Delta_{\phi^4}=4+\sum_{n=1}^\infty\delta_{\phi^4}^{(n)}\epsilon^n+O(\epsilon^3)\,,\quad C_{\phi^4}=\sum_{n=0}^\infty C_{\phi^4}^{(n)}\epsilon^{2+n}\,.
\ee
We will take $\ell'=0,2$ derivative conditions at $r=0,1$. To remove the $\alpha_0$ dependence in the non-derivative conditions, we will subtract the $\ell'=0,r=0$ equation from every non-derivative equation. We will work with $\ell'=0$, $r=1$  subtracted conditions and $\ell'=2$, $r=0,1$ equations. The explicit expressions for contact terms lead to the conclusion that they will contribute at a higher order in epsilon.
Expanding these equations to $O(\epsilon^3)$, in addition to the free theory solution, we can solve order by order to find  the following:
\begin{eqnarray}
\Delta_\phi&=&1-\frac{\e}{2}+\frac{\e^2}{108}+(109+72\delta_{\phi^4}^{(1)})\frac{\e^3}{11664}+O(\e^4)\,,\\
\Delta_0&=&2-\frac{2\e}{3}+(19+18\delta_{\phi^4}^{(1)})\frac{\e^2}{162}+O(\e^3)\,,\\
C_0 &=&2-\frac{2\e}{3}-2(17+9\delta_{\phi^4}^{(1)})\frac{\e^2}{81}+\frac{\e^3}{486}(-\frac{95}{6}+12\delta_{\phi^4}^{(1)}-972\delta_{0}^{(3)}+63 \zeta(3))\,,\\
C_2&=&\frac{1}{3}-\frac{11\e}{36}+\frac{37\e^2}{486}+\frac{451\e^3}{52488}+O(\e^4)\,,\\
C_{\phi^4}&=&\frac{1}{54}+O(\e)\,.
\end{eqnarray}
Thus we can solve consistently in terms of one unknown, which we have parametrised in terms of the $\phi^4$ anomalous dimension. Curiously the stress tensor OPE coefficient squared, $C_2$, has no dependence on the $\phi^4$ anomalous dimension up to this order.  The $O(\e^3)$ OPE coefficient for $C_2$ was first computed in \cite{Gopakumar:2016wkt, Gopakumar:2016cpb} and an independent, indirect consistency check based on diagrammatic calculations was carried out in \cite{Dey:2016mcs}. 
Using the CFT derivation of $\delta_{\phi^4}^{(1)}=0$ which follows from eq.(\ref{phin}), we find agreement with the Wilson-Fisher results, with new results for OPE coefficients. This gives a rigorous derivation of the $\e$-expansion results of \cite{Gopakumar:2016wkt, Gopakumar:2016cpb, Gopakumar:2018xqi}. In these papers, the $O(\e^3)$ anomalous dimensions of the double field operators of general even spins were derived and found to perfectly agree with the diagrammatic approach. The $O(\e^3)$ OPE coefficients were also derived, which were new. 
Defining the central charge $c_T$ as
\be
c_T=\frac{d^2\Dphi^2}{(d-1)^2 C_2}\,,
\ee
we find  \cite{Gopakumar:2016wkt, Gopakumar:2016cpb}
\be \label{cT}
\frac{c_T}{c_{free}}=1-\frac{5\e^2}{324}-\frac{233\e^3}{8748}+O(\e^4)\,.
\ee
Including the $O(\e^3)$ result improves the agreement with the numerical bootstrap result of \cite{fracnum} as indicated in the figure. 

\begin{figure}
\centering
\includegraphics[width=0.5\textwidth]{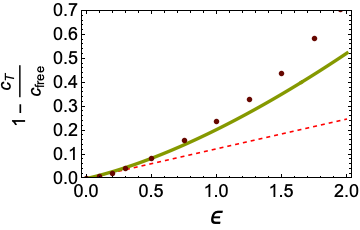}
\caption{Behaviour of $c_T$ in fractional dimensions. The solid points are numerical bootstrap data \cite{fracnum}. The red dashed line is the $O(\e^2)$ result while the green solid line is the $O(\e^3)$ result. Figure adapted from \cite{Gopakumar:2016cpb}. }
    \label{fig:convreg}
\end{figure}

Note that to compute the $O(\e^3)$ term in $C_0$ we would need $\delta_0^{(3)}$.  This has not been possible so far using bootstrap techniques. It is also possible to show that the at $O(\e)$, $C_{\phi^4}=\frac{1}{54}-\frac{47}{1458}\e$ using the $\ell'=0, r=1$ derivative condition, where no other operator information, apart from those listed above, is needed for this calculation. Using similar techniques, one can further derive the anomalous dimension of the unique spin-2, twist-4 operator and its OPE coefficient \cite{Gopakumar:2018xqi} using similar techniques.
\subsection{Bootstrap results at $O(\e^4)$ and open questions}
In this section, we will summarize what results have been possible at $O(\e^4)$. In \cite{Alday:2017zzv}, it was observed that if one assumed ``pure transcendentality'' of the position space basis of the conformal correlator, then the known $O(\e^4)$ anomalous dimensions for the ${\mathcal O}_\ell$ operators were reproduced. Using this and assuming the $O(\e^3)$ anomalous dimension of $\phi^2$, the $O(\e^4)$ contribution in eq.(\ref{cT}) was shown to be
\be \label{cT4}
\frac{c_T}{c_{free}}{\bigg |}_{\e^4}=-\left(\frac{100651}{3779136}-\frac{55}{2916}\zeta_3\right)\,.
\ee
It will be remarkable if the same assumptions can calculate higher-order anomalous dimensions and OPE coefficients. 
In  \cite{Carmi:2020ekr}, the OPE coefficient result for $\phi^2$ at $O(\e^4)$ is worked out. This needs the $\delta_0^{(3)}, \delta_0^{(4)}$ as well as certain information about twist-4 operators. One finds
\be
C_0=2-\frac{2\e}{3}-\frac{34}{81}\e^2+\frac{1863\zeta(3)-611}{4374}\e^3+\chi_4 \e^4+O(\e^5)\,,
\ee
where $\chi_4\approx -0.656398$ was determined numerically. It was found that putting $\e=1$ made the agreement with the 3d-Ising model worse. This is expected since the $\epsilon$ expansion is asymptotic, as we have pointed out earlier. In the same paper, averaged OPE coefficients for twist-4 operators up to $O(\e^3)$ have also been derived numerically using sum rules arising from the Polyakov conditions.

\subsubsection*{Open questions}
\begin{enumerate}
\item
 As we have explained above, in addition to the Polyakov conditions, there are also the so-called locality constraints (null or odd-spin constraints). No systematic study has been carried out about what role these constraints play in the analytic bootstrap. This appears to be an immediate problem to address. In \cite{Gopakumar:2021dvg}, these locality constraints played an important role in establishing the equivalence between the sum rules in \cite{Penedones:2019tng, Caron-Huot:2020adz}. Preliminary analysis at $O(\e^2)$ carried out in \cite{Gopakumar:2021dvg} showed that these results are consistent with the locality constraints. However, it is unclear if these constraints can be used to yield additional analytic information. 
\item
An important open question is to derive the ``pure transcendentality'' ansatz used in \cite{Alday:2017zzv} and understand this in Mellin space. This may point the way forward to connect with higher-order results without resorting to studying mixed correlators\footnote{The dispersion relation technology for mixed correlators is not well developed. See the recent work \cite{Trinh:2021mll} for the fixed-$t$ dispersive sum rules.}. The position space dispersion relation considered in \cite{Carmi:2019cub, Bissi:2019kkx} may be a good starting point to examine this ansatz.
\item 
The crossing symmetric dispersion relation also connects with a fascinating area of mathematics called Geometric Function Theory \cite{Haldar:2021rri, Raman:2021pkf}. One of the primary outcomes of the S-matrix bootstrap is concerned is a derivation of two-sided bounds on the Taylor expansion coefficients of the flat space scattering amplitudes at low energies. These bounds arise from the so-called Bieberbach-Rogosinski type inequalities for typically real univalent functions. It will be fascinating to study these in the context of CFTs.

\item A systematic study of higher-order $O(1/N)$ results in the context of $O(N)$ models should be possible using the same framework. Attempts using the Polyakov bootstrap were made in \cite{Dey:2016mcs}. Further progress using large spin perturbation theory was made in \cite{Alday:2019clp}. Some preliminary attempts using the crossing symmetric dispersion was made in \cite{Kaviraj:2021cvq}. Unlike the epsilon expansion, going to the second subleading order in $1/N$ is challenging since one needs to resum the contribution of an infinite number of operators. A recent review on critical $O(N)$ CFTs is \cite{Henriksson:2022rnm}. $O(M)\times O(N)$ critical CFTs have been reviewed in \cite{Henriksson:2020fqi} while CFTs with $MN$ global symmetry has been reviewed in \cite{Henriksson:2021lwn}.

\item On a conceptual level, these initial successes in analytically producing the first few orders in perturbation theory (and the related success of producing the OPE coefficients, which are otherwise hard in the diagrammatic approach) still leave a lot to be explored. The immediate criticism is that the Feynman diagram approach, at least in principle, gives an algorithm to compute up to very high orders in perturbation theory. While this may be difficult in practice, the algorithm is well understood and is implementable. The same level of understanding has not been reached with the analytic bootstrap methods so far. At a technical level, if the orthogonality of the Mack polynomials was understood, it may facilitate progress in this direction.
\end{enumerate}

\newpage
\markboth{10\quad BOOTSTRAPPING TREE-LEVEL CORRELATORS: POSITION SPACE}{}
\section{Bootstrapping tree-level correlators: Position space}\label{Sec:PositionSpace}
Correlation functions in superconformal CFTs are mapped to scattering amplitudes in AdS by the AdS/CFT correspondence. They are the most basic observables in the theory, from which we can extract various data. However, the standard diagrammatic expansion method for computing amplitudes is accompanied by enormous technical difficulties, making it extremely cumbersome to use in practice.\footnote{See, {\it e.g.}, \cite{DHoker:1999pj,Arutyunov:2000py,Arutyunov:2002ff,Arutyunov:2002fh,Arutyunov:2003ae} for early progress using this approach.} A better strategy is to use symmetries and consistency conditions to fix the correlators, {\it i.e.}, to {\it bootstrap} them. Such a bootstrap strategy does not rely on the explicit details of the effective Lagrangians, which are usually very complicated, and works directly with the ``on-shell'' observables. This makes the bootstrap methods   extremely efficient, and has produced many impressive results which are impossible to obtain by using the traditional strategy. Moreover, the bootstrap methods also help to manifest remarkable hidden simplicity of holographic correlators which is obscured by the diagrammatic expansion. We will review some of these bootstrap methods and their results in Section \ref{Sec:PositionSpace}-\ref{Sec:loops}.

Our main focus in this review is maximally superconformal theories, {\it i.e.}, theories with sixteen Poincar\'e supercharges. Non-maximally superconformal theories will be briefly dicussed in Section \ref{Subsec:supergluons}. For $d>2$, there are only three such maximally superconformal  theories
\begin{itemize}
\item 4d $\mathcal{N}=4$ super Yang-Mills (SYM) theory, dual to IIB string theory in $AdS_5\times S^5$, with a superconformal symmetry group $PSU(2,2|4)$.
\item 3d $\mathcal{N}=8$ Aharony-Bergman-Jafferis-Maldacena (ABJM) theory \cite{Aharony:2008ug}, dual to M-theory in $AdS_4\times S^7$, with a superconformal symmetry group $OSp(8|4)$.
\item 6d $\mathcal{N}=(2,0)$ theory, dual to M-theory in $AdS_7\times S^4$, with a superconformal symmetry group $OSp(8^*|4)$.
\end{itemize}
We will focus on the supergravity limit, where the central charge of the theory is taken to be large. In the case of string theory, the string length needs to be further set to zero. In this limit, all the single-particle states in AdS are supergravity fields, and belong to the so-called $\frac{1}{2}$-BPS multiplets of the superconformal group which are annihilated by half of the supercharges.\footnote{There are other types of multiplets in the theory, but they appear as ``bound states'' in the supergravity limit. For classifications of superconformal multiplets, see, {\it e.g.}, \cite{Dolan:2002zh,Buican:2016hpb,Cordova:2016emh}.} We will focus on the correlators of the superconformal primaries of these $\frac{1}{2}$-BPS multiplets, which are the {\it super gravitons}. The method which we will present in this section applies to both the 4d and 6d theories, and serves as a proof of principle for the bootstrap strategy. The method relies on special properties of the spectrum as we will explain at the end of Section \ref{Subsec:positionspace}. A more powerful method which applies to all three theories will be reviewed in Section \ref{Sec:MellinSpace}. Theories with non-maximal superconformal symmetry will be discussed in Section \ref{Subsec:supergluons}, where we will focus on the correlators of AdS {\it super gluons}.

This section is organized as follows. In Section \ref{Subsec:scfkinematics} we set the stage by discussing the superconformal kinematics of four-point correlators. This discussion will also be used in later sections on holographic correlators. In Section \ref{Subsec:positionspace} we review a bootstrap method for computing four-point functions in position space. In Section \ref{Subsec:5ptfunctions} we review how a similar  strategy can be used to compute five-point functions.

\subsection{Superconformal kinematics}\label{Subsec:scfkinematics}
The  R-symmetry groups of the three maximally superconformal theories all have the form of $SO(\mathtt{d})$, as listed below 
\begin{eqnarray}
\nonumber &&d=3:\;\; \mathtt{d}=8\;,\;\; SO(8)\;,\\
&&d=4:\;\; \mathtt{d}=6\;,\;\; SO(6)\simeq SU(4)\;,\\
\nonumber &&d=6:\;\; \mathtt{d}=5\;,\;\; SO(5)\simeq Sp(4)\;.
\end{eqnarray} 
We focus on the $\frac{1}{2}$-BPS operators which are the superconformal primaries of the $\frac{1}{2}$-BPS multiplets. They are labelled by an integer $k=2,3,\ldots$, and transform in the rank-$k$ symmetric traceless representation of the $SO(\mathtt{d})$ R-symmetry group. The $k=2$ multiplet is special because the multiplet contains the stress tensor operator as a superconformal descendant, and the multiplet is known as the stress tensor multiplet. Thanks to superconformal symmetry, the conformal dimensions of the $\frac{1}{2}$-BPS operators are determined by their R-symmetry representations
\begin{equation}
\Delta_k=\epsilon k
\end{equation}
where\footnote{The $\epsilon$ we defined here should not be confused with the $\epsilon$ used in the epsilon expansion sections.} 
\begin{equation}
\epsilon=\frac{d-2}{2}\;.
\end{equation}
We will denote these operators as $\mathcal{O}^{I_1\ldots I_k}(x)
$ where $I_i=1,\ldots \mathtt{d}$ are vector indices of $SO(\mathtt{d})$.  However, it is more convenient to keep track of the R-symmetry indices by contracting them with auxiliary {\it null} vectors $t_i^I$
\begin{equation}
\mathcal{O}_k(x,t)=\mathcal{O}^{I_1\ldots I_k}(x)t_{I_1}\ldots t_{I_k}\;.
\end{equation}
Here null means that the vectors $t_i^I$ satisfy 
\begin{equation}
t_i\cdot t_i=0\;,
\end{equation}
and this property ensures that the R-symmetry indices are projected to the symmetric traceless representation. The four-point functions 
\begin{equation}
G_{k_1k_2k_3k_4}(x_i,t_i)=\langle \mathcal{O}_{k_1}(x_1,t_1) \mathcal{O}_{k_2}(x_2,t_2)  \mathcal{O}_{k_3}(x_3,t_3)  \mathcal{O}_{k_4}(x_4,t_4) \rangle
\end{equation}
are then functions not only of the spacetime coordinates $x_i$ but also of the internal coordinates $t_i$. It is also clear that $G_{k_1k_2k_3k_4}$ can only depend on $t_i$ as a polynomial of the $SO(\mathtt{d})$ invariants $t_{ij}=t_i\cdot t_j$, and satisfies the following relation
\begin{equation}
G_{k_1k_2k_3k_4}(x_i,\xi_i t_i)=\prod_i \zeta_i^{k_i} G_{k_1k_2k_3k_4}(x_i,t_i)\;,
\end{equation}
under independent rescaling of each null polarization vector. 

Superconformal symmetry imposes strong constraints on the form of the correlators. Let us first consider the consequence of covariance under the bosonic subgroups, namely, the conformal group $SO(d+1,1)$ and the R-symmetry group $SO(\mathtt{d})$. The bosonic symmetries allow us to write the four-point functions as functions of two pairs of cross ratios. To make this statement more precise, we can consider without loss of generality, the following ordering of external weights $k_1\leq k_2\leq k_3\leq k_4$. We further need to distinguish two cases 
\begin{equation}\label{twocases}
k_1+k_4\geq k_2+k_3 \;\; \text{(case I)}\;,\quad\quad\;\; k_1+k_4< k_2+k_3 \;\; \text{(case II)}\;.
\end{equation}
We can extract a kinematic factor as follows
\begin{equation}\label{GandcalG}
G_{k_1k_2k_3k_4}(x_i,t_i)=\prod_{i<j}\left(\frac{t_{ij}}{x_{ij}^{2\epsilon}}\right)^{\gamma^0_{ij}}\left(\frac{t_{12}t_{34}}{x_{12}^{2\epsilon}x_{34}^{2\epsilon}}\right)^{\mathcal{E}}\mathcal{G}_{k_1k_2k_3k_4}(U,V;\sigma,\tau)\;.
\end{equation}
This factor takes care of the covariant transformation properties. Then the correlator becomes a function of the invariant cross ratios
\begin{equation}
U=\frac{x_{12}^2x_{34}^2}{x_{13}^2x_{24}^2}\;,\;\;\quad V=\frac{x_{14}^2x_{23}^2}{x_{13}^2x_{24}^2}\;,\;\;\quad \sigma=\frac{t_{13}t_{24}}{t_{12}t_{34}}\;,\;\;\quad\tau=\frac{t_{14}t_{23}}{t_{12}t_{34}}\;.
\end{equation}
Here $x_{ij}=x_i-x_j$, and $\mathcal{E}$ is defined to be the {\it extremality}
\begin{equation}
\mathcal{E}=\frac{k_1+k_2+k_3-k_4}{2} \;\;\; \text{(case I)}\;,\;\;\quad\quad \mathcal{E}=k_1 \;\;\; \text{(case II)}\;,
\end{equation}
which measures the complexity of the correlator. The exponents in the factor of (\ref{GandcalG}) are given by
\begin{eqnarray}
&&\gamma_{12}^0=\gamma_{13}^0=0\;,\;\; \gamma_{34}^0=\frac{\kappa_s}{2}\;,\; \; \gamma_{24}^0=\frac{\kappa_u}{2}\;,\\
\nonumber && \gamma_{14}^0=\frac{\kappa_t}{2}\,,\;\; \gamma_{23}^0=0\;\;\text{(I)}\;,\;\;\;\;\gamma_{14}^0=0\,,\;\; \gamma_{23}^0=\frac{\kappa_t}{2}\;\;\text{(II)}\;,
\end{eqnarray}
where
\begin{equation}\label{kappastu}
\kappa_s\equiv|k_3+k_4-k_1-k_2|\;,\;\; \kappa_t\equiv|k_1+k_4-k_2-k_3|\;,\;\; \kappa_u\equiv|k_2+k_4-k_1-k_3|\;.
\end{equation}
Recall that the null vectors appear in $G_{k_1k_2k_3k_4}$ as a polynomial of $t_{ij}$. The kinematic factor extracted in (\ref{GandcalG}) then ensures that $\mathcal{G}_{k_1k_2k_3k_4}$ is a degree-$\mathcal{E}$ polynomial in $\sigma$ and $\tau$. 

This is as far as we can go by exploiting only the bosonic subgroups. However, the fermionic generators of the superconformal group impose additional constraints and relate the dependence on the conformal and R-symmetry cross ratios. These constraints are known as the {\it superconformal Ward identities}, and take the same form in different spacetime dimensions \cite{Dolan:2004mu}
\begin{equation}\label{scfWardid}
(z\partial_z-\epsilon \alpha\partial_\alpha)\mathcal{G}(z,\bar{z};\alpha,\bar{\alpha})\big|_{\alpha=1/z}=0\;. 
\end{equation}
Here we have made a convenient change of variables 
\begin{equation}
U=z \bar{z}\;,\quad V=(1-z)(1-\bar{z})\;,\quad \sigma=\alpha\bar{\alpha}\;,\quad \tau=(1-\alpha)(1-\bar{\alpha})\;.
\end{equation}
Moreover, it is understood that in (\ref{scfWardid}) we need to first act on $\mathcal{G}(z,\bar{z};\alpha,\bar{\alpha})$ with the differential operator before setting $\alpha=1/z$. We also note that $\mathcal{G}(z,\bar{z};\alpha,\bar{\alpha})$ is invariant separately under $z\leftrightarrow\bar{z}$ and $\alpha\leftrightarrow\bar{\alpha}$. Therefore we can make these replacements in (\ref{scfWardid}) and obtain in total a set of four identities. While the superconformal Ward identities (\ref{scfWardid}) take a universal form for all spacetime dimensions, their solutions can look drastically different. It was shown in  \cite{Dolan:2004mu} that the solutions generally involve differential operators which are determined by superconformal symmetry. The solutions are simpler for even $d$, while for odd $d$ the differential operators are non-local. The case of $d=4$ is particularly simple where the differential operator becomes an algebraic factor.

Let us notice that the superconformal Ward identities (\ref{scfWardid}) have amusing implications. For $d=4$, it is clear that  (\ref{scfWardid}) implies that the correlator with the special R-symmetry configuration $\bar{\alpha}=1/\bar{z}$ is independent of $\bar{z}$
\begin{equation}\label{4dtwisted}
\partial_{\bar{z}}\mathcal{G}^{({\rm 4d})}(z,\bar{z};\alpha,1/\bar{z})=0\;.
\end{equation}
Further setting $\alpha=1/z$ implies that the correlator $\mathcal{G}^{({\rm 4d})}(z,\bar{z};1/z,1/\bar{z})$ is topological 
\begin{equation}
\partial_z\mathcal{G}^{({\rm 4d})}(z,\bar{z};1/z,1/\bar{z})=\partial_{\bar{z}}\mathcal{G}^{({\rm 4d})}(z,\bar{z};1/z,1/\bar{z})=0\;.
\end{equation}
Similarly, setting $\alpha=\bar{\alpha}=1/\bar{z}$  for $d=6$ leads to
\begin{equation}
\partial_{\bar{z}}\mathcal{G}^{({\rm 6d})}(z,\bar{z};1/\bar{z},1/\bar{z})=0\;,
\end{equation}
and shows that $\mathcal{G}^{({\rm 6d})}(z,\bar{z};1/\bar{z},1/\bar{z})$ also becomes holomorphic. For $d=3$, we need to set instead $z=\bar{z}=1/\alpha$ and this gives
\begin{equation}
\partial_{z}\mathcal{G}^{({\rm 3d})}(z,z;\alpha,1/z)=0\;.
\end{equation}
This equation tells us the correlator $\mathcal{G}^{({\rm 3d})}(z,z;\alpha,1/z)$ is topological on the $z=\bar{z}$ slice which corresponds to the configuration of all four operators on a line. These interesting holomorphic and topological properties can be understood more generally from the chiral algebra construction \cite{bllprv13,brv14} and the topological twisting \cite{Drukker:2009sf,Chester:2014mea,Beem:2016cbd}. In these constructions, certain protected operators are restricted to a two dimensional complex plane or a one dimensional line (except for one twisting introduced in \cite{Drukker:2009sf}).  Moreover, the operators are required to have special position-dependent R-symmetry polarizations known as {\it twists}, which corresponds to setting $\bar{\alpha}=1/\bar{z}$, {\it etc} in the four-point correlator. Such configurations of twisted operators preserve a certain amount of supersymmetry, which renders the correlation functions holomorphic or topological. Note that in general these conditions are weaker than the superconformal Ward identities (\ref{scfWardid}) and follow from them as corollaries. However, they generalize easily to arbitrary $n$-point correlators with $n\geq 5$, where the superconformal Ward identities have not been obtained in the literature.\footnote{In the simpler setup of 1d SCFTs, an interesting conjecture of their multi-point generalizations based on perturbative calculations was given in \cite{Barrat:2021tpn}.} We will see in a five-point function example in Section \ref{Subsec:5ptfunctions} that the conditions following from these constructions still impose sufficiently nontrivial constraints.

\subsection{The position space method}\label{Subsec:positionspace}
Traditionally, holographic correlators are computed by using the standard method of Feynman diagrams (which are known as the Witten diagrams in AdS). This procedure requires us to first expand the AdS supergravity effective action to a certain order to obtain the needed vertices (cubic and quartic in the case of tree-level four-point functions). Using these vertices and AdS propagators one constructs all possible Witten diagrams from these vertices, and computes the correlator as the sum of these diagrams. Unfortunately, this seemingly straightforward algorithm is very cumbersome to use. In particular, expanding the effective action to derive vertices is an enormously complicated task. The only case where the general quartic vertices have been worked out is IIB supergravity on $AdS_5\times S^5$, and the results occupied 15 pages \cite{Arutyunov:1999fb}. Moreover, the diagrams involved proliferate as we consider correlators with higher external weights, which soon exceeds our practical computational power. Meanwhile, we should notice that this brute force algorithm makes no use of the large amount of symmetry in the theories. In particular, the superconformal Ward identities (\ref{scfWardid}) were never  exploited in this procedure. These identities appear to be highly nontrivial from the diagrammatic expansion perspective because they are not satisfied by each individual diagram. Therefore, a natural question is whether one can use superconformal symmetry to facilitate the computation of holographic correlators so that we do not have to go through all the steps. 

Concretely, to implement this idea we can proceed as follows. We start with an ansatz which is the linear combination of all possible exchange and contact diagrams. However, instead of using the correct coefficient for each diagram, which would be calculable with the precise vertices, we will leave the coefficients as unfixed parameters. We then impose the superconformal Ward identities, and see if all coefficients get fixed. This always turns out to be possible (up to an overall coefficient which we can fix by using protected CFT data), and gives a unique answer for the holographic correlator. The use of superconformal symmetry greatly improves the traditional algorithm, as it bypasses the most difficult step of expanding the complicated supergravity effective action. This strategy was first proposed in \cite{Rastelli:2016nze,Rastelli:2017udc}, where it was dubbed the {\it position space method}.

Let us demonstrate this method by computing the stress tensor multiplet ($k_i=2$) four-point function in $AdS_5\times S^5$ IIB supergravity. 
The ansatz
\begin{equation}\label{positionansatz}
\mathcal{G}_{\rm anstaz}(U,V;\sigma,\tau)=\mathcal{A}_s+ \mathcal{A}_t+\mathcal{A}_u+\mathcal{A}_{\rm con}
\end{equation}
is separated into exchange contributions and a contact contribution. In each channel, there are only three fields that can be exchanged: the super graviton $s$ itself with $\Delta=2$ and in the $\mathbf{20}'$ representation of $SO(6)$, a spin-1 graviphoton field $V_\mu$ with $\Delta=3$ and in $\mathbf{15}$ of the R-symmetry group, and a spin-2 graviton field $\varphi_{\mu\nu}$ with $\Delta=4$ and singlet under R-symmetry. This comes as a result of the selection rules and the fact that all the supergravity fields reside in $\frac{1}{2}$-BPS multiplets, as we will explain in more detail towards the end of this subsection. In particular, these three exchanged fields all live in the same stress tensor multiplet (see Appendix \ref{App:multipletcomponents} for details). Therefore, the s-channel exchange is given by the following linear combination
\begin{equation}
\mathcal{A}_s=\lambda_s Y_{\mathbf{20}'}(\sigma,\tau) \mathcal{W}_{2,0}(U,V)+\lambda_v Y_{\mathbf{15}}(\sigma,\tau) \mathcal{W}_{3,1}(U,V)+\lambda_g Y_{\mathbf{1}}(\sigma,\tau) \mathcal{W}_{4,2}(U,V)
\end{equation}
where $\lambda_s$, $\lambda_v$, $\lambda_g$ are parameters to be fixed. Here $Y_{\mathbf{R}}(\sigma,\tau)$ are the R-symmetry polynomials characterizing the exchanged R-symmetry representation
\begin{equation}
Y_{\mathbf{1}}(\sigma,\tau)=1\;,\quad Y_{\mathbf{15}}(\sigma,\tau)=\sigma-\tau\;,\quad Y_{\mathbf{20}'}(\sigma,\tau)=\sigma+\tau-\frac{1}{3}\;.
\end{equation}
They can be obtained from solving the R-symmetry two-particle quadratic Casimir equation, and are the analogue of conformal blocks. Moreover, $\mathcal{W}_{\Delta,\ell}$ are the exchange Witten diagrams. In general, these exchange diagrams are difficult to evaluate in position space in terms of simple functions. However, it is pointed out in \cite{DHoker:1999mqo} that when the conformal twist, {\it i.e.}, $\Delta-\ell$, of the exchanged field satisfies 
\begin{equation}\label{spectrumcondition}
\Delta_1+\Delta_2-(\Delta-\ell)=2\mathbb{Z}_+\;,\quad \text{or}\quad \Delta_3+\Delta_4-(\Delta-\ell)=2\mathbb{Z}_+\;,
\end{equation}
the exchange Witten diagrams can be expressed as a {\it finite} sum of the $D$-functions (or $\bar{D}$-functions when written as functions of cross ratios, see (\ref{dbar}) for the definition). A $D$-function $D_{\Delta_1\Delta_2\Delta_3\Delta_4}$ represents a contact diagram where the external operators have dimensions $\Delta_i$. Using the formulae in \cite{DHoker:1999mqo} (which will also be reviewed in Appendix \ref{App:WittenDiagrams}), we have 
\begin{eqnarray}
\nonumber\mathcal{W}_{2,0}&=&\frac{\pi^2}{8}U\bar{D}_{1122}\;,\\
\mathcal{W}_{3,1}&=&\frac{\pi^2}{8}U(\bar{D}_{1223}-\bar{D}_{2123}+\bar{D}_{2132}-V\bar{D}_{1232})\;,\\
\nonumber \mathcal{W}_{4,2}&=&\frac{2\pi^2}{3}U(3(\bar{D}_{2123}+\bar{D}_{2132}-\bar{D}_{3133})-2\bar{D}_{1122})\;.
\end{eqnarray}
The other two channels are related to the s-channel by crossing symmetry
\begin{equation}\label{Atucrossing}
\begin{split}
\mathcal{A}_t(U,V;\sigma,\tau)=&\big(\frac{U\tau}{V}\big)^2\mathcal{A}_s(V,U;\sigma/\tau,1/\tau)\;,\\
\mathcal{A}_u(U,V;\sigma,\tau)=&(U\sigma)^2\mathcal{A}_s(1/U,V/U;1/\sigma,\tau/\sigma)
\end{split}
\end{equation}
where $\bar{D}$-functions are mapped to themselves under crossing with explicit relations given in (\ref{Dbarpermutation}). Finally, the contact part $\mathcal{A}_{\rm con}$ contains at most two contracted derivatives and all possible R-symmetry structures. The two-derivative requirement is to match the two-derivative structure of the supergravity action.\footnote{While this expectation is clear in flat space, it is less obvious in AdS. Superficially, the effective action contains four-derivatives terms as well \cite{Arutyunov:1999fb}. However, these terms cancel and there are no intrinsic four-derivative interactions \cite{Arutyunov:2017dti}.} It is convenient to also write it as the sum of three channels 
\begin{equation}\label{Acons}
\mathcal{A}_{\rm con}=\mathcal{A}_{{\rm con},s}+\mathcal{A}_{{\rm con},t}+\mathcal{A}_{{\rm con},u}
\end{equation}
with 
\begin{equation}
\mathcal{A}_{{\rm con},s}=\underbrace{\big(\!\!\sum_{0\leq a+b\leq 2} \!\!\! c_{ab}\sigma^a\tau^b\big)2\pi^2U^2(2\bar{D}_{2222}-\bar{D}_{2233}-U\bar{D}_{3322})}_{\text{2-derivative}}+\underbrace{\big(\!\!\sum_{0\leq a+b\leq 2} \!\!\! \tilde{c}_{ab}\sigma^a\tau^b\big)\pi^2U^2\bar{D}_{2222}}_{\text{0-derivative}}\;.
\end{equation}
The other two terms $\mathcal{A}_{{\rm con},t,u}$ are related to $\mathcal{A}_{{\rm con},s}$ in the same way as (\ref{Atucrossing}). Here $c_{ab}$ and $\tilde{c}_{ab}$ are symmetric because the s-channel contribution is invariant under $1\leftrightarrow 2$. In fact, the parametrization in (\ref{Acons}) is redundant. One can use $\bar{D}$-function identities to show that the zero-derivative contribution in the crossing symmetric $\mathcal{A}_{\rm con}$ can be absorbed into the two-derivative one (see Appendix B of \cite{Rastelli:2017udc} for details). Therefore we can set $\tilde{c}_{ab}=0$ without loss of generality. 

This gives us the most general ansatz for the stress tensor four-point function. To impose the superconformal Ward identities, we need to decompose the ansatz into a convenient basis of functions. This is done by exploiting two important properties of $D$-functions. First, the $D$-functions obey the ``weight-shifting'' relations\footnote{Equivalent formulae written in terms of $\bar{D}$-functions are collected in Appendix \ref{App:WittenDiagrams}.}
\begin{equation}\label{Dws}
D_{\Delta_1\ldots \Delta_i+1\ldots \Delta_j+1\ldots \Delta_n}=\frac{d-\sum_{a=1}^n\Delta_a}{2\Delta_i\Delta_j}\frac{\partial}{\partial x_{ij}^2}D_{\Delta_1\ldots \Delta_n}\;,
\end{equation} 
which allows us to reach every $\bar{D}$-function in the ansatz from $\bar{D}_{1111}$ with the action of differential operators. Second, the ``fundamental'' $\bar{D}_{1111}$ function is also known as the one-loop box integral, and is evaluated to be \cite{Usyukina:1992jd}
\begin{equation}
\bar{D}_{1111}\equiv\Phi(z,\bar{z})=\frac{1}{z-\bar{z}}\left(2{\rm Li}_2(z)-2{\rm Li}_2(\bar{z})+\log(z\bar{z})\log\big(\frac{1-z}{1-\bar{z}}\big)\right)\;.
\end{equation}
From this expression, it is easy to verify the following differential recursion relations
\begin{equation}\label{Phidiffrecur}
\begin{split}
\partial_z\Phi(z,\bar{z})=&-\frac{\Phi(z,\bar{z})}{z-\bar{z}}+\frac{\log U}{(z-1)(z-\bar{z})}-\frac{\log V}{z(z-\bar{z})}\;,\\
\partial_{\bar{z}}\Phi(z,\bar{z})=&\frac{\Phi(z,\bar{z})}{z-\bar{z}}-\frac{\log U}{(\bar{z}-1)(z-\bar{z})}+\frac{\log V}{\bar{z}(z-\bar{z})}\;.
\end{split}
\end{equation}
These two properties of $\bar{D}$-functions tell us that the ansatz (\ref{positionansatz}) can be expanded in the basis formed by $\Phi$, $\log U$, $\log V$ and $1$ 
\begin{equation}\label{ansatzinbasis}
\mathcal{G}_{\rm ansatz}=R_\Phi\, \Phi+R_{U}\log U+R_{V}\log V+R_1\;,
\end{equation}
where the coefficients $R_{\Phi,U,V,1}$ are {\it rational} functions of $z$ and $\bar{z}$. It is then straightforward to impose the superconformal Ward identity (\ref{scfWardid}), which can be cast in the same form as (\ref{ansatzinbasis}) upon using (\ref{Phidiffrecur}) again. Requiring the rational coefficient functions to vanish gives rise to a system of linear equations for the parameters in the ansatz, of which the solution reads 
\begin{equation}\label{sol2222}
\begin{split}
&\lambda_s=\xi\;,\quad \lambda_v=-\frac{1}{2}\xi\;,\quad \lambda_g=\frac{1}{16}\xi\;,\\
& c_{00}=\frac{1}{32}\xi\;,\quad c_{01}=-\frac{1}{8}\xi\;,\quad c_{02}=\frac{1}{32}\xi\;,\quad c_{11}=-\frac{1}{16}\xi\;.
\end{split}
\end{equation}
Note that there is a remaining unfixed coefficient $\xi$, because the  superconformal Ward identities (\ref{scfWardid}) are linear and homogenous. However, it can be determined in terms of the central charge as the exchanged multiplet is the stress tensor multiplet. More conveniently, we can look at the holomorphic twisted correlator (\ref{4dtwisted}), which is independent of the 't Hooft coupling \cite{bllprv13}. Substituting the solution (\ref{sol2222}) into the ansatz (\ref{positionansatz}), we find that the twisted correlator is indeed a holomorphic function\footnote{See \cite{Behan:2021pzk} for a systematic discussion of chiral algebra correlators in the holographic context.}
\begin{equation}
\mathcal{G}_{\rm ansatz}(z,\bar{z};\alpha,1/\bar{z})=-\frac{\pi^2\xi(\alpha^2z^2-2\alpha z^2+2\alpha z-z)}{8N^2(z-1)}\;.
\end{equation} 
By matching with the twisted correlator in the free theory (where correlators are  computed by simple Wick contractions), we find 
\begin{equation}
\xi=\frac{32}{N^2\pi^2}\;.
\end{equation}

This strategy straightforwardly generalizes to four-point functions of operators with higher Kaluza-Klein levels. In each exchange channel, there are only finitely many supergravity fields that can appear. The finite number of exchanged fields is dictated by two selection rules: the selection rule following from R-symmetry\footnote{It is just the requirement that the R-symmetry irreducible representation carried by the exchanged field should be contained in the tensor product of representations both of $\mathcal{O}_{k_1}$, $\mathcal{O}_{k_2}$ and of $\mathcal{O}_{k_3}$, $\mathcal{O}_{k_4}$ (say in the s-channel). The R-symmetry representations of component fields in a $\frac{1}{2}$-BPS multiplet are listed in Appendix \ref{App:multipletcomponents}.}, and the requirement that the conformal twists of the exchanged fields satisfy inequalities ({\it e.g.}, in the s-channel) $\Delta_1+\Delta_2<\Delta-\ell$ and $\Delta_3+\Delta_4<\Delta-\ell$. The R-symmetry selection rule already ensures $\Delta_1+\Delta_2\leq\Delta-\ell$ and $\Delta_3+\Delta_4\leq\Delta-\ell$ because the spectrum of the supergravity states is determined by their R-symmetry charges. The latter requirement merely forbids the possibility where the bound is saturated  (also known as being {\it extremal} in the literature). That such couplings are forbidden
comes from the fact that extremal three-point Witten diagram integrals are divergent, and would otherwise lead to an inconsistent infinite effective action. 

Note that the position space method relies crucially on the fact that the ansatz can be decomposed in a basis spanned by $\Phi$, $\log U$, $\log V$, $1$, with rational coefficient functions, as in (\ref{ansatzinbasis}). This requires the exchange diagrams to be written as a finite sum of $\bar{D}$-functions, which only happens in theories with  special spectra guaranteeing the conditions (\ref{spectrumcondition}). IIB supergravity on $AdS_5\times S^5$ and eleven dimensional supergravity on $AdS_7\times S^4$ are two examples where such conditions are satisfied. By contrast, eleven dimensional supergravity on $AdS_4\times S^7$ does not share this property, and therefore the position space method does not apply to this case. 

Finally, we note that while the position space method bypasses the formidable computations to extract the vertices, it also becomes cumbersome when we apply it to more general correlators. The number of $\bar{D}$-functions in the ansatz quickly increases, making the coefficient functions in (\ref{ansatzinbasis}) increasingly complicated. Nevertheless, the position space method taught us an important lesson, namely, holographic correlators can be completely fixed by symmetries. In Section \ref{Sec:MellinSpace} we apply this lesson  in Mellin space, where the correlators have much simpler analytic structure, and show how all tree-level correlators can be obtained in a closed form. However, before we do that let us first look at an application of this lesson to higher-point functions.
 
\subsection{Higher-point correlators}\label{Subsec:5ptfunctions}
The above position space method can also be extended to bootstrap higher-point correlation functions. In \cite{Goncalves:2019znr}, the five-point function of $k_i=2$ super gravitons in $AdS_5\times S^5$ was computed. In this subsection, we briefly review this result and outline its strategy. 

\begin{figure}
\centering
\includegraphics[width=0.6\textwidth]{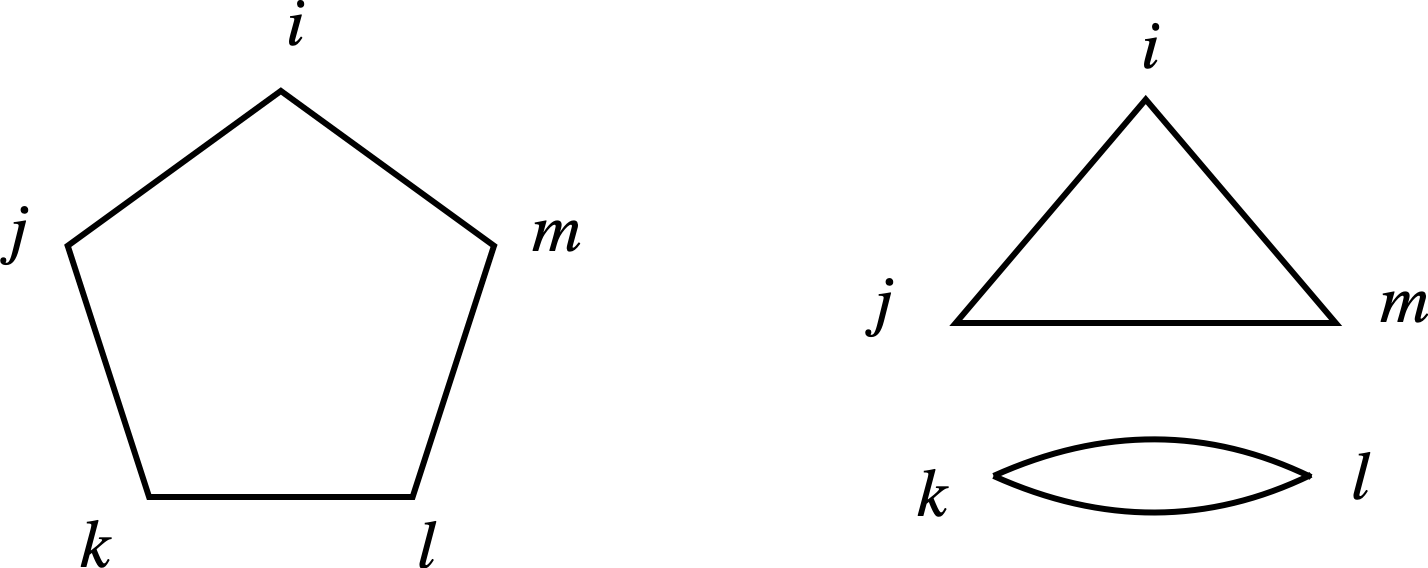}
\caption{Independent R-symmetry structures of the $k_i=2$ five-point function.}
    \label{fig:5ptRstruct}
\end{figure}

The five-point function of the $k_i=2$ operator 
\begin{equation}
G_5(x_i,t_i)=\langle \mathcal{O}_2(x_1,t_1) \mathcal{O}_2(x_2,t_2) \mathcal{O}_2(x_3,t_3) \mathcal{O}_2(x_4,t_4) \mathcal{O}_2(x_5,t_5)\rangle
\end{equation}
has more complicated structures compared to the four-point function. It has 22 independent R-symmetry structures which correspond to the Wick contractions in Figure \ref{fig:5ptRstruct}. The pentagon type contraction gives $t_{ij}t_{jk}t_{kl}t_{lm}t_{mi}$, while the type with a triangle gives $t_{ij}t_{jm}t_{mi}t_{kl}^2$. Exploiting conformal symmetry and R-symmetry, we can write $G_5$ as a function of cross ratios. There are five conformal cross ratios
\begin{equation}
V_1=\frac{x_{12}^2x_{34}^2}{x_{13}^2x_{24}^2}\;,\;\; V_2=\frac{x_{14}^2x_{23}^2}{x_{13}^2x_{24}^2}\;,\;\; V_3=\frac{x_{14}^2x_{35}^2}{x_{13}^2x_{45}^2}\;,\;\; V_4=\frac{x_{15}^2x_{34}^2}{x_{13}^2x_{45}^2}\;,\;\; V_5=\frac{x_{12}^2x_{35}^2}{x_{13}^2x_{25}^2}\;,
\end{equation}
and similarly five cross ratios for R-symmetry. On the other hand, a  comprehensive study of the superconformal properties of higher-point correlators has not been performed in the literature. Therefore, superconformal Ward identities, which played a central role in bootstrapping four-point functions, have not been written down for five-point functions. Nevertheless, two highly nontrivial superconformal constraints are known in the literature from supersymmetric twisting.

\begin{itemize}
\item {\bf Chiral algebra twist.} The first constraint comes from the chiral algebra twist \cite{bllprv13}. To perform this twist, we restrict all the operators on a two-dimensional plane parameterized by the complex coordinates $(z,\bar{z})$. Furthermore, we restrict the $SO(6)$ R-symmetry group to $SO(4)$ by setting the last two components of the six dimensional auxiliary vector $t^I$ to zero. The resulting four-dimensional vector $t^\mu$ can be written as a pair of two-component spinors $v^a$ and $\bar{v}^{\dot{a}}$ as $t^\mu=\sigma^\mu_{a\dot{a}}v^a\bar{v}^{\dot{a}}$ where $\sigma^\mu_{a\dot{a}}$ are the Pauli matrices. Using the scaling degree of freedom of the null vector, we can normalize the spinors as 
$v=(1,y)$, $\bar{v}=(1,\bar{y})$. The chiral algebra twist corresponds to setting $\bar{y}_i=\bar{z}_i$ for each operator. One can then show that there exists a nilpotent supercharge preserved by this configuration, and the twisted operators are in its cohomology class. Moreover, the twisted translations, {\it i.e.}, transformations changing $\bar{z}_i$ while maintaining $\bar{y}_i=\bar{z}_i$, are exact with respect to this supercharge. It then follows from a standard argument that the correlator after twisting is independent of the anti-holomorphic coordinates $\bar{z}_i$. Furthermore, it can be shown that the twisted correlator is independent of the coupling \cite{bllprv13} and thus takes the same value as in the free theory. Therefore, the chiral algebra yields the following condition 
\begin{equation}\label{5ptcat}
G_5(z_i,\bar{z}_i;v_i,\bar{v}_i=\bar{z}_i)=G_{5,{\rm free}}(z_i,\bar{z}_i;v_i,\bar{v}_i=\bar{z}_i)\;.
\end{equation}
Note that before twisting the correlator $G_5$ is a complicated function. Even when restricted on the plane, it contains both holomorphic and anti-holomorphic dependence. On the other hand, the RHS of the above identity is a simple rational function of $z_i$ and $v_i$, which can be computed by performing Wick contractions in the free theory. Therefore, (\ref{5ptcat}) imposes highly nontrivial constraints on the five-point function. 

\item {\bf Drukker-Plefka twist.} The other important constraint is given by the topological twist introduced in \cite{Drukker:2009sf}. In this case, the operators are allowed to be inserted at generic locations $x_i\in \mathbb{R}^4$. However, their $SO(6)$ polarizations are fixed by their positions
\begin{equation}
t=\big(2ix^1,2ix^2,2ix^3,2ix^4,i(1-x\cdot x),1+x\cdot x\big)\;.
\end{equation}
It was shown in \cite{Drukker:2009sf} that this configuration preserves two supercharges. Moreover, twisted translations and the exactly marginal deformation are exact with respect to the supercharges. As a result, the twisted correlator is a coupling-independent constant which can be computed in the free theory
\begin{equation}
G_5(x_i,t_{ij}=x_{ij}^2)=\frac{20\sqrt{2}}{N}+\frac{48\sqrt{2}}{N^3}\;.
\end{equation}
Here the two terms on the RHS correspond to the disconnected and connected contributions respectively. The connected term is relevant for the supergravity computation in this subsection. 
\end{itemize}

\begin{figure}[h]
\centering
\includegraphics[width=\textwidth]{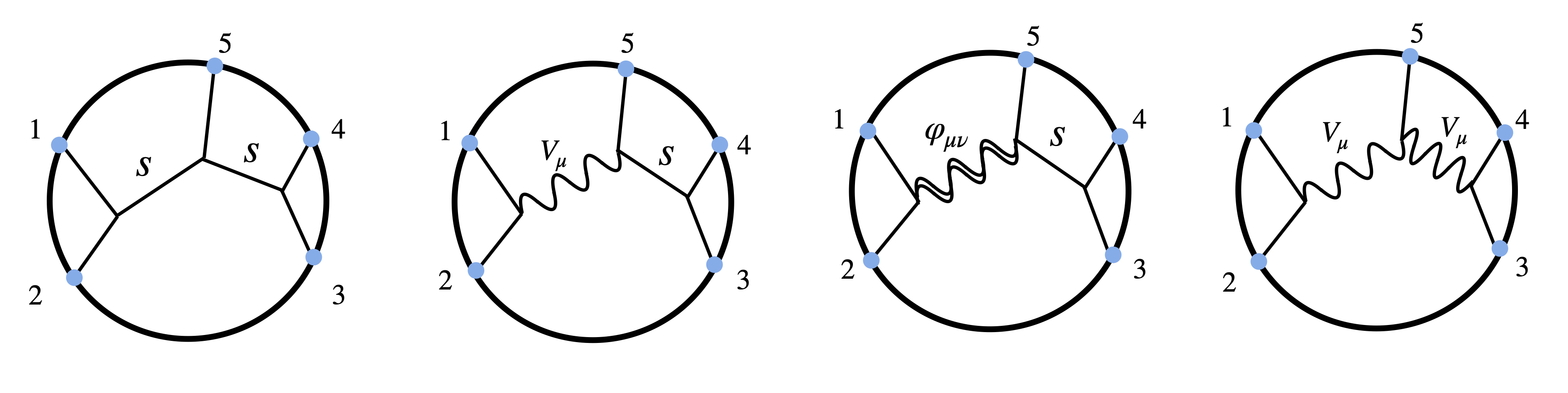}
\caption{Double-exchange Witten diagrams.}
    \label{fig:5ptde}
\end{figure}

\begin{figure}[h]
\centering
\includegraphics[width=0.85\textwidth]{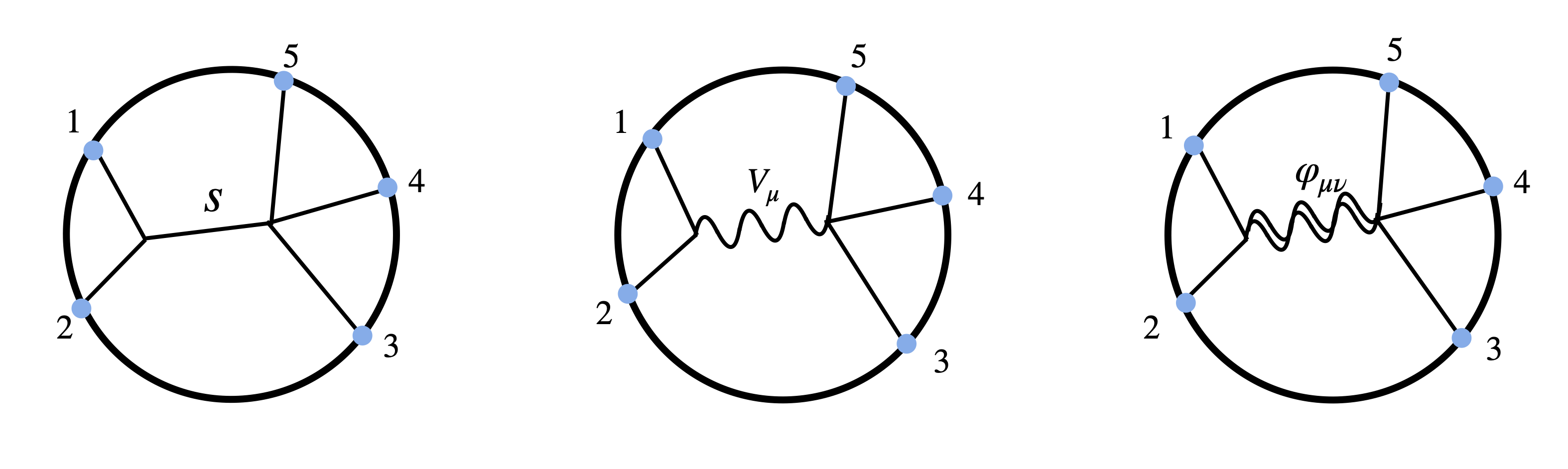}
\caption{Single-exchange Witten diagrams.}
    \label{fig:5ptse}
\end{figure}

To bootstrap this correlator, we start by making an ansatz which includes all the possible exchange diagrams and contact diagrams. There are two types of exchange diagrams, double-exchange diagrams and single-exchange diagrams, depicted in Figure \ref{fig:5ptde} and Figure \ref{fig:5ptse} correspondingly. There are three fields that can appear in the exchanges: the $k_i=2$ scalar field $s$, the graviphoton field $V_\mu$ and the graviton field $\varphi_{\mu\nu}$. Like in the four-point function case, each bosonic Witten diagram is multiplied with an R-symmetry polynomial. For the double-exchange diagrams, these R-symmetry polynomials are linear combinations of the 22 monomial R-symmetry structures such that they  are the simultaneous eigenfunctions of the R-symmetry quadratic Casimir in the two exchange channels (the 12 channel and 34 channel for the diagrams in Figure \ref{fig:5ptde}). All possible double-exchange diagrams have been listed in Figure \ref{fig:5ptde}, and we note that they exclude certain diagrams because of R-symmetry selection rules. For example, we cannot have a double-exchange diagram where both internal lines are gravitons. The cubic vertex including the external leg 5 would violate R-symmetry because gravitons are uncharged.  For single-exchange diagrams, the R-symmetry polynomials are only eigenfunction of one Casimir equation, and therefore are not unique. We should include in the ansatz all allowed solutions to the R-symmetry Casimir equation. Finally, we also include a quintic contact contribution in the ansatz. This contact term contains all 22 R-symmetry structures and contact Witten diagrams with zero and two derivatives.  

\begin{figure}[h]
\centering
\includegraphics[width=0.25\textwidth]{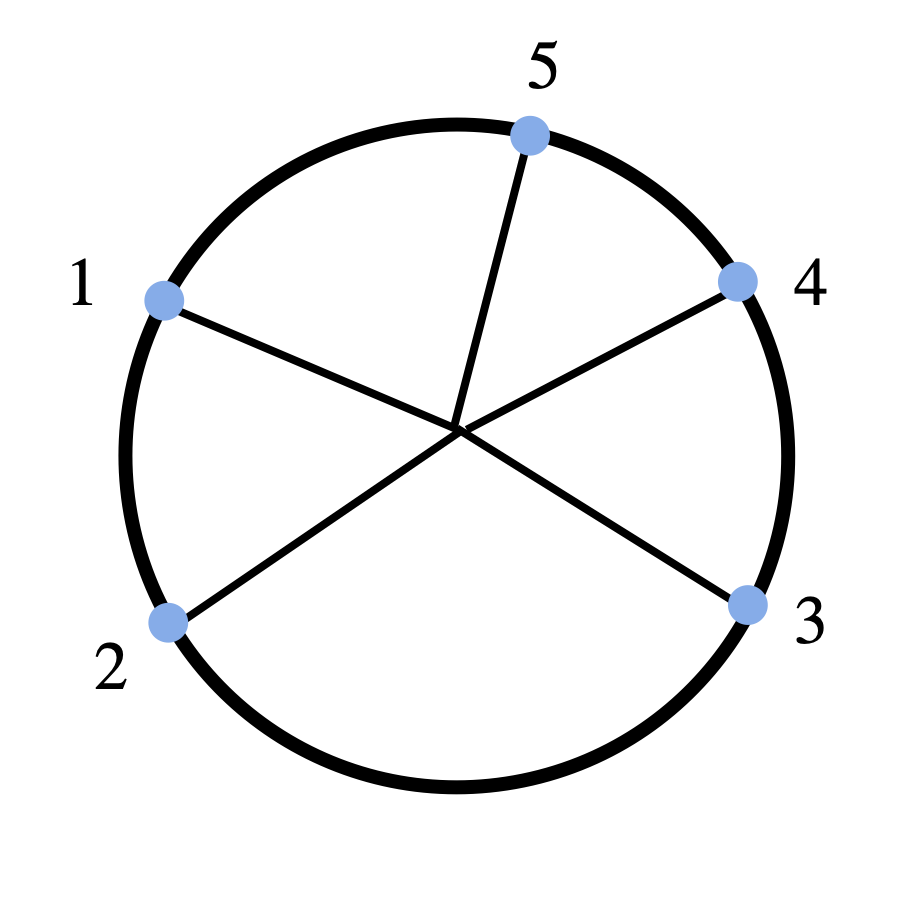}
\caption{Quintic contact Witten diagrams.}
    \label{fig:5ptcon}
\end{figure}

To implement the superconformal twists, we need to evaluate the ansatz and express it in a useful form. The contact diagrams are the simplest. For example, the zero-derivative contact diagram is just
\begin{equation}
D_{22222}\;,
\end{equation}
and the two-derivative contact diagram (with derivatives on legs 1 and 2) is 
\begin{equation}
4(D_{22222}-2x_{12}^2D_{33222})\;.
\end{equation}
To evaluate the exchange diagrams, we need to extend the result of \cite{DHoker:1999mqo} for four-point exchange diagrams which expresses them in terms of a truncated sum of $D$-functions. It turns out that the truncation only relies on half of the diagram, namely, the integral involving two bulk-to-boundary propagators and a bulk-to-bulk propagator. We can express this integrated cubic vertex as a sum of contact vertices with just two bulk-to-boundary propagators. We collect such integrated vertex identities in \ref{Subapp:ivi},  and they are sufficient for computing all five-point exchange Witten diagrams. For example, a special case of the identity which involves the scalar bulk-to-bulk propagator is 
\begin{equation}
\int \frac{d^5z}{z_0^5} G^2_{B\partial}(z,x_1)G^2_{B\partial}(z,x_2) G^2_{BB}(z,w)=\frac{1}{4}x_{12}^{-2}G^1_{B\partial}(z,x_1)G^1_{B\partial}(z,x_2)\;,
\end{equation}
where $G^\Delta_{B\partial}(z,x)$ and $G^\Delta_{BB}(z,x)$ are bulk-to-boundary and bulk-to-bulk propagators respectively. Using this identity on the vertex joining 1, 2 and the vertex joining 3, 4 in the first diagram in Figure \ref{fig:5ptde}, we find that the scalar double-exchange diagram evaluates to
\begin{equation}
\frac{1}{16}x_{12}^{-2}x_{34}^{-2}D_{11112}\;.
\end{equation}
Similarly, the scalar single-exchange diagram in \ref{fig:5ptse} is 
\begin{equation}
\frac{1}{4}x_{12}^{-2}D_{11222}\;,
\end{equation}
when the quartic vertex in the diagram contains no derivatives. All the diagrams encountered in the five-point function computation can be evaluated in this way, and the results are expressed as linear combinations of $D$-functions. An important feature of these results is that all the $D$-functions can be generated from $D_{11112}$ (and its permutations) by using the differential weight-shifting relations (\ref{Dws}). This is clear, for example, for the contact and exchange Witten diagrams we computed above. Furthermore, the function $D_{11112}$ is known in the amplitude literature as the pentagon integral and has been well studied \cite{Bern:1992em,Bern:1993kr}. It can be expressed as the linear combination of five one-loop box diagrams $D_{1111}$ with rational coefficients of the coordinates, where each box function involves four of the five points (see (\ref{Df5pt}) for the explicit expression). From the differential recursion relations (\ref{Phidiffrecur}), we then know that the ansatz can be expressed in terms of polylogarithmic functions with transcendental degrees 2 and lower. Although computationally it still requires quite some heavy lifting to implement the two superconformal twisting constraints, conceptually it is very straightforward and is similar to the four-point function case.

\begin{figure}[h]
\centering
\includegraphics[width=0.8\textwidth]{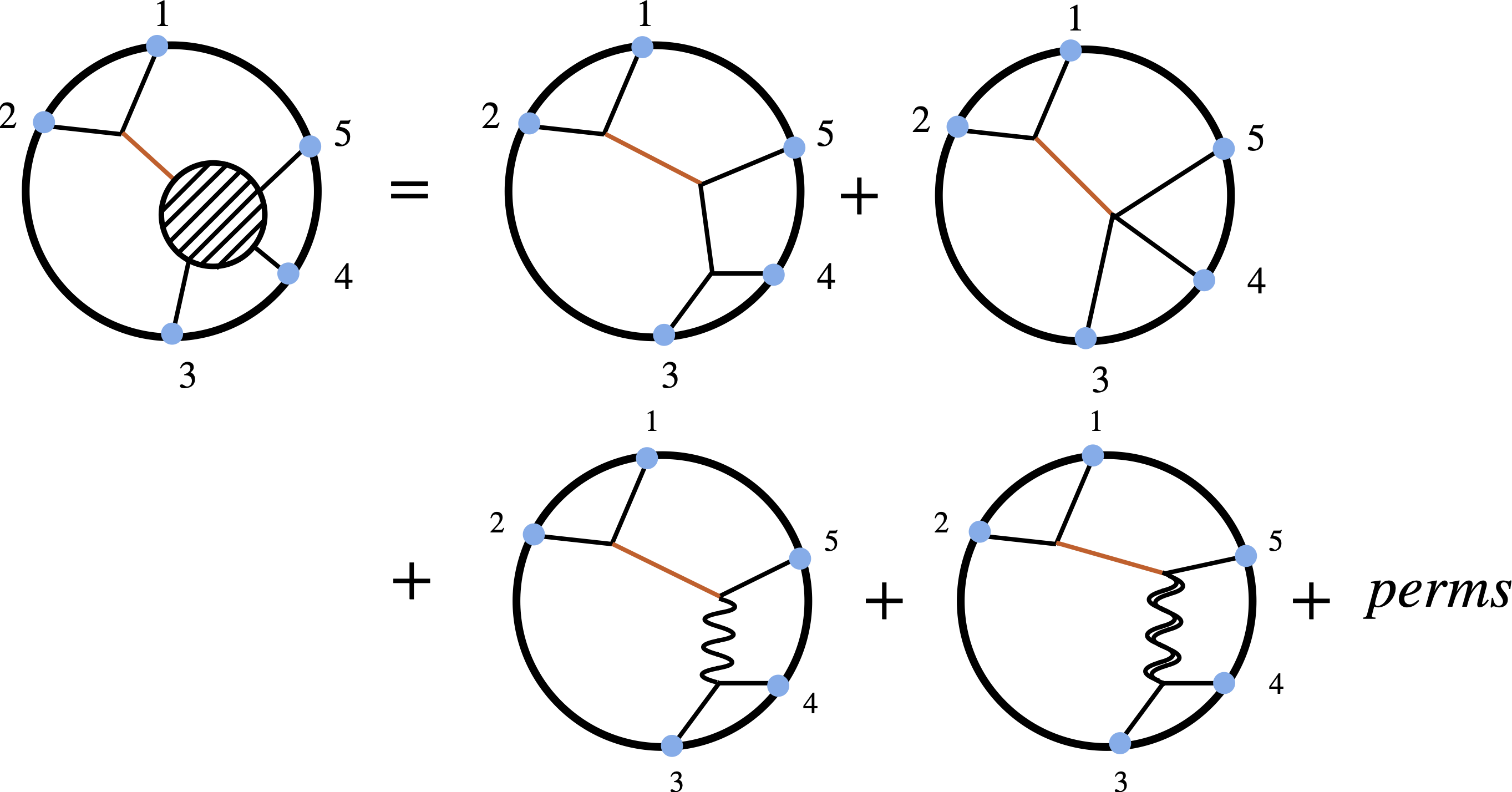}
\caption{Factorization on a scalar propagator. Here {\it perm} denotes similar diagrams obtained by permuting the external legs 3, 4, 5.}
    \label{fig:5ptfacs}
\end{figure}

\begin{figure}[h]
\centering
\includegraphics[width=0.8\textwidth]{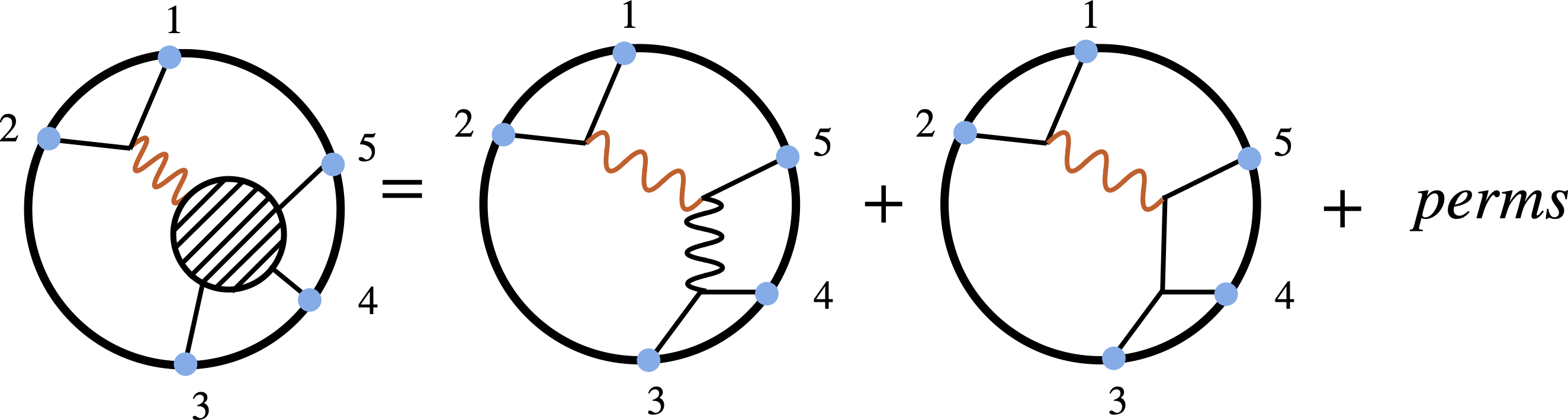}
\caption{Factorization on a graviphoton propagator.}
    \label{fig:5ptfacv}
\end{figure}

\begin{figure}[h]
\centering
\includegraphics[width=0.8\textwidth]{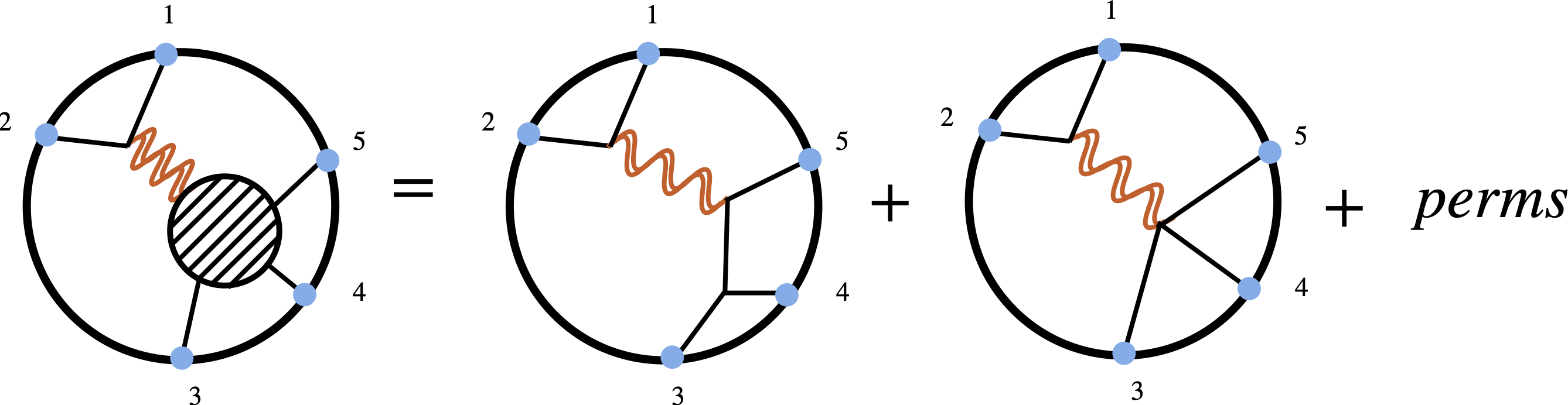}
\caption{Factorization on a graviton propagator.}
    \label{fig:5ptfacg}
\end{figure}

Another important constraint on the five-point function is its consistency with lower-point correlation functions. We can group the exchange Witten diagrams in the ansatz into several sets. In each set we have the bulk-to-bulk propagator of a certain field which connects two external legs on one side and three remaining legs on the other side. This is illustrated in  Figure \ref{fig:5ptfacs}, \ref{fig:5ptfacv} and \ref{fig:5ptfacg} for the scalar, graviphoton, and the graviton fields respectively. When we cut open the internal propagator and place the ends on the boundary of AdS, we obtain a three-point function $\langle \mathcal{O}_2\mathcal{O}_2\mathcal{O}_{\rm int}\rangle$ and a four-point function $\langle \mathcal{O}_{\rm int}\mathcal{O}_2\mathcal{O}_2\mathcal{O}_2\rangle$ where $\mathcal{O}_{\rm int}$ denotes the operator dual to the internal field. This is the idea of factorization in flat space, and it has a concrete realization in Mellin space for holographic correlators \cite{Goncalves:2014rfa}. When we look at the pole of the Mellin amplitude associated with the internal propagator, the residue can be expressed in terms of the Mellin amplitudes of the three-point function and the four-point function. Moreover, since the graviphoton and graviton fields are superconformal descendants of the $k=2$ scalar field, correlators involving these spinning fields are related to the scalar four-point function by differential relations determined by superconformal symmetry \cite{Belitsky:2014zha}. These spinning correlators were explicitly computed in \cite{Goncalves:2019znr}. 

These conditions, two superconformal constraints from supersymmetric twisting and factorization in Mellin space, form a system of complementary constraints. For example, the factorization condition is agnostic of the five-point contact interaction diagrams which do not have any poles. On the other hand, the supersymmetric twisting constraints can only be satisfied when the contact terms are included. Together they impose strong constraints on the super graviton five-point function, and was shown to uniquely fix the correlator in \cite{Goncalves:2019znr}. Although the expression of the five-point function is a bit complicated in position space, its Mellin amplitude turns out to be quite simple and resembles a  flat-space amplitude of massive particles. We refer the reader to \cite{Goncalves:2019znr} for details of how the constraints fix the correlator and explicit expressions of the amplitude. Finally, let us point out that the method we reviewed in this section can also be used to bootstrap the super gluon five-point function in $AdS_5$ \cite{Alday:2022lkk}.

\newpage
\markboth{11\quad BOOTSTRAPPING TREE-LEVEL CORRELATORS: MELLIN SPACE}{}
\section{Bootstrapping tree-level correlators: Mellin space}\label{Sec:MellinSpace}
In Section \ref{Sec:PositionSpace} we showed how to bootstrap holographic correlators in position space. While the algorithm works in the same way for correlators of higher weights, the implementation becomes more and more cumbersome as the external conformal dimensions are increased. Therefore using this method to find a closed form formula for general correlators does not seem very feasible. In this section, we introduce alternative methods in Mellin space which allow us to obtain general correlators with arbitrary Kaluza-Klein weights. Crucially, these methods exploit the simple analytic structure of holographic correlators in Mellin space, which allows us to extend our intuition of flat-space scattering amplitudes. We first look at the case of IIB supergravity in $AdS_5\times S^5$ in Section \ref{Subsec:AdS5Mellin}, and review the method of \cite{Rastelli:2016nze,Rastelli:2017udc}. In this approach the task of computing correlators can be translated into solving an algebraic bootstrap problem in Mellin space which is formulated by imposing symmetry constraints and consistency conditions. This algebraic problem can be solved in general. The solution is elegantly simple and gives all tree-level four-point amplitudes of super gravitons with arbitrary Kaluza-Klein levels. Unfortunately, this method implicitly relies on special features of the $AdS_5\times S^5$ theory, and is not as effective for $AdS_7\times S^4$. Moreover, it does not apply to $AdS_4\times S^7$. Nevertheless, in the latter two cases superconformal symmetry is still constraining enough to uniquely fix super graviton correlators. This suggests the existence of a universal method which exploits superconformal symmetry in a dimension-independent way and treats all three backgrounds on an equal footing. Such a method was developed in \cite{Alday:2020lbp,Alday:2020dtb}, building on earlier work \cite{Zhou:2017zaw}, as we will review in Section \ref{Subsec:Mellinscfwi} and Section \ref{Subsec:MRV}. In Section \ref{Subsec:Mellinscfwi} we first explain how the position space superconformal Ward identities can be exploited in Mellin space and translated into a system of difference equations. Then in Section \ref{Subsec:Mellinscfwi} we examine a special kinematic limit where the Mellin amplitudes drastically simplify and can be easily computed. Having solved the correlators in this limit, we can then obtain amplitudes in generic kinematic configurations by using symmetries. This method was first developed in \cite{Alday:2020lbp,Alday:2020dtb} for computing super graviton amplitudes in maximally superconformal theories. However, with small modifications it can also be used to compute tree-level super gluon amplitudes in a variety of non-maximally superconformal theories in different spacetime dimensions, as we will briefly discuss in Section \ref{Subsec:supergluons}.

\subsection{A Mellin bootstrap problem for $AdS_5\times S^5$ IIB supergravity}\label{Subsec:AdS5Mellin}
In Section \ref{Subsec:positionspace} we have seen that the superconformal Ward identities (\ref{scfWardid}) play a central role in bootstrapping tree-level four-point functions. One might wonder if it is possible to directly solve these differential constraints, which will then automatically take the consequence of superconformal symmetry into account. While for generic spacetime dimensions the solutions to (\ref{scfWardid}) are quite complicated \cite{Dolan:2004mu}, in $d=4$ the answer is rather simple. One can show that the four-point functions split into two parts \cite{Eden:2000bk,Nirschl:2004pa}
\begin{equation}\label{solscfwi4d}
G_{k_1k_2k_3k_4}=G_{{\rm free},k_1k_2k_3k_4}+{\rm R}\, H_{k_1k_2k_3k_4}\;,
\end{equation}
where $G_{{\rm free},k_1k_2k_3k_4}$ is the four-point correlator in the free $\mathcal{N}=4$ SYM theory and $H_{k_1k_2k_3k_4}$ is the {\it reduced} correlator containing all the dynamical information. The factor ${\rm R}$ is determined by superconformal symmetry to be 
\begin{equation}
{\rm R}=t_{12}^2t_{34}^2x_{13}^4x_{24}^4(1-z\alpha)(1-\bar{z}\alpha)(1-z\bar{\alpha})(1-\bar{z}\bar{\alpha})\;.
\end{equation}
Since ${\rm R}$ carries nontrivial weights under conformal and R-symmetry transformations, the conformal dimensions and R-symmetry charges of the reduced correlator $H_{k_1k_2k_3k_4}$ are shifted
\begin{equation}
\text{conformal dimensions: }k_i\to k_i+2\;,\quad\quad \text{R-symmetry charges: }k_i\to k_i-2\;.
\end{equation}
Compared to the full correlator $G_{k_1k_2k_3k_4}$, the reduced correlator $H_{k_1k_2k_3k_4}$ is generally much simpler. For example, $G_{2222}$ contains six independent R-symmetry structures corresponding to the Wick contractions
\begin{equation}
t_{12}^2t_{34}^2\;,\quad t_{13}^2t_{24}^2\;,\quad t_{14}^2t_{23}^2\;,\quad t_{12}t_{23}t_{34}t_{14}\;,\quad t_{13}t_{23}t_{24}t_{14}\;,\quad t_{12}t_{24}t_{34}t_{13}\;.
\end{equation} 
By contrast, $H_{2222}$ is independent of $t_i$ and therefore has only one R-symmetry structure. 

We now translate the solution (\ref{solscfwi4d}) into Mellin space. From the full correlator $G_{k_1k_2k_3k_4}$, we define the Mellin amplitude $\mathcal{M}_{k_1k_2k_3k_4}$ in the standard way
\begin{equation}\label{defMfull}
G_{k_1k_2k_3k_4}=\int_{-i\infty}^{i\infty}\frac{dsdt}{(4\pi i)^2} {\rm K}(x_{ij}^2;s,t) \mathcal{M}_{k_1k_2k_3k_4}(s,t;t_{ij}) \Gamma_{\{k_i\}}(s,t)\;.
\end{equation}
Here to manifest Bose symmetry, we wrote the correlator without extracting the kinematic factor in contrast to what we did in (\ref{GandcalG}). The factor ${\rm K}(x_{ij}^2;s,t)$ is defined by
\begin{equation}
{\rm K}(x_{ij}^2;s,t)=(x_{12}^2)^{\frac{s-k_1-k_2}{2}}(x_{34}^2)^{\frac{s-k_3-k_4}{2}}(x_{14}^2)^{\frac{t-k_1-k_4}{2}}(x_{23}^2)^{\frac{t-k_2-k_3}{2}}(x_{13}^2)^{\frac{u-k_1-k_3}{2}}(x_{24}^2)^{\frac{u-k_2-k_4}{2}}\;,
\end{equation}
with $s+t+u=k_1+k_2+k_3+k_4$, and 
\begin{equation}
\Gamma_{\{k_i\}}(s,t)=\Gamma[\tfrac{k_1+k_2-s}{2}]\Gamma[\tfrac{k_3+k_4-s}{2}]\Gamma[\tfrac{k_1+k_4-t}{2}]\Gamma[\tfrac{k_2+k_3-t}{2}]\Gamma[\tfrac{k_1+k_3-u}{2}]\Gamma[\tfrac{k_2+k_4-u}{2}]\;.
\end{equation}
Similarly, we define the {\it reduced} Mellin amplitude $\widetilde{\mathcal{M}}_{k_1k_2k_3k_4}$ from the reduced correlators $H_{k_1k_2k_3k_4}$
\begin{equation}\label{defMreduced}
H_{k_1k_2k_3k_4}=\int_{-i\infty}^{i\infty}\frac{dsdt}{(4\pi i)^2} \widetilde{{\rm K}}(x_{ij}^2;s,t) \widetilde{\mathcal{M}}_{k_1k_2k_3k_4}(s,t;t_{ij}) \widetilde{\Gamma}_{\{k_i\}}(s,t)\;,
\end{equation}
where 
\begin{equation}
\widetilde{{\rm K}}(x_{ij}^2;s,t)=(x_{12}^2)^{\frac{s-k_1-k_2}{2}}(x_{34}^2)^{\frac{s-k_3-k_4}{2}}(x_{14}^2)^{\frac{t-k_1-k_4}{2}}(x_{23}^2)^{\frac{t-k_2-k_3}{2}}(x_{13}^2)^{\frac{\tilde{u}-k_1-k_3}{2}}(x_{24}^2)^{\frac{\tilde{u}-k_2-k_4}{2}}\;,
\end{equation}
\begin{equation}
\widetilde{\Gamma}_{\{k_i\}}(s,t)=\Gamma[\tfrac{k_1+k_2-s}{2}]\Gamma[\tfrac{k_3+k_4-s}{2}]\Gamma[\tfrac{k_1+k_4-t}{2}]\Gamma[\tfrac{k_2+k_3-t}{2}]\Gamma[\tfrac{k_1+k_3-\tilde{u}}{2}]\Gamma[\tfrac{k_2+k_4-\tilde{u}}{2}]\;,
\end{equation}
and $s+t+\tilde{u}=k_1+k_2+k_3+k_4-4$. Note that the shift $\tilde{u}=u-4$ is needed because $H_{k_1k_2k_3k_4}$ has shifted conformal dimensions relative to $G_{k_1k_2k_3k_4}$. Moreover, we should note that Bose symmetry, which permutes $s$, $t$, $u$ in (\ref{defMfull}), now permutes $s$, $t$, $\tilde{u}$ in (\ref{defMreduced}).

Once including the factor ${\rm R}$, the combination ${\rm R}\, H_{k_1k_2k_3k_4}$ has the same weights as $G_{k_1k_2k_3k_4}$ and therefore should have the same Mellin representation (\ref{defMfull}). This leads us to interpret ${\rm R}$ as a difference operator in Mellin space. We note that ${\rm R}$ is a polynomial in $x_{ij}^2$, and multiplicative $x_{ij}^2$ monomials outside of the inverse Mellin transformation can be absorbed into the $\widetilde{{\rm K}}(x_{ij}^2;s,t)$ factor by shifting $s$ and $t$. More precisely, let us write 
\begin{equation}
\frac{{\rm R}}{x_{13}^4x_{24}^4}=t_{12}^2t_{34}^2\bigg(\tau+(1-\sigma-\tau)V+(\tau^2-\tau-\sigma\tau)U+(\sigma^2-\sigma-\sigma\tau)UV+\sigma V^2+\sigma\tau U^2\bigg)\;.
\end{equation}
Comparing (\ref{defMfull}) and (\ref{defMreduced}), we find that each monomial $U^mV^n$ in the RHS becomes a difference operator $\widehat{U^mV^n}$ which acts as 
\begin{equation}
\widehat{U^mV^n}\circ \widetilde{\mathcal{M}}_{k_1k_2k_3k_4}(s,t;t_{ij})=\frac{\widetilde{\Gamma}_{\{k_i\}}(s-2m,t-2n)}{\Gamma_{\{k_i\}}(s,t)} \widetilde{\mathcal{M}}_{k_1k_2k_3k_4}(s-2m,t-2n;t_{ij})\;.
\end{equation}
This substitution defines an operator $\mathbb{R}$ acting on the reduced Mellin amplitude
\begin{equation}
\mathbb{R}\circ\widetilde{\mathcal{M}}_{k_1k_2k_3k_4}\;.
\end{equation}
Finally, it can be argued that the free correlator $G_{{\rm free},k_1k_2k_3k_4}$ does not contribute to the Mellin amplitudes \cite{Rastelli:2016nze,Rastelli:2017udc}. When we multiply with the factor ${\rm R}$, the contours in the inverse Mellin transformations are also shifted. In bringing the contours to the correct ones, we encounter situations where the contours are pinched at poles with a {\it vanishing} Mellin amplitude. The ``zero times infinity'' contribution coming from contour pinching gives rise to rational terms which together become precisely the free correlator $G_{{\rm free},k_1k_2k_3k_4}$.  We will not keep track of the contours in this review. Therefore, as far as the Mellin amplitudes are concerned we can ignore the free correlators. To see how the free correlator is explicitly reproduced in an example with $k_i=2$, see \cite{Rastelli:2017udc}. All in all, the solution to the superconformal Ward identity (\ref{solscfwi4d}) implies the following difference relation 
\begin{equation}\label{MandMt}
\mathcal{M}_{k_1k_2k_3k_4}=\mathbb{R}\circ\widetilde{\mathcal{M}}_{k_1k_2k_3k_4}\;,
\end{equation}
which compactly packages the full Mellin amplitudes in terms of the reduced Mellin amplitudes. 

Note that (\ref{MandMt}) has only exploited superconformal symmetry. The Mellin amplitudes $\mathcal{M}_{k_1k_2k_3k_4}$ further need to satisfy a number of other consistency conditions in order to be physical. First of all, the Mellin amplitude should have Bose symmetry. This requirement means that the Mellin amplitude is invariant under exchanging external particle labels, which also permutes the Mandelstam variables. Secondly, the Mellin amplitude is local. It has simple poles at locations corresponding to the twist of exchanged single-trace particles, and the residues at these poles are polynomial in the other independent Mandelstam variable. Finally, the high energy limit of the Mellin amplitude with $s,t,u\to\infty$ at the same rate is proportional to the flat-space scattering amplitude of IIB super gravity. The latter grows linearly in energy. Therefore the high-energy growth of the Mellin amplitude must also have the same linear behavior. 

These three conditions together with (\ref{MandMt}) formulate a highly constraining bootstrap problem. For example, it is not difficult to convince oneself that for $k_i=2$ the reduced Mellin amplitude can only be proportional to 
\begin{equation}
\frac{1}{(s-2)(t-2)(\tilde{u}-2)}\;,
\end{equation}
in order to be compatible with all the above conditions. On the other hand, this reformulation of the problem places correlators with any choice of $k_i$ on the same footing, which makes it possible to find a general solution in one go. After studying a few explicit examples, \cite{Rastelli:2016nze,Rastelli:2017udc} found the following ansatz for the reduced Mellin amplitudes
\begin{equation}\label{solAdS5}
\widetilde{\mathcal{M}}_{k_1k_2k_3k_4}=\prod_{i<j}t_{ij}^{\gamma_{ij}^0}(t_{12}t_{34})^{\mathcal{E}}\sum_{{\tiny\begin{split}{}&i+j+k=\mathcal{E}-2,\\{}&0\leq i,j,k\leq \mathcal{E}-2\end{split}}} \frac{a_{ijk}\sigma^i\tau^j}{(s-s_M+2k)(t-t_M+2j)(\tilde{u}-u_M+2i)}\;,
\end{equation}
where $a_{ijk}$ are unknown parameters, and
\begin{equation}
\begin{split}
s_M={}&\min\{k_1+k_2, k_3+k_4\}-2\;,\\
t_M={}&\min\{k_1+k_4,k_2+k_3\}-2\;,\\
u_M={}&\min\{k_1+k_3, k_2+k_4\}-2\;.
\end{split}
\end{equation}
Imposing the bootstrap conditions, we find that $a_{ijk}$ are uniquely fixed up to an overall constant 
\begin{equation}
a_{ijk}=\frac{\mathcal{N}_{k_1k_2k_3k_4}}{i!j!k!(i+\frac{\kappa_u}{2})!(j+\frac{\kappa_t}{2})!(k+\frac{\kappa_s}{2})!}\;,
\end{equation}
where $\kappa_{s,t,u}$ were defined in (\ref{kappastu}). The remaining constant is fixed to be \cite{Aprile:2018efk}
\begin{equation}
\mathcal{N}_{k_1k_2k_3k_4}=\frac{2}{N^2}\sqrt{k_1k_2k_3k_4}\;.
\end{equation} 
There are many ways to compute this normalization factor. In \cite{Aprile:2018efk} it was obtained by requiring the correlator to vanish in a light-like limit. Alternatively, one can use (\ref{MandMt}) to convert the reduced Mellin amplitudes into full Mellin amplitudes, from which one can extract three-point function coefficients of $\frac{1}{2}$-BPS operators and match with the known result \cite{Lee:1998bxa}. The reduced Mellin amplitudes (\ref{solAdS5}) reproduced all explicit examples computed in the literature \cite{Arutyunov:2000py,Arutyunov:2002fh,Arutyunov:2003ae,Berdichevsky:2007xd,Uruchurtu:2008kp,Uruchurtu:2011wh}, and also confirmed a conjecture for the $k_i=p$ correlators \cite{Dolan:2006ec}. The result was further checked in many examples in \cite{Arutyunov:2018neq,Arutyunov:2018tvn} where the traditional method was simplified to increase computational power. 

The remarkably simple reduced Mellin amplitudes (\ref{solAdS5}) have further interesting hidden structures. In \cite{Caron-Huot:2018kta}, it was pointed out that they can be unified into a single object in terms of a  hidden conformal symmetry in higher dimensions. Taking the simplest reduced correlator $H_{2222}(x_{ij}^2)$, which is a function of spacetime coordinates $x_{ij}^2$ only, one can construct a generating function $\mathbf{H}=H_{2222}(x_{ij}^2-t_{ij})$ by replacing its arguments $x_{ij}^2$ with the ten dimensional distances $x_{ij}^2-t_{ij}$. Then all the other reduced correlators $H_{k_1k_2k_3k_4}$ can be obtained by Taylor expanding $\mathbf{H}$ in $t_{ij}$ and collecting the allowed R-symmetry structures. Similar hidden structures have also been found in other theories defined on conformally flat backgrounds such as IIB supergravity on $AdS_3\times S^3\times K3$ \cite{Rastelli:2019gtj,Giusto:2020neo}, certain 4d $\mathcal{N}=2$ SCFTs containing super gluons on $AdS_5\times S^3$ \cite{Alday:2021odx}, and hypermultiplets on $AdS_2\times S^2$ \cite{Abl:2021mxo}. However, these properties are not shared by eleven dimensional supergravity on $AdS_7\times S^4$ and $AdS_4\times S^7$. Moreover, the approach reviewed in this subsection is also not suitable for these two backgrounds. For $AdS_7\times S^4$ one  can set up a similar algebraic bootstrap problem \cite{Rastelli:2017ymc}, and case by case one can show that the solution is unique. However, the analytic structure of the reduced Mellin amplitudes turns out to be much more obscure, which makes it difficult to find the general solution. For $AdS_4\times S^7$, the situation is even worse. The position space solution to the superconformal Ward identities involves non-local differential operators which are unclear how to interpret in Mellin space. Therefore, one needs to have a different strategy to deal with these cases, as we will explain in the next two subsections.

\subsection{Superconformal Ward identities for Mellin amplitudes}\label{Subsec:Mellinscfwi}
As was pointed out in the previous subsection, the Mellin space method of \cite{Rastelli:2016nze,Rastelli:2017udc} is specific to the $AdS_5\times S^5$ background. While the superconformal Ward identities (\ref{scfWardid}) have the same form in any spacetime dimension $d$, the form of the solutions are rather sensitive to the value of $d$. In order to have a bootstrap method in Mellin space which can be applied to any spacetime dimensions, we must return to the superconformal Ward identities themselves. In this subsection, we explain how we can translate the superconformal Ward identities into Mellin space, following \cite{Zhou:2017zaw,Alday:2020dtb}. 

We start by recalling that Mellin amplitudes are defined from $\mathcal{G}_{k_1k_2k_3k_4}(U,V;\sigma,\tau)$ as 
\begin{equation}\label{defManyd}
\mathcal{G}_{k_1k_2k_3k_4}(U,V;\sigma,\tau)=\int_{-i\infty}^{i\infty}\frac{dsdt}{(4\pi i)^2}U^{\frac{s}{2}-a_s}V^{\frac{t}{2}-a_t}\mathcal{M}_{k_1k_2k_3k_4}(s,t;\sigma,\tau)\Gamma_{\{k_i\}}\;,
\end{equation}
where $a_s=\frac{\epsilon}{2}(k_1+k_2)-\epsilon\mathcal{E}$, $a_t=\frac{\epsilon}{2}\min\{k_1+k_4,k_2+k_3\}$. The Gamma function factor $\Gamma_{\{k_i\}}$ also depends on $\epsilon$
\begin{equation}
\Gamma_{\{k_i\}}(s,t)=\Gamma[\tfrac{\epsilon(k_1+k_2)-s}{2}]\Gamma[\tfrac{\epsilon(k_3+k_4)-s}{2}]\Gamma[\tfrac{\epsilon(k_1+k_4)-t}{2}]\Gamma[\tfrac{\epsilon(k_2+k_3)-t}{2}]\Gamma[\tfrac{\epsilon(k_1+k_3)-u}{2}]\Gamma[\tfrac{\epsilon(k_2+k_4)-u}{2}]\;,
\end{equation}
and $s+t+u=\epsilon\sum_{i=1}^4 k_i\equiv \epsilon \Sigma$. The major obstacle of implementing the superconformal Ward identities (\ref{scfWardid}) in Mellin space is that the variables $z$ and $\bar{z}$ appear asymmetrically. By contrast, in the Mellin representation (\ref{defManyd}) $z$ and $\bar{z}$ appear only in the combination $U=z\bar{z}$, $V=(1-z)(1-\bar{z})$, and therefore $z\leftrightarrow\bar{z}$ is a symmetry. If one were to express $z$, $\bar{z}$ in terms of $U$ and $V$ in (\ref{scfWardid}), one would encounter square roots which are difficult to make sense of in Mellin space. However, this issue can be resolved if we take the linear combination of (\ref{scfWardid}) and the equation with the replacement $z\leftrightarrow\bar{z}$, as we now explain.

To begin, we write $z\partial_z$ in the superconformal Ward identity
\begin{equation}\label{scfWardidb1}
(z\partial_z-\epsilon \alpha\partial_\alpha)\mathcal{G}_{k_1k_2k_3k_4}(z,\bar{z};\alpha,\bar{\alpha})\big|_{\alpha=1/z}=0
\end{equation}
as
\begin{equation}
z\partial_z=U\partial_U-\frac{z}{1-z}V\partial_V\;.
\end{equation}
In Mellin space, $U\partial_U$ and $V\partial_V$ have simple multiplicative actions
\begin{equation}
U\partial_U\to\left(\frac{s}{2}-a_s\right)\times\;,\quad\quad V\partial_V\to\left(\frac{t}{2}-a_t\right)\times\;.
\end{equation}
On the other hand, $z$ is difficult to interpret due to the aforementioned square root. To proceed, we expand the Mellin amplitude $\mathcal{M}_{k_1k_2k_3k_4}$ in powers of $\alpha$
\begin{equation}
\mathcal{M}_{k_1k_2k_3k_4}(s,t;\alpha,\bar{\alpha})=\sum_{q=0}^{\mathcal{E}}\alpha^q \mathcal{M}_{k_1k_2k_3k_4}^{(q)}(s,t;\bar{\alpha})\;.
\end{equation}
In terms of the component amplitudes $\mathcal{M}_{k_1k_2k_3k_4}^{(q)}(s,t;\bar{\alpha})$, (\ref{scfWardidb1}) becomes
\begin{equation}
\sum_{q=0}^{\mathcal{E}}\left((1-z)z^{\mathcal{E}-q}\left(\frac{s}{2}-a_s-q\right)-z^{\mathcal{E}-q+1}\left(\frac{t}{2}-a_t\right)\right)\mathcal{M}_{k_1k_2k_3k_4}^{(q)}(s,t;\bar{\alpha})=0\;.
\end{equation}
We can obtain an inequivalent identity by replacing $z\to\bar{z}$ and get 
\begin{equation}
\sum_{q=0}^{\mathcal{E}}\left((1-\bar{z})\bar{z}^{\mathcal{E}-q}\left(\frac{s}{2}-a_s-q\right)-\bar{z}^{\mathcal{E}-q+1}\left(\frac{t}{2}-a_t\right)\right)\mathcal{M}_{k_1k_2k_3k_4}^{(q)}(s,t;\bar{\alpha})=0\;.
\end{equation}
Taking the sum and difference of these two relations we arrive at
\begin{equation}\label{WIpm}
\sum_{q=0}^{\mathcal{E}}\left((\zeta_\pm^{\mathcal{E}-q}-\zeta_\pm^{\mathcal{E}-q+1})\left(\frac{s}{2}-a_s-q\right)-\zeta_\pm^{\mathcal{E}-q+1}\left(\frac{t}{2}-a_t\right)\right)\mathcal{M}_{k_1k_2k_3k_4}^{(q)}(s,t;\bar{\alpha})=0\;,
\end{equation}
where 
\begin{equation}
\zeta_+^{(n)}=z^n+\bar{z}^n\;,\quad \quad \zeta_-^{(n)}=\frac{z^n-\bar{z}^n}{z-\bar{z}}\;.
\end{equation}
Importantly, each $\zeta_\pm^{(n)}$ is a polynomial of $U$ and $V$
\begin{equation}
\begin{split}
\zeta_+^{(n)}=&2^{1-n}\sum_{k=0}^{\lfloor n/2 \rfloor}\binom{n}{2k}\,\left((1+U-V)^2-4U\right)^k\,(1+U-V)^{n-2k}\;,\\
\zeta^{(n)}_-=&2^{1-n}\,\sum_{k=0}^{\lfloor n/2 \rfloor}
\binom{n}{2k+1}\,\left((1+U-V)^2-4U\right)^k\,(1+U-V)^{n-2k-1}\;.
\end{split}
\end{equation}
Every monomial $U^mV^n$ can be easily interpreted in Mellin space as a difference operator $\mathbb{O}_{mn}$ which acts as 
\begin{equation}
\mathbb{O}_{mn}\circ \mathcal{M}_{k_1k_2k_3k_4}^{(q)}(s,t;\bar{\alpha})=\frac{\Gamma_{\{k_i\}}(s-2m,t-2n)}{\Gamma_{\{k_i\}}(s,t)} \mathcal{M}_{k_1k_2k_3k_4}^{(q)}(s-2m,t-2n;\bar{\alpha})\;.
\end{equation}
Then (\ref{WIpm}) becomes difference constraints for the Mellin amplitudes which constitute the Mellin space version of the superconformal Ward identities. 

Let us now apply the Mellin space superconformal Ward identities in a concrete example. We compute the $k_i=2$ four-point amplitude of eleven dimensional super gravity on $AdS_4\times S^7$, which is beyond the reach of the position space method and the Mellin space method of \cite{Rastelli:2016nze,Rastelli:2017udc}. Similar to the position space method, we start with an ansatz in Mellin space which is the linear combination of all possible exchange and contact diagrams. In this example, we have three exchanged fields in each channel: a scalar field dual to $\mathcal{O}_2$ with $\Delta=1$ and $SO(8)$ Dynkin label $[2,0,0,0]$, a vector field with $\Delta=2$ and representation $[0,1,0,0]$, and the graviton field which has $\Delta=3$ and is neutral under $SO(8)$ R-symmetry. They are determined by the same selection rules as in the $AdS_5\times S^5$ case. Therefore, the s-channel exchange part of the ansatz is 
\begin{equation}\label{AdS4keq2s}
\mathcal{M}_s=\lambda_s Y_{[2,0,0,0]}\mathcal{M}_{1,0}+\lambda_v Y_{[0,1,0,0]}\mathcal{M}_{2,1}+\lambda_g Y_{[0,0,0,0]}\mathcal{M}_{3,2}\;,
\end{equation}
where 
\begin{equation}
Y_{[2,0,0,0]}=\sigma+\tau-\frac{1}{4}\;,\quad Y_{[0,1,0,0]}=\sigma-\tau\;,\quad Y_{[0,0,0,0]}=1\;,
\end{equation}
and the Mellin amplitudes of the bosonic Witten diagrams $\mathcal{M}_{\Delta,\ell}$ are
\begin{eqnarray}\label{bosonicM}
\nonumber \mathcal{M}_{1,0}&=&\sum_{m=0}^\infty-\frac{2 \Gamma \left(m+\frac{1}{2}\right)}{\pi ^{7/2} \Gamma (m+1)(s-1-2m)}=\frac{\Gamma \left(\frac{1}{2}-\frac{s}{2}\right)}{\pi ^{5/2} \Gamma \left(1-\frac{s}{2}\right)}\;,\\
\mathcal{M}_{2,1}&=&\sum_{m=0}^\infty\frac{16 \Gamma \left(m+\frac{1}{2}\right) (u-t)}{\pi ^{7/2} (2 m+1) \Gamma (m+1)(s-1-2m)}=\frac{8(t-u)}{s}\left(\mathcal{M}_{1,0}-\frac{1}{\pi^2}\right)\;,\\
\nonumber \mathcal{M}_{3,2}&=&\sum_{m=0}^\infty -\frac{64 \Gamma \left(m+\frac{1}{2}\right)^2 \left(t^2-6 t (u-1)+u (u+6)-8\right)}{3 \pi ^{7/2} \Gamma (m+1) \Gamma \left(m+\frac{5}{2}\right)(s-1-2m)}\\
\nonumber &=&\frac{128 \left(t^2-6 t u+6 t+u^2+6 u-8\right)}{3 s (s+2)}\left(\mathcal{M}_{1,0}-\frac{s+4}{4 \pi ^2}\right)\;.
\end{eqnarray}
The expressions of these bosonic exchange Mellin amplitudes can be found in, {\it e.g.}, Appendix B of \cite{Alday:2020dtb}. The exchange contributions $\mathcal{M}_t$, $\mathcal{M}_u$ in the t- and u-channel are obtained from $\mathcal{M}_s$ by crossing
\begin{equation}
\mathcal{M}_t(s,t;\sigma,\tau)=\tau^2 \mathcal{M}_s(t,s;\tfrac{\sigma}{\tau},\tfrac{1}{\tau})\;,\quad \mathcal{M}_u(s,t;\sigma,\tau)=\sigma^2\mathcal{M}_s(u,t;\tfrac{1}{\sigma},\tfrac{\tau}{\sigma})\;.
\end{equation}
Finally, we also include a contact term $\mathcal{M}_{\rm con}$ which is a polynomial linear in $s$, $t$ and of degree 2 in $\sigma$, $\tau$ with unknown coefficients. The Mellin ansatz is 
\begin{equation}\label{AdS4keq2}
\mathcal{M}_{2222}=\mathcal{M}_s+\mathcal{M}_t+\mathcal{M}_u+\mathcal{M}_{\rm con}\;.
\end{equation} 
It is now straightforward to impose the Mellin space superconformal Ward identities as described above. It turns out that the identities with the $+$ sign are already sufficient to fix all the unknowns up to an overall constant, and the $-$ sign identities impose no further constraints. We find that  
\begin{equation}\label{lambdas}
\lambda_v=-\frac{1}{8}\lambda_s\;,\quad \lambda_g=\frac{3}{512}\lambda_s\;,
\end{equation}
and the contact term is fixed to be
\begin{equation}
\mathcal{M}_{\rm con}=\frac{\lambda_s}{16\pi^2}(-s \sigma ^2-16 \tau  (-\sigma  s+s+4 \sigma +t)+s-\sigma ^2 t+8 \sigma  (\sigma +2 t-8)+(t+4) \tau ^2+4)\;.
\end{equation}
The remaining parameter $\lambda_s$ can be fixed by using the known three-point function of $\mathcal{O}_2$ to be
\begin{equation}
\lambda_s=-\frac{3\pi}{\sqrt{2}N^{\frac{3}{2}}}\;,
\end{equation}  
where $N$ is the number of M2 branes. 

This bootstrap method in Mellin space is quite powerful. Although we demonstrated the method here in the $AdS_4\times S^7$ case, it applies in the exact same way to $AdS_5\times S^5$ and $AdS_7\times S^4$ backgrounds. Moreover, we can analogously consider four-point amplitudes of operators with higher Kaluza-Klein levels. Imposing the Mellin space superconformal Ward identities also fixes all exchange coefficients and contact terms up to an overall constant factor. However, as the external weights are increased there are more and more unknowns and the complexity of the calculation also increases. Therefore, without a better understanding of the structure of the Mellin amplitudes it is a bit difficult to obtain all the amplitudes using this method. We will study the properties of these Mellin amplitudes in detail in the next subsection, and point out several remarkable simplifying features which are dictated by symmetries. With this improved understanding, we explain how to derive all tree-level super graviton four-point amplitudes in all maximally superconformal theories. 

\subsection{The MRV method and super graviton amplitudes in all maximal SCFTs}\label{Subsec:MRV}
In this subsection, we review the method of \cite{Alday:2020lbp,Alday:2020dtb} which applies to all theories with maximal superconformal symmetry. We will use the $AdS_4\times S^7$ four-point amplitude computed in the previous subsection as a concrete example to show various properties, and we will demonstrate the new method by re-deriving the result. 

Although this $k_i=2$ example is the simplest correlator in the $AdS_4\times S^7$ theory, its Mellin amplitude is already quite complicated. However, as we will see below, there are special limits where the Mellin amplitude drastically simplifies. In \cite{Alday:2020lbp,Alday:2020dtb}, a notion called {\it maximally R-symmetry violating} (MRV) was introduced. In the MRV limit, we restrict ourselves to the special kinematic configuration where the R-symmetry polarization vectors of operator 1 and 3 are parallel, {\it i.e.}, $t_1=t_3$.\footnote{More precisely, this limit is the u-channel MRV limit. One can also define MRV limits in the other two channels.} In terms of the R-symmetry cross ratios, this amounts to setting  $\sigma=0$ and $\tau=1$. This slice of the Mellin amplitude (\ref{AdS4keq2}), defined to be the MRV amplitude, now has a particularly simple structure
\begin{equation}\label{AdS4keq2MRV}
\begin{split}
&\mathbf{MRV}_{2222}(s,t)=\mathcal{M}_{2222}(s,t;0,1)\\
&=\frac{3(u-2)(u-4)}{\sqrt{2\pi}N^{\frac{3}{2}}}\sum_{m=0}^\infty \frac{1}{m!\Gamma \left(\frac{1}{2}-m\right)^2 \Gamma \left(m+\frac{5}{2}\right)} \bigg( \frac{1}{s-1-2m}+\frac{1}{t-1-2m}\bigg)\;.
\end{split}
\end{equation}
There are two important features in this expression worth noticing. First, the MRV amplitude does not have any poles in the u-channel. Only the poles in the s- and t-channels are present. Second, the MRV amplitude contains a factor of zeros in $u$, and the locations of the zeros correspond to the double-trace long operators with low-lying twists.   

Physically, these two features can be understood as follows. Setting $t_1=t_3=t_u$ implies that the only R-symmetry representation which can propagate through the u-channel is the rank-4 symmetric traceless representation. This is easy to see from the fact that the Wick contraction of $t_1$ with $t_3$ is zero and it leads to
\begin{equation}
t_1^{\{I_1}t_1^{I_2\}}t_3^{\{I_3}t_3^{I_4\}}=t_u^{\{I_1}t_u^{I_2}t_u^{I_3}t_u^{I_4\}}\;.
\end{equation}
On the other hand, we do not have any exchanged single-trace particles in the $k_i=2$ four-point amplitude which transform in the rank-4 symmetric traceless representation. Recall we mentioned in Section \ref{Subsec:positionspace} that couplings of single-trace fields with weights satisfying $k_a+k_b=k_c$ are known as extremal. Three-point Witten diagrams associated with such a coupling diverge. In order to keep the effective action finite extremal couplings must vanish and all the exchanges are sub-extremal. This explains why we do not see any poles in the u-channel associated with single-trace particle exchanges. Relatedly, the fact that only the rank-4 symmetric traceless representation is exchanged in the u-channel also implies that certain long operators are decoupled in the MRV limit. This has to do with the structure of the long superconformal multiplets, in which the superconformal descendants have varying R-symmetry representations. In order for all R-symmetry representations of the entire multiplet to fit inside the $k_i=2$ four-point correlator, the superconformal primaries of the long multiplets must be  R-symmetry singlets. Therefore, in the MRV limit we do not see the exchange of such super primaries. The visible super descendants, which are in the rank-4 symmetric traceless representation, have conformal twists larger than their super primaries by 4. Since all long operators in the supergravity limit are double-trace, the minimal conformal twists of the super primaries is 2. Long operators with twist 2 and 4 are missing in the MRV limit. The decoupling of such operators is manifested in the Mellin amplitudes as zeros.  

The presence of the zeros turns out to be a very useful fact. In fact, this property is satisfied by the exchange amplitude of each single-trace super multiplet in each channel. It turns out that we can use the zeros to fix the relative coefficients of the component fields, {\it i.e.}, the ratios of $\lambda_s$, $\lambda_v$ and $\lambda_g$, without using the superconformal Ward identities. However, before we are able to do this, we need to take a small detour to explain an obstruction. Let us first notice that the s-channel part of the MRV amplitude has a better Regge behavior than the bosonic exchange amplitudes (\ref{bosonicM}). Here the relevant Regge limit is the u-channel Regge limit, in which we take $s\to\infty$ keeping $u$ fixed. The s-channel part of the MRV amplitude scales as
\begin{equation}
\mathbf{MRV}^{(s)}=\frac{3(u-2)(u-4)}{\sqrt{2\pi}N^{\frac{3}{2}}}\sum_{m=0}^\infty \frac{1}{m!\Gamma \left(\frac{1}{2}-m\right)^2 \Gamma \left(m+\frac{5}{2}\right)}\frac{1}{s-1-2m}\sim \frac{1}{s}\;.
\end{equation}
Such a Regge behavior is expected because the numerator at each pole has degree 2 and is saturated by the two zero factors in $u$. By contrast, the s-channel spin-$\ell$ exchange amplitude scales as 
\begin{equation}
\mathcal{M}_{\Delta,\ell}\sim \frac{1}{s^{1-\ell}}\;.
\end{equation}
Naively, this would seem to be a contradiction as the MRV amplitude is a collection of bosonic Mellin amplitudes. However, we note that we can improve the bosonic Mellin amplitudes by adding contact terms. We do this by simply replacing $t$ in the summands in (\ref{bosonicM}) by $3-u-2m$, which follows from solving 
\begin{equation}\label{mvst}
s+t+u=4\;,\quad\quad s=1+2m\;.
\end{equation}
Note that we are allowed to use the pole value of $s$ as in the second condition when focusing on the residues of the Mellin amplitude. Apparently, the new bosonic amplitudes 
\begin{eqnarray}\label{bosonicMimp}
\nonumber \mathcal{M}_{1,0}^{\rm imp}=&&\mathcal{M}_{1,0}\\
\mathcal{M}_{2,1}^{\rm imp}=&&\sum_{m=0}^\infty\frac{16 \Gamma \left(m+\frac{1}{2}\right) (2u-3+2m)}{\pi ^{7/2} (2 m+1) \Gamma (m+1)(s-1-2m)}\;,\\
\nonumber \mathcal{M}_{3,2}^{\rm imp}=&&\sum_{m=0}^\infty -\frac{64 \Gamma \left(m+\frac{1}{2}\right)^2 \left(4 m^2+8 m (2 u-3)+8 u^2-24 u+19\right)}{3 \pi ^{7/2} \Gamma (m+1) \Gamma \left(m+\frac{5}{2}\right)(s-1-2m)}\;,
\end{eqnarray}
have the improved $s^{-1}$ Regge behavior as $\mathbf{MRV}^{(s)}$. The differences between the original and Regge-improved exchange amplitudes are contact terms as promised
\begin{equation}
\mathcal{M}_{2,1}-\mathcal{M}_{2,1}^{\rm imp}=\frac{8}{\pi^2}\;,\quad\quad  \mathcal{M}_{3,2}-\mathcal{M}_{3,2}^{\rm imp}=\frac{32 (7 s+8 t-20)}{3 \pi ^2}\;.
\end{equation}
We now take the following modified combination of s-channel exchange amplitudes
\begin{equation}\label{AdS4keq2sp}
\mathcal{M}'_s=\lambda_s Y_{[2,0,0,0]}\mathcal{M}_{1,0}^{\rm imp}+\lambda_v Y_{[0,1,0,0]}\mathcal{M}_{2,1}^{\rm imp}+\lambda_g Y_{[0,0,0,0]}\mathcal{M}_{3,2}^{\rm imp}
\end{equation}
and require that it has zeros at $u=2$ and $u=4$ when $\sigma=0$, $\tau=1$. It is most straightforward to impose the zeros at every pole, and we find this condition fixes the exchange coefficients to be (\ref{lambdas}). However, using the Regge-improved Witten diagrams also comes with a problem: it breaks the s-channel Bose symmetry, {\it i.e.}, $\mathcal{M}'_s$ is not invariant under the $1\leftrightarrow 2$ exchange
\begin{equation}
\mathcal{M}'_s(s,t;\sigma,\tau)\neq \mathcal{M}'_s(s,u;\tau,\sigma)\;.
\end{equation}
To understand this issue better, let us write $\mathcal{M}'_s$ as a sum over simple poles
\begin{equation}
\mathcal{M}'_s=\lambda_s \sum_{m=0}^\infty \frac{\Gamma \left(m+\frac{1}{2}\right)^2}{\pi ^{7/2} \Gamma (m+1) \Gamma \left(m+\frac{5}{2}\right)}\times \frac{L_m(u;\sigma,\tau)}{s-1-2m}\;,
\end{equation}
where 
\begin{equation}
L_m(u;\sigma,\tau)=(2 m+3) \tau  (u-2)-(2 m+u-1) ((2 m+3) \sigma +u-2)\;.
\end{equation}
Clearly, the absence of the $t$ variable, which is mapped to $u$ under $1\leftrightarrow 2$ crossing, is responsible for the breaking of  Bose symmetry. A simple fix is to reverse (\ref{mvst}), and set $m=\frac{1}{2}(3-t-u)$ in $L_m$. This leads to a new s-channel multiplet exchange amplitude
\begin{equation}\label{Nfackeq2}
M_s=\lambda_s \sum_{m=0}^\infty \frac{\Gamma \left(m+\frac{1}{2}\right)^2}{\pi ^{7/2} \Gamma (m+1) \Gamma \left(m+\frac{5}{2}\right)}\times \frac{N(t,u;\sigma,\tau)}{s-1-2m}\;,
\end{equation}
with symmetric numerators
\begin{equation}
N(t,u;\sigma,\tau)=-(t-2) (\sigma  (t+u-6)-u+2)-\tau  (u-2) (t+u-6)\;,
\end{equation}
which differs from $\mathcal{M}'_s$ and $\mathcal{M}_s$ by contact terms. Importantly, we note that this prescription leading to $L_m\to N$ preserves the desired behavior at the MRV limit
\begin{equation}
M_s(s,t;0,1)=\mathbf{MRV}^{(s)}(s,t)\;.
\end{equation}
So far, we have only focused on the exchange part of the correlator which involves poles. There are in principle also contact terms which are regular. However, there is an interesting surprise in choosing $M_s$ to be the s-channel exchange amplitude. Using crossing symmetry, we obtain $M_t$ and $M_u$ as
\begin{equation}
M_t(s,t;\sigma,\tau)=\tau^2 M_s(t,s;\tfrac{\sigma}{\tau},\tfrac{1}{\tau})\;,\quad M_u(s,t;\sigma,\tau)=\sigma^2M_s(u,t;\tfrac{1}{\sigma},\tfrac{\tau}{\sigma})\;.
\end{equation}
Remarkably, we find that the full $k_i=2$ Mellin amplitude can be written in terms of just the exchange amplitudes
\begin{equation}
\mathcal{M}_{2222}=M_s+M_t+M_u\;,
\end{equation}
without any additional contact terms! Physically, this result implies that the contact interactions are not intrinsic.\footnote{Such a property would be important for the existence of Britto-Cachazo-Feng-Witten type relations \cite{Britto:2005fq} which recursively construct higher-point amplitudes from lower-point ones.} Compared to (\ref{AdS4keq2}), our new prescription of constructing multiplet exchange amplitudes has the clear advantage that it automatically absorbs the contact terms.  

The above appealing features of the $k_i=2$ correlator in fact generalize to correlators with arbitrary Kaluza-Klein weights, and also to correlators in the $AdS_5\times S^5$ and $AdS_7\times S^4$ backgrounds. Exploiting these features, we get a universal new method which can be used to efficiently compute all tree-level four-point correlators in all maximally superconformal theories. We will only outline the procedures here, and refer the reader  to the original papers \cite{Alday:2020lbp,Alday:2020dtb} for details and explicit expressions of general correlators. The method works as follows.
\begin{enumerate}
\item We start with an ansatz similar to (\ref{AdS4keq2s}) for the s-channel exchange amplitude of each multiplet. For a generic multiplet with Kaluza-Klein level $p\geq 4$, there are six component fields with Lorentz spins up to 2 (see Appendix \ref{App:multipletcomponents}).
\item We improve the u-channel Regge behavior in each bosonic exchange Mellin amplitude $\mathcal{M}_{\Delta,\ell}$ by eliminating the $t$ variable in the numerators in favor of $u$ and $m$, which is possible because of the condition $s+t+u=\epsilon \Sigma$ and the fact that poles are located at $s=\epsilon p+2m$.  
\item We impose the condition that in the MRV limit, the multiplet exchange amplitude should have two zeros at $u=\epsilon\max\{k_1+k_3,k_2+k_4\}$ and $u=\epsilon\max\{k_1+k_3,k_2+k_4\}+2$. This fixes the exchange coefficients of each component field in the multiplet up to an overall factor which can be chosen to be the exchange coefficient of the super primary. Using the known three-point function coefficients of the $\frac{1}{2}$-BPS operators, this overall factor can be computed. 
\item We now restore the s-channel Bose symmetry which has been lost when using the Regge-improved exchange Witten diagrams. It turns out that every numerator at each simple pole contains the same kinematic factor which includes all the Mandelstam variable dependence and is a degree-2 polynomial in $u$. We reverse the substitution performed in step 2, and replace $m$ in this factor by $\frac{1}{2}(\epsilon(\Sigma-p)-t-u)$ to obtain a new kinematic factor.   
\item Finally, we add up all the multiplet exchange amplitudes in three channels which are compatible with the selection rules. These include the R-symmetry selection rule and the requirement that cubic couplings are non-extremal. {\it A priori}, we should also include the most general contact terms in the ansatz which have all possible R-symmetry structures and are linear in the Mandelstam variables. However, by solving the Mellin space superconformal Ward identities we find all such additional contact terms vanish. The four-point amplitudes are therefore purely made of the multiplet exchange amplitudes. 
\end{enumerate}

We conclude this subsection with a few comments. In Step 3 we fixed the overall coefficient of each multiplet by using the known three-point functions. This is convenient but not necessary. We could leave these coefficients undetermined and then use the superconformal Ward identities to fix them up to an overall constant. The remaining overall constants for different correlators are not independent because they can be interpreted as the product of two three-point function coefficients. The same three-point function coefficients can appear in different correlators. By considering a system of mixed correlators ({\it e.g.}, correlators of the form $\langle ppqq \rangle$) we can reduce all remaining constants to the overall factor which appears in the $k_i=2$ factor. The latter is determined by the central charge of the theory. 

The other comment is related to the factorized structure mentioned in Step 4. Each multiplet exchange amplitude contains a degree-2 polynomial of $t$ and $u$ which is independent of $m$ ({\it e.g.}, it is the factor $N(s,t;\sigma,\tau)$ in (\ref{Nfackeq2}) for $k_i=2$ in $AdS_4\times S^7$). This factor can be brought outside of the Mellin transformation as a differential operator. The remaining dependence on the Mandelstam variables is a sum over simple poles with constant numerators, and resembles a scalar exchange Mellin amplitude. More precisely, one can show that it can be written as the linear combination of three $AdS_{d+1}$ scalar exchange amplitudes with internal dimensions $\epsilon p$, $\epsilon p+2$ and $\epsilon p+4$. For example, (\ref{Nfackeq2}) can be written as       
\begin{equation}
M_s=\lambda_s N(s,t;\sigma,\tau) \left(-\frac{2}{3}\mathcal{M}_{1,0}^{AdS_4}+\frac{1}{30}\mathcal{M}_{3,0}^{AdS_4}-\frac{3}{11200}\mathcal{M}_{5,0}^{AdS_4}\right)\;.
\end{equation}
Remarkably, we can also write the sum over simple poles as a single scalar exchange diagram with internal dimension $\epsilon p$ but in a {\it lower} dimensional $AdS_{d-3}$ space! For example, the reader can check that (\ref{Nfackeq2}) can also be expressed as\footnote{Note that the spacetime dimension $d$ is treated here as a formal parameter.} 
\begin{equation}
M_s=-\frac{3}{2} \lambda_s N(s,t;\sigma,\tau) \mathcal{M}_{1,0}^{AdS_0}\;.
\end{equation}
In fact, this emergent dimensional reduction structure was observed in \cite{Behan:2021pzk} for all four-point correlators in all maximally supersymmetry CFTs, and appears to be related to the Parisi-Sourlas supersymmetry \cite{Parisi:1979ka}. As was shown in  \cite{Kaviraj:2019tbg}, Parisi-Sourlas supersymmetry gives rise to dimensional reduction relations for conformal blocks as a kinematic consequence. These identities can further be lifted into AdS space as reduction formulae for exchange Witten diagrams \cite{Zhou:2020ptb} (see Appendix \ref{Subapp:recur} for details), which underlie the dimensional reduction structure mentioned above. However, at this moment there is still no understanding of the physical meaning of the observed structure.

\subsection{Super gluon four-point amplitudes}\label{Subsec:supergluons}
The MRV method reviewed in the previous subsection can also be used to efficiently compute super gluon scattering amplitudes in AdS. In many holographic CFTs with non-maximal superconformal symmetry, the theories contain a sector of states which decouple from the rest of the theories in the large central charge limit. The single-trace states in this sector have at most Lorentz spin 1. Therefore, the sector is not gravitational, and we have a supersymmetric gauge theory of gluons and their super partners. 

More precisely, these superconformal CFTs usually preserve eight Poincar\'e supercharges. In the holographic dual descriptions of these SCFT$_d$ there are singular loci of the form $AdS_{d+1}\times S^3\subset AdS_{d+1}\times_{(w)} X$,\footnote{Here $w$ denotes the possibility of a warped product.} where $X$ is the full internal space of dimension $9-d$ or $10-d$ depending on whether it is a string or M theory. On these loci, there are localized degrees of freedom which are organized by supersymmetry into a vector multiplet. The vector multiplet transforms in the adjoint representation of some global symmetry group $G_F$ which depends on the theory and is interpreted as a gauge group from the bulk perspective. Performing a Kaluza-Klein reduction on $S^3$, we get an infinite tower of states with spins up to 1 and they are organized into $\frac{1}{2}$-BPS multiplets of the non-maximal superconformal algebra. We refer to the scalar super primaries as the super gluons. By contrast, gravity lives in the total space $ AdS_{d+1}\times_{(w)} X$ and the Kaluza-Klein reduction on the internal space $X$ gives rise to gravitons and their super partners with varying masses. Importantly, the self-couplings of the super gluons are much stronger than the couplings of super gluons with super gravitons by powers of the central charge, in the limit where the central charge is large. Therefore, in the large central charge limit we can decouple gravity when considering the leading contribution to the super gluon correlators.

The $SO(4)$ isometry group of $S^3$ can be written as $SU(2)_R\times SU(2)_L$. The first $SU(2)_R$ factor is an R-symmetry group, while the second $SU(2)_L$ group is a global symmetry. The super gluons have $SU(2)$ spins $j_R=\frac{k}{2}$, $j_L=\frac{k-2}{2}$, with $k=2,3,\ldots$, and their conformal dimensions are determined by their Kaluza-Klein levels $k$ to be $\Delta=\epsilon k$. The tree-level four-point amplitudes of super gluons are similar to the super graviton amplitudes at the level of the diagrammatic structure. They contain exchange contributions of super gluon multiplets (the relevant bosonic component fields are tabulated in Appendix \ref{App:multipletcomponents}), and possibly also contact contributions. In \cite{Alday:2021odx}, all super gluon four-point functions with arbitrary Kaluza-Klein levels $k_i$ were systematically computed in a variety of theories with $d=3,4,5,6$. Here we will only demonstrate the computational method in the so-called E-string theory \cite{Ganor:1996mu,Seiberg:1996vs} which has $d=6$, and point out various interesting features. In this theory, the flavor group $G_F$ is fixed to be $E_8$. Furthermore, we will only consider the simplest $k_i=2$ correlator \cite{Zhou:2018ofp}, and refer the reader to \cite{Alday:2021odx} for a comprehensive treatment. 

Compared to the super graviton case, the main difference here is that super gluons carry flavor (or color, from the bulk perspective) indices and the correlator
\begin{equation}
G^{A_1A_2A_3A_4}(x_i;v_i)=\langle \mathcal{O}^{A_1}_2(x_1;v_1) \mathcal{O}^{A_2}_2(x_2;v_2) \mathcal{O}^{A_3}_2(x_3;v_3) \mathcal{O}^{A_4}_2(x_4;v_4) \rangle
\end{equation}
has several independent flavor structures. Here $A=1,\ldots, {\rm dim}G_F$, and we have contracted the super gluon operators with two-component auxiliary spinors $v^a$, $a=1,2$, to keep track of the R-symmetry polarizations
\begin{equation}
\mathcal{O}^{A}_2(x;v)=\mathcal{O}^{A;a_1a_2}(x)v^{b_1}v^{b_2}\epsilon_{a_1b_1}\epsilon_{a_2b_2}\;.
\end{equation}
The number of independent flavor structures corresponds to the number of representations that appear in the tensor product of two adjoint representations of $G_F$. We can concretely discuss these flavor structures by decomposing the correlator into different flavor channels using projectors
\begin{equation}
G^{A_1A_2A_3A_4}=\sum_{a\in {\rm adj}\otimes {\rm adj}} {\rm P}^{A_1A_2|A_3A_4}_a G_a\;.
\end{equation}
Here ${\rm P}^{A_1A_2|A_3A_4}_a$ are s-channel projectors, and represent the exchange of irreducible flavor representations $a$ in this channel. For example, the projectors associated with exchanging the identity representation and the adjoint representations are
\begin{equation}\label{P1Padj}
{\rm P}^{A_1A_2|A_3A_4}_{\bf 1}=\frac{1}{{\rm dim}(G_F)}\delta^{A_1A_2}\delta^{A_3A_4}\;,\quad {\rm P}^{A_1A_2|A_3A_4}_{\rm adj}=\frac{1}{\psi^2 h^\vee}f^{A_1A_2A_5}f^{A_5A_3A_4}\;,
\end{equation} 
where $h^\vee$ is the dual Coxeter number, $\psi^2$ is the length squared of the longest root, and $f^{ABC}$ are the structure constants of the flavor group. These projectors satisfy
\begin{equation}
{\rm P}^{A_1A_2|A_3A_4}_a=(-1)^{{\rm R}_a}{\rm P}^{A_2A_1|A_3A_4}_a\;,\quad {\rm P}^{A_1A_2|A_3A_4}_a={\rm P}^{A_3A_4|A_1A_2}_a\;,
\end{equation}
where ${\rm R}_a$ is $0$ for symmetric representations and $1$ for antisymmetric representations. Moreover, they are idempotent 
\begin{equation}\label{idempo}
{\rm P}^{A_1A_2|A_3A_4}_a{\rm P}^{A_4A_3|A_5A_6}_b=\delta_{ab}{\rm P}^{A_1A_2|A_5A_6}_a\;,
\end{equation}
and contracting external indices gives a delta function for the exchanged representations 
\begin{equation}\label{Pdimform}
{\rm P}^{A_1A_2|A_3A_4}_a{\rm P}^{A_1A_2|A_3A_4}_b=\delta_{ab}{\rm dim}({\rm R}_a)\;.
\end{equation}
We will also encounter the situation where we cross a t- or u-channel representation into the s-channel. This is accomplished by the flavor crossing matrices which are the overlaps of projectors in two channels
\begin{equation}\label{defFtu}
({\rm F}_t)_{a}{}^{a'}\equiv \frac{1}{{\rm dim}({\rm R}_a)}{\rm P}^{A_3A_2|A_1A_4}_{a} {\rm P}^{A_1A_2|A_3A_4}_{a'}\;,\;\; ({\rm F}_u)_{a}{}^{a'}\equiv \frac{1}{{\rm dim}({\rm R}_a)}{\rm P}^{A_4A_2|A_3A_1}_{a} {\rm P}^{A_1A_2|A_3A_4}_{a'}\;.
\end{equation}
For $E_8$, these crossing matrices explicitly read \cite{Chang:2017xmr}
\begin{equation} 
{\rm F}_t=\left(
\begin{array}{ccccc}
 \frac{1}{248} & \frac{125}{8} & \frac{3375}{31} & 1 & \frac{245}{2} \\
 \frac{1}{248} & -\frac{3}{8} & \frac{27}{31} & \frac{1}{5} & -\frac{7}{10} \\
 \frac{1}{248} & \frac{1}{8} & \frac{23}{62} & -\frac{1}{30} & -\frac{7}{15} \\
 \frac{1}{248} & \frac{25}{8} & -\frac{225}{62} & \frac{1}{2} & 0 \\
 \frac{1}{248} & -\frac{5}{56} & -\frac{90}{217} & 0 & \frac{1}{2} \\
\end{array}
\right)\;,\quad {\rm F}_u=\left(
\begin{array}{ccccc}
 \frac{1}{248} & \frac{125}{8} & \frac{3375}{31} & -1 & -\frac{245}{2} \\
 \frac{1}{248} & -\frac{3}{8} & \frac{27}{31} & -\frac{1}{5} & \frac{7}{10} \\
 \frac{1}{248} & \frac{1}{8} & \frac{23}{62} & \frac{1}{30} & \frac{7}{15} \\
 -\frac{1}{248} & -\frac{25}{8} & \frac{225}{62} & \frac{1}{2} & 0 \\
 -\frac{1}{248} & \frac{5}{56} & \frac{90}{217} & 0 & \frac{1}{2} \\
\end{array}
\right)\;
\end{equation}
where  the representations in each column are $a=\bf{1},\bf{3875},\bf{27000},\bf{248}\;({\rm adj}), \bf{30380}$ from top to bottom. The first three representations are symmetric, while the last two are anti-symmetric. 

We are now ready to bootstrap super gluon four-point amplitude. We similarly start with an ansatz which consists of exchange contributions in three channels and possible contact terms
\begin{equation}
\mathcal{M}^{A_1A_2A_3A_4}(s,t;\alpha)=\mathcal{M}_s^{A_1A_2A_3A_4}+\mathcal{M}_t^{A_1A_2A_3A_4}+\mathcal{M}_u^{A_1A_2A_3A_4}+\mathcal{M}_{\rm con}^{A_1A_2A_3A_4}\;.
\end{equation}
We have extracted an overall factor $(v_1\cdot v_2)^2(v_3\cdot v_4)^2$, where $(v_i\cdot v_j)\equiv v_i^av_j^b\epsilon_{ab}$ , from the correlator so that we can write it as a degree-2 polynomial in the $SU(2)_R$ R-symmetry cross ratio
\begin{equation}
\alpha=\frac{(v_1\cdot v_3)(v_2\cdot v_4)}{(v_1\cdot v_2)(v_3\cdot v_4)}\;.
\end{equation}
In each channel, we have two exchanged fields: the super gluon itself with $\Delta=4$ and $SU(2)_R$ spin 1, and a vector field with $\Delta=5$ and neutral under $SU(2)_R$. Since the super gluon multiplet is in the adjoint representation, the exchange amplitudes in each channel contain a single flavor structure, namely
\begin{equation}
\mathcal{M}_s^{A_1A_2A_3A_4}=\mathtt{c}_s M_s\;,\quad \mathcal{M}_t^{A_1A_2A_3A_4}=\mathtt{c}_t M_t\;,\quad \mathcal{M}_u^{A_1A_2A_3A_4}=\mathtt{c}_u M_u
\end{equation}
where 
\begin{equation}
\mathtt{c}_s=f^{A_1A_2B}f^{BA_3A_4}\;,\quad \mathtt{c}_t=f^{A_1A_4B}f^{BA_2A_3}\;,\quad \mathtt{c}_u=f^{A_1A_3B}f^{BA_4A_2}
\end{equation}
are proportional to the projectors ${\rm P}^{A_1A_2|A_3A_4}_{\rm adj}$, ${\rm P}^{A_3A_2|A_1A_4}_{\rm adj}$, ${\rm P}^{A_4A_2|A_3A_1}_{\rm adj}$. Crossing symmetry relates $M_{s,t,u}$ as 
\begin{equation}
M_t=(\alpha-1)^2\big(M_s\big|^{\{s,t,u\}\to\{t,u,s\}}_{\alpha\to\frac{1}{1-\alpha}}\big)\;,\quad M_u=(-\alpha)^2\big(M_s\big|^{\{s,t,u\}\to\{u,s,t\}}_{\alpha\to\frac{\alpha-1}{\alpha}}\big)\;.
\end{equation}
On the other hand, the contact term ansatz $\mathcal{M}_{\rm con}^{A_1A_2A_3A_4}$ should include all flavor structures. Therefore, in each flavor channel we should write down a regular amplitude which is degree 2 in $\alpha$ and independent of Mandelstam variables. The latter is because we expect that the quartic vertices do not  include any derivatives. 

At this point, we have two choices. We could proceed by simply writing $M_s$ as the linear combination of the scalar and vector exchange Mellin amplitudes as in (\ref{AdS4keq2s}), and then imposing the superconformal Ward identities. The $SU(2)_R$ polynomials associated with spin $j_R$ are given by $Y_{j_R}=\frac{(j_R!)^2}{(2j_R)!}P_{j_R}(2\alpha-1)$. The superconformal Ward identities are almost identical to the maximally superconformal case (\ref{scfWardid}), except that now we do not have the cross ratio $\bar{\alpha}$ and there is an independent identity for each flavor channel. Therefore, we decompose the Mellin amplitude ansatz in the s-channel flavor channels, and impose the Mellin space superconformal Ward identities (\ref{WIpm})  in each channel with $\bar{\alpha}$ set to zero. That is how the computation was performed in \cite{Zhou:2018ofp}, and one finds that solving the superconformal Ward identities leads to a unique answer, modulo an overall constant. However, written in this way the structure of the amplitude is quite obscure. The contact terms in different flavor channels are not vanishing. Note that one can also repeat this exercise in  other theories where one has multiple choices of the flavor group (such as in the 5d Seiberg exceptional theories \cite{Seiberg:1996bd}). The expressions of the contact terms depend on the gauge group chosen. Therefore it is not obvious if the {\it whole} amplitudes depend sensitively on $G_F$. The other choice is to extend the MRV method to super gluon amplitudes. Remarkably, as we will show below, using the MRV method leads to a very nice form of the answer which does not have explicit contact terms, and is furthermore agnostic about $G_F$. This allows us to have a much clearer understanding of the structure of super gluon correlators.  

In analogy with the super graviton case, we can define the (u-channel) MRV limit by requiring the polarization spinors of particle 1 and 3 to be identical $v_1=v_3$. Translated in terms of the cross ratio, the MRV limit corresponds to the slice with $\alpha=0$. In the MRV limit, only the $j_R=2$ representation can propagate in the u-channel. As a result of superconformal symmetry, we find the MRV amplitude should have the following two features similar to the maximally superconformal case:
\begin{itemize}
\item There are no poles in the u-channel.
\item The MRV amplitude contains a factor of zero in $u$ at the minimal double-trace twist location $u=8$. 
\end{itemize}
Note that compared to the super graviton case, here we have only one zero in the MRV limit instead of two. This is because we have less supersymmetry and long multiplets are smaller in size. We can now follow the same procedure to determine the multiplet exchange amplitude. We use the Regge-improved bosonic Mellin amplitudes (\ref{bosonicMimp}) in the exchange ansatz. Imposing the zero at every pole fixes the ratio of exchange coefficients. Finally, we restore the s-channel Bose symmetry using the prescription given in Section \ref{Subsec:MRV} and we find 
\begin{equation}\label{Msgluon}
M_s=\frac{(\alpha  (t+u-16)-u+8)}{4N^2}\left(\frac{1}{s-4}+\frac{1}{3(s-6)}\right)\;.
\end{equation}
Here we have also fixed the overall coefficient using the flavor current central charge  and $N$ is the number of M5 branes in the brane construction of this theory. Using these multiplet exchange amplitudes, one can check that superconformal Ward identities force the additional contact terms to be zero. The $k_i=2$ four-point amplitude is therefore simply
\begin{equation}\label{gluon4pttree}
\mathcal{M}^{A_1A_2A_3A_4}(s,t;\alpha)=\mathtt{c}_s M_s+\mathtt{c}_t M_t+\mathtt{c}_u M_u\;.
\end{equation}
The MRV method applies similarly to massive correlators with $k_i>2$, as well as to other super gauge theories in other spacetime dimensions. In all these theories, we find that the amplitudes always have the form of (\ref{gluon4pttree}) and contain only exchange contributions. 

Let us end this section with a few comments. The Parisi-Soulars-like dimensional reduction structure we showed in the previous subsection can also be found in the super gluon amplitudes. For example, the multiplet amplitude (\ref{Msgluon}) can be written as
\begin{equation}
M_s=\frac{(\alpha  (t+u-16)-u+8)}{4N^2}\left(-\frac{1}{12}\mathcal{M}_{4,0}^{AdS_7}+\frac{1}{90}\mathcal{M}_{6,0}^{AdS_7}\right)\;.
\end{equation} 
However, we can also express the sum of the two $AdS_7$ scalar exchange diagrams as a single $AdS_5$ scalar exchange diagram 
\begin{equation}
M_s=\frac{(\alpha  (t+u-16)-u+8)}{4N^2}\left(-\frac{1}{12}\mathcal{M}_{4,0}^{AdS_5}\right)\;.
\end{equation} 
More generally, the exchange amplitude of a multiplet with weight $k$ can be written as a degree-1 polynomial in the Mandelstam variables times the linear combination of two $AdS_{d+1}$ scalar exchange amplitudes with internal dimensions $\epsilon k$ and $\epsilon k+2$. By using the dimensional reduction formulae of Witten diagrams, the two scalar exchange amplitudes can be rewritten as a single scalar exchange amplitude of dimension $\epsilon k$ in $AdS_{d-1}$. Recall that in the maximally superconformal case, correlators in $AdS_{d+1}$ can be expressed in terms of scalar exchange diagrams in $AdS_{d-3}$. Therefore, the number of reduced spacetime dimensions is correlated with the amount of supersymmetry. Just as in the super graviton case, this Parisi-Sourlas dimensional reduction structure is found in all super gluon amplitudes in all $AdS_{d+1}\times S^3$ backgrounds. 

Another interesting feature of the amplitude can be seen by writing (\ref{gluon4pttree}) as
\begin{equation}
\mathcal{M}^{A_1A_2A_3A_4}=\mathtt{c}_s\mathtt{n}_s \mathcal{M}_{4,0}^{AdS_5,(s)}+\mathtt{c}_t\mathtt{n}_t \mathcal{M}_{4,0}^{AdS_5,(t)}+\mathtt{c}_u\mathtt{n}_u \mathcal{M}_{4,0}^{AdS_5,(u)}\;,
\end{equation}
where $\mathcal{M}_{4,0}^{AdS_5,(s)}=\mathcal{M}_{4,0}^{AdS_5}$, and $\mathcal{M}_{4,0}^{AdS_5,(t,u)}$ are the scalar exchange amplitudes in the t- and u-channel. This form of the amplitude resembles the flat-space tree-level gluon scattering amplitude, and we can think of $\mathcal{M}_{4,0}^{AdS_5,(s)}$ as the flat-space scalar propagator $1/s$. By the Jacobi identity, the color structures satisfy
\begin{equation}
\mathtt{c}_s+\mathtt{c}_t+\mathtt{c}_u=0\;.
\end{equation} 
Interestingly, we can check by straightforward calculation that the kinematic numerators also satisfy a similar relation
\begin{equation}
\mathtt{n}_s+\mathtt{n}_t+\mathtt{n}_u=0\;.
\end{equation}
This gives an AdS extension of the flat-space color-kinematic duality \cite{Bern:2008qj}. The same structure is also found for the massless $k_i=2$ super gluon amplitudes in other spacetime dimensions \cite{Alday:2021odx}. On the other hand, replacing the color factors with the kinematic factors does not give the super graviton amplitudes, unlike what happens in the flat-space double copy relation \cite{Bern:2010ue}. Nevertheless, the flat-space prescription of double copy does work at the level of {\it reduced} Mellin amplitudes for a set of $AdS_5$ theories \cite{Zhou:2021gnu}. It was shown that there is a different realization of the color-kinematic duality for SYM on $AdS_5\times S^3$ in terms of the reduced Mellin amplitudes. In this realization, all four-point amplitudes for IIB supergravity on $AdS_5\times S^5$ and bi-adjoint scalars on $AdS_5\times S^1$, both massless and massive, can be obtained from the super gluon amplitudes in the same way as in the flat-space case. Searching for AdS extensions of flat-space amplitude properties is an interesting line of research, although such a program is still in its infancy. For recent developments, see \cite{Farrow:2018yni,Lipstein:2019mpu,Eberhardt:2020ewh,Roehrig:2020kck,Armstrong:2020woi,Albayrak:2020fyp,Alday:2021odx,Jain:2021qcl,Zhou:2021gnu,Diwakar:2021juk,Sivaramakrishnan:2021srm,Herderschee:2022ntr,Alday:2022lkk,Cheung:2022pdk}.

\newpage
\markboth{12\quad BOOTSTRAPPING LOOP-LEVEL HOLOGRAPHIC CORRELATORS}{}
\section{Bootstrapping loop-level holographic correlators}\label{Sec:loops}
In this section we  discuss how to compute one-loop level correlators in full-fledged holographic models by incorporating the techniques discussed in Section \ref{Sec:largeN}. A major complexity that arises in these supersymmetric theories is the so-called operator mixing, and this complexity requires us to modify the techniques of Section \ref{Sec:largeN}. While conceptually not difficult to digest, a full discussion of the details of unmixing would require a great deal of additional technical knowledge and goes beyond the scope of this section. Therefore, we will keep this part of the discussion as schematic as possible, with the goal being only to explain the problem and to outline its solution. The main focus of this section is the explicit computation of the one-loop amplitude after taking the solution of the mixing problem as an input. The procedures of this calculation will be explained in detail. We will discuss the case of 4d $\mathcal{N}=4$ SYM which is dual to IIB supergravity on $AdS_5\times S^5$. To keep the discussion pedagogical, we will only consider the simplest correlator $\langle \mathcal{O}_2\mathcal{O}_2\mathcal{O}_2\mathcal{O}_2\rangle$.

\subsection{The mixing problem}
As discussed in Section \ref{Sec:largeN}, in order to construct the one-loop answer, it is sufficient to know the tree-level anomalous dimension $\gamma_{n,\ell}^{(1)}$ and the OPE coeffient of leading order $a^{(0)}_{n,\ell}$. Generically, to extract this piece of information it is enough to consider the correlator with the same external operators at order $N^{0}$ and $N^{-2}$ and decompose it in conformal blocks. However, this algorithm rests on the assumption that the intermediate operators are unique, meaning that there is a single operator with the same quantum numbers. 

For $\mathcal{N}=4$ SYM at strong coupling, this is not the case and the above algorithm needs modifications. In particular, the unprotected superconformal primaries that appear in $\langle \mathcal{O}_2\mathcal{O}_2\mathcal{O}_2\mathcal{O}_2\rangle$ as intermediate operators are singlets under the $SU(4)$ R-symmetry. Schematically, these operators are linear combinations of double-trace operators of the form $\left[\mathcal{O}_2\mathcal{O}_2\right]_{n,\ell}$, $\left[\mathcal{O}_3\mathcal{O}_3\right]_{n-1,\ell}$, $\dots$ $\left[\mathcal{O}_{2+n}\mathcal{O}_{2+n}\right]_{0,\ell}$. Each operator is neutral under R-symmetry, and has the same spin $\ell$ and bare conformal twist $2n+4$. Therefore, in general there is mixing among all such double-trace operator. This adds an extra layer of complication to the loop-level computation. However, we can still use the method of Section \ref{Sec:largeN} once we solve the mixing problem. More specifically, we need to diagonalize the dilatation operator and apply the squaring of anomalous dimension on each eigenstate. To unmix operators with conformal twist $2n+4$,  one needs to consider the family of four-point functions $\langle {\cal O}_2{\cal O}_2{\cal O}_p{\cal O}_p\rangle$  for $p=2,\dots, 2+n$. At order $N^0$ and $N^{-2}$ respectively, we extract from each correlator the averages 
 \begin{equation}
\langle a^{(0)}\rangle=\sum_{I=1}^{1+n} c_{22,I}c_{pp,I}\;.
 \end{equation}
 and 
 \begin{equation}
 \langle a^{(0)}\gamma^{(1)}\rangle=\sum_{I=1}^{1+n} c_{22,I}c_{pp,I} \gamma_I^{(1)}
 \end{equation}
where $c$ are the three-point function. Notice that for simplicity, we removed the spin and conformal dimension labels. Here we have also chosen a normalization in which these eigenstates $\Sigma^I$ are orthonormal, {\it i.e.}, $\langle \Sigma^{I} \Sigma^{J}\rangle =\delta^{IJ}$. This gives the mixing matrices which are diagonalized in \cite{Aprile:2017bgs, Alday:2017xua}. The explicit results which are relevant for our purposes are recorded in the next subsection. Let us also mention that to compute one-loop correlators with higher Kaluza-Klein weights, we need to consider more general tree-level correlators. Diagonalizing these more complicated mixing matrices is discussed in \cite{Aprile:2017xsp,Aprile:2017bgs, Aprile:2018efk} and in general the spectrum still has remaining degeneracy.\footnote{For the singlet sector the degeneracy is lifted completely to order $1/N^2$, see \cite{Aprile:2018efk}.}

\subsection{Super graviton one-loop amplitude in $AdS_5\times S^5$}
In the previous subsection, we explained how to solve the mixing problem. With the data from its solution we can proceed to compute the leading logarithmic singularity. Knowing it allows us to fix the full correlator. However, for $AdS_5\times S^5$ correlators we can also define the reduced correlators. At tree level, we have also seen in Section \ref{Subsec:AdS5Mellin} that using them leads to a lot of simplifications, as they automatically take into account superconformal symmetry. For this reason, we will continue to work with the reduced correlator at one-loop level. Note that in Section \ref{Sec:largeN} we explained the principle of the calculation in position space. However, the computation of loop-level amplitudes is particularly simple in Mellin space and we find a very compact answer in this representation. Therefore, for pedagogical purpose, we will only give below a review of the Mellin method of  \cite{Alday:2018kkw,Alday:2019nin}. There are other complementary approaches such as the position space method \cite{Aprile:2017bgs,Aprile:2017qoy,Aprile:2019rep} and a method based on the Lorentzian inversion formula \cite{Alday:2017vkk,Caron-Huot:2018kta}. We will briefly comment on these approaches in Section \ref{Sec:openproblemscorrelators}, but will refer the reader to these references for details.

The one-loop leading logarithmic singularity $\mathcal{H}^{(2)}|_{\log^2U}(U,V)$ is singular in the small $V$ limit, and has the form
\begin{equation}\label{HlognV}
\mathcal{H}^{(2)}|_{\log^2U}(U,V)=f_2(U,V)\log^2V+f_1(U,V)\log V+f_0(U,V)\;,
\end{equation}
where the coefficient functions $f_i(U,V)$ are regular in $U,V\to 0$. This structure is expected from crossing symmetry as $\{\log^2U, \log U,1\}$ are mapped to $\{\log^2V, \log V,1\}$ under $U\leftrightarrow V$. Closed form expressions of these functions can be found in \cite{Aprile:2017bgs,Caron-Huot:2018kta}, but it is not necessary for our purpose. Instead we will only need them order by order in the power expansion with respect to $U$
\begin{equation}
f_i(U,V)=U^2(f_i^{(0)}(V)+Uf_i^{(1)}(V)+\ldots)\;.
\end{equation}
Let us explain how to compute them from the conformal block decomposition of the leading logarithmic singularity\footnote{Note there is a shift of 4 in the  dimension of the conformal block. This is because we are looking at the reduced correlator. See \cite{Nirschl:2004pa} for details.}
\begin{equation}\label{lls1loop}
\mathcal{H}^{(2)}|_{\log^2U}(U,V)=\frac{1}{2}\sum_n\sum_{\ell\text{ even}}\langle a^{(0)}_{n,\ell}(\gamma^{(1)}_{n,\ell})^2\rangle U^{-2}G_{8+2n+\ell,\ell}(z,\bar{z})\;.
\end{equation}
Here $G_{\tau,\ell}(z,\bar{z})$ are the conformal blocks with the extra power of $U^{\frac{\tau}{2}}$
\begin{equation}
    G_{\tau+\ell,\ell}(z,\bar{z})=(z\bar{z})^{\frac{\tau}{2}}g_{\tau,\ell}(z,\bar{z})\;,
\end{equation}
compared to the one used in Section \ref{Sec:largespin}. For reader's convenience, we have reproduced the average $\langle a^{(0)}_{n,\ell}(\gamma^{(1)}_{n,\ell})^2\rangle$ from \cite{Aprile:2017bgs}
\begin{equation}
\langle a^{(0)}_{t-2,\ell}(\gamma^{(1)}_{t-2,\ell})^2\rangle=\sum_{i=1}^{t-1}C^{(0)}_{t,\ell}R_{t,\ell,i}a_{t,i}\;,
\end{equation}
where 
\begin{eqnarray}
\nonumber &&C^{(0)}_{t,\ell}=\frac{2((t+\ell+1)!)^2(t!)^2(\ell+1)(2t+\ell+2)}{(2t)!(2t+2\ell+2)!}\;,\\
&&R_{t,\ell,i}=\frac{2^{1-t}(2\ell+3+4i)(\ell+i+1)_{t-i-1}(t+\ell+4)_{i-1}}{(\frac{5}{2}+\ell+i)_{t-1}}\;,\\
\nonumber &&a_{t,i}=\frac{2^{1-t}(2+2i)!(t-2)!(2t-2i+2)!}{3(i-1)!(i+1)!(t+2)!(t-i-1)!(t-i+1)!}\;.
\end{eqnarray}
Notice that conformal blocks can be power-expanded in $U$ and $(1-V)$. In particular, the leading $U$ power of $G_{8+2n+\ell,\ell}(z,\bar{z})$ is $U^{4+n}$. As a result, for a fixed power of $U$ in the leading logarithmic singularity, there are only finitely values of $n$ that can contribute. On the other hand, the expansion of the conformal blocks in $(1-V)$ consists of finitely many terms of the form
\begin{equation}
(1-V)^{a+\ell}{}_2F_1(A,B;C;1-V)\;,
\end{equation}
with the minimal power of $(1-V)$ controlled by the spin of the exchanged operator. Therefore, we can truncate the sum over $\ell$ in (\ref{lls1loop}) when obtaining the coefficients $B^{(n)}_i$ of $(1-V)^i$ with small $i$ since they are not affected by the large spins. With the help of $\mathtt{Mathematica}$ we can easily find the general formula for the coefficients $B^{(n)}_i$ as a function of $i$ from a few low-lying values. Performing the infinite sum over $i$ gives $f^{(n)}_i(V)$
\begin{equation}
\sum_{i=0}^\infty B^{(n)}_i (1-V)^i=f^{(n)}_2(V)\log^2V+f^{(n)}_1(V)\log V+f^{(n)}_0(V)\;.
\end{equation}
For example, we have\footnote{We do not keep track of the overall normalization of the correlator in this subsection.}
\begin{eqnarray}
\nonumber &&f^{(0)}_2(V)=\frac{96(V^2+4 V+1)}{(V-1)^6}\;,\quad f^{(0)}_1(V)=-\frac{288 (V+1)}{(V-1)^5}\;,\quad f^{(0)}_0(V)=0\;,\\
\nonumber &&f^{(1)}_2(V)=\frac{48 (5 V^3+37 V^2+37 V+5)}{(V-1)^8}\;,\quad f^{(1)}_1(V)=-\frac{144 (7 V^2+22 V+7)}{(V-1)^7}\;,\\
&& f^{(1)}_0(V)=\frac{576 (V+1)}{(V-1)^6}\;,\\
\nonumber &&f^{(2)}_2(V)=\frac{48 (59 V^4+706 V^3+1494 V^2+706 V+59)}{7 (V-1)^{10}}\;,\\
\nonumber &&f^{(2)}_1(V)=-\frac{144 (101 V^3+627 V^2+627 V+101) }{7 (V-1)^9}\;,\quad f^{(2)}_0(V)=\frac{384 (5 V^2+14 V+5)}{(V-1)^8}\;.
\end{eqnarray}

Now let us focus on $f_2^{(n)}(V)$ which multiplies $U^{2+2n}\log^2U\log^2V$. To produce such terms from the Mellin representation we must have simultaneous cubic poles at $s=4+2n$ and $t=4+2m$. Note that the Gamma function factor in the definition
\begin{equation}
\mathcal{H}=\int\frac{dsdt}{(4\pi i)^2}U^{\frac{s}{2}}V^{\frac{t}{2}-2}\widetilde{\mathcal{M}}(s,t)\Gamma^2[\frac{4-s}{2}]\Gamma^2[\frac{4-t}{2}]\Gamma^2[\frac{4-\tilde{u}}{2}]\;,\quad s+t+\tilde{u}=4\;,
\end{equation}
already provides double poles at these locations. Therefore, the reduced Mellin amplitude must contain a pair of simultaneous pole. Our minimal assumption is 
\begin{equation}
\widetilde{\mathcal{M}}(s,t)\supset \sum_{n,m=0}^\infty \frac{c_{mn}}{(s-4-2n)(t-4-2m)}\;,
\end{equation}
with {\it constant} $c_{mn}$ coefficients which are independent of the Mandelstam variables. These coefficients are picked up by taking residues in the Mellin representation at $s=4+2n$, $t=4+2m$ and can be determined by comparing with the Taylor expansion coefficients of $f_2^{(n)}(V)$. In practice, we can proceed by first finding an expression for the coefficients for fixed $n$ and then obtain a list of these functions as we increase $n$. It is not difficult to find from these data points that the general expression is given by a symmetric function
\begin{equation}
c_{mn}=\frac{ p^{(6)}(m,n)}{5(m+n-1)_5}\;,
\end{equation}
where $p^{(6)}(m,n)$ is a degree 6 polynomial 
\begin{equation}
\begin{split}
p^{(6)}(m,n)={}&32\big(15 m^4 n^2+25 m^4 n+12 m^4+30 m^3 n^3+120 m^3 n^2+114 m^3 n+36 m^3\\
{}&+15 m^2 n^4+120 m^2 n^3+216 m^2 n^2+77 m^2 n-8 m^2+25 m n^4+114 m n^3\\
{}&+77 m n^2-76 m n-40 m+12 n^4+36 n^3-8 n^2-40 n\big)\;.
\end{split}
\end{equation}

Let us now go back to check the assumption that $c_{mn}$ are constants. By crossing symmetry, the Mellin amplitude should also contain simultaneous poles in $s$, $\tilde{u}$ and $t$, $\tilde{u}$. Therefore, our minimal assumption corresponds to the following expression 
\begin{equation}\label{M1loop}
\begin{split}
\widetilde{\mathcal{M}}(s,t)={}& \sum_{n,m=0}^\infty c_{mn}\bigg(\frac{1}{(s-4-2n)(t-4-2m)}+\frac{1}{(s-4-2n)(\tilde{u}-4-2m)}\\
{}&\quad\quad\quad\quad\quad +\frac{1}{(t-4-2n)(\tilde{u}-4-2m)}\bigg)\;.
\end{split}
\end{equation}
To check it, we again take the residue at $s=4+2n$ and select the term proportional to $\log^2U$ corresponding to the leading logarithmic singularity. Notice that it receives contributions only from the first two pairs of simultaneous poles in (\ref{M1loop}). We then perform the sum over $m$ and compute the residue of $t$ at $t=4+2\mathbb{Z}_{\geq 0}$. We find that $f_2(U,V)$ in (\ref{HlognV}) is matched by construction. However, very nontrivially, both $f_1(U,V)$ and $f_0(U,V)$ are also fully reproduced. This tells us that there cannot be {\it single poles} of the form $1/(s-4-2n)$ in (\ref{M1loop}). They do not modify the $\log^2U\log^2V$ coefficients but can change the $\log^2U\log^pV$ coefficient functions for $p=0,1$. By crossing symmetry we also rule out the existence of single poles in the other Mandelstam variables. Therefore, the only ambiguities are regular terms which correspond to contact interactions. These contact terms can be fixed by looking at the flat-space limit of the reduced Mellin amplitude (see \cite{Alday:2018kkw,Alday:2019nin} for details). Therefore, we conclude that (\ref{M1loop}) is the full one-loop amplitude. 

The same computational strategy also applies to one-loop correlators with higher Kaluza-Klein weights and a closed form expression for all correlators of the form $\langle22pp\rangle$ was obtained in \cite{Alday:2019nin}. It also can be used to compute one-loop super gluon amplitudes on $AdS_5\times S^3$ \cite{Alday:2021ajh} where the amplitudes have the same structure of simultaneous poles (see also \cite{Behan:2022uqr} for related work on super gluon one-loop correlators in the so called S-fold theories). 
\newpage
\markboth{13\quad HOLOGRAPHIC CORRELATORS: OTHER ASPECTS AND OPEN PROBLEMS}{}
\section{Holographic correlators: Other aspects and open problems}\label{Sec:openproblemscorrelators}
\subsection{Other developments}
Our review of holographic correlators in top-down models is unfortunately incomplete in many regards, and we had to omit many other interesting topics. Here we list a few of these research directions which we did not have the chance to discuss and we will refer the interested reader to the original references for details. 

\vspace{0.5cm}
\noindent{\it Other approaches to loop correlators.} 
\vspace{0.1cm}

In addition to the Mellin space method reviewed here, there are other approaches to loop correlators. The computations of one-loop correlators can also be performed in position space, as they were first obtained \cite{Aprile:2017bgs,Aprile:2017qoy,Aprile:2019rep}. Schematically, this is achieved by formulating an ansatz in terms of certain basis functions which are ladder integrals with maximal transcendental degree 4. One then fixes this ansatz by matching with the double-discontinuity and by imposing a few physical conditions. Recently, it was shown in \cite{Huang:2021xws} that the position space method can be extended to two loops with the help of an educated ansatz based on the hidden conformal symmetry for $AdS_5\times S^5$. In the simpler 1d case of half-BPS line defects, the application of a variation of the position method has also led to results at three loops \cite{Ferrero:2021bsb}. In addition to the position space approach, another complementary method for $AdS_5\times S^5$ loop correlators is to use the Lorentzian inversion formula \cite{Caron-Huot:2017vep}. This approach also gives a nice rederivation of the results of \cite{Rastelli:2016nze,Rastelli:2017udc} at tree level. Note that the double-discontinuity vanishes for crossed-channel double-trace operators.  This is particularly convenient for tree-level correlators as there are only finitely many single-trace operators which contribute. For details of this method, see \cite{Alday:2017vkk,Caron-Huot:2018kta}. 

\vspace{0.5cm}
\noindent{\it Higher derivative corrections.} 
\vspace{0.1cm}

In this review, we have mostly focused on the two-derivative supergravity limit. However, one can also consider string theory and M-theory corrections in the low-energy limit which are manifested at tree level as higher-derivative contact terms. In general, superconformal symmetry and the flat-space limit are not constraining enough to completely fix these contact terms \cite{Goncalves:2014ffa}. However, by using additional independent input, such as constraints from the chiral algebra and the supersymmetric localization, one can fix the remaining parameters in some types of correlators up to a certain order in the expansion \cite{Chester:2018aca,Chester:2018dga,Binder:2018yvd,Binder:2019jwn,Binder:2019mpb,Chester:2019pvm}. Moreover, in $AdS_5\times S^5$ by exploiting the hidden conformal symmetry and making ansatz based on patterns observed in double-trace anomalous dimensions, one is able to make further progress and deal with more general correlators \cite{Drummond:2019odu,Drummond:2020dwr,Abl:2020dbx,Aprile:2020mus}. String and M-theory corrections can also be considered at the level of loop amplitudes. Works in this direction include \cite{Alday:2018pdi,Drummond:2019hel,Alday:2020tgi,Drummond:2020uni,Alday:2021ymb}. Another interesting limit in $AdS_5\times S^5$ corresponds to taking $N$ to infinity with $g_{\rm YM}$ fixed in the dual 4d $\mathcal{N}=4$ SYM. Note that this is not the usual 't Hooft limit, which has fixed $g^2_{\rm YM} N$, but it is still amenable to the supersymmetric localization aforementioned. By studying correlators in this limit, one can access $SL(2,\mathbb{Z})$ properties of superstring amplitudes and perform non-perturbative precision tests of AdS/CFT. For works in this direction, see \cite{Binder:2019jwn,Chester:2019jas,Chester:2020dja,Chester:2020vyz,Green:2020eyj,Dorigoni:2021bvj,Dorigoni:2021rdo}.

\vspace{1.5cm}
\noindent{\it Tree-level correlators from the heavy-heavy-light-light limit.}
\vspace{0.1cm}

Another independent approach to compute holographic correlators in $AdS_3$ was developed in \cite{Giusto:2018ovt,Giusto:2019pxc,Giusto:2020neo} based on earlier works \cite{Bombini:2017sge,Galliani:2017jlg}, which obtains correlators of four light operators from a heavy-heavy-light-light limit. In the latter case, heavy operators can be described by a classical supergravity solution and the light operators correspond to fluctuations on this background. It was shown that it is possible to take a formal ``light'' limit to obtain all-light correlators. The limit is smooth at the level of correlators even though the end point is outside the regime of validity.

\vspace{0.5cm}
\noindent{\it Correlators of multi-particle states.}
\vspace{0.1cm}

In the standard AdS/CFT duality, we consider fluctuations of the supergravity fields which correspond to single-particle states. The amplitudes we obtain are thus all for such single-particle states.  However, it is also interesting to consider AdS scattering amplitudes involving ``bound states''. From the CFT perspective, these are multi-trace operators. In principle, one can obtain such correlators by taking OPE limits of correlators of single-trace operators. However, it would also be beneficial to avoid taking this detour and obtain these correlators more directly. Some progress has been made in this direction. In \cite{Bissi:2021hjk} a systematic analysis was carried out for the superconformal kinematics of 4d $\mathcal{N}=4$ SYM four-point functions containing $\frac{1}{4}$-BPS double-trace operators. This paves the way for a future bootstrap strategy. On the other hand, in $AdS_3$ it was shown in \cite{Ceplak:2021wzz} that the approach of \cite{Bombini:2017sge,Galliani:2017jlg} can be extended to compute tree-level four-point functions with two single-trace operators and two multi-trace operators. A particularly interesting feature pointed out in \cite{Ceplak:2021wzz} is that the tree-level multi-trace correlators necessarily involve building block functions in position space which are generalizations of the $D$-functions.

\subsection{Open problems}
Clearly, there are many open problems in this modern program of holographic correlators. Here we will outline some of the major research avenues.\footnote{See also \cite{Gopakumar:2022kof} for a complementary discussion on related topics.} 

\begin{itemize}
\item {\bf Constructive methods and higher multiplicities:} Most of the methods reviewed here are bootstrap in nature. However, it would also be useful to explore alternative methods which are more constructive. For example, is there an AdS analogue of the flat-space Britto-Cachazo-Feng-Witten recursion relation \cite{Britto:2005fq} which would allow us to recursively build higher-point correlators from lower-point ones?\footnote{See \cite{Raju:2011mp,Raju:2010by} for early progress in this direction.} Such constructive methods would not only provide crosschecks for the bootstrap results, but could also greatly facilitate the study of correlators with higher multiplicities. 
\item {\bf Spinning correlators:} The majority of past work in the literature has focused on correlators of scalar operators. This is partly because they preserve more supersymmetry (when they are superconformal primaries), and also due to the fact that the kinematics is simpler. However, it is also important to look into holographic correlators of operators with spins as they provide more direct analogies with gluon and graviton amplitudes in flat space. Much work is still needed to make progress in this direction. On the one hand, it is necessary to find convenient formalisms to simplify the kinematics.\footnote{For example, it would be useful to generalize the spinning Mellin formalism \cite{Goncalves:2014rfa} to include arbitrarily many spinning operators. Other promising options include the AdS spinor-helicity formalism \cite{Maldacena:2011nz,Nagaraj:2018nxq,Nagaraj:2019zmk,Binder:2020raz,Nagaraj:2020sji} and weight-shifting operators \cite{Karateev:2017jgd,Costa:2018mcg,Binder:2020raz}.} On the other, one also needs to develop efficient computational techniques which do not rely on supersymmetry. 
\item {\bf Color-kinematic duality and double copy:} While there has been much evidence for color-kinematic duality and double copy in AdS space at tree level, a systematic understanding is still currently lacking. In particular, it would be very interesting to find a concrete realization for $n$-point tree-level correlators of bosonic Yang-Mills and Einstein-Hilbert gravity in AdS. On the other hand, the existence of a double copy relation at loop levels is currently a completely open question. In flat space, the double copy relation can be manifested in the Cachazo-He-Yuan (CHY) formalism \cite{Cachazo:2013hca,Cachazo:2013iea}. The investigation in AdS may be similarly facilitated by exploiting such a representation. The study of CHY-like representations for holographic correlators has been initiated in \cite{Eberhardt:2020ewh,Roehrig:2020kck}.
\item {\bf Higher loops:} Higher-loop correlators are another research area which has not been sufficiently explored. The leading logarithmic term has been constructed in \cite{Bissi:2020wtv,Bissi:2020woe} for the stress tensor multiplet four-point function in $AdS_5\times S^5$ IIB supergravity at any loop order. But with these results only a particular part of the dual AdS amplitude can be reconstructed. A major challenge in fully computing higher-loop correlators is to determine the higher-trace operator contributions, which can be extracted from higher-point tree-level correlators. This problem is particularly severe in full-fledged holographic models where there is an internal manifold, as in the $AdS_5\times S^5$ example. One has to glue together infinitely many tree-level correlators to obtain loop-level correlators even in the simplest case.

\item {\bf Stringy correlators:} Current technologies only allow us to compute the correlators in expansions of $\alpha'$. As discussed in the previous subsection, one can fix the corrections order by order by using additional input from the SCFT. However, it would be great to go beyond that. An extremely interesting open problem is to devise a strategy to obtain all-order results and find the analogue of Virasoro-Shapiro amplitude in AdS. 

\item {\bf Quantum gravity:} It has often been advocated that via AdS/CFT one can gain insight into quantum gravity from the dual CFT. Holographic correlators provide a concrete way to exploit this duality, and computing loops in AdS allows us to access the quantum regime of gravity. However, it is not fully clear what precise lessons we can learn from studying holographic correlators. It would be very interesting to have a sharp answer to this question.
\end{itemize}

\newpage
\markboth{14\quad FURTHER READING}{}
\section{Further reading}\label{Sec:omissions}
In order to keep the discussion manageable, and also as a result of our limited competence in this vast and rapidly evolving subject, we have focused on only a few selected topics in this review and left out many exciting research directions. In this concluding section, we give a brief discussion on further reading material on related topics.

\begin{itemize}
    \item {\it Modular bootstrap:} Despite the early success of the classification program of rational CFTs, progress in learning about the general landscape of 2d conformal field theories is extremely hard to make. However, nontrivial constraints can be extracted from the modular invariance of the torus partition function. The invariance condition allows one to derive rigorous bounds on the CFT spectrum, much like the bootstrap equation for four-point functions. This is the ``modular bootstrap'' program pioneered \cite{Hellerman:2009bu} by Hellerman and pursued by many others \cite{Hellerman:2010qd,Keller:2012mr,Friedan:2013cba,Hartman:2014oaa,Benjamin:2016fhe,Collier:2016cls,Cardy:2017qhl,Cho:2017fzo,Dyer:2017rul,Anous:2018hjh,Afkhami-Jeddi:2019zci,Mukhametzhanov:2019pzy,Benjamin:2019stq,Mukhametzhanov:2020swe,Pal:2020wwd,Benjamin:2020zbs,Dymarsky:2020bps,Lin:2021udi,Benjamin:2021ygh, Mukhi:2019xjy}. The application of this program goes beyond delineating the allowed region of consistent 2d CFTs. Via the AdS/CFT correspondence, modular bootstrap bounds allow us to address the fundamental question whether ``pure'' AdS gravity exists as a consistent quantum theory. Moreover,  there also exists a precise connection between the modular bootstrap and the sphere packing problem in Euclidean geometry \cite{Hartman:2019pcd}.

     \item {\it Axiomatic CFT}: While it is commonly accepted that in Euclidean signature, CFT correlators satisfy well-defined rules, the status of their Lorentzian counterpart is less clear. This question has been explored in great detail recently in \cite{slavaaxiom}. The main finding is that QFT axioms like the Osterwalder-Schrader and Wightman axioms follow from the Euclidean CFT axioms which rely on unitarity, reality constraints and convergence of the OPE. In particular, an independent derivation of the Wightman axioms for CFTs from the Euclidean axioms was provided.  While the results we have included in this review, which rely on lightcone bootstrap are correct as they find support using numerical bootstrap, it will be worthwhile to establish these results in a mathematically rigorous manner building on the theorems of \cite{Kravchuk:2020scc, slavaaxiom}.
    
    \item {\it S-matrix bootstrap}:
    The success of the numerical methods exploring the conformal bootstrap has spurred a new research direction in the numerical studies of the S-matrix bootstrap \cite{smat1,smat2,smat3, Guerrieri:2021ivu, Guerrieri:2021tak, He:2021eqn}. The directions that have been explored include trying to constrain the space of pion S-matrices using unitarity and crossing symmetry. It is not clear so far, what selection rules need to be imposed to zoom in on the QCD answer. Analytic progress in this regard is still in its infancy. It will be fascinating to see if the analytic methods in the conformal bootstrap have a role to play in the S-matrix bootstrap endeavor as well. In fact, we expect cross-fertilization between ideas in the two programs. As an example, certain positivity properties of Gegenbauer polynomials are very well studied and play a crucial role in the S-matrix bootstrap program. The analogous properties for the Mack polynomials are hard to establish analytically but can be checked numerically and appear to hold. We expect such interplay to be fruitful for both programs.
    
    \item {\it Bounds on EFTs}: In the context of low energy effective field theories, fascinating progress has happened in providing two-sided bounds on Wilson coefficients \cite{pos1, simon1, simon2} over the last year or so. The main technical tools used here are positivity of the partial wave amplitude and crossing symmetry using the fixed-$t$ dispersion relation. Alternatively, as in \cite{Sinha:2020win}, one uses a crossing symmetric dispersion relation and demands locality in addition to positivity. Since Mellin space dispersion relations exist for CFT correlators, in principle, the same study can be carried out for such correlators and one could try to see for instance, how big the space of allowed theories in the Wilson coefficient space is and where interesting CFTs like the 3d Ising model sit. Some preliminary studies keeping the AdS/CFT correspondence in mind have been carried out recently in \cite{simon3, kundu}. It will be worthwhile developing these ideas further. For instance, large spin perturbation theory techniques focus on a universal sector of CFTs. What role does this universal sector play in the space of Wilson coefficients?

    \item {\it Cosmological correlators:} In this review we mostly discussed applications to AdS physics, in particular in the context of the AdS/CFT correspondence. However, there is a very interesting line of research to study correlators in dS spaces, where one can study late time correlators. A promising approach is  to use techniques motivated by the bootstrap philosophy. Here the aim is to explore cosmological correlators from the consistency conditions of the CFT, using weight shifting operators, the structure of singularities and the Mellin space approach \cite{Arkani-Hamed:2018kmz,Baumann:2019oyu,Baumann:2020dch,Sleight:2019hfp,Sleight:2020obc,Sleight:2021iix,Gomez:2021qfd,Sleight:2021plv, Hogervorst:2021uvp}. 
    
    \item {\it Large charge expansion:} Another arena for analytic studies is the regime of large charge $Q$ in presence of a global symmetry, where $Q$ is a generic quantum number. This sector is usually not attainable to bootstrap techniques (for an exception, see \cite {Jafferis:2017zna}). In particular, it is possible to use the power of symmetries (conformal symmetry, and possibly supersymmetry) to write a large charge expansion systematically \cite{Araujo:2021sjv,Hellerman:2021qzz,Orlando:2021usz,Dondi:2021buw,Orlando:2019skh,Banerjee:2017fcx,Hellerman:2015nra}. This approach complements the conformal bootstrap techniques, but in spirit it is driven by the same guiding principles. For a review on this topic, see \cite{Gaume:2020bmp}.
    
    \item{\it CFT-like theories:} The philosophy and techniques of the analytic conformal bootstrap can also be extended to theories which are CFT-like. An important class of such theories is the theory of logarithmic CFTs. Unlike in the usual CFTs where two-point functions obey power laws, the two-point functions in logarithmic CFTs contain logarithms which are caused by reducible but indecomposable representations. These logarithmic CFTs  have a vast range of applications from critical percolations to systems with quenched disorder. For works in this direction, see \cite{Ghezelbash:1997cu,Creutzig:2013hma,Hogervorst:2016itc,Gorbenko:2020xya,Nivesvivat:2020gdj}. One can also consider theories with modified conformal symmetries. Examples include nonrelativistic theories with Schr\"odinger, Carrollian and Galilean symmetries \cite{Nishida:2007pj,Bagchi:2009my,Golkar:2014mwa,Goldberger:2014hca, Bagchi:2017cpu,Pal:2018idc,Chen:2019hbj, Gupta:2020dtl,Chen:2020vvn,Chen:2021xkw,Hao:2021urq}.  
    
    \item{\it Supersymmetric theories:} When we discussed supersymmetric CFTs in this review, we have mostly focused on the limit where the holographic duals are weakly coupled local theories. For instance, in the canonical example of 4d $\mathcal{N}=4$ SYM, such a limit corresponds to large ranks and strong 't Hooft coupling. However, it is also interesting to go away from this limit. This can be achieved by combining both analytic and numerical bootstrap techniques, as was initiated in the seminal works \cite{Beem:2013qxa,Chester:2014mea,Beem:2014zpa,Beem:2015aoa}. The chiral algebra or the topological quantum mechanics structure in these SCFTs imposes a baby version of the bootstrap which can be solved analytically. The solution is then used as an input in the full-fledged bootstrap problem to obtain bounds on the operator spectrum and OPE coefficients via numerics. For research along this direction, see \cite{Bobev:2015jxa,Beem:2015aoa,Liendo:2015ofa,Lemos:2015awa,Lemos:2015orc,Ramirez:2016lyk,Lin:2016gcl,Lemos:2016xke,Cornagliotto:2017dup,Chang:2017xmr,Bobev:2017jhk,Chang:2017cdx,Cornagliotto:2017snu,Liendo:2018ukf,Li:2018mdl,acp19,Chang:2019dzt,Gimenez-Grau:2020jrx,Chester:2021aun}.

    \item{\it Integrability methods for conformal blocks:} Conformal blocks are one of the key ingredients when discussing the conformal bootstrap. In even integer spacetime dimensions and for four external scalar operators they have been found in a closed form in  \cite{Dolan:2000ut}. In particular, conformal blocks are eigenfunctions of the quadratic Casimir of the conformal group.
    More recently, in \cite{Isachenkov:2016gim} another interesting way of studying them has been discussed. The key observation is that the conformal Casimir equation can be mapped into an eigenvalue problem for a Calogero-Sutherland Hamiltonian, which is integrable. This has been shown to be valid in any number of spacetime dimensions. This program has been further carried out for spinning external operators \cite{Schomerus:2017eny,Schomerus:2016epl,Isachenkov:2017qgn}, for blocks in defect CFTs \cite{Isachenkov:2018pef,Buric:2020zea}, for superconformal blocks \cite{Buric:2019rms, Buric:2020buk} and for conformal blocks of multi-point correlators \cite{Buric:2020dyz,Buric:2021ywo, Buric:2021ttm,Buric:2021kgy}.

    \item{\it Conformal colliders bound:} In \cite{Hofman:2008ar} Hofman and Maldacena found universal bounds on scale anomaly coefficients, which can be read off from correlators of the stress tensors, as a consequence of causality. Similarly, three-point functions of gravitons have been studied in \cite{Camanho:2014apa}, and the anomaly coefficients have been related to the dimension of the lightest operator appearing in the OPE. Due to the intrinsic Lorentzian nature of these bounds, an approach to find and study them is through lightcone approaches applied to four-point correlators and imposing analyticity. These ideas have been introduced in \cite{Hartman:2015lfa}, and other developments in  \cite{Hartman:2016dxc,Chowdhury:2018uyv,Hofman:2016awc,Meltzer:2017rtf, Cordova:2017dhq,Manenti:2019kbl}. 
    
    \item{\it Other explorations:} Let us conclude by listing out a few other interesting areas that we did not cover in this review. An area that we did not touch upon in detail is the analysis of correlators which involve two heavy and two light scalar primary operators in $d>2$ \cite{andrei1,liambh1,Fitzpatrick:2019zqz,andrei2,Karlsson:2020ghx,Li:2020dqm} using the so-called ``Conformal Regge Bootstrap''. In the context of the AdS/CFT correspondence, one can think of the heavy field as a black hole and the light field as a particle scattering off the black hole. By analysing the crossed channel, one can make predictions for the anomalous dimension of a double-trace operator arising from the combination of a heavy and a light field. This OPE data then can be checked against the AdS/CFT calculation involving phase shifts in a black hole background. Impressive agreements have been reported in \cite{andrei1,andrei2}. Another interesting direction is the research related to celestial amplitudes \cite{Kapec:2014opa, Pasterski:2016qvg,Pasterski:2021raf}. Here, scattering of massless particles in four dimensions is mapped to the celestial sphere and the information of four dimensional scattering is encoded in a potentially interesting two dimensional CFT living on the celestial sphere. It would be interesting to see if analytic bootstrap techniques, such as large spin perturbation theory and dispersion relations, can play a role in clarifying properties of this 2d CFT. 
\end{itemize}

\section*{Acknowledgements} We thank Fernando Alday, Faizan Bhat, Parijat Dey, Giulia Fardelli, Pietro Ferrero, Rajesh Gopakumar, Apratim Kaviraj, Himanshu Khanchandani, Andrea Manenti, Maria Nocchi, Slava Rychkov, Jacopo Sisti, Alexander Soderberg, Ahmadullah Zahed for comments on certain portions of the review. We thank all our collaborators over the last several years who have contributed to our understanding of much of the material presented in this review. The work of A.B. is supported by Knut and Alice Wallenberg Foundation under grant KAW 2016.0129 and by VR grant 2018-04438. A.S. acknowledges support from MHRD, Govt. of India, through a SPARC grant P315 and from DST through the SERB core grant CRG/2021/000873. The work of X.Z. is supported by starting funds from University of Chinese Academy of Sciences (UCAS) and from the Kavli Institute for Theoretical Sciences (KITS), and the part of work carried out in Princeton Center for Theoretical Science was supported by Simons Foundation Grant No. 488653.
\newpage
\begin{appendices}
\markboth{A\quad EPSILON EXPANSION: SOME DETAILS}{}
\section{Epsilon expansion: Some details}
\subsection{Conventions}\label{secConv}
In section \ref{Sec:epsilonbootstrap}, we use
\be\label{NRdef}
\begin{split}
\mathcal{N}_{\Delta, \ell}&=\frac{2^{\ell}(\Delta+\ell-1) \Gamma^{2}(\Delta+\ell-1) \Gamma(\Delta-h+1)}{\Gamma(\Delta-1) \Gamma^{4}\left(\frac{\Delta+\ell}{2}\right) \Gamma^{2}\left(\Delta_{\phi}-\frac{\Delta-\ell}{2}\right) \Gamma^{2}\left(\Delta_{\phi}-\frac{2 h-\Delta-\ell}{2}\right)}\,,\\ \mathcal{R}_{\D,\ell}^{(k)}&=\frac{\Gamma^{2}\left(\frac{\Delta+\ell}{2}+\Delta_{\phi}-h\right)\left(1+\frac{\Delta-\ell}{2}-\Delta_{\phi}\right)_{k}^{2}}{k ! \Gamma(\Delta-h+1+k)}\,.
\end{split}
\ee
Notice that $\mathcal{N}_{\Delta,\ell}$ has zeros when $\Delta=2\Dphi+2n+\ell$, {\it i.e.,} the GFF values. Finding a general form for conformal blocks in Mellin space is a formidable challenge, see e.g. \cite{Dolan:2011dv, Chen:2019gka}. A suitable form for the Mack polynomial that we will use can be found in \cite{Gopakumar:2018xqi, Gopakumar:2021dvg}
\be
P^{}_{\D,\ell}(s_1,s_2)=\sum_{m=0}^\ell\sum_{n=0}^{\ell-m}\mu^{(\D,\ell)}_{n,m}~~\left(\frac{\D-\ell}{2}-s_1-\frac{2\Dphi}{3}\right)_{m}~~\left(\frac{\Dphi}{3}-s_2\right)_{n}\,,
\ee
where 
\be\label{Ap:mu}
\begin{split}
&\mu^{(\D,\ell)}_{n,m}
=\frac{2^{-\ell} \ell! (-1)^{m+n} (h+\ell-1)_{-m} \left(\frac{\bar\tau }{2}-m\right)_m (\bar\tau -1)_{n-\ell} \left(\frac{\tau}{2}+n\right)_{\ell-n} \left(\frac{\tau}{2}+m+n\right)_{\ell-m-n} \,}{m! n! (\ell-m-n)!}\\
&\times {}_4F_3\left(-m,-h+\frac{\tau }{2}+1,-h+\frac{\tau}{2}+1,n+\Delta -1;\frac{\bar\tau}{2}-m,\frac{\tau}{2}+n,-2 h+\tau +2;1\right)\,.
\end{split}
\ee
Here $\tau=\Delta-\ell, \bar\tau=\Delta+\ell$ and $(a)_b\equiv \Gamma(a+b)/\Gamma(a)$ is the Pochhammer symbol.
We will further use the following "very well poised" ${}_7F_6$ hypergeometric function which shows up in our calculations. 
\begin{eqnarray}\label{Wdef}
&&W(a;b,c,d,e,f)\equiv \nonumber\\
&&{}_7F_6\bigg{(}\begin{matrix} a, & 1+\frac{1}{2}a, & b, & c, & d, & e, & f\\ ~&\frac{1}{2}a,& 1+a-b, & 1+a-c, & 1+a-d, & 1+a-e, & 1+a-f\end{matrix};1\bigg{)}\nonumber\\
&=&\frac{\G(1+a-b)\G(1+a-c)\G(1+a-d)\G(1+a-e)\G(1+a-f)}{\G(1+a)\G(b)\G(c)\G(d)\G(1+a-c-d)\G(1+a-b-d)\G(1+a-b-c)\G(1+a-e-f)}\nonumber\\
&\times&\frac{1}{2\pi i}\int_{-i\infty}^{i\infty}d\sigma\, \frac{\G(-\sigma)\G(1+a-b-c-d-\sigma)\G(b+\sigma)\G(c+\sigma)\G(d+\sigma)\G(1+a-e-f+\sigma)}{\G(1+a-e+\sigma)\G(1+a-f+\sigma)}\,.\nonumber\\
\end{eqnarray}
Here we have $a_\ell=1-\Dphi+(\Delta-\ell)/2$ and 
\begin{eqnarray}\label{params}
a&=&\ell'+2(a_\ell+m+s_1+\frac{2\Dphi}{3}-1)\,,\quad b=e=a_\ell+m,\nonumber\\ c&=&d=a_\ell+m+s_1+\frac{2\Dphi}{3}-1\,,\quad f=2(s_1-\frac{\Dphi}{3})+h+m+\ell'-\ell\,.
\end{eqnarray}
For the above expression to be finite, we need $4a-2(b+c+d+e+f-2)>0$. When this is not satisfied, we need to analytically continue the expression (for instance in the epsilon expansion for non-zero spins in the $t$-channel, this condition is not respected for $m=\ell$).
\subsection{Lengthy formulas}\label{seclen}
\be
q_{\ell'}^{diss}(s_1)=-\frac{(s_1-\frac{\D_\phi}{3})}{\G^2(\frac{\D_\phi}{3}-s_1+1)} \frac{2^{-\ell'}\G(2s_1+2\ell+\frac{4\Dphi}{3})}{\G^2(s_1+\frac{2\Dphi}{3})\G^2(s_1+\ell+\frac{2\Dphi}{3})} \ {}_3F_2\bigg[\begin{matrix} -\ell,2s_1+\ell-1+\frac{4\Dphi}{3},s_1-\frac{\Dphi}{3}\\
s_1+\frac{2\Dphi}{3}, s_1+\frac{2\Dphi}{3}
\end{matrix};1\bigg]\,,
\ee
is the contribution from the identity operator which we have added by hand. Here, we have
\begin{eqnarray}\label{qsgen}
q^{(s)}_{\D, \ell' |\ell}(s_1) &=& \sum_{m,n} \mu_{m,n}^{(\ell)}(\frac{\D-\ell}{2}-s_1-\frac{2\Dphi}3{})_m\chi^{(n)}_{\ell'}(s_1)
 \frac{\Gamma^2 \left(\frac{\Delta+\ell }{2}+\Delta_\phi -h\right)}{(\frac{\Delta-\ell}{2}-s_1-\frac{2\Dphi}{3}) 
 \Gamma (\Delta-h +1)} \,\\
&\times & _3F_2\left[\begin{matrix}\frac{\Delta -\ell}{2}-s_1-\frac{2\Dphi}{3},1+\frac{\Delta-\ell }{2}-\Delta_\phi ,1+\frac{\Delta-\ell }{2}-\Delta_\phi \\1+\frac{\Delta-\ell }{2}-s_1-\frac{2\Dphi}{3},\Delta-h +1\end{matrix};1\right]\,,\nonumber 
\end{eqnarray}
where
\be\label{schi1}
\chi_{\ell'}^{(n)}\!(s_1)=(-1)^{\ell'}2^{-\ell'}\frac{\G(2s_1+2\ell'+\frac{4\Dphi}{3})\G(s_1+\frac{2\Dphi}{3}+n)^2}{\ell'! \G(\ell'+s_1+\frac{2\Dphi}{3})^2\G(2s_1+n+\frac{4\Dphi}{3})}\,{}\frac{(-n)_{\ell'}}{(2s_1+n+\frac{4\Dphi}{3})_{\ell'}}\,.
\ee
The crossed-channel expression is given by:
\begin{eqnarray}\label{qcoefft2s}
&& q^{(t)}_{\D, \ell' |\ell}(s_1)  = \frac{2^{-\ell'}}{\ell'!}\frac{\G(2s_1+2\ell'+\frac{4\Dphi}{3})}{\G^2(s_1+\ell'+\frac{2\Dphi}{3})\G(a_\ell)} \frac{\Gamma \left(2\Delta_\phi +\ell -h\right)}{(a_\ell+\ell+2\Delta_{\phi}-h-1)} \,, \nonumber\\
&\times & \sum_{p=0}^{\ell'}\sum_{n=0}^{\ell}\sum_{m =0}^{\ell-n} \mu_{m,n}^{(\ell)}(\frac{\Dphi}{3}-s_1)_n \frac{\G^2(s_1+\frac{2\Dphi}{3}+m+a_\ell-1)}{\G(2s_1+\frac{4\Dphi}{3}+p+m+a_\ell-1)}\frac{(-\ell')_p(2s_1+\frac{4\Dphi}{3}+\ell'-1)_p}{p!} \nonumber \\ 
& \times & \int_{0}^{1}  dy \,
 y^{s_1+\frac{2\Dphi}{3}-1}(1-y)^{a_\ell-1}{}_2F_1[1, a_\ell, a_\ell+\ell+(2\Delta_{\phi}-h);y]\nonumber \\ &&~~~~~\times {}_2F_1[s_1+\frac{2\Dphi}{3}+p,s_1+\frac{2\Dphi}{3}+m+a_\ell-1, 2s_1+\frac{4\Dphi}{3}+p+m+a_\ell-1;1-y]  \nonumber\,,\\
\end{eqnarray}
where $a_\ell=1+\frac{\D-\ell}{2}-\Dphi$.
Remarkably, this admits a closed form expression in terms of ${}_7F_6$ hypergeometric functions.

\begin{eqnarray}\label{qtchann}
q^{(t)}_{\D,\ell'|\ell}(s_1)&=&\sum_{n=0}^\ell\sum_{m=0}^{\ell-n} (-1)^{\ell'+m}2^{-\ell'}\mu_{m,n}^{(\ell)} (\frac{\Dphi}{3}-s_1)_n (a_\ell)_m^2 \G(2s_1+2\ell'+\frac{4\Dphi}{3})\nonumber \\ &\times& \G^2(d)\G(\frac{a}{2})\G(a+1)\G^2(1+a-f-b)\tilde W(a;b,c,d,e,f)\,,
\end{eqnarray}
 the parameters $a,b,c$ etc. are given in eq.(\ref{params}) and the $\tilde W$ is the regularized version of a special (``very well poised") ${}_7F_6$ hypergeometric function as defined in eq.(\ref{Wdef}). For $\ell>\ell'$ there are a finite set of terms that need to be added to the above expression \cite{fgsz}. We will use eq.(\ref{qcoefft2s}) for performing calculations.

\newpage
\markboth{B\quad BOSONIC COMPONENTS OF $\frac{1}{2}$-BPS MULTIPLETS}{}
\section{Bosonic components of $\frac{1}{2}$-BPS multiplets}\label{App:multipletcomponents}
In this appendix, we give more details of the $\frac{1}{2}$-BPS multiplets which correspond to supergravity and supersymmetric gauge theory fields in AdS. We only keep the relevant bosonic fields which can appear in four-point functions. But the multiplets themselves contain more components. For a comprehensive discussion on superconformal multiplets in various spacetime dimensions and the complete set of superconformal descendants, see \cite{Cordova:2016emh}. The tables below for the bosonic components are reproduced from \cite{Alday:2020dtb,Behan:2021pzk}, and we now explain these tables.

\vspace{0.5cm}
\noindent {\bf The cases with sixteen Poincar\'e supercharges (maximally superconformal)}
\vspace{0.3cm}

{\begin{center}
 \begin{tabular}{||c| c | c | c | c | c | c ||} 
 \hline
component field & $s_p$ & $A_{p,\mu}$ & $\varphi_{p,\mu\nu}$ &  $C_{p,\mu}$ & $t_p$ & $r_p$ \\ [0.5ex] 
 \hline\hline
Lorentz spin $\ell$ & 0 & 1 & 2 & 1& 0 & 0\\ 
 \hline
conformal dimension $\Delta$ & $\epsilon p$ & $\epsilon p+1$ & $\epsilon p+2$ & $\epsilon p+3$ & $\epsilon p+4$ & $\epsilon p+2$ \\
 \hline
$d_1$ &  $p$ & $p-2$ & $p-2$ & $p-4$ & $p-4$ & $p-4$ \\ 
\hline $d_2$ &  $0$ & $2$  & $0$  & $2$ & $0$ & $4$  \\ [0.5ex] 
 \hline
\end{tabular}
\end{center}}

For supergravity theories with maximal superconformal symmetry, the supergravity fields are all organized into such $\frac{1}{2}$-BPS multiplets. The parameter $\epsilon=\frac{d-2}{2}$ takes value $\frac{1}{2}$, $1$, $2$, corresponding to the three maximal superconformal cases.  The multiplets are labelled by an integer $p$ which corresponds to the Kaluza-Klein level with $p=2,3,\ldots$. The lowest value $p=2$ corresponds to the stress tensor multiplet, and the massless graviton field $\varphi_{2,\mu\nu}$ is dual to the stress tensor. In the table, the quantum numbers $d_1$, $d_2$ are associated with the R-symmetry representation of the component fields. They parameterize the Dynkin labels of the R-symmetry groups as follows
\begin{equation}
SO(5):\;\; [d_1,d_2]\;,\quad\quad SU(4):\;\; [\tfrac{d_2}{2},d_1,\tfrac{d_2}{2}]\;,\quad\quad SO(8):\;\; [d_1,\tfrac{d_2}{2},0,0]\;.
\end{equation} 
Note that for $p<4$, some of the $d_1$ values in the table are negative. In this case the corresponding components are absent from the multiplet. Therefore, a generic multiplet with $p\geq 4$ contains six bosonic fields which can be exchanged in the four-point function. For $p=2,3$ the multiplets are extra-short and contain only three such fields. 

\vspace{0.5cm}
\noindent {\bf The cases with eight Poincar\'e supercharges}
\vspace{0.3cm}

{\begin{center}
 \begin{tabular}{||c| c | c | c ||} 
 \hline
component field & $s^I_p$ & $A^I_{p,\mu}$ & $r^I_p$ \\ [0.5ex] 
 \hline\hline
Lorentz spin $\ell$ & 0 & 1 &  0\\ 
 \hline
conformal dimension $\Delta$ & $\epsilon p$ & $\epsilon p+1$ & $\epsilon p+2$ \\
 \hline
$SU(2)_R$ spin $j_R$ &  $\frac{p}{2}$ & $\frac{p}{2}-1$ & $\frac{p}{2}-2$  \\ [0.5ex] 
 \hline
 $SU(2)_L$ spin $j_L$ &  $\frac{p-2}{2}$ & $\frac{p-2}{2}$ & $\frac{p-2}{2}$  \\ [0.5ex] 
 \hline
\end{tabular}
\end{center}}

The cases with eight Poincar\'e supercharges are relevant for our discussion of super gluons on $AdS_{d+1}\times S^3$, and all super gluon fields as well as their superconformal descendants reside in the $\frac{1}{2}$-BPS multiplets. Multiplets with different Kaluza-Klein levels are labelled by the integer $p=2,3,\ldots$. The lowest value $p=2$ corresponds to the flavor current multiplet.  As in the maximally superconformal cases, fields with negative $SU(2)$ quantum numbers are absent. 

\newpage
\markboth{C\quad PROPERTIES OF WITTEN DIAGRAMS}{}
\section{Properties of Witten Diagrams}\label{App:WittenDiagrams}
In this appendix we review several properties of tree-level Witten diagrams which are useful in applications to holographic correlators and in various analytic conformal bootstrap methods. In Appendix \ref{Subapp:dfun}, we focus on the contact Witten diagrams, {\it i.e.}, the $D$-functions. In Appendix \ref{Subapp:ivi}, we review vertex identities obtained from integrating out an internal bulk-to-bulk propagator. These identities express integrated cubic vertices with a bulk-to-bulk propagator to contact vertices with only bulk-to-boundary propagators, and are useful for computing higher-point exchange diagrams. In Appendix \ref{Subapp:eomid}, we review how the boundary two-particle Casimir equation translates to the equation of motion identity in the bulk, which relates exchange Witten diagrams and contact Witten diagrams. We also discuss several applications of this identity.  In Appendix \ref{Subapp:ivi} we discuss recursion relations satisfied by Witten diagrams, and demonstrate some general properties in a few explicit examples.

\subsection{$D$-functions}\label{Subapp:dfun}
The $D$-functions are a class of special functions defined as 
\begin{equation}\label{defDf}
D_{\Delta_1\ldots \Delta_n}(x_i)=\int \frac{d^d\vec{z}dz_0}{z_0^{d+1}}\prod_{i=1}^nG^{\Delta_i}_{B\partial}(z,x_i)\;,\quad G^{\Delta_i}_{B\partial}(z,x_i)=\left(\frac{z_0}{z_0^2+(\vec{z}-\vec{x}_i)^2}\right)^{\Delta_i}\;,
\end{equation}
which are $n$-point contact Witten diagrams in $AdS_{d+1}$ with no derivatives. Contact diagrams with derivatives can also be expressed as $D$-functions with shifted weights by using the identity
\begin{equation}
\nabla^\mu G^{\Delta_1}_{B\partial} \nabla_\mu G^{\Delta_2}_{B\partial}=\Delta_1\Delta_2(G^{\Delta_1}_{B\partial}G^{\Delta_2}_{B\partial}-2x_{12}^2G^{\Delta_1+1}_{B\partial}G^{\Delta_2+1}_{B\partial})\;.
\end{equation}
It is convenient to write the $D$-functions as functions of cross ratios by extracting a kinematic factor. For $n=4$, one defines the $\bar{D}$-functions as 
\begin{equation}\label{dbar}
\frac{ \prod_{i=1}^4\Gamma(\Delta_i)}{\Gamma(\frac{1}{2}\Sigma_\Delta-\frac{1}{2}d)}\frac{2}{\pi^{\frac{d}{2}}}D_{\Delta_1\Delta_2\Delta_3\Delta_4}(x_i)=\frac{(x_{14}^2)^{\frac{1}{2}\Sigma_\Delta-\Delta_1-\Delta_4}(x^2_{34})^{\frac{1}{2}\Sigma_\Delta-\Delta_3-\Delta_4}}{(x^2_{13})^{\frac{1}{2}\Sigma_\Delta-\Delta_4}(x^2_{24})^{\Delta_2}}\bar{D}_{\Delta_1\Delta_2\Delta_3\Delta_4} (U,V)\, ,
\end{equation}
where $\Sigma_\Delta=\sum_{i=1}^n\Delta_i$.

We can also represent the $D$-functions using the Feynman parameter representation
\begin{equation}
D_{\Delta_1\ldots \Delta_n}(x_i)=\frac{\pi^{\frac{d}{2}}\Gamma(\frac{1}{2}\Sigma_\Delta-\frac{1}{2}d)\Gamma(\frac{1}{2}\Sigma_\Delta)}{2\prod_i\Gamma(\Delta_i)}\int \prod_j \frac{d\alpha}{\alpha_j}\alpha_j^{\Delta_j}\frac{\delta(\sum_j\alpha_j-1)}{(\sum_{k<l}\alpha_k\alpha_l x_{kl}^2)^{\frac{1}{2}\Sigma_\Delta}}\;.
\end{equation}
From this representation, it is clear that we have the following derivative relations relating $D$-functions with different weights
\begin{equation}\label{DWS}
D_{\Delta_1\ldots \Delta_i+1\ldots \Delta_j+1\ldots \Delta_n}(x_i)=\frac{d-\Sigma_\Delta}{2\Delta_i\Delta_j}\frac{\partial}{\partial x_{ij}^2}D_{\Delta_1\ldots \Delta_n}(x_i)\;.
\end{equation} 
For $n=4$, we can further rewrite the relations in terms of $
\bar{D}$-functions and derivatives of $U$ and $V$
\begin{equation}\label{DbarWS}
\begin{split}
\bar{D}_{\Delta_1+1,\Delta_2+1,\Delta_3,\Delta_4}={}&-\partial_U \bar{D}_{\Delta_1,\Delta_2,\Delta_3,\Delta_4}\;,\\
\bar{D}_{\Delta_1,\Delta_2,\Delta_3+1,\Delta_4+1}={}&(\Delta_3+\Delta_4-\tfrac{1}{2}\Sigma_\Delta-U\partial_U )\bar{D}_{\Delta_1,\Delta_2,\Delta_3,\Delta_4}\;,\\
\bar{D}_{\Delta_1,\Delta_2+1,\Delta_3+1,\Delta_4}={}&-\partial_V \bar{D}_{\Delta_1,\Delta_2,\Delta_3,\Delta_4}\;,\\
\bar{D}_{\Delta_1+1,\Delta_2,\Delta_3,\Delta_4+1}={}&(\Delta_1+\Delta_4-\tfrac{1}{2}\Sigma_\Delta-V\partial_V )\bar{D}_{\Delta_1,\Delta_2,\Delta_3,\Delta_4}\;,\\
\bar{D}_{\Delta_1,\Delta_2+1,\Delta_3,\Delta_4+1}={}&(\Delta_2+U\partial_U+V\partial_V )\bar{D}_{\Delta_1,\Delta_2,\Delta_3,\Delta_4}\;,\\
\bar{D}_{\Delta_1+1,\Delta_2,\Delta_3+1,\Delta_4}={}&(\tfrac{1}{2}\Sigma_\Delta-\Delta_4+U\partial_U+V\partial_V )\bar{D}_{\Delta_1,\Delta_2,\Delta_3,\Delta_4}\;.
\end{split}
\end{equation}
Another set of useful identities arise from the invariance of (\ref{defDf}) under permutations of operators. This gives the identities
\begin{equation}\label{Dbarpermutation}
\begin{split}
\bar{D}_{\Delta_1\Delta_2\Delta_3\Delta_4} (U,V)=&V^{-\Delta_2}\bar{D}_{\Delta_1\Delta_2\Delta_4\Delta_3} (U/V,1/V)\\
=&V^{\Delta_4-\frac{1}{2}\Sigma_\Delta}\bar{D}_{\Delta_2\Delta_1\Delta_3\Delta_4} (U/V,1/V)\\
=&\bar{D}_{\Delta_3\Delta_2\Delta_1\Delta_4} (V,U)\\
=&V^{\Delta_1+\Delta_4-\frac{1}{2}\Sigma_\Delta}\bar{D}_{\Delta_2\Delta_1\Delta_4\Delta_3} (U,V)\\
=&U^{\Delta_3+\Delta_4-\frac{1}{2}\Sigma_\Delta}\bar{D}_{\Delta_4\Delta_3\Delta_2\Delta_1} (U,V)\;.
\end{split}
\end{equation}
There are other identities of $\bar{D}$-functions which are not used in this review, but can be found in, {\it e.g.}, Appendix D of \cite{Arutyunov:2002fh}.

Let us now focus on two special $D$-functions which played important roles in the position space computation of holographic correlators. The first case is $n=4$ and $\Delta_i=1$. The $\bar{D}$-function is the well known scalar one-loop box integral in four dimensions, and evaluates to \cite{Usyukina:1992jd}
\begin{equation}
\bar{D}_{1111}\equiv\Phi(z,\bar{z})=\frac{1}{z-\bar{z}}\left(2{\rm Li}_2(z)-2{\rm Li}_2(\bar{z})+\log(z\bar{z})\log\big(\frac{1-z}{1-\bar{z}}\big)\right)\;.
\end{equation}
From this expression, we find the following differential recursion relations
\begin{equation}\label{Phidiffrecurapp}
\begin{split}
\partial_z\Phi(z,\bar{z})=&-\frac{\Phi(z,\bar{z})}{z-\bar{z}}+\frac{\log U}{(z-1)(z-\bar{z})}-\frac{\log V}{z(z-\bar{z})}\;,\\
\partial_{\bar{z}}\Phi(z,\bar{z})=&\frac{\Phi(z,\bar{z})}{z-\bar{z}}-\frac{\log U}{(\bar{z}-1)(z-\bar{z})}+\frac{\log V}{\bar{z}(z-\bar{z})}\;.
\end{split}
\end{equation}
These relations imply that any $\bar{D}$-function obtained from $\bar{D}_{1111}$ by using the ``weight-shifting'' operators in (\ref{DbarWS}) can be written as a linear combination of the basis functions $\Phi(z,\bar{z})$, $\log U$, $\log V$ and $1$, with rational coefficients in $z$ and $\bar{z}$. This property was critical for the position space method reviewed in Section \ref{Subsec:positionspace}.

Another important case is $n=5$ with $\Delta_1=\Delta_2=\Delta_3=\Delta_4=1$ and $\Delta_5=2$, which is relevant for the computation of five-point functions in $AdS_5\times S^5$. This $D$-function can also be evaluated in terms of the scalar one-loop box integral \cite{Bern:1992em,Bern:1993kr}
\begin{equation}\label{Df5pt}
D_{11112}=\frac{4\pi^2}{x_{14}^2x_{35}^2x_{25}^2}\sum_{i=1}^5\frac{\eta_{i5}\Phi^{(i)}}{N_5}\;.
\end{equation}
Here $N_5$ and $\eta_{i5}$ are defined via a matrix $\rho$
\begin{equation}
\rho=N_5\eta^{-1}\;, \quad\quad N_5=2^{4}\det \rho\;,
\end{equation}
where
\begin{equation}
\rho=\left(\begin{array}{ccccc}0 & V_4 & 1 & 1 & V_3 \\V_4 & 0 & V_5 & 1 & 1 \\1 & V_5 & 0 & V_1 & 1 \\1 & 1 & V_1 & 0 & V_2 \\V_3 & 1 & 1 & V_2 & 0\end{array}\right)\;,
\end{equation}
with
\begin{equation}
V_1=\frac{x_{25}^2 x_{34}^2}{x_{24}^2 x_{35}^2}\;,\quad V_2=\frac{x_{31}^2 x_{45}^2}{x_{35}^2 x_{14}^2}\;,\quad V_3=\frac{x_{24}^2 x_{15}^2}{x_{14}^2 x_{25}^2}\;,\quad V_4=\frac{x_{12}^2 x_{35}^2}{x_{25}^2 x_{13}^2}\;,\quad V_5=\frac{x_{14}^2 x_{23}^2}{x_{13}^2 x_{24}^2}\;.
\end{equation}
The function $\Phi^{(i)}$ is the scalar one-loop box diagram where the point $i$ is omitted from the set of five. For example, we have
\begin{equation}
\Phi^{(5)}=\Phi(z,\bar{z})\big|_{z\bar{z}=V_1V_4}^{(1-z)(1-\bar{z})=V_5}\;.
\end{equation}
Starting from (\ref{Df5pt}) one can then use the relation (\ref{DWS}) and the differential recursion relations of the box function (\ref{Phidiffrecurapp}) to compute more complicated five-point $D$-functions.

\subsection{Integrated vertex identities}\label{Subapp:ivi}
\begin{figure}
\centering
\includegraphics[width=0.8\textwidth]{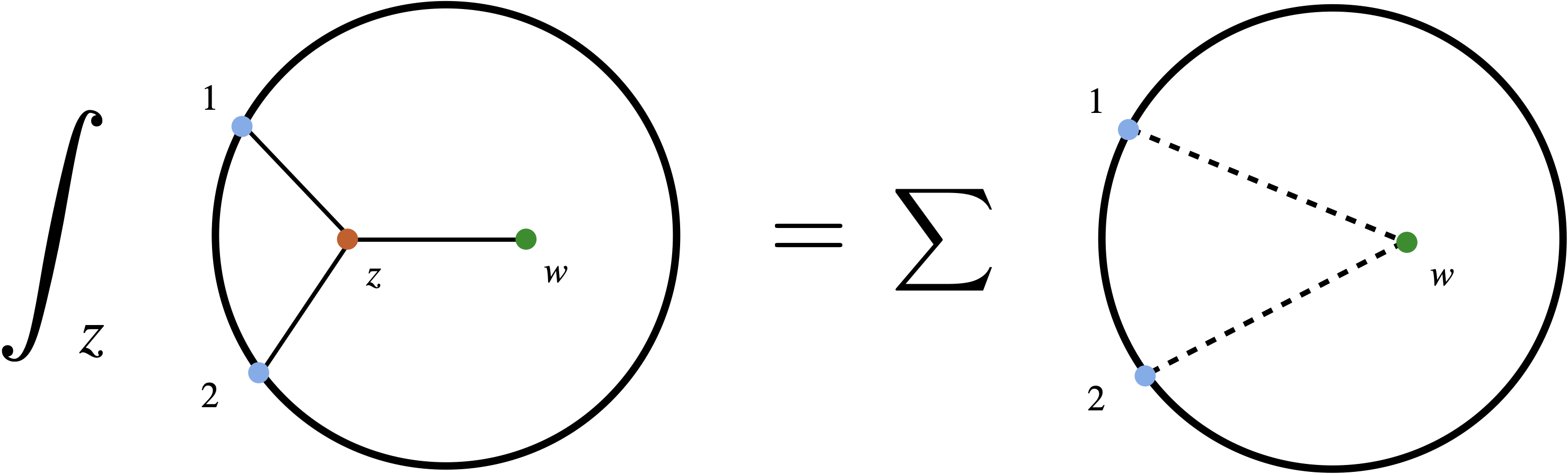}
\caption{Integrated vertex identities. On the LHS we have a cubic vertex connecting two scalar bulk-to-boundary propagators and one bulk-to-bulk propagator of dimension $\Delta$ and spin $\ell$. The bulk point $z$ is integrated over. When $\Delta_1+\Delta_2-\Delta+\ell\in 2\mathbb{Z}_+$, the integral can be written as a finite sum of contact vertices as on the RHS.}
    \label{fig:appavertex}
\end{figure}

It was shown in \cite{DHoker:1999mqo} that when the external and internal quantum numbers satisfy certain relations, the internal bulk-to-bulk propagator can be integrated out and the four-point exchange Witten diagram can be written as a finite sum of $D$-functions. However, this truncation into finitely many $D$-functions relies only on half of the diagram, {\it i.e.}, the integral on the LHS of Figure \ref{fig:appavertex}. We can extract from the results of  \cite{DHoker:1999mqo} identities for the integrated cubic vertex which express them as a sum of contact vertices as illustrated in Figure \ref{fig:appavertex}. These integrated vertex identities are very useful for computing higher-point exchange Witten diagrams. For illustration, we will explain in detail how truncation happens for the scalar case and extract the corresponding vertex identity.  The cases with spinning internal propagators are similar, and we will only record the results.

In the scalar case, we consider the integral
\begin{equation}\label{Ascalar}
A(x_1,x_2,w)=\int \frac{d^{d+1}z}{z_0^{d+1}}G^{\Delta_1}_{B\partial}(z,x_1)G^{\Delta_2}_{B\partial}(z,x_2)G^{\Delta}_{BB}(z,w)\;.
\end{equation}
The scalar bulk-to-bulk propagator is explicitly given by
\begin{equation}
G^\Delta_{BB}=\tilde{C}_\Delta  (2u^{-1})^\Delta {}_2F_1\big(\Delta,\Delta-\frac{d}{2}+\frac{1}{2};2\Delta-d+1;-2u^{-1}\big)\;,
\end{equation}
where
\begin{equation}
\tilde{C}_\Delta=\frac{\Gamma(\Delta)\Gamma(\Delta-\frac{d}{2}+\frac{1}{2})}{(4\pi)^{\frac{d+1}{2}}\Gamma(2\Delta-d+1)}\;,
\end{equation}
and
\begin{equation}
u=\frac{(z-w)^2}{2z_0w_0}\;.
\end{equation}
It satisfies the equation of motion identity 
\begin{equation}\label{seom}
(-\square+m^2)G_{BB}^\Delta(u)=\delta(z,w)\;,
\end{equation}
where $m^2=\Delta(\Delta-d)$.
To evaluate the integral (\ref{Ascalar}), it is convenient to first simplify it by performing a translation
\begin{equation}
x_1\to 0\;,\quad\quad x_2\to x_{21}=x_2-x_1\;,
\end{equation} 
and then a conformal inversion
\begin{equation}
x'_{21}=\frac{x_{21}}{|x_{21}|^2}\;,\quad z'=\frac{z}{z^2}\;,\quad w'=\frac{w}{w^2}\;.
\end{equation}
The integral becomes 
\begin{equation}
A(x_1,x_2,w)=(x_{12})^{-2\Delta_2}I(w'-x'_{12})\;,
\end{equation}
where 
\begin{equation}
I(w)=\int \frac{d^{d+1}z}{z_0^{d+1}}G^\Delta_{BB}(u)z_0^{\Delta_1}\left(\frac{z_0}{z^2}\right)^{\Delta_2}\;.
\end{equation}
The scaling behavior of $I(w)$ under $w\to\lambda w$ and Poincar\'e invariance dictate that $I(w)$ must take the form
\begin{equation}
I(w)=w_0^{\Delta_1-\Delta_2}f(t)\;,
\end{equation}
with $t$ defined by
\begin{equation}
t=\frac{w_0^2}{w^2}\;.
\end{equation}
On the other hand, we can act on $I(w)$ with the operator $(-\square+m^2)$ and use the equation of motion (\ref{seom}). This operator collapses the bulk-to-bulk propagator to a delta function, and leads to the following differential equation for $f(t)$ 
\begin{equation}
4t^2(t-1)f''+4t((\Delta_{12}+1)t-\Delta_{12}+\frac{d}{2}-1)f'+(\Delta_{12}(d-\Delta_{12})+m^2)f=t^{\Delta_2}\;,
\end{equation}
where $\Delta_{12}=\Delta_1-\Delta_2$. This differential equation is further accompanied by two boundary conditions 
\begin{enumerate}
\item The function $f(t)$ is smooth as $t\to 1$. This can be seen from the fact that $I(w)$ is regular at $\vec{w}=0$.
\item The function $f(t)\sim t^{\frac{\Delta-\Delta_{12}}{2}}$ as $t\to 0$. We can see this from the $w_0\to 0$ limit of $I(w)$ where it behaves as $I(w)\sim w_0^\Delta$.
\end{enumerate}
Remarkably, the differential equation has a polynomial solution when a certain relation is obeyed by $\Delta_1$, $\Delta_2$ and $\Delta$. Assuming that $f(t)$ has the representation 
\begin{equation}
f(t)=\sum_k a_k t^k\;,
\end{equation}
we find the following recursion relation for $a_k$ from the differential equation
\begin{equation}
a_{k-1}=\frac{(k-\frac{\Delta}{2}+\frac{\Delta_{12}}{2})(k-\frac{d}{2}+\frac{\Delta}{2}+\frac{\Delta_{12}}{2})}{(k-1)(k-1+\Delta_{12})}a_k\;.
\end{equation}
We can consistently set $a_k=0$ when $k\geq \Delta_2$ so that the maximal value of $k$ is $k_{\rm max}=\Delta_2-1$ and its coefficient is
\begin{equation}
a_{\Delta_2-1}=\frac{1}{4(\Delta_1-1)(\Delta_2-1)}\;.
\end{equation}
Note that when $\Delta_1+\Delta_2-\Delta$ is a positive even integer, the series $f(t)$ truncates at a minimal value $k_{\rm min}=\frac{1}{2}(\Delta-\Delta_{12})$. It is easy to check that the polynomial solution satisfies both boundary conditions. We can now undo the inversion and translation and write the solution for $A(x_1,x_2,w)$ as 
\begin{equation}
A(x_1,x_2,w)=\sum_{k=k_{\rm min}}^{k_{\rm max}} a_k\,(x_{12}^2)^{k-\Delta_2}\, G_{B\partial}^{k+\Delta_1-\Delta_2}(w,x_1)G_{B\partial}^k(w,x_2)\;.
\end{equation}
This is the integrated vertex identity for an internal scalar bulk-to-bulk propagator. 

With some extra work, one can similarly obtain the integrated vertex identities for vector fields and gravitons. We give the answer below, which is taken from \cite{Goncalves:2019znr}. For simplicity, we will consider the case where the two external weights are equal $\Delta_1=\Delta_2=\Delta_{\rm ext}$. 

We first look at the case of exchanging a vector field with dimension $\Delta$. When $\Delta=d-1$, the vector field is a massless gauge field. But we will not restrict $\Delta$ to this special value. We consider the coupling of the vector field to a conserved current
\begin{equation}
A^\mu(x_1,x_2,w)\equiv \int \frac{d^{d+1}z}{z_0^{d+1}}  \left(G_{B\partial}^{\Delta_{\rm ext}}(z,x_1) {\overset{\leftrightarrow}{\nabla}}_\nu G_{B\partial}^{\Delta_{\rm ext}}(z,x_2)\right) G_{BB}^{\Delta,1,\mu\nu}(z,w)\;,
\end{equation}
where $G_{BB}^{\Delta,1,\mu\nu}(z,y)$ is the vector bulk-to-bulk propagator. This integral can be evaluated as a sum of contact vertices 
\begin{equation}
A^\mu(x_1,x_2,w)=-\sum_{k=k_{\rm min}}^{k_{\rm max}} \frac{a_k}{2k}(x_{12}^2)^{-\Delta_{\rm ext}+k} g^{\mu\nu}(w) \left(G_{B\partial}^{k}(w,x_1) {\overset{\leftrightarrow}{\nabla}}_\nu G_{B\partial}^{k}(w,x_2)\right)\;,
\end{equation}
where
\begin{equation}
   \begin{split}
      k_{\rm min} = {}& \frac{d-2}{4}+\frac{1}{4}\sqrt{(d-2)^2+4(\Delta-1)(\Delta-d+1)}\;, \\
      k_{\rm max}=   {}& \Delta_{\rm ext}-1\;, \\
      a_{k-1}= {}& \frac{2k(2k+2-d)-(\Delta-1)(\Delta-d+1)}{4(k-1)k}a_k\;,\\
      a_{\Delta_{\rm ext}-1}={}&\frac{1}{2(\Delta_{\rm ext}-1)}\;.
   \end{split}
\end{equation}
Truncation requires that $k_{\rm max}-k_{\rm min}$ is a non-negative integer.

Now we consider the case of gravitons which have dimension $d$. The cubic integral is given by
\begin{equation}
\begin{split}
{}&A^{\mu\nu}(x_1,x_2,w)=\int \frac{d^{d+1}z}{z_0^{d+1}} G^{\Delta=d,\ell=2,\;\mu\nu;\rho\sigma}_{BB}(z,w)\times\bigg(\nabla_{\rho} G^{\Delta_{\rm ext}}_{B\partial}(z,x_1) \nabla_{\sigma} G^{\Delta_{\rm ext}}_{B\partial}(z,x_2)\\
{}&\quad\quad-\frac{1}{2}g^{\rho\sigma}(z)(\nabla^\kappa G^{\Delta_{\rm ext}}_{B\partial}(z,x_1)  \nabla_\kappa G^{\Delta_{\rm ext}}_{B\partial}(z,x_2)+m^2 G^{\Delta_{\rm ext}}_{B\partial}(z,x_1) G^{\Delta_{\rm ext}}_{B\partial}(z,x_2))\bigg)\;.
\end{split}
\end{equation}
Using the result of \cite{DHoker:1999mqo}, we find that this integral reduces to the following sum of contact vertices
\begin{equation}
\begin{split}
A^{\mu\nu}(x_1,x_2,w)={}&\sum_{k=k_{\rm min}}^{k_{\rm max}}a_k (x_{12}^2)^{-\Delta_{\rm ext}+k}\bigg( \frac{g^{\mu\nu}(y)}{d-1}G^{k}_{B\partial}(w,x_1) G^{k}_{B\partial}(w,x_2)\\
+{}&\frac{1}{k(k+1)}\left(\nabla^\mu \nabla^\nu G^{k}_{B\partial}(w,x_1)+k g^{\mu\nu}(w)G^{k}_{B\partial}(w,x_1)\right)G^{k}_{B\partial}(w,x_2)\bigg)\;,
\end{split}
\end{equation}
where
 \begin{equation}
   \begin{split}
      k_{\rm min} = {}& \frac{d}{2}-1\;, \\
      k_{\rm max}=   {}& \Delta_{\rm ext}-1\;, \\
      a_{k-1}= {}& \frac{k+1-\frac{d}{2}}{k-1}a_k\;,\\
      a_{\Delta_{\rm ext}-1}={}&-\frac{\Delta_{\rm ext}}{2(\Delta_{\rm ext}-1)}\;.
   \end{split}
\end{equation}
Note that in deriving the integrated cubic vertex $A^{\mu\nu}$, we have assumed that the rest part of the exchange diagram (or a sum of diagrams) is a conserved current.  

\subsection{Equation of motion identities}\label{Subapp:eomid}
Let us start by considering the scalar exchange Witten diagram defined by the integral
\begin{equation}\label{Wscalar}
W_{\Delta,0}=\int \frac{d^{d+1}z}{z_0^{d+1}}\frac{d^{d+1}w}{w_0^{d+1}} G_{B\partial}^{\Delta_1}(z,x_1)G_{B\partial}^{\Delta_2}(z,x_2) G_{BB}^{\Delta}(z,w) G_{B\partial}^{\Delta_3}(w,x_3)G_{B\partial}^{\Delta_4}(w,x_4)\;,
\end{equation}
which can be written as 
\begin{equation}
W_{\Delta,0}=\int \frac{d^{d+1}w}{w_0^{d+1}} A(x_1,x_2,w) G_{B\partial}^{\Delta_3}(w,x_3) G_{B\partial}^{\Delta_4}(w,x_4)\;,
\end{equation}
using $A(x_1,x_2,w)$ defined in (\ref{Ascalar}). Note that $A(x_1,x_2,w)$ is conformally invariant. Therefore, we have  
\begin{equation}
(L_1^{AB}+L_2^{AB}+\mathcal{L}_w^{AB})A(x_1,x_2,w)=0\;,
\end{equation}
where $L_{1,2}^{AB}=-L_{1,2}^{BA}$, $A,B=0,1,\ldots,d+1$ are the conformal generators acting on $x_1$ and $x_2$, and $\mathcal{L}_w^{AB}$ is the isometry generator of $AdS_{d+1}$. From this identity we get
\begin{equation}
{\rm Cas}\circ A(x_1,x_2,w)= -\frac{1}{2}\mathcal{L}_w^{AB}\mathcal{L}_{w,AB} A(x_1,x_2,w)=\square_w A(x_1,x_2,w)\;,
\end{equation}
where ${\rm Cas}=-\frac{1}{2}(L_1^{AB}+L_2^{AB})(L_{1,AB}+L_{2,AB})$ is the two-particle quadratic conformal Casimir with respect to points 1 and 2. Using the equation of motion of the bulk-to-bulk propagator (\ref{seom}), we find 
\begin{equation}
\big({\rm Cas}-\Delta(\Delta-d)\big) A(x_1,x_2,w)=G_{B\partial}^{\Delta_1}(w,x_1)G_{B\partial}^{\Delta_2}(w,x_2)\;.
\end{equation}
Inserting it into (\ref{Wscalar}), we arrive at 
\begin{equation}
\big({\rm Cas}-\Delta(\Delta-d)\big) W_{\Delta,0}=D_{\Delta_1\Delta_2\Delta_3\Delta_4}\;,
\end{equation}
which shows the action of the Casimir operator turns an exchange Witten diagram into a contact Witten diagram. The above derivation can be extended to spinning particle exchange diagrams. In general, for an exchange Witten diagram exchanging a particle of dimension $\Delta$ and spin $\ell$ we have 
\begin{equation}\label{eomidgen}
\big({\rm Cas}-C_{\Delta,\ell}\big) W_{\Delta,\ell}=W_{\rm con}\;,
\end{equation}
where $C_{\Delta,\ell}$ is the Casimir eigenvalue 
\begin{equation}
C_{\Delta,\ell}=\Delta(\Delta-d)+\ell(\ell+d-2)\;,
\end{equation}
and $W_{\rm con}$ is a finite collection of contact diagrams containing no more than $2\ell-2$ derivatives and depends on the cubic vertices.  

The equation of motion identity (\ref{eomidgen}) of exchange Witten diagrams has many uses. For example, one can use it to obtain the Mellin amplitudes of exchange Witten diagrams \cite{Costa:2012cb}. Extracting a kinematic factor, we can write $W_{\Delta,\ell}$ in terms of cross ratios 
\begin{equation}
W_{\Delta,\ell}=\frac{1}{(x_{12}^2)^{\frac{\Delta_1+\Delta_2}{2}}(x_{34}^2)^{\frac{\Delta_3+\Delta_4}{2}}}\left(\frac{x_{14}^2}{x_{24}^2}\right)^a\left(\frac{x_{14}^2}{x_{13}^2}\right)^b \mathcal{W}_{\Delta,\ell}(U,V)\;,
\end{equation}
with $a=\frac{1}{2}(\Delta_2-\Delta_1)$, $b=\frac{1}{2}(\Delta_3-\Delta_4)$. The Casimir operator acts on $\mathcal{W}_{\Delta,\ell}(U,V)$ as 
\begin{equation}
\begin{split}
{\rm Cas}=&2(V^{-1}-UV^{-1}-1)V\frac{\partial}{\partial V}\left(V\frac{\partial}{\partial V}+a+b\right)+U\frac{\partial}{\partial U}\left(2U\frac{\partial}{\partial U}-d\right)\\
&-2(1+U-V)\left(U\frac{\partial}{\partial U}+V\frac{\partial}{\partial V}+a\right)\left(U\frac{\partial}{\partial U}+V\frac{\partial}{\partial V}+b\right)\;,
\end{split}
\end{equation}
which is easy to interpret as a difference equation in the Mellin representation 
\begin{equation}
\begin{split}
\mathcal{W}_{\Delta,\ell}=&\int_{-i\infty}^{i\infty}\frac{dsdt}{(4\pi i)^2}U^{\frac{s}{2}}V^{\frac{t}{2}-\frac{\Delta_2+\Delta_3}{2}}\mathcal{M}_{\Delta,\ell}(s,t)\,\Gamma(\tfrac{\Delta_1+\Delta_2-s}{2})\Gamma(\tfrac{\Delta_3+\Delta_4-s}{2})\\
&\quad\quad\quad\times \Gamma(\tfrac{\Delta_1+\Delta_4-t}{2})\Gamma(\tfrac{\Delta_2+\Delta_3-t}{2}) \Gamma(\tfrac{\Delta_1+\Delta_3-u}{2})\Gamma(\tfrac{\Delta_2+\Delta_4-u}{2})\;,
\end{split}
\end{equation}
by using the following replacement 
\begin{equation}
\begin{split}
& U\frac{\partial}{\partial U}\to \frac{s}{2}\times\;,\quad\quad\quad V\frac{\partial}{\partial V}\to \left(\frac{t}{2}-\frac{\Delta_2+\Delta_3}{2}\right)\times\;, \\
& U^mV^n \to \mathbb{D}_{mn}\;,\\
& \mathbb{D}_{mn}\circ f(s,t)=f(s-2m,t-2n)\left(\tfrac{\Delta_1+\Delta_2-s}{2}\right)_m \left(\tfrac{\Delta_3+\Delta_4-s}{2}\right)_m \left(\tfrac{\Delta_1+\Delta_4-t}{2}\right)_n\\
&\quad\quad\quad\quad\quad\quad\quad \times \left(\tfrac{\Delta_2+\Delta_3-t}{2}\right)_n \left(\tfrac{s+t-\Delta_1+\Delta_3}{2}\right)_{-m-n}\left(\tfrac{s+t-\Delta_2+\Delta_4}{2}\right)_{-m-n}\;.
\end{split}
\end{equation}
Recall that the Mellin amplitude of an exchange diagram has the form\footnote{Here we focus on the polar part of the amplitude and have chosen a specific regular term. Note also that such Mellin amplitudes have improved $u$-channel Regge behavior $1/s$.} 
\begin{equation}
\mathcal{M}_{\Delta,\ell}(s,t)=\sum_{m=0}^\infty \frac{Q_{\ell,m}(u)}{s-(\Delta-\ell)-2m}\;,
\end{equation}
where $Q_{\ell,m}(u)$ is a degree-$\ell$ polynomial of $u$. We can fix $Q_{\ell,0}(u)$ by first taking the residue for $s$ at $s=\Delta-\ell$ and then requiring the $t$ integral to produce the collinear conformal block with dimension $\Delta$ and spin $\ell$. Then the difference equation following from (\ref{eomidgen}) allows us to recursively obtain all $Q_{\ell,m}(u)$ with $m>0$ from $Q_{\ell,0}(u)$.

As another application, let us prove that the difference of two exchange Witten diagrams with opposite quantizations ({\it i.e.}, with conformal dimension $\Delta$ versus $d-\Delta$) is proportional to the conformal partial wave
\begin{equation}
\mathcal{W}_{\Delta,\ell}-\mathcal{W}_{d-\Delta,\ell}\propto \Psi_{\Delta,\ell}\;.
\end{equation}
The conformal partial wave $\Psi_{\Delta,\ell}$ is defined to be the linear combination of a conformal block and its shadow such that it is single-valued in Euclidean space ({\it i.e.}, when $\bar{z}=z^*$).\footnote{Note that each conformal block is not single-valued.} To prove this relation, we act on the combination $\mathcal{W}_{\Delta,\ell}-\mathcal{W}_{d-\Delta,\ell}$ with the operator $({\rm Cas}-C_{\Delta,\ell})$. The contact term on the RHS of (\ref{eomidgen}) does not distinguish the two quantizations, and therefore 
\begin{equation}
\big( {\rm Cas}-C_{\Delta,\ell}\big)  (\mathcal{W}_{\Delta,\ell}-\mathcal{W}_{d-\Delta,\ell})=0\;.
\end{equation}
This equation tells us that the double-trace conformal blocks in each exchange Witten diagram have been precisely cancelled, and the difference is a linear combination of the single-trace conformal blocks with dimensions $\Delta$ and $d-\Delta$. On the other hand, single-valuedness is obvious. It follows from the fact that each exchange Witten diagram is single-valued. 

Finally, let us mention that the equation of motion identity also implies efficient recursion relations that can be used to obtain the crossed channel conformal block decomposition coefficients of exchange Witten diagrams or conformal partial waves. The latter is related to the crossing kernel (also known as the $6j$ symbol) of the conformal group. The idea is that the equation of motion turns an exchange Witten diagram into contact Witten diagrams (or a conformal partial wave into zero) which can be easily decomposed into conformal blocks in the crossed channel. On the other hand, the conformal Casimir operator acts nicely on crossed channel conformal blocks, and its action can be expressed as a linear combination of finitely many conformal blocks with shifted dimensions and spins. This gives rise to relations among the crossed channel conformal block decomposition coefficients which can be recursively solved. We refer the reader to \cite{Zhou:2018sfz} for details of this recursive approach. For other approaches to this problem, see \cite{Hogervorst:2017sfd,Sleight:2018epi,Liu:2018jhs,Sleight:2018ryu}.

\subsection{Recursion relations}\label{Subapp:recur}
It is well known that conformal blocks satisfy various intricate recursion relations (see, {\it e.g.}, \cite{Dolan:2003hv,Dolan:2011dv}). These recursion relations are very useful for studying the properties of conformal blocks and for performing conformal block decomposition for conformal correlators. Exchange Witten diagrams are intuitively very similar to conformal blocks. They contain a single-trace conformal block which is associated with the exchange of a particle in AdS. But at the same time they also contain infinitely many double-trace conformal blocks which are two-particle states. Because of the infinitely many conformal blocks involved, at first sight it seems rather unlikely that  similar recursion relations can exist for Witten diagrams. However, it was pointed out in \cite{Zhou:2020ptb} that their existence is always {\it guaranteed}. There is an intimate connection between the recursion relations of conformal blocks and Witten diagrams, and one can easily generate Witten diagram recursion relations from known recursion relations of conformal blocks.

Let us demonstrate this correspondence in the simplest situation where the conformal block recursion relations have the form of a linear combination of conformal blocks with constant coefficients. The prime examples in this category are the dimensional reduction formulae. It was found in \cite{Hogervorst:2016hal} that a $d$-dimensional conformal block can be expressed in terms of infinitely many $(d-1)$-dimensional ones 
\begin{equation}
g^{(d)}_{\Delta,\ell}=\sum_{n=0}^\infty\sum_{j} A_{n,j}\, g^{(d-1)}_{\Delta+2n,j}\;,\quad j=\ell\;,\ell-2\;,\ldots\;, \ell\;\text{mod}\;2\;.
\end{equation}
On the other hand, for conformal blocks in $d$ and $d-2$ dimensions it is possible to find a relation with finitely many terms  \cite{Kaviraj:2019tbg}
\begin{equation}
g^{(d-2)}_{\Delta,\ell}=g^{(d)}_{\Delta,\ell}+c_{2,0}g^{(d)}_{\Delta+2,\ell}+c_{1,-1}g^{(d)}_{\Delta+1,\ell-1}+c_{0,-2}g^{(d)}_{\Delta,\ell-2}+c_{2,-2}g^{(d)}_{\Delta+2,\ell-2}\;.
\end{equation}
Here $A_{n,j}$ and $c_{i,j}$ are numerical constants whose explicit expressions are not important for our discussion and can be found in \cite{Hogervorst:2016hal} and \cite{Kaviraj:2019tbg}. To obtain recursion relations for Witten diagrams from these identities, a simple prescription was pointed out in \cite{Zhou:2020ptb}. One just needs to replace the conformal blocks $g^{(D)}_{\Delta,\ell}$ by the corresponding $AdS_{D+1}$ exchange Witten diagrams $W^{AdS_{D+1}}_{\Delta,\ell}$ which contain $g^{(D)}_{\Delta,\ell}$ as the single-trace conformal block.\footnote{The normalization is chosen such that the single-trace conformal block appears with the unit coefficient.} Note that for spin $\ell\geq 1$, one has multiple choices for the contact terms in the exchange diagrams. Therefore, one further needs to choose appropriate contact terms in order for the identities to hold. But such choices turn out to always exist as we shall see. Let us first write down the corresponding Witten diagram relations following from the above prescription
\begin{equation}
W^{AdS_{d+1}}_{\Delta,\ell}=\sum_{n=0}^\infty\sum_{j} A_{n,j}\, W^{AdS_d}_{\Delta+2n,j}\;,\quad j=\ell\;,\ell-2\;,\ldots\;, \ell\;\text{mod}\;2\;,
\end{equation}
\begin{equation}
W^{AdS_{d-1}}_{\Delta,\ell}=W^{AdS_{d+1}}_{\Delta,\ell}+c_{2,0}W^{AdS_{d+1}}_{\Delta+2,\ell}+c_{1,-1}W^{AdS_{d+1}}_{\Delta+1,\ell-1}+c_{0,-2}W^{AdS_{d+1}}_{\Delta,\ell-2}+c_{2,-2}W^{AdS_{d+1}}_{\Delta+2,\ell-2}\;,
\end{equation}
where we have left the choice of the contact terms implicit. Note that identities of the second kind are responsible for the Parisi-Sourlas dimensional reduction structure found in the super graviton and super gluon correlators (with $\ell=0$), as we mentioned in Section \ref{Subsec:MRV} and \ref{Subsec:supergluons}. To understand why this simple prescription works and also to see how to choose the contact terms, it is most convenient to go to the Mellin space. We recall that the Mellin amplitude of an exchange Witten diagram is a sum over simple poles plus a polynomial regular term. On the other hand, under conformal block decomposition, an exchange Witten diagram contains a single-trace conformal block and infinitely many double-trace conformal blocks. The single-trace conformal block is determined by the singular terms, and is produced when we take the residues at these simple poles. By contrast, the double-trace conformal blocks in the Witten diagram are produced when we take residues at the poles of the Gamma function factor. Note that crucially there is no freedom left to change the double-trace conformal blocks once the Mellin amplitude is determined. With this observation, it is easy to see why this prescription gives the correct answer. The original conformal block recursion relation, which yields the equality of the single-trace conformal blocks, guarantees that the singular part of the Mellin amplitudes are the same on both sides. The remaining task is to match the polynomial terms, which are the sums of the contact terms in the exchange Witten diagrams. Since each spin-$\ell$ exchange Witten diagram can accommodate a contact term which is a degree-$(\ell-1)$ polynomial, clearly this is always possible. 

In Mellin space the existence of these Witten diagram relations is almost obvious following the above reasoning. However, from the position space perspective such identities are rather remarkable, as they require intricate cancellations of infinitely many double-trace conformal blocks. One can also take these Witten diagram identities and decompose them in the crossed channel. These identities then give rise to highly nontrivial relations which constrain the crossed channel conformal block decomposition coefficients of exchange Witten diagrams. 

In the above, we have only discussed the simplest scenario. More generally, conformal block relations may have cross ratio dependence in their linear combination coefficients. Such relations also induce Witten diagram relations although sometimes additional correction terms are needed. The simplest example in this class is the Casimir equation for conformal blocks
\begin{equation}
\big({\rm Cas}-C_{\Delta,\ell}\big) g_{\Delta,\ell}=0\;.
\end{equation}
It is mapped to the equation of motion identity for exchange Witten diagrams encountered in the previous subsection
\begin{equation}
\big({\rm Cas}-C_{\Delta,\ell}\big) W_{\Delta,\ell}=W_{\rm con}\;.
\end{equation}
That we can generate Witten diagram relations from conformal block relations in the more general case is essentially guaranteed by the same fact as before, namely, the double-trace conformal blocks are fully determined by the Mellin amplitudes. However, an important difference to note is the extra term on the RHS. This contact term cannot be absorbed by redefining the contact part in the exchange diagram.\footnote{It is easiest to convince oneself of this fact in the example which has $\ell=0$. The scalar exchange Witten diagram admits no contact term. However, the equation of motion identity has a zero-derivative contact term on the RHS.} This represents a general feature when the coefficients of the conformal blocks are no longer just constants. These cross ratio dependent coefficients translate into difference operators in Mellin space. Such operators generically shift the simple poles of the Mellin amplitude. But at the same time they can also generate new poles or multiply the Mellin amplitudes by polynomials. As a result, whenever this happens we need to add a finite number of extra exchange Witten diagrams or contact Witten diagrams in order to match the Mellin amplitudes. Many examples of such relations were given in \cite{Zhou:2020ptb} and were verified by explicit computations. However, they are a bit too technical to be included here and we will not discuss this further. The interested reader can read \cite{Zhou:2020ptb} for more details.

\end{appendices}

\newpage

\setlength{\bibsep}{4pt plus 0.2ex}

{\small
\bibliography{refs}
\bibliographystyle{utphys}
}
\end{document}